\documentclass[onecolumn,showpacs,floatfix,superscriptaddress,amsmath,amssymb]{revtex4}
\usepackage{mathrsfs}
\usepackage{txfonts}
\usepackage{amssymb}
\usepackage{graphicx}
\usepackage{hyperref}
\usepackage{ulem}
 \usepackage{overpic}
 \usepackage{psfrag}
\usepackage{tabularx}
\usepackage{array}
\usepackage{placeins}
\usepackage{subfigure}
\newcommand{\be}{\begin{equation}}
\newcommand{\ee}{\end{equation}}

\newcommand{\PreserveBackslash}[1]{\let\temp=\\#1\let\\=\temp}

\newcommand{\ket} [1] {| #1 \rangle}

\newcommand\dbar{{\mathchar22\mkern-12mu d}}
\newcommand{\tabincell}[2]{\begin{tabular}{@{}#1@{}}#2\end{tabular}}
\makeatletter

\begin{document}

\title{Fidelity mechanics: analogues of the four thermodynamic laws and Landauer's principle}
\author{Huan-Qiang Zhou}
\altaffiliation{hqzhou@cqu.edu.cn}
\affiliation{Centre for Modern Physics, Chongqing University, Chongqing 400044, The People's Republic of China}
\author{Qian-Qian Shi}
\affiliation{Centre for Modern Physics, Chongqing University, Chongqing 400044, The People's Republic of China}
\author{Yan-Wei Dai}
\affiliation{Centre for Modern Physics, Chongqing University, Chongqing 400044, The People's Republic of China}\vspace{10pt}

\begin{abstract}
Fidelity mechanics is formalized as a framework to investigate quantum critical phenomena in quantum many-body systems. This is achieved by introducing fidelity temperature to properly quantify quantum fluctuations, which, together with fidelity entropy and fidelity internal energy, constitute three basic state functions in fidelity mechanics, thus enabling us to formulate analogues of the four thermodynamic laws and Landauer's principle at zero temperature.  Fidelity flows are defined and may be interpreted as an alternative form of renormalization group flows.  Thus, both stable and unstable fixed points are characterized in terms of fidelity temperature and fidelity entropy: divergent fidelity temperature for unstable fixed points and zero fidelity temperature and (locally) maximal fidelity entropy for stable fixed points.  In addition, an inherently fundamental role of duality is clarified, resulting in a canonical form of the Hamiltonian in fidelity mechanics.  Dualities, together with symmetry groups and factorizing fields, impose the constraints on a fidelity mechanical system, thus shaping fidelity flows from an unstable fixed point to a stable fixed point.

A  detailed analysis of fidelity mechanical state functions is presented for the quantum XY model, the transverse field quantum Ising chain in a longitudinal field, the spin-$1/2$ XYZ model and the XXZ model in a magnetic field, with help of classical simulations of quantum many-body systems in terms of a tensor network algorithm in the context of matrix product states. Rich physics is unveiled even for these well-studied models.  First, for the quantum XY chain, the disorder circle is interpreted as a separation line between two different types of fidelity flows, with one type of fidelity flows starting from unstable fixed points with central charge $c=1$, and the other type of fidelity flows starting from unstable fixed points with central charge $c=1/2$.  Both types of fidelity flows end at the same stable fixed point, where fidelity entropy reaches its local maximum.  Another remarkable feature is that fidelity temperature is zero on the disorder circle, as it should be, since no quantum fluctuations exist in a factorized state.  However, fidelity temperature is not well-defined at the Pokrovsky-Talapov transition point. In fact, it ranges from 0 to $\infty$, depending on how it is approached.  Second, for the quantum Ising chain with transverse field $\lambda$ and longitudinal field $h$, there are stable fixed points at $(0,0)$, $(0,\infty)$, $(\infty, 0)$, and $(1, \infty)$. The existence of stable fixed points $(0,0)$ and $(\infty, 0)$ is protected by the $Z_2$ symmetry when $h=0$, whereas the existence of stable fixed points $(0, \infty)$ and $(1, \infty)$ may be interpreted as a consequence of the variation of the symmetry group with $\lambda$: $U(1)$ for $\lambda =0$, and none for  $\lambda \neq 0$, when $h\neq 0$. Third, for the quantum spin-$1/2$ XYZ model, five different dualities have been identified, which enable us to reproduce the ground state phase diagram. Fourth, for the quantum XXZ model in a magnetic field, at the phase boundary between the critical XY phase and  the antiferromagnetic phase, fidelity temperature is not well-defined, ranging from a finite value to $\infty$.  That is,  a quantum phase transition at this phase boundary is an intermediate case interpolating between a Kosterlitz-Thouless transition  and a Pokrovsky-Talapov transition, which represents a new universality class.

Fidelity flows are irreversible, as follows from the second law in fidelity mechanics. We also present an argument to justify why the thermodynamic, psychological/computational and cosmological arrows of time should align with each other in the context of fidelity mechanics, with the psychological/computational arrow of time  being singled out as a master arrow of time.

\end{abstract}

\maketitle

\section{introduction}~\label{intro}

Quantum critical phenomena \cite{Sachdev, Wen, ortiz} arise from the
cooperative behavior in quantum many-body systems. Conventionally,
there are two categories of theories to describe these fascinating physical phenomena.
One is adapted from Landau's spontaneous symmetry-breaking (SSB) theory~\cite{Landau}, with
a symmetry-broken phase characterized by the nonzero values
of a local order parameter.  A SSB occurs in a system when
its Hamiltonian enjoys a certain symmetry, whereas the
ground state wave functions do not preserve it~\cite{Anderson, Coleman}. The implication of a SSB is twofold: first, a system has stable
and degenerate ground states, each of which breaks the symmetry
of the system; second, the symmetry breakdown results
from random perturbations.
The other  is  Wilson's renormalization group (RG) theory~\cite{RG1,RG2,Michael,jean,RG3},
originated from the field-theoretic approach to classical many-body systems.  However, this so-called Landau-Ginzburg-Wilson paradigm suffers from a few fundamental problems:
first, ubiquitous topologically ordered states occur beyond SSB order~\cite{XGWen, Kane, SCZhang}; second, even for SSB ordered states, only local order parameters are not enough to quantify quantum fluctuations;
third, proliferation of an unlimited number of irrelevant coupling constants occurs in various RG schemes, which makes it impractical to work out RG flows from unstable fixed points to stable fixed points; fourth, intrinsic irreversibility
i.e., information loss, along RG flows is baffling, due to the fact that a number of high energy degrees of freedom are discarded during the construction of the effective Hamiltonian.
As such, a full characterization of quantum critical phenomena is still lacking.

The latest advances in our understanding of quantum critical phenomena originate
from a perspective of  fidelity
\cite{Zanardi,Zhou1,Zhou2,Zanardi2,fid1,fid2}, a basic notion in quantum information
science.  In Refs.~\cite{Zhou1,Zhou2}, it has been argued that ground state
fidelity per site is fundamental, in the sense that it
may be used to characterize quantum phase transitions (QPTs), regardless of what type of
internal order is present in quantum many-body states.
In other words, ground state fidelity per
site is able to describe QPTs arising from  symmetry-breaking order
and/or topological order \cite{note1}. This has been further confirmed in
Refs.~\cite{Zhao,Wang}, where topologically ordered states in
the Kitaev model on a honeycomb lattice~\cite{Kitaev} and the
Kosterlitz-Thouless (KT) phase transition~\cite{Berezinskii, KT} are investigated.
The argument is solely based on the basic postulate of quantum
mechanics on quantum measurements, which implies that
two nonorthogonal quantum states are not reliably distinguishable
\cite{Nielsen}.  Moreover, even for quantum many-body systems
with symmetry-breaking order, it is advantageous to
adopt ground state fidelity per  site instead of using
the conventional local order parameters, due to the fact that it
is model-independent, although one may systematically derive
local order parameters from tensor network representations~\cite{DMRG,TN1,TN2}
of  ground state wave functions
by investigating the reduced density matrices for local
areas on an infinite-size lattice \cite{Zhou3}.
In fact, a systematic scheme to study  quantum critical phenomena in
the context of the fidelity approach to quantum critical phenomena consists of three stpdf,
as advocated in Ref.~\cite{Zhou3}: first, map out the ground state
phase diagram by evaluating ground state fidelity per
site; second, derive local order parameters (if any) from
the reduced density matrices for a representative ground state
wave function in a given phase; third, characterize any phase
without any long-range order in terms of non-local order parameters.

An intriguing question is to ask whether or not the fidelity approach provides a full characterization of quantum critical phenomena, in the sense that it is not only able to signal critical points/unstable fixed points,
but also offers a way to locate stable fixed points. Moreover,  it has to clarify in what sense  a quantum many-body system flows from unstable fixed points to stable fixed points in the coupling parameter space, which may be understood as a flow discarding irrelevant information along the way.  Given this as our main goal, our work is motivated by a few specific questions:\\

(i) There is a long-standing folklore, pointing towards a similarity between critical points and black holes~\cite{folklore}. Given that both critical points and black holes originate from singularities, there should be a way to clarify a formal similarity between QPTs and black holes.\\

(ii) RG flows from an unstable fixed point to a stable fixed point are irreversible. This is relevant to Zamolodchikov's $c$-theorem and Cardy's $a$-theorem~\cite{ctheorem,cardy1,cardy2,atheorem}, which may be regarded as the adaptation of the renowned Boltzmann's H theorem to the RG setting.  In real space RG theories, such as Kadanoff block spins as well as other coarse-graining RG schemes, high energy degrees of freedom are discarded. Therefore, RG flows seem irreversible in a similar sense to the situations described by Boltzmann's H theorem, where the physical time is replaced by a RG parameter~\cite{gaite}. Thus, it is desirable to see if there is any intrinsic explanation for the irreversibility from the fidelity perspective.\\

(iii) As Landauer first noted~\cite{Landauer}, at finite temperature $T$,  in order to erase one bit of information, we need to do the minimum work $w$:  $w=k_BT\ln{2}$, with $k_B$ being the Boltzmann constant.  At zero temperature, do we still need to do any sort of minimum work to erase one bit of information?\\

(iv) During the construction of an effective Hamiltonian along any RG flow, an unlimited number of irrelevant coupling constants proliferate. In practice, this prevents access to a stable fixed point.  It is tempting to see if there is any deep reason underlying this observation.\\

(v) A proper definition of QPTs is still lacking. Traditionally, the ground state energy density is used as an indicator to signal a critical point, but fails in many situations~\cite{Wolf}.  Instead, some exotic indicators, such as entanglement entropy~\cite{Amico}, topological entanglement entropy~\cite{Preskill,xgwen1} and ground state fidelity per site~\cite{Zhou1,Zhou2}, are introduced to signal QPTs, due to recent progress in our understanding of quantum critical phenomena.  Therefore, it is important to find a proper criterion to define QPTs.\\

In this work, we aim to answer these questions. This is achieved by introducing fidelity temperature to quantify quantum fluctuations present in a given ground state wave function for a quantum many-body system, which exhibits QPTs. As it turns out,   fidelity temperature, together with fidelity entropy and fidelity internal energy,  offer us  a proper basis to describe QPTs, both continuous and discontinuous.  As a consequence,  this allows us to formulate analogues of the four thermodynamic laws and Landauer's principle.  As illustrations, we  discuss the quantum XY model, the transverse field quantum Ising chain in a longitudinal field, the spin-$1/2$ quantum XYZ model and the quantum XXZ model in a magnetic field, with help of classical simulations of quantum many-body systems in terms of a tensor network algorithm in the context of matrix product states~\cite{DMRG,TN1,TN2}. For each model, a canonical form of the Hamiltonian is introduced via dualities.
Rich physics is unveiled even for  these well-studied models. First, for the quantum XY chain,  the disorder circle is interpreted as a separation line between two different types of fidelity flows, with one type of fidelity flows starting from unstable fixed points with central charge $c=1$, and the other type of fidelity flows starting from unstable fixed points with central charge $c=1/2$.  Both types of fidelity flows end at the same stable fixed point $(0,1)$, at which fidelity entropy reaches its local maximum.  Another remarkable feature is that fidelity temperature is zero on the disorder circle, as it should be, since no quantum fluctuations exist in a factorized state. However, fidelity temperature is not well-defined at  the Pokrovsky-Talapov (PT) transition point~\cite{PT,PT1}, ranging from 0 to $\infty$, depending on how it is approached.  Second, for the quantum Ising chain with transverse field $\lambda$ and longitudinal field $h$,  there are stable fixed points at $(0,0)$, $(0,\infty)$, $(\infty, 0)$, and $(1, \infty)$. The existence of stable fixed points $(0,0)$ and $(\infty, 0)$ is protected by the $Z_2$ symmetry when $h=0$, whereas the existence of stable fixed points $(0, \infty)$ and $(1, \infty)$ may be interpreted as a consequence of the variation of the symmetry group with $\lambda$: $U(1)$ for $\lambda =0$, and none for $\lambda \neq 0$, when $h\neq 0$.  Third, for the spin-$1/2$ XYZ model, five different dualities have been identified, which enable us to reproduce the ground state phase diagram for the quantum XYZ chain. Fourth, for the XXZ model in a magnetic field, at the phase boundary between the XY critical phase and  the antiferromagnetic (AF) phase, fidelity temperature is not well-defined, ranging from a finite value to $\infty$.  That is,  a QPT at this phase boundary is an intermediate case interpolating between a KT transition  and a PT transition, which represents a new universality class, different from both the KT and the PT transitions.

The layout of this work is as follows. Section~\ref{intro} is an introduction, describing our motivations to formalize a full characterization of quantum critical phenomena in the context of fidelity. In particular, five questions are raised regarding the current status of theoretical investigations into quantum critical phenomena.
In Section~\ref{fmsut}, we  first define a fidelity mechanical system and its environment, thus attaching a physical meaning to the present, the past and the future, with information storage as a key ingredient,
and then introduce three fidelity mechanical state functions, i.e., fidelity entropy, fidelity temperature, and fidelity  internal energy, with an analogue of Landauer's principle at zero temperature as a basic requirement from the internal logical consistency. In Section~\ref{duality}, a canonical form of the Hamiltonian in fidelity mechanics is discussed, thus unveiling an inherently fundamental role of duality in fidelity mechanics.  In Section~\ref{qxymodel},  we present fidelity mechanical state functions for the quantum XY chain, which is a typical example for continuous QPTs.  In Section~\ref{tfismodel},  fidelity mechanics is discussed for the transverse field quantum Ising chain in a longitudinal field,
which exhibits a first-order QPT.  In Section~\ref{xyzmodel},  fidelity mechanics is discussed for the spin-$1/2$ XYZ chain, thus offering a prototype for the role of duality in fidelity mechanics.  In Section~\ref{xxzhmodel}, an analysis of fidelity mechanical state functions is presented for the quantum spin-$1/2$ XXZ chain in a magnetic field,  which enables us to unveil intermediate cases between the KT transitions and the PT transitions. Here, we stress that, except for the quantum XY chain,  a tensor network algorithm~\cite{TN1,TN2} in terms of a matrix product state representation has been exploited to simulate quantum many-body systems in these illustrative examples, thus making it possible to numerically compute ground state fidelity per site and in turn fidelity mechanical state functions. In Section~\ref{answer}, we answer the questions raised in the Introduction, and define fidelity flows as an alternative form of RG flows.  Moreover, an argument is presented to justify why the thermodynamic, psychological/computational and cosmological arrows of time should align with each other in the context of fidelity mechanics, with the psychological/computational arrow of time  being singled out as a master arrow of time.  The last Section~\ref{conclusion} is devoted to concluding remarks.

Some complementary details are also presented in Appendices.   In Appendix~\ref{pinch}, we introduce ground state fidelity per  site and define a pinch point as an intersection point between two singular lines on a fidelity surface.  As typical examples, the transverse field quantum Ising chain and the Kitaev model on a honeycomb are presented to illustrate QPTs arising from symmetry-breaking order and topological order, respectively.
In Appendix~\ref{fidelityfromTN}, we describe an efficient scheme to compute ground state fidelity per site in the context of matrix product states. In Appendix~\ref{itebd}, we summarize the infinite time-evolving block decimation algorithm~\cite{TN2}, which is efficient to generate a matrix product state representation of ground state wave functions for quantum many-body systems in one spatial dimension.
In Appendix~\ref{gedensity}, we describe an efficient means to numerically identify (unentangled) separable states in the context of tensor networks~\cite{TN1,TN2}.  In Appendix~\ref{timearrows}, arrows of time are discussed, with focus on the thermodynamic, psychological/computational  and cosmological arrows of time.  It is argued that fidelity mechanics may be regarded as an attempt to understand the psychological/computational arrow of time in the context of quantum many-body systems.  Appendix~\ref{threetheorems} recalls three theorems in quantum information science, which are used to justify our definition of a fidelity mechanical system and its environment.
In Appendix~\ref{connectionUT}, we establish a relation between an unknown function, as a defining factor for fidelity internal energy, and fidelity temperature.
In Appendix~\ref{xychain}, mathematical details are discussed about fidelity entropy, fidelity temperature and fidelity internal energy for the quantum XY chain.  In Appendix~\ref{ishchain}, explicit expressions for fidelity entropy, fidelity temperature and fidelity internal energy are presented for the transverse field quantum Ising chain in a longitudinal field.  In Appendix~\ref{xyzchain},  mathematical details for fidelity entropy, fidelity temperature and fidelity internal energy are discussed for the quantum XYZ model.
In Appendix~\ref{fictitious}, a fictitious parameter $\sigma$  is introduced to address different choices of a dominant control parameter in a given regime for  quantum many-body systems. As demonstrated, information encoded in this fictitious parameter arising from different choices of a dominant control parameter is irrelevant in fidelity mechanics.
In Appendix~\ref{shift}, the meaning of a canonical form of the Hamiltonian is clarified through relating duality with a shift operation in the Hamiltonian. Therefore,  any artificial choice of the definition of duality is irrelevant, as long as the identification of unstable and stable fixed points is concerned.
In  Appendix~\ref{dual}, dualities are presented for the quantum XYZ models on both bipartite and non-bipartite lattices.
In Appendix~\ref{dualityinkitaev},  dualities are presented for the quantum spin-$s$ Kitaev model on a honeycomb lattice.
In Appendix~\ref{scaling}, a scaling behavior of fidelity entropy near a critical point is discussed.
In Appendix~\ref{gaussion},  a scaling analysis is presented for the quantum XY chain near a line of the Gaussian critical points.

\section{Fidelity mechanics: Basic state functions}~\label{fmsut}

\subsection{Preliminaries}~\label{fmsuta}

Consider a quantum many-body system on a lattice described by a Hamiltonian $H(x_1,x_2)$, with $x_1$ and $x_2$ being two coupling parameters~\cite{twoton}.  It is necessary to determine its ground state phase diagram, in addition to symmetries, dualities and factorizing fields.  As is well known, lines of critical/transition points divide the parameter space into different phases, which may be characterized in terms of local order parameters for symmetry-breaking ordered phases and non-local order parameters for topologically ordered phases, respectively. In contrast,  symmetries, dualities and factorizing fields furnish characteristic lines in the parameter space.  Here, we define an intersection point between two or more characteristic lines, arising from symmetries, dualities and factorizing fields, as a characteristics point.  As a consequence,  a peculiar type of characteristic lines appears, originating from a multi-critical point and ending at a characteristic point. This happens, only if no characteristic line, arising from symmetries, dualities and factorizing fields, exists at this multi-critical point.

The ground state phase diagram may be mapped out by evaluating ground state fidelity per site.  As demonstrated in Refs.~\cite{Zhou1,Zhou2,Zhao,Wang,Zhou3}, ground state fidelity per site is able to signal QPTs arising from symmetry-breaking order and/or topological order.  Here, we restrict ourselves to briefly recall  the definition of ground state fidelity per site (cf. Appendix~\ref{pinch} for more details). For two ground states $|\psi (x_1,x_2) \rangle$ and $|\psi (y_1,y_2) \rangle$, fidelity is a measure of the similarity between them.  Mathematically, it is defined as the absolute value of their overlap: $F(x_1,x_2; y_1,y_2)=|\langle\psi(y_1,y_2)|\psi (x_1,x_2)\rangle|$. Here, we stress that $y_1$ and $y_2$ should be understood as different values of the {\it same} coupling parameters as $x_1$ and $x_2$, respectively.
In the thermodynamic limit, any two ground states are always distinguishable (orthogonal).  That is, fidelity between these two states vanishes.  However,
for a large but finite lattice, $F_N(x_1,x_2; y_1,y_2)$ scales as $d^N(x_1,x_2; y_1,y_2)$, with $N$ being the total number of lattice sites, and $d(x_1,x_2; y_1,y_2)$  being a scaling parameter.
In the thermodynamic limit, ground state fidelity per site $d(x_1,x_2; y_1,y_2)=\lim_{N\rightarrow\infty}{F_N^{1/N}}(x_1,x_2; y_1,y_2)$ is well-defined.  For an efficient scheme to compute ground state fidelity per site in the context of the time-evolving block decimation in terms of matrix product states, we refer to Appendix~\ref{fidelityfromTN} and Appendix~\ref{itebd}.

Close scrutiny should be performed about symmetries of  the Hamiltonian under investigation. Generically,  the symmetry group of the Hamiltonian varies with the coupling parameters.  In particular, a $U(1)$ symmetry occurs when one coupling parameter is infinite in value.  A characteristic line with a specific symmetry group often arises from a QPT belonging to different universality class from others.  Characteristic lines also arise from dualities~\cite{ortiz}, which are defined via a local or nonlocal unitary transformation.  Another type of characteristic lines comes from factorizing fields~\cite{factorizing,factorizing1}. In addition to various analytical approaches, there is an efficient numerical means to identify factorizing fields for quantum many-body systems in the context of tensor networks~\cite{geometricGE} (also cf. Appendix~\ref{gedensity}).

Given that characteristic lines separate a given quantum phase into different regimes, we need to clarify physical reasons underlying this separation. In our scenario, all the ground states in a given phase share the same relevant information, with their distinguishability fully attributed to the fact that irrelevant information encoded in different ground states is different (cf. Appendix~\ref{pinch} for more details). However, this does not categorize any different types of irrelevant information that are possible in a given phase, which in turn may be traced back to critical points belonging to different universality classes.  Actually, it is the four different types of characteristic lines that separate a given phase into different regimes, making it possible to attach a certain type of irrelevant information to each regime. That is, there is a one-to-one correspondence between a regime and a type of irrelevant information in a given phase. In addition, this also applies to characteristic lines themselves: different types of irrelevant information are attached to different characteristic lines.

To proceed, we  introduce a {\it dominant} control parameter $x$ and an {\it auxiliary} control parameter $\tau$ to replace the original coupling parameters $x_1$ and $x_2$ such that there is a one-to-one correspondence between ($x$, $\tau$) and ($x_1$, $x_2$) in a specific regime.  Therefore, the Hamiltonian $H(x_1,x_2)$ is re-parametrized as $H(x,\tau)$.  As a dominant control parameter, $x$ has to satisfy the conditions: (i) as a function of $x$, the ground state energy density $e(x,\tau)$ is monotonic with increasing $x$ for a fixed $\tau$; (ii) the range of $x$ is {\it finite}.  Generically, $x$ starts from a critical/transition point $x_c/x_d$ for continuous/discontinuous QPTs, and ends at a characteristic line/point.  Such a characteristic line itself starts from a multi-critical point, and describes a QPT belonging to a universality class different from what $x_c/x_d$ belongs to.  As a consequence,  our emphasis is on irrelevant information instead of relevant information encoded in ground state wave functions for quantum many-body systems.  This is in contrast to local order parameters in Landau's SSB theory, but resembles real space RG theories that merely manipulate high energy degrees of freedom.

From now on, if the auxiliary control parameter $\tau$ is fixed, then we shall drop $\tau$ in the Hamiltonian $H(x,\tau)$ to keep the notation simple.

\subsection{A fidelity mechanical system and its environment}~\label{fmsutb}

For a quantum many-body system described by a Hamiltonian $H(x)$, if we treat $x$ as a parameter varying with time $t$, then the time evolution is subject to the time-dependent Schr\"{o}dinger equation, which is invariant under the time-reversal operation. In particular, as the adiabatic theorem~\cite{Messiah} tells, if $x$ is slowly varying, then the system remains in a ground state, if it is initially in a ground state, unless a critical point is crossed~\cite{Campo}. However,  everyday experience teaches us that we remember the past, but not the future. This so-called psychological/computational arrow of time distinguishes the past from the future (for a brief summary about arrows of time, cf.  Appendix~\ref{timearrows})~\cite{timearrow}.  A fundamental issue is to understand the ensuing consequences resulted from information storage, i.e., recording information encoded in the past states in media.  As it turns out, information storage is a key ingredient in fidelity mechanics.

\begin{figure}
  \centering
\includegraphics[angle=0,totalheight=5.8cm]{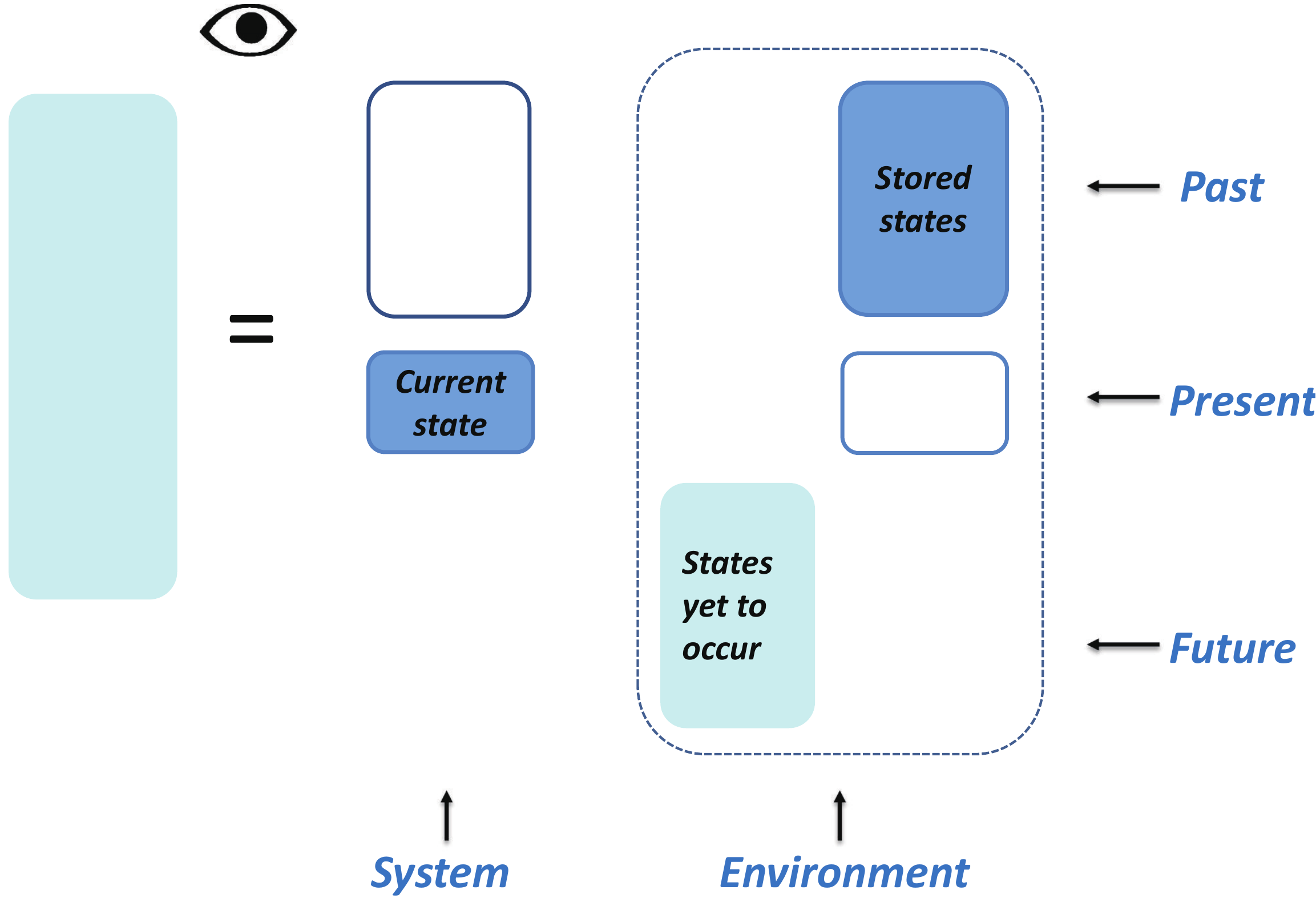}
\caption{A fidelity mechanical system and its environment.  A fidelity mechanical system is defined to be the current state stored in a medium.  An environment consists of the past states, which are stored in other media, and any possible states yet to occur in the future, which are simply left blank in media.  Here, the current state, the past states and the future states are associated with a quantum many-body system described by the Hamiltonian $H(x)$, with $x$ being a dominant control parameter, meaning that the ground state energy density $e(x)$ is monotonic with increasing $x$ and the range of $x$ is finite.  The present lies exactly at the intersection of  the past and the future. Note that an outside observer, as an information processor,  is equipped with a quantum copier tailored to a collection of mutually orthogonal states generated via a time evolution.  Thus, a certain amount of information is extractible by comparing the current state with the past states, both of which are stored in media.}\label{ego}
\end{figure}

An outsider observer,  as an information processor, is equipped with a quantum copier tailored to a collection of mutually orthogonal states generated via a time evolution.  Note that the no-cloning theorem does not rule out the possibility for copying a set of mutually orthogonal states~\cite{orthogonality} (cf. Appendix~\ref{threetheorems} for more details about the no-cloning, no-deleting and no-hiding theorems).  This enables us to turn quantum states at different instants, which arise from a time evolution, into quantum states at the same instant, recorded in media, via quantum copying. This is in sharp contrast to the case when one considers a quantum few-body system. For the latter, it is impossible to set up such a quantum copier~\cite{difference}.

Now we are ready to define a fidelity mechanical system and its environment. A fidelity mechanical system is defined to be the current state stored in a medium.  An environment consists of the past states, which are stored in other media, and any possible states yet to occur in the future, which are simply left blank in media.  The present lies exactly at the intersection of  the past and the future.  A pictorial representation for a fidelity mechanical system and its environment is presented in Fig.~\ref{ego}.

Now we turn to a description of a state for a given fidelity mechanical system.  For this purpose, we introduce a quantum mechanical equivalent of the relaxation time scale in thermodynamics~\cite{Kittel}, which tells how much time a non-equilibrium state needs to adjust to an equilibrium state.   From the adiabatic theorem, one knows that, as long as the inverse of the gap is small enough, a quantum system starting its evolution from a ground state remains in the ground state. However, if it is driven at finite rate, then it will be excited. In fact, the inverse of the gap acts as a quantum mechanical equivalent of the relaxation time scale~\cite{Damski}.  Therefore, it is plausible to regard an adiabatic evolution as an analogue of a quasi-static process in thermodynamics. In fact, a fidelity mechanical system, with the current state stored in a medium being a ground state, is in equilibrium with its environment, with all the past states being ground states. Accordingly,  a fidelity mechanical system, with the current state stored in a medium being a low-lying state,  takes
at least as much time as required by a quantum mechanical equivalent of the relaxation time scale,  in order to return to an equilibrium state with its environment.
This allows us to define basic state functions, i.e., fidelity entropy, fidelity temperature and fidelity internal energy for a fidelity mechanical system.

\subsection{Fidelity entropy, fidelity temperature and fidelity internal energy: continuous quantum phase transitions}~\label{fmsutc}

For a given fidelity mechanical system, which is in equilibrium with its environment,  an important question is to quantify what amount of information may be recovered from the environment, due to information storage that makes information encoded in the past states available. Here, it is proper to clarify what type of information we are trying to extract. In fact, we may categorize information into two different types: (i)  information encoded in a given state, which may be quantified in terms of, e.g., entanglement entropy. In this case,  {\it only} one state is concerned, with quantum correlations at different spatial locations involved. So this type of information is {\it spatial};  (ii) information extractible by comparing the current state with the past states, both of which are stored in media. Thus,  different states at different temporal instants are involved. So this type of information is {\it temporal}. In fidelity mechanics,  we solely deal with information of the second type.

For a continuous QPT, fidelity entropy $S(x)$ is defined to characterize the uncertainty accumulated from  a critical point $x_c$ to $x$:
\begin{equation}
S(x)=-2\int_{x_c}^{x}{\ln{d(x,y)}d{y}} +S_0. \label{entropy}
\end{equation}
Here, $d(x,y)$ is ground state fidelity per site for two ground states $|\psi (x) \rangle$ and $|\psi (y) \rangle$,  $S_0$ is an additive constant, representing residual fidelity entropy at a critical point.  Fidelity entropy $S(x)$ quantifies the amount of information that is extractible from comparing the current state at $x$ with the stored states at $y$ in the past. Actually, there is an interpretation for the first term in the definition of  fidelity entropy $S(x)$ in terms of Shanon entropy, if one regards the squared fidelity between two quantum states as a probability.

We assume $e(x)$ is always negative for any $x$.  Given fidelity entropy $S(x)$, we need to define, in a consistent way, fidelity internal energy $U(x)$ and fidelity temperature $T(x)$~\cite{fidtem}. Indeed,  we define fidelity temperature $T(x)$ as $T(x)={\partial{U(x)}}/{\partial{S(x)}}$, which implies that  no fidelity work is involved, when $x$ is varied.  This amounts to stipulating a rule that separates an increment of fidelity internal energy $\Delta U(x)$ into an increment of fidelity heat $\Delta Q(x)$ and an increment of fidelity work $\Delta W(x)$, with $\Delta Q(x) = T(x) \Delta S(x)$~\cite{separation}.

Suppose the Hamiltonian $H(x)$  is defined through a Hamiltonian density $h(x)$ acting locally on the Hilbert space for a translation-invariant quantum many-body system.  With the translational invariance in mind, we have $\langle \psi(y)|h(x)|\psi(x)\rangle=e(x)\langle \psi(y)|\psi(x)\rangle$, with $e(x)$ being the ground state energy density.  Given $\langle \psi(y)|\psi(x)\rangle$ scales exponentially with $N$, it is reasonable to postulate that the dependence of fidelity internal energy $U(x)$ on the ground state energy density $e(x)$ should be logarithmic.  Hence, we define fidelity internal energy $U(x)$ as follows
\begin{equation}
U(x)= \mp \ln{(\frac{e(x)}{e(x_c)})}V(x) + U_0,\label{internalenergy}
\end{equation}
where $V(x)$ is a quantity, as a function of $x$,  yet to be determined consistently, and $U_0$ is an additive constant.  Here, $\mp$ is introduced to ensure that $V(x)$ is positive: $-/+$ corresponds to monotonically increasing/decreasing $e(x)$ with increasing $x$, respectively.  It is proper to remark that fidelity entropy $S(x)$ and fidelity internal energy $U(x)$, as defined, should be understood as fidelity entropy per site and fidelity internal energy per site, respectively.

Given two undefined quantities $V(x)$ and fidelity temperature $T(x)$, we really need another constraint. As it turns out, such a constraint occurs in the guise of an analogue of Landauer's principle at zero temperature: at zero temperature, due to quantum fluctuations, a certain amount of fidelity work needs to be done to erase any information.  Logically, the internal consistency ascertains that the minimum fidelity work to be done to erase one bit of information must be $w(x)=  \pm T(x) \ln{2}$,  with $T(x)$ being fidelity temperature quantifying quantum fluctuations and $+/-$ corresponding to increasing/decreasing $e(x)$ with increasing $x$, respectively.

In Appendix~\ref{connectionUT},  a key relation between fidelity temperature $T(x)$ and $V(x)$  is derived from an analogue of Landauer's principle at zero temperature:\begin{equation}
T(x)=- \frac{\partial{V(x)}} {\partial{x}} \equiv -V_x(x).
\end{equation}
Here, $V(x)$ must be monotonically decreasing with increasing $x$, in order to guarantee positive fidelity temperature $T(x)$.  Combining this relation with the definition of fidelity temperature $T(x)$:
\begin{equation}
T(x)=\frac{\partial{U(x)}/\partial{x}}{\partial{S(x)}/\partial{x}},
\label{temperature}
\end{equation}
we have
\begin{equation}
T(x) =\mp \frac{(\ln{(e(x)/e(x_c))})_xV(x)+\ln{e(x)/e(x_c)}V_x(x)}{S_x(x)}.
\label{temperature}
\end{equation}
Here,  $S_x(x) \equiv \partial{S(x)}/\partial{x}$  and $(\ln{(e(x)/e(x_c))})_x\equiv \partial{\ln{(e(x)/e(x_c))}}/\partial{x}$.  This implies
\begin{equation}
V_x(x)=\alpha (x) ~V (x),
 \label{alpha}
\end{equation}
where $\alpha (x)$ is defined to be

\begin{equation}
\alpha (x)= \pm \frac{(\ln{(e(x)/e(x_c))})_x}{S_x(x) \mp \ln{(e(x)/e(x_c))}}.
\label{alphapm}
\end{equation}
Here, we emphasize that $\alpha (x)$ is singular when a critical point $x_c$ is approached. Therefore,  Eq. (\ref{alpha}) is a singular first-order differential equation. It plays a fundamental role in fidelity mechanics. In fact, once it is solved, we are able to determine fidelity internal energy $U(x)$ and fidelity temperature $T(x)$.

\subsection{Fidelity entropy, fidelity temperature and fidelity internal energy: discontinuous quantum phase transitions}~\label{fmsutd}

For first-order (discontinuous) QPTs, some modifications are needed. For a quantum many-body system undergoing a first-order QPT at a transition point $x_d$, fidelity entropy $S(x)$ is defined as
\begin{equation}
S(x)=-2\int_{x_d}^{x}{\ln{d(x,y)}dy} +S_0,
\label{entropyd}
\end{equation}
where $S_0$ is residual fidelity entropy at  a first-order transition point $x_d$.
However, in order to keep consistency with the fact that fidelity temperature $T(x)$ is finite at a first-order transition point $x_d$, an additional parameter $\kappa$ is introduced in  fidelity internal energy $U(x)$:
\begin{equation}
U(x)= \mp [\ln \kappa + \ln{(\frac{e(x)}{e(x_d)})}]V(x) + U_0.
\label{energyd}
\end{equation}
Here,   $V(x)>0$ is an undetermined function of $x$, $U_0$ is an additive constant, and $-/+$ corresponds to monotonically increasing/decreasing of $e(x)$  with increasing $x$, respectively.  Note that fidelity temperature $T(x)$ is again determined by $T(x)=-V_{x}$, since the same argument still applies to first-order (discontinuous) QPTs.  In fact, $V(x)$ must be monotonically decreasing with increasing $x$, in order to guarantee positive fidelity temperature $T(x)$.  Combining the definition of  fidelity temperature  $T(x)=\partial{U(x)}/\partial{S(x)}=\partial{U(x)}/\partial{x}/\partial{S(x)}/\partial{x}$ with $T(x)=-V_{x}(x)$, we have
\begin{equation}
V_x(x)=\alpha_d (x) ~V (x),
 \label{alphaf}
\end{equation}
where $\alpha_d (x)$ is defined to be
\begin{equation}
\alpha_d (x)=\pm \frac{(\ln{e(x)/{e(x_d)})}_{x}}{S_{x}(x)\mp(\ln \kappa +\ln{(e(x)/e(x_d))})}.
\label{alphad0}
\end{equation}
Note that, in contrast to continuous QPTs, $\alpha_d (x)$ is regular when a transition point $x_d$ is approached. Therefore,  Eq. (\ref{alphaf}) is a  regular first-order differential equation. Once it is solved, we are able to determine fidelity internal energy $U(x)$ and fidelity temperature $T(x)$.

\subsection{A contribution to fidelity entropy from rescaling in the ground state energy density }~\label{fmsute}

For a quantum many-body system, the ground state phase diagram exhibits a diversity of distinct phases, which in turn involve different regimes.
A dominant control parameter $x$ has been introduced via a one-to-one correspondence between $(x_1,x_2)$ and $(x, \tau)$ in a specific regime. However, it is impossible for one single dominant control parameter to work in all the regimes.  This might be due to the fact that either the ground state energy density $e(x)$ is not monotonic as a function of $x$, or the range of $x$ is not finite. Therefore,
there is another possible contribution to fidelity entropy $S(x)$, which arises from rescaling in the Hamiltonian $H(x) \rightarrow H'(x')$, such that the ground state energy density $e(x)$ becomes $e'(x')$: $e(x)=k(x) \; e'(x')$,  where $x' $ is a monotonic function of $x$ and $k(x)>0$ is a rescaling factor, which has to be a monotonic function of $x$. It is convenient to introduce $k'(x')$ such that $k'(x')=k(x)$, which implicitly defines $x$ as a monotonic function of $x'$.
A few remarks are in order.  First,  the rescaled ground state energy $e'(x')$ must be monotonic as a function of $x'$, although $e(x)$ itself is not necessary to be monotonic as a function of $x$.  Second,  there should be an $x'_0$ such that $k'(x'_0)=1$, and $k'(x')$ is monotonic as a function of $x'$. Third, the range of $x'$ has to be {\it finite}, although the range of $x$ may be finite or infinite, depending on a specific regime. In fact, if the range of $x$ is infinite, then rescaling  is necessary to ensure that the range of a properly chosen dominant control parameter $x'$ is finite. However, even if the range of $x$ is finite, rescaling is still needed in certain occasions, since $e(x)$ may be not monotonic with respect to $x$. Fourth, rescaling occurs when duality exists, as discussed in Section~\ref{duality}. In this case, a unitary transformation is involved. However,  there exist other situations, for which it is also necessary to rescale the Hamiltonian $H(x)$.  Detailed treatments to specific examples are presented in, e.g., Sections~\ref{qxymodel}, ~\ref{tfismodel} and~\ref{xyzmodel}.

Given $e'(x')$ is monotonically increasing/decreasing with increasing $x'$, fidelity entropy $S_\phi '(x')$, fidelity temperature $T_\phi '(x')$ and fidelity internal energy $U_\phi '(x')$ for the rescaled Hamiltonian $H'(x')$ follow from our discussions above, with the replacements $x \rightarrow x'$, $x_c \rightarrow x'_c$, $x_d \rightarrow x'_d$, $y \rightarrow y'$, $e(x) \rightarrow e'(x')$,  and $V(x) \rightarrow V'(x')$.  For continuous QPTs, we have $S_\phi '(x')=-2\int_{x'_c}^{x'}{\ln{d(x',y')}d{y'}} +S'_{\phi 0}$, $T_\phi '(x')=-V'_{x'}(x')$, and $U_\phi '(x')= \mp \ln{(e'(x')/e'(x'_c))}V'(x') + U'_{\phi 0}$,  where $V'_{x'}(x')=\alpha (x')\;V' (x')$, with $\alpha' (x')= \pm (\ln{(e'(x')/e'(x'_c))})_{x'} /(S'_{\phi x'}(x') \mp \ln{(e'(x')/e'(x'_c))})$. For discontinuous QPTs, we have $S_\phi '(x')=-2\int_{x'_d}^{x'}{\ln{d(x',y')}d{y'}} +S'_{\phi 0}$, $T_\phi '(x')=-V'_{x'}(x')$, and $U_\phi '(x')=\mp (\ln \kappa + \ln{(e'(x')/e'(x'_d))})V'(x') + U'_{\phi 0}$,  where $V'_{x'}(x')=\alpha (x')\;V' (x')$, with $\alpha' (x')= \pm (\ln{(e'(x')/e'(x'_d))})_{x'} /(S'_{\phi x'}(x') \mp (\ln \kappa + \ln{(e'(x')/e'(x'_d))}))$. Here, a subscript $\phi$ has been used to indicate that no contribution to fidelity entropy from rescaling in the Hamiltonian is considered.

We turn to fidelity entropy $S(x)$, fidelity temperature $T(x)$ and fidelity internal energy $U(x)$ for the Hamiltonian $H(x)$. Fidelity temperature $T(x)$ must be identical to $T_\phi '(x')$, given fidelity temperature quantifies quantum fluctuations present in a ground state wave function, which remains the same under rescaling. In addition, fidelity internal energy  $U(x)$ must remain the same as $U'_\phi (x')$, since we are dealing with the same Hamiltonian, up to a rescaling factor $k'(x')$. The latter causes an extra uncertainty
$\Delta k'(x') / k'(x')$.  Therefore, the only change arising from rescaling is an additive contribution to fidelity entropy $S(x)$, with fidelity temperature  $T(x)$ and fidelity internal energy  $U(x)$ left intact: $S(x) = S_\phi(x)+S_\sigma(x)$, $T(x) =T_\phi(x)$ and $U(x) = U_\phi (x)$, where $S_\phi (x) \equiv S'_\phi (x')$, $T_\phi (x) \equiv T'_\phi (x')$, and $U_\phi (x) \equiv U'_\phi (x')$.  Here,  $S_\sigma(x)$ represents scaling entropy, defined as $S_\sigma (x) \equiv S'_\sigma (x')$, where $S'_\sigma (x') \equiv \pm \ln k'(x')$ up to a constant, with  $\pm$  being determined to ensure that $d S'_\sigma (x')/d x' >0$. That is,  scaling entropy $S'_\sigma(x')$ is monotonically increasing with increasing $x'$.  Physically, this is due to the fact that the presence of a rescaling factor $k'(x')$  amounts to a variation of an energy scale with $x'$, which makes a contribution to fidelity entropy, with the variation of scaling entropy $S'_\sigma(x')$  being proportional to  $\Delta k'(x') / k'(x')$.  The latter represents the uncertainty due to the variation of energy scales.   Accordingly, the increment of scaling entropy $S'_\sigma (x')$ is compensated by a certain amount of fidelity work $\Delta W'(x')$ such that $0 = T'_\phi(x') \Delta S'_\sigma (x') +\Delta W'(x')$, if $x'$ is changed to $x'+\Delta x'$.  Combining with $\Delta U'_\phi (x') = T'_\phi (x') \Delta S'_\phi(x')$, we have

\begin{equation}
\Delta U(x) = T(x) \Delta  S(x) +\Delta W(x).
\end{equation}
Here, $\Delta W(x)$ is identified as $\Delta W'(x')$.  The same argument is also applicable to discontinuous QPTs.

\subsection{Shifts in fidelity temperature and fidelity internal energy: the continuity requirements}~\label{fmsutf}

Up to the present we focused on a dominant control parameter $x/x'$ for a quantum many-body system described by the Hamiltonian $H(x)/H'(x')$, depending on whether or not rescaling in the Hamiltonian is needed. Now we turn to the role of an auxiliary control parameter $\tau$, which results in the continuity requirements.

As demonstrated, characteristic lines divide a given phase into different regimes in the ground state phase diagram. 
As a result of dualities, not all the regimes are independent; we refer to all the independent regimes as principal regimes, which actually represent the underlying physics for a given quantum many-body system. If a line of critical points is involved as a boundary in a given regime, then one may categorize principal regimes into different classes:  (i) the ranges of  the two coupling parameters $x_1$ and $x_2$ in a given regime are finite; (ii) the range of one of the two coupling parameters $x_1$ and $x_2$ in a given regime is finite,  with the line of critical points is not finite in extent; (iii) the range of one of the two coupling parameters $x_1$ and $x_2$ in a given regime is finite,  with the line of critical points is finite in extent; (iv) the ranges of  the two coupling parameters $x_1$ and $x_2$ in a given regime are not finite. Otherwise, only a line of discontinuous phase transition points occurs, which ends at an isolated critical point.  This leads to a different class of principal regimes: (v) the ranges of  the two coupling parameters $x_1$ and $x_2$ in a given regime are not finite, with an isolated point as a critical point.
Typically, a principal regime in class (i) is enclosed by boundaries,  consisting of one line of critical/transition points, one characteristic line originating from a critical/transition point, and one characteristic line originating from a multi-critical point;  a principal regime in class (ii) is enclosed by boundaries,  consisting of one line of critical/transition points, one characteristic line originating from a critical/transition point, and one characteristic line originating from a multi-critical point at infinity; a principal regime in class (iii) is enclosed by boundaries,  consisting of one line of critical/transition points, one characteristic line originating from a critical/transition point, and one characteristic line originating from a multi-critical point, with these two characteristic lines converging at a characteristic point at infinity;  a principal regime in class (iv) is enclosed by boundaries,  consisting of one line of critical/transition points, one characteristic line originating from a critical/transition point, and one characteristic line originating from a multi-critical point at infinity, with these two characteristic lines converging at a characteristic point at infinity; a principal regime in class (v) is enclosed by boundaries,  consisting of a critical point and two characteristic lines originating from the same critical point, characterized by different symmetry groups, due to the variation of the symmetry group with coupling parameters.  As for a choice of a dominant control parameter in a principal regime, our convention is that a dominant control parameter for a principal regime in classes (i) and (ii) originates from a critical/transition point and ends at a characteristic line originating from a multi-critical point, with an auxiliary control parameter being finite in extent for a principal regime in class (i) and infinite in extent for a principal regime in class (ii);  that a dominant control parameter for a principal regime in classes (iii) and (iv) originates from a critical/transition point and ends at a characteristic point at infinity,  with an auxiliary control parameter being finite in extent for a principal regime in class (iii) and infinite in extent for a principal regime in class (iv); and that a dominant control parameter for a principal regime in class (v) originates from a critical point and ends at a characteristic line at infinity,  with an auxiliary control parameter being finite in extent. Note that scaling in the ground state energy is necessary for a principal regime in classes (iii), (iv) and (v) to ensure that the range of a dominant control parameter is finite. In addition, scaling in the ground state energy is often needed on a characteristic line for a principal regime in classes (i) and (ii), either to ensure that the ground state energy density is monotonic with an increasing dominant control parameter, or to keep consistency with the requirements from duality and factorizing fields.

A few remarks are in order. (1) Two characteristic lines, as boundaries in a principal regime, belong to different universality classes.
(2) A multi-critical point at infinity arises  when one of the two coupling parameters $x_1$ and $x_2$ in a given regime is infinite in value, with an extra  $U(1)$ symmetry at this multi-critical point.  This appears to be a result of duality, if a self-dual point does not describe a critical point. (3) A characteristic point at infinity arises when one of the two coupling parameters $x_1$ and $x_2$ in a given regime is infinite in value, with an extra  $U(1)$ symmetry at this characteristic point.  In particular, a factorized ground state occurs at this characteristic point. (4) A characteristic line at infinity arises when the two coupling parameters $x_1$ and $x_2$ in a given regime are infinite in value in proportionality, with an extra  $U(1)$ symmetry at this characteristic line, if a factorized state occurs as a ground state.   Here,  we emphasize that, for a quantum many-body system,  if one of the coupling parameters  $x_1$ and $x_2$ is infinite in value, then there are two possibilities:  it accommodates either a trivial factorized ground state or a multi-critical point - a fact that remains unnoticed in the conventional theories. A point that deserves to be mentioned is that, when we speak of a multi-critical point at infinity or a characteristic point at infinity, we are referring to the original coupling parameters $x_1$ and $x_2$ instead of dominant and auxiliary control parameters, given the range of a dominant control parameter is, by definition, finite.  Related to this is that a characteristic point at infinity is occasionally referred to as a characteristic line, since such a characteristic point at infinity should be regarded as a point as a result of identification, given the Hamiltonian is essentially the same on a characteristic line located at infinity (at most up to a local unitary transformation). 

 With a proper choice of a dominant control parameter in a principal regime as well as on its characteristic lines,  we are able to determine fidelity entropy, according to the definition in Eq.~(\ref{entropy}). The strategy is to start from characteristic lines for a principal regime in class (i) if any, or from characteristic lines for a principal regime in class (iii).  Since fidelity entropy is {\it relative}, in a sense that it is only determined up to an additive constant,  one may choose residual fidelity entropy at a critical point, which is an originating point along a characteristic line, to be zero.  Then,  residual fidelity entropy at a multi-critical point follows from the continuity requirement for fidelity entropy at a characteristic point - an intersection point between the two characteristic  lines. Hence, fidelity entropy at a characteristic line originating from a multi-critical point follows.  With this in mind, we may determine residual fidelity entropy at a line of critical/transition points from the continuity requirement for fidelity entropy at a characteristic line originating from a multi-critical point. 
Hence, fidelity entropy in this principal regime is determined.
Once this is done, we move to a principal regime in class (ii) if any, and repeat the procedure to determine residual fidelity entropy at other critical/transition lines. Afterwards, fidelity entropy for a principal regime in classes (iii) and (iv) as well as along their characteristic lines follows from Eq.~(\ref{entropy}), with determined residual entropy at critical/transition lines, which are shared as boundaries with principal regimes for classes (i) and (ii).   This ensures that fidelity entropy is {\it continuous} on boundaries between different regimes for each phase in the ground state phase diagram. Note that fidelity entropy may not be continuous if a characteristic line is crossed, which describes a QPT belonging to a different universality class.

In order to determine fidelity temperature and fidelity internal energy on a characteristic line, we need to solve a singular first-order differential equation (\ref{alpha})  for continuous QPTs and a regular first-order differential equation  (\ref{alphaf}) for discontinuous QPTs.  This results in an integration constant $V_0/V'_0$, depending on whether or not rescaling in the Hamiltonian is needed on this characteristic line.  Afterwards, we need to shift fidelity temperature, accompanied by a shift in fidelity internal energy to ensure the continuity requirements for fidelity temperature and fidelity internal energy at a characteristic point.   To proceed, we distinguish two different situations: (i) no rescaling in the Hamiltonian is needed on a characteristic line;  (ii) rescaling in the Hamiltonian is needed on a characteristic line.

(i) If no rescaling in the Hamiltonian is needed on a characteristic line,  a shift in fidelity temperature $T(x) \rightarrow T(x)-T_0$ is performed, accompanied by a shift in fidelity internal energy $U (x)\rightarrow U(x)- T_0S(x)$, in order to ensure the continuity requirement for fidelity temperature at a characteristic point: it must be zero, since a factorized state occurs as a ground state.  Generically,  fidelity temperature is zero at any factorized fields.  Hence, $T_0$ simply represents fidelity temperature at a characteristic point, evaluated from a dominant control parameter on a characteristic line originating from a critical point on a line of critical points.  Then, we refer to the shifted fidelity temperature $T(x)-T_0$ as fidelity temperature $T_f(x)$ and the shifted fidelity internal energy  $U(x)- T_0 \; S(x)$ as fidelity internal energy $U_f(x)$, respectively, with fidelity entropy $S(x)$ left intact.  That is, $S_f(x) \equiv S(x)$, $T_f (x)\equiv T(x)-T_0$, and  $U_f(x)\equiv U(x)- T_0 \; S(x)$.

(ii) If rescaling in the Hamiltonian is needed on a characteristic line, we need to combine it with rescaling in the ground state energy density by taking scaling entropy into account.  In fact,  a shift  is performed to fidelity temperature  $T_\phi'(x')\rightarrow T'_\phi(x')-T_0$, accompanied by a shift in fidelity internal energy $U_\phi'(x')\rightarrow U'_{\phi}(x')- T_0 \; S'_{\phi}(x')$, with fidelity entropy $S_\phi'(x')$ left intact.  This is necessary to ensure the continuity requirement for fidelity temperature at a characteristic point: it must be zero, since a factorized state occurs as a ground state.  Hence, $T_0$ simply represents fidelity temperature at a characteristic point, evaluated from a dominant control parameter on a characteristic line originating from a critical point on a line of critical points. That is, $S'_{\phi f}(x') \equiv S'_{\phi}(x') $, $T'_{\phi f}(x') \equiv T'_\phi(x')-T_0$, and  $U'_{\phi f}(x') \equiv U'_{\phi}(x')- T_0 \; S'_{\phi}(x')$.
Therefore, combined with scaling entropy $S'_{\sigma}(x')$, fidelity entropy $S_f(x)$ takes the form: $S_f(x) = S_{\phi f}(x)+S_{\sigma f}(x)$, with $S_{\phi f}(x)\equiv S'_{\phi f}(x')$ and $S_{\sigma f}(x) \equiv S'_{\sigma f}(x')$.
Fidelity temperature $T_f(x)$ and fidelity internal energy $U_f(x)$ take the form:  $T_f(x) = T_{\phi f}(x)$ and $U_f(x) =
U_{\phi f}(x)$, with  $T_{\phi f}(x) \equiv T'_{\phi f}(x')$ and $U_{\phi f}(x) \equiv U'_{\phi f}(x')$.

Let us now discuss  fidelity temperature and fidelity internal energy in a principal regime for continuous QPTs.   Suppose  we have solved a singular first-order differential equation (\ref{alpha}), with a chosen dominant control parameter in this principal regime.  This results in an integration constant $V_0/V'_0$, depending on whether or not rescaling in the Hamiltonian is performed in this principal regime.  To proceed, we need to distinguish different situations, depending on whether or not rescaling in the Hamiltonian is needed in a principal regime and/or on its characteristic line originating from a multi-critical point.
Here, we restrict ourselves to consider three different situations: 
(i) no rescaling in the Hamiltonian is needed in a principal regime and on its characteristic line; 
(ii) rescaling in the Hamiltonian is needed both in a principal regime and on its characteristic line;
(iii) rescaling in the Hamiltonian is needed on its characteristic line, but not in a principal regime.

(i) In order to ensure that the continuity requirement for fidelity temperature is satisfied, a shift in fidelity temperature  $T(x) \rightarrow T(x)-T_0$ is performed, accompanied by a shift in fidelity internal energy $U (x)\rightarrow U(x)- T_0S(x)$.  Generically, $T_0 \equiv T_m-T_t$,  where $T_m$ represents fidelity temperature at a  chosen point on a characteristic line originating from a critical point on a multi-critical point, evaluated from a dominant control parameter $x$ in this principal regime, whereas $T_t$ represents fidelity temperature at the same point, which is determined from a dominant control parameter along this characteristic line itself.  Therefore, we refer to the shifted fidelity temperature $T(x)-T_0$ as fidelity temperature $T_f(x)$ and the shifted fidelity internal energy  $U(x)- T_0 \; S(x)$ as fidelity internal energy $U_f(x)$, respectively, with fidelity entropy $S(x)$ left intact. That is, $S_f(x) \equiv S(x)$, $T_f (x)\equiv T(x)-T_0$, and  $U_f(x)\equiv U(x)- T_0 \; S(x)$.

(ii) In order to ensure that the continuity requirement for fidelity temperature is satisfied, a shift in fidelity temperature $T_\phi'(x')\rightarrow T'_\phi(x')-T_0$ is performed, accompanied by a shift in fidelity internal energy $U_\phi'(x')\rightarrow U'_{\phi}(x')- T_0 \; S'_{\phi}(x')$, with fidelity entropy $S_\phi'(x')$ left intact. Generically, $T_0 \equiv T_m-T_t$, where $T_m$ represents fidelity temperature at a  chosen point on a characteristic line originating from a critical point on a multi-critical point, evaluated from a dominant control parameter $x'$ in this principal regime, whereas $T_t$ represents fidelity temperature at the same point, which is determined from a dominant control parameter along this characteristic line itself. That is, $S'_{\phi f}(x') \equiv S'_{\phi}(x') $, $T'_{\phi f}(x') \equiv T'_\phi(x')-T_0$, and  $U'_{\phi f}(x') \equiv U'_{\phi}(x')- T_0 \; S'_{\phi}(x')$.
Therefore, combined with scaling entropy $S'_{\sigma}(x')$, fidelity entropy $S_f(x)$ takes the form: $S_f(x) = S_{\phi f}(x)+S_{\sigma f}(x)$, with $S_{\phi f}(x)\equiv S'_{\phi f}(x')$ and $S_{\sigma f}(x) \equiv S'_{\sigma f}(x')$.
Fidelity temperature $T_f(x)$ and fidelity internal energy $U_f(x)$ take the form:  $T_f(x) = T_{\phi f}(x)$ and $U_f(x) =
U_{\phi f}(x)$, with $T_{\phi f}(x) \equiv T'_{\phi f}(x')$ and $U_{\phi f}(x) \equiv U'_{\phi f}(x')$.

(iii) If rescaling in the ground state energy at a characteristic line, which appears to be a boundary in this principal regime,  is performed,  then  a contribution from scaling entropy to residual fidelity entropy at a line of critical points exists.  Therefore, we need to introduce residual scaling entropy $S_{\sigma 0}$, which is necessary to satisfy the continuity requirement for $S_f(x)$.
That is, we have $S_f(x)=S(x)+S_{\sigma 0}$.
In order to ensure that the continuity requirement for fidelity temperature is satisfied, a shift in fidelity temperature  $T(x) \rightarrow T(x)-T_0$ is performed, accompanied by a shift in fidelity internal energy $U (x)\rightarrow U(x)- T_0S(x)$.  Here, $T_0 \equiv T_m-T_t$,  where $T_m$ represents fidelity temperature at a chosen point on a characteristic line originating from a multi-critical point, evaluated from a dominant control parameter $x$ in this principal regime, whereas $T_t$ represents fidelity temperature at the same point, which is determined from a dominant control parameter along this characteristic line itself.  Therefore, we refer to the shifted fidelity temperature $T(x)-T_0$ as fidelity temperature $T_f(x)$ and the shifted fidelity internal energy $U(x)- T_0 \; S(x)$ as fidelity internal energy $U_f(x)$, respectively, with fidelity entropy $S(x)$ left intact.  Therefore, in this principal regime, we have $T_f (x)\equiv T(x)-T_0$, and  $U_f(x)\equiv U(x)- T_0 \; S(x)$.

Once this is done, an additive constant $U_0/U'_{\phi 0}$ in fidelity internal energy and an integration constant $V_0/V'_0$ are determined from the continuity requirements for fidelity internal energy $U(x)/U'_\phi(x')$ on a characteristic line originating from a multi-critical point, with an extra condition that fidelity internal energy $U(x)/U'_\phi(x')$ is zero at a critical point.  This extra condition will be justified in  the next subsection.

This also applies to discontinuous QPTs, with modifications that an additive constant $U_0/U'_{\phi 0}$ in fidelity internal energy and an integration constant $V_0/V'_0$ are determined from the continuity requirements for fidelity internal energy on a characteristic line originating from a multi-critical point and at a first-order transition point $x_d$, and that
an extra parameter $\kappa$ is determined from the continuity requirement for fidelity temperature at a first-order transition point $x_d$.  Note that $x_d$ itself must be on a line of first-order transition points, which ends at a critical point. Therefore,
fidelity temperature at a first-order transition point is evaluated from a critical point, with a dominant control parameter along a characteristic line, which is identified as a line of first-order transition points, due to symmetry and/or duality.

Once fidelity mechanical state functions in all the principal regimes are determined,  we are able to determine fidelity mechanical state functions in all the other regimes as a result of dualities (cf. Section~\ref{fmsute} for a more thorough discussion about dualities). 
Specific examples to illustrate how to implement our prescription for quantum many-body systems may be found in Appendix~{\ref{xychain}} for the quantum XY chain, in Appendix~{\ref{ishchain}} for the transverse field quantum Ising chain in a longitudinal field, and in Appendix~{\ref{xyzchain}} for the quantum XYZ chain, respectively.

\subsection{Generic remarks}~\label{fmsutg}

We are able to draw some consequences from our argument above, combining with discussions about duality in Section~\ref{duality}. First, residual fidelity entropy $S_0$ depends on choices of a dominant control parameter in a given regime, thus it does not reflect information encoded in ground state wave functions at critical points. That is, it is {\it extrinsic}, in the sense that it is impossible to determine it from the Hamiltonian itself at a critical point. Second, there are a lower bound and an upper bound for fidelity internal energy $U(x)$ if no rescaling in the Hamiltonian is needed~\cite{lowerbound}. This is due to the fact that the range of a dominant control parameter $x$ is finite.  The same statement is also valid for fidelity internal energy $U'_\phi(x')$ if rescaling in the Hamiltonian is performed.  As a convention, we always choose the lower bound to be zero. Third, at a critical point, fidelity internal energy $U(x)$  must be zero, if no rescaling in the Hamiltonian is needed, thus leading to the requirement that $U_0 = T_0 \; S_0$.  Physically, this is a consequence of the fact that, at a critical point, it is impossible to extract any relevant information from discarding irrelevant information, since any relevant information is covered up by irrelevant information.  Mathematically, at a critical point, fidelity internal energy becomes $U_0 -T_0 \; S_0$, which has to satisfy $U_0 -T_0 \; S_0 \geq 0$, due to the convention that the lower bound is zero. However, if it takes a positive value, then it is impossible to guarantee that fidelity internal energy $U(x)$ is monotonically increasing with increasing $x$. That is, the internal logical consistency demands that fidelity internal energy $U(x)$  must be zero at a critical point. The same reasoning applies to fidelity internal energy $U'_\phi(x')$,
if rescaling in the Hamiltonian is performed.  Fourth, for a given quantum many-body system, fidelity internal energy $U(x)$, if no rescaling in the Hamiltonian is needed, or $U'_\phi(x')$, if rescaling in the Hamiltonian is performed, takes the same value at all stable fixed points.  This reflects the fact that the Hamiltonians are unitarily equivalent at all stable fixed points.  Fifth, fidelity temperature is zero at factorizing fields, given that no quantum fluctuations are present in factorized states.

In addition,  characteristic lines intersect with each other at a characteristic point in the parameter space.  Such a characteristic point is identified as a stable fixed point,  as follows from our discussion about fidelity flows in Section~\ref{answer}: a stable fixed point is characterized by zero fidelity temperature and (local) maximal fidelity entropy.
Instead, a critical point is identified as an unstable fixed point, which is characterized by divergent fidelity temperature, as a result from the fact that  $\alpha (x)$, defined in Eq. (\ref{alphapm}), diverges at a critical point. Here, we remark that a characteristic point at infinity will be labeled in terms of the two original coupling parameters, with one of them being infinite in value.  As it turns out, it is necessary to keep the other finite coupling parameter in labeling a characteristic point at infinity, given that the symmetry group varies with the two coupling parameters.

Note that different choices are allowed for  a dominant control parameter in a given regime.  However, different choices lead to different fidelity mechanical state functions.  Therefore, one may raise a concern whether or not it is possible to extract any sensible physics from our formalism. This concern has been addressed in Appendix~\ref{fictitious}. As argued, any  two different sets of fidelity entropy, fidelity temperature and fidelity internal energy, resulted from two different choices, are related to each other via introducing a fictitious parameter $\sigma$. Actually, information encoded in $\sigma$ arising from different choices of a dominant control parameter for a given regime is {\it irrelevant}, in the sense that both stable and unstable fixed points remain the same. Physically, this is due to the fact that the constraints imposed by symmetries, dualities and factorizing fields are {\it rigid}, meaning that  there is no flexibility in choosing a dominant control parameter on such a characteristic line, although it is still allowed to choose different parametrization variables on a characteristic line, subject to the condition  that, for any two parametrization variables, one must be a monotonically increasing function of the other, and vice versa. Although this does change fidelity mechanical state functions, but does not change where fidelity temperature diverges or becomes zero, and does not change where fidelity entropy takes a (local) maximum.

In practice, we may take advantage of this freedom to properly choose a dominant control parameter $x$  such that numerical computation is more efficient, when we exploit a tensor network algorithm~\cite{DMRG,TN1,TN2} to simulate quantum many-body systems.

We emphasize that, once fidelity mechanical state functions are determined in all the regimes, we have to transform back to the original coupling parameters $x_1$ and $x_2$, according to one-to-one correspondence between ($x_1, x_2$) and ($x, \tau$). As a convention, we use a subscript $f$ to indicate fidelity mechanical state functions, with the original coupling parameters as their arguments, for a specific quantum many-body system, unless otherwise stated.

\section{Duality and a canonical form of the Hamiltonian}~\label{duality}

For a quantum many-body system described by a given  Hamiltonian,  in principle we are able to determine both the ground state energy density and ground state wave functions analytically and/or numerically. From the latter, ground state fidelity per site may be computed straightforwardly.  In addition,  the ground state phase diagram may be read off from singular behavior in ground state fidelity per site.  In practice, we may determine fidelity mechanical quantities simply following our formalism.

However, given the ground state energy density is involved in our formalism, an important question remains to be addressed: what form of the Hamiltonian should be chosen, given the Hamiltonian is mathematically determined up to a multiplying factor and an additive constant?
As is well known, for a given Hamiltonian $H(x)$, the physics itself does not change under the operations of rescaling and shift: $H^*(x) = g H(x)+b$, with $g>0$ and $b$ being real numbers.   Rescaling or shift in the Hamiltonian  $H(x)$ generates rescaling or shift in the ground state energy density.  Therefore, our question can be reshaped as, what is a canonical form of the Hamiltonian $H$ among equivalent Hamiltonians, related via $H^* (x) = g H(x)+b$, in fidelity mechanics?  The answer rests on a well-known notion: duality.

Duality is nothing but a unitary mapping between quantum Hamiltonians that preserves the quasi-local character of their interaction terms (see, e.g., Ref.~\cite{ortiz}). Mathematically, this corresponds to $H(x) = k'(x') U H' (x')U^{\dagger}$, where $H'(x')$ is the Hamiltonian unitarily equivalent to $H(x)$, $U$ is a unitary operator, and $k'(x')>0$ is monotonic as a function of $x'$, with $x'$ in turn being monotonic as a function of $x$.   Of special interests are dualities, which are unitary mappings conserving the form of the Hamiltonian operator $H(x)$. That is,  $H(x) = k'(x') U H(x')U^{\dagger}$.  In this work, we only refer to this type of unitary mappings as dualities.  In other words, duality is one of the possible realizations for rescaling in the Hamiltonian, with the only difference that duality always involves a nontrivial unitary transformation.
A prototypical example is the transverse field quantum Ising chain.  Physically, duality allows us to relate the weak-coupling regime to the strong-coupling regime.

It is important to note that duality leaves no room for the Hamiltonian $H^* (x)$ but  a constant multiplying factor. That is, for a fixed $g$, there is only one value of $b$ such that duality exists in the corresponding Hamiltonian.  Therefore, one may choose a specific form of the Hamiltonian $H$ among equivalent Hamiltonians, related via $H^*(x) = g H(x)+b$, up to a constant multiplying factor.  This form is a {\it canonical form} of the Hamiltonian in fidelity mechanics, meaning that it only makes sense to adopt the ground state energy density $e(x)$ from a canonical form of the Hamiltonian in  Eqs.~(\ref{alphapm}) and  (\ref{alphad0}) to determine fidelity mechanical state functions.  We remark that, generically,  duality is lacking in a given Hamiltonian. However, we are still able to define a canonical form in such a case; there are three ways to do so.  First, for a Hamiltonian depending on more than one coupling parameters, we may find a special case that hosts duality, thus enabling us to determine a specific value of $b$, as happens for the quantum XY chain. For this model, the transverse field quantum Ising chain as a special case does host duality. Second, for a Hamiltonian without any coupling parameter (except for a global multiplying constant as an energy scale), we may introduce more coupling constants by embedding a given Hamiltonian into  a more general Hamiltonian with more than one coupling parameters, and try to see if there is any special case to host duality. This happens to the quantum XXX model and the XXZ model, which may be extended to the quantum XYZ model. The latter hosts duality, as discussed in Appendix~\ref{dual}. Third, an established canonical form of a given Hamiltonian may be exploited to justify a canonical form of a related Hamiltonian, which reduces to the given Hamiltonian in some limit. This happens to the t-J model, as it reduces to the quantum XXX model at half filling.

We stress that the presence of a constant multiplying factor does not change fidelity entropy $S(x)$, fidelity temperature $T(x)$ and fidelity internal energy $U(x)$, as long as it is kept to be a constant.
In contrast, extra attention needs to be paid to the shift operation $H^*(x) = H(x)+b$. Suppose $H(x)$ is in a canonical form, then, generically, we have $e(x) <0$.  Thus, fidelity entropy $S(x)$, fidelity temperature $T(x)$ and fidelity internal energy $U(x)$ follows accordingly, with $V(x)$ determined from the singular first-order differential equation~(\ref{alpha}) for continuous QPTs and the regular first-order differential equation (\ref{alphaf}) for first-order QPTs. Under the shift, the ground state energy density $e(x)$ becomes $e^*(x)=e(x)+b$, but ground state wave functions remain the same. Therefore,  fidelity entropy $S(x)$ and fidelity temperature $T(x)$ are left intact, since the former only depends on ground state wave functions and the latter quantifies quantum fluctuations present in ground state wave functions. However, fidelity internal energy $U(x)$ depends explicitly on $b$. That is, one may attribute any change in fidelity internal energy to fidelity work needed to be done to perform the shift.
Three possible cases need to be distinguished, depending on $b$:
(i) for a chosen $b$, $e(x)+b <0$ for all possible values of $x$; (ii) for a chosen $b$, there is some $x_0$ such that $e(x_0)+b = 0$;
(iii) for a chosen $b$, $e(x)+b >0$ for all possible values of $x$.  In all cases, fidelity internal energy $U(x)$  becomes $U'(x)=\mp \ln{|e(x)+b|/|e(x_c)+b|}V+U'_0$, with $-/+$ corresponding to monotonically increasing/decreasing $e(x)$ with $x$, respectively, as this is the only choice to keep consistency with the requirement from an analogue of Landauer's principle at zero temperature, in addition to the fact that $U'(x) \rightarrow U(x)$, when $b \rightarrow 0$.  Thus, fidelity work, needed to be done to  shift from $0$ to $b$, is $W (x)=\mp (\ln{|e(x)+b|/|e(x)|}-  \ln{|e(x_c)+b|/|e(x_c)|} V + U_0'-U_0$, since both fidelity temperature and fidelity entropy remain the same during the shift operation.  There are three regimes to be identified with three different choices of $b$. In case (i), $U'(x)-U'_0$ is monotonically increasing with increasing $x$. In case (ii), $U'(x)-U'_0$ is either monotonically increasing with $x$ from $x_c$ to $x_0$, then decreasing with $x$ beyond $x_0$, or monotonically decreasing with increasing $x$ from $x_c$ to $x_0$, then increasing with increasing $x$ beyond $x_0$,  corresponding to monotonically increasing/decreasing $e(x)$ with $x$, respectively. In case (iii), $U'(x)-U'_0$ is monotonically decreasing with increasing $x$. In particular,  in order to locate the exact value of $x_0$ such that $e(x_0)+b=0$,  the amount of fidelity work to be done is divergent: $W(x) \rightarrow \infty$. This quantifies an observation that an infinite amount of work is necessary to achieve unlimited accuracy, when one is trying to find a solution $x_0$ to an algebraic equation $e(x_0)+b = 0$.

In addition,  duality, by definition, results in extra rescaling in the ground state energy density.  Hence, for a dual Hamiltonian, rescaling only introduces an extra contribution to fidelity entropy arising from rescaling, with fidelity temperature and fidelity internal energy left intact, as discussed in Section~\ref{fmsut}.

Our discussion up to now leaves an impression that a canonical form of a given Hamiltonian seems to occupy a unique position in fidelity mechanics. However, this is not true, since
the definition of a canonical form of the Hamiltonian depends on the definition of duality, which in turn depends on a shift operation in the Hamiltonian.  As argued in Appendix~\ref{shift}, information encoded in a shift parameter $b$ in different canonical forms of the Hamiltonian arising from different definitions of duality is {\it irrelevant}, in the sense that
both stable and unstable fixed points remain the same for different canonical forms of the Hamiltonian resulted from different definitions of duality.

Now we are ready to justify our assumption about the ground state energy density $e(x)$ that it is negative for all $x$, made in Section~\ref{fmsut}. For a given Hamiltonian $H(x)$, if  $e(x)$ is not always negative, then it should be shifted  to ensure that it is negative. Then fidelity entropy, fidelity temperature and fidelity internal energy may be determined, following our formalism in Section~\ref{fmsut}. As argued in Appendix~\ref{shift}, we are able to assign the shifted Hamiltonian to be a canonical form of the Hamiltonian according to a specific definition of duality. In order to shift it back to the original Hamiltonian, we resort to our discussions above in case (ii) or case (iii):
if $e(x)$ changes its sign, then it is the case (ii); if $e(x)$ is positive for all $x$, then it is the case (iii). This allows us to determine fidelity internal energy, with fidelity entropy and fidelity temperature left intact. As such, our assumption that $e(x)$ is negative for all $x$, made in Section~\ref{fmsut}, does not prevent us from investigating any quantum many-body system in the context of fidelity mechanics.

In passing, we remark that duality is ubiquitous for quantum many-body systems, as shown in Appendices~\ref{dual} and \ref{dualityinkitaev}, respectively, for the quantum XYZ models on both bipartite and non-bipartite
lattices and for the quantum Kitaev model on a honeycomb lattice, with arbitrary spin $s$.

\section{Quantum XY chain - a typical example for continuous quantum phase transitions}~\label{qxymodel}

\begin{figure}
  \centering
\includegraphics[totalheight=4.5cm]{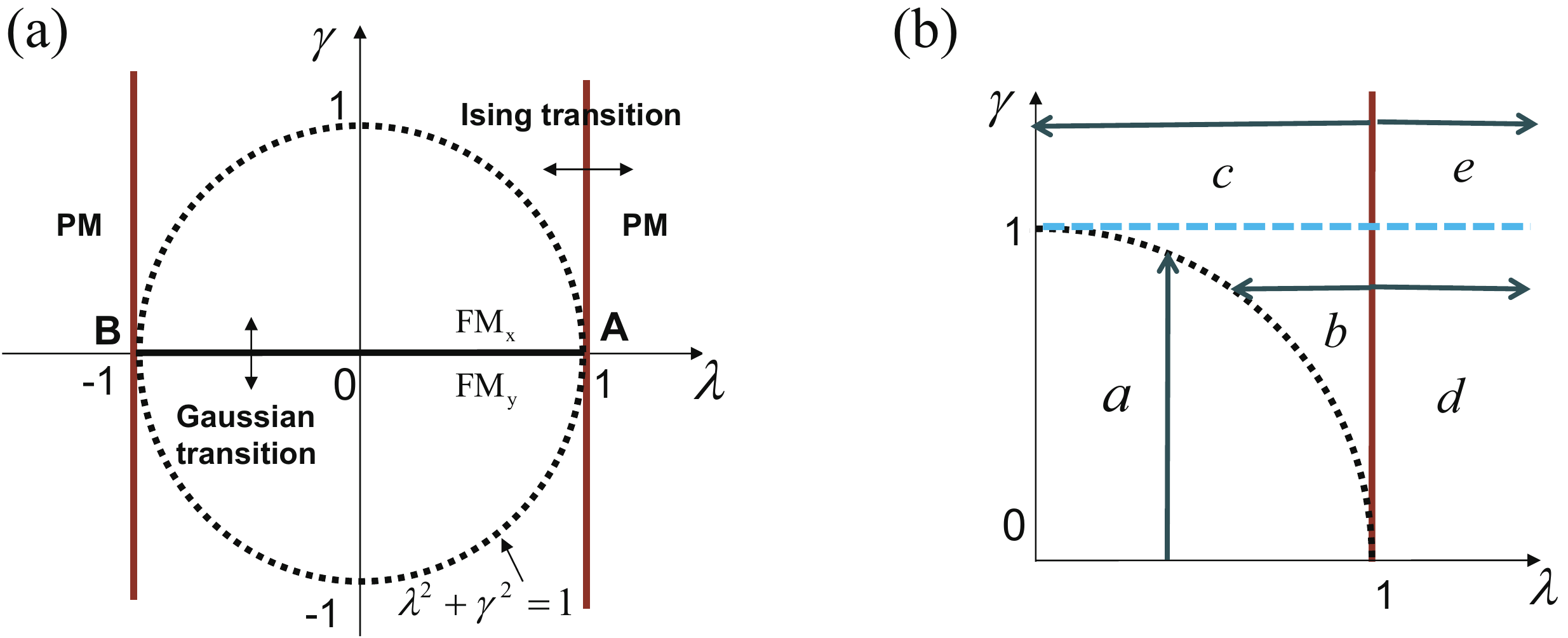}
\caption{ (a) Ground state phase diagram for the quantum XY chain. There is a marked difference between the regimes inside and outside the disorder circle.  Indeed, as claimed~\cite{Illuminati},  long-range entanglement driven order exists inside the disorder circle.  However, we stress that the same order must also exist on the line $\lambda =0$, $\gamma \geq 1$, due to the presence of duality between  $\lambda =0$, $\gamma \geq 1$ and $\lambda =0$, $\gamma \leq 1$.
(b)  Choices of a dominant control parameter in five different principal regimes: (1) for regime $a$,
a dominant control parameter is chosen to be $\gamma$,  starting from $\gamma=\gamma_c = 0$ up to the disorder circle, with $\lambda$ fixed; (2)  for regime $b$ or $c$, a dominant control parameter is chosen to be  $1-\lambda$, starting from $\lambda=\lambda_c =1$ up to the disorder circle or $\lambda=0$, with $\gamma$ fixed; (3) for regime $d$ or $e$, a dominant control parameter is chosen to be  $1-1/\lambda$, starting from $\lambda=\lambda_c =1$ up to $\lambda=\infty$, with $\gamma$ fixed. This choice is to keep consistency with duality for the transverse field quantum Ising chain, which corresponds to the quantum XY chain with $\gamma =1$.}
\label{XYphase}
\end{figure}

The Hamiltonian for the quantum XY chain takes the form
\begin{equation}
H(\lambda, \gamma)=-\sum_{i}{(\frac{1+\gamma}{2}\sigma_i^x\sigma_{i+1}^x+\frac{1-\gamma}{2}\sigma_i^y\sigma_{i+1}^y+\lambda \sigma_{i}^{z})}, \label{xyham}
\end{equation}
where $\sigma_i^{\beta}$ are the Pauli matrices at site $i$, with $\beta=x,y,z$, $\gamma$ is the anisotropic parameter, and $\lambda$ is the transverse field.
The model is exactly solvable~\cite{Lieb, pfeuty, Liebmattis}, with its ground state phase diagram shown in Fig. ~\ref{XYphase}. For $\gamma \neq 0$, the Hamiltonian (\ref{xyham}) possesses a $Z_2$ symmetry group, defined by $\sigma^x_i\leftrightarrow - \sigma^x_i$,
$\sigma^y_i\leftrightarrow - \sigma^y_i$ and $\sigma^z_i \leftrightarrow  \sigma^z_i$. For $\gamma = 0$, the symmetry group becomes $U(1)$.  In addition, a $U(1)$ symmetry occurs at two isolated points: $\lambda=0$ and $\gamma = \pm 1$, as well as at a multi-critical point when  $\gamma$ is infinite in value and at a characteristic point when $\lambda$  is infinite in value.

The ground state phase diagram may be read off from the singularities exhibited in ground state fidelity per site (cf. Appendix~\ref{pinch} for the transverse field quantum Ising chain as a special case).  The system undergoes QPTs when the two lines of critical points: $\lambda=1$, with $\gamma \neq 0$, and $\gamma=0$, with $-1<\lambda<1$,  are crossed in the thermodynamic limit.
For  a fixed $\gamma$,  the model is driven to cross a critical point at $\lambda=1$ from an ordered ferromagnetic (FM) phase to a disordered paramagnetic (PM) phase, which is a QPT, belonging to the Ising universality, characterized in terms of  central charge $c=1/2$ in conformal field theory.  Specifically, for $\gamma>0$ ($\gamma<0$),
when  $ \lambda \in (-1,1)$,  the system is in the ferromagnetic order along the $x~(y)$ direction, labeled as ${\rm FM}_x/{\rm FM}_y$ in Fig. ~\ref{XYphase}. For a fixed $\lambda \in (-1,1)$,  the system is driven through a Gaussian critical point at $\gamma=0$, with central charge $c=1$.  For $\gamma=0$, a PT transition from a critical phase to a fully polarized phase occurs at $\lambda=\pm1$, protected by the symmetry group $U(1)$.  Note that, at two multi-critical points $(\pm 1,0)$,  denoted as A and B in Fig.~\ref{XYphase}, dynamic critical exponent $z$ is $z = 2$. Hence, the underlying field theories are not conformally invariant.

 An interesting feature of the model is the  disorder circle: $\lambda^2+\gamma^2=1$, characterized by the fact that ground states on the circle are factorized states~\cite{factorizing,factorizing1}.    As demonstrated in Ref.~\cite{Wolf}, the model on the disorder circle is unitarily equivalent to a spin- $1/2$ model with three-body interactions, with ground states being restricted to matrix product states, and the bond dimension being equal to two.  Therefore,
 we have to treat transition points at $(\pm 1, 0)$ as an exotic QPT, given that the ground state energy density  is a constant on the disorder circle, even at the transition points $(\pm 1, 0)$. In addition, there is a marked difference between the regimes inside and outside the disorder circle.  Indeed, as claimed~\cite{Illuminati}, long-range entanglement driven order exists in the disordered regime.  However, we stress that the same order must also exist on the line $\lambda =0$ with $\gamma \geq 1$, due to the presence of duality between $\lambda =0$, $\gamma \geq 1$ and $\lambda =0$, $0<\gamma \leq 1$.  In fact, there are two dualities along the lines $\gamma=1$ and $\lambda=0$, as discussed in Appendix~\ref{xychain}.  We remark that a factorized state also occurs when $\lambda$ is infinite in value and that the presence of duality along $\lambda=0$ implies a multi-critical point $(0, \infty)$ at infinity. Accordingly, the Hamiltonian (\ref{xyham}) is, by definition, in a canonical form.

We may restrict ourselves to the Hamiltonian (\ref{xyham}) with $\lambda \geq 0$ and $\gamma \geq 0$,  since the Hamiltonian (\ref{xyham})  is symmetric with respect to $\gamma \leftrightarrow - \gamma$  and $\lambda \leftrightarrow - \lambda$. Meanwhile,  the consideration of the dualities and factorizing fields allows us to separate the whole region with $\gamma>0$ and $\lambda>0$ into five different principal regimes: (a)  the regime inside the disorder circle, with $0<\lambda<1$ and $0<\gamma<\sqrt{1-\lambda^2}$; (b) the regime outside the disorder circle, with $0<\lambda<1$ and $\sqrt{1-\lambda^2}<\gamma<1$; (c) the regime with $0<\lambda<1$ and $\gamma>1$; (d) the regime with $\lambda>1$ and $0<\gamma<1$; (e) the regime with $\lambda>1$ and $\gamma>1$.
In each regime, we may choose a dominant control parameter, as long as such a choice is consistent with the constraints imposed by the symmetry groups, dualities and factorizing fields, meaning that any choice has to respect all the boundaries between different regimes. Here, our choice is: (1) for regime $a$,
a dominant control parameter is chosen to be $\gamma$, starting from $\gamma=\gamma_c = 0$ up to the disorder circle, with $\lambda$ fixed; (2)  for regime $b$ or $c$, a dominant control parameter is chosen to be  $1-\lambda$, starting from $\lambda=\lambda_c = 1$ up to the disorder circle or $\lambda=0$, with $\gamma$ fixed; (3) for regime $d$ or $e$, a dominant control parameter is chosen to be $1-1/\lambda$, starting from $\lambda=\lambda_c = 1$ up to $\lambda=\infty$, with $\gamma$ fixed. This choice is to keep consistency with duality for the transverse field quantum Ising chain, which corresponds to the quantum XY chain with $\gamma =1$.

It is numerically confirmed that fidelity entropy $S_{\phi}(\lambda, \gamma)$  scales as $\gamma ^{ \nu +1}$ near a critical line: $\lambda \in (-1, 1)$ for $\gamma =0$,  and scales as $|1-\lambda|^{\nu +1}$ near a critical line: $\lambda=1$ for $\gamma \neq 0$, respectively. Here, $\nu$ is the critical exponent for correlation length. In both cases, $\nu =1$  (cf. Appendix~\ref{scaling}).

\begin{figure}
\centering
     \includegraphics[angle=0,totalheight=4.5cm]{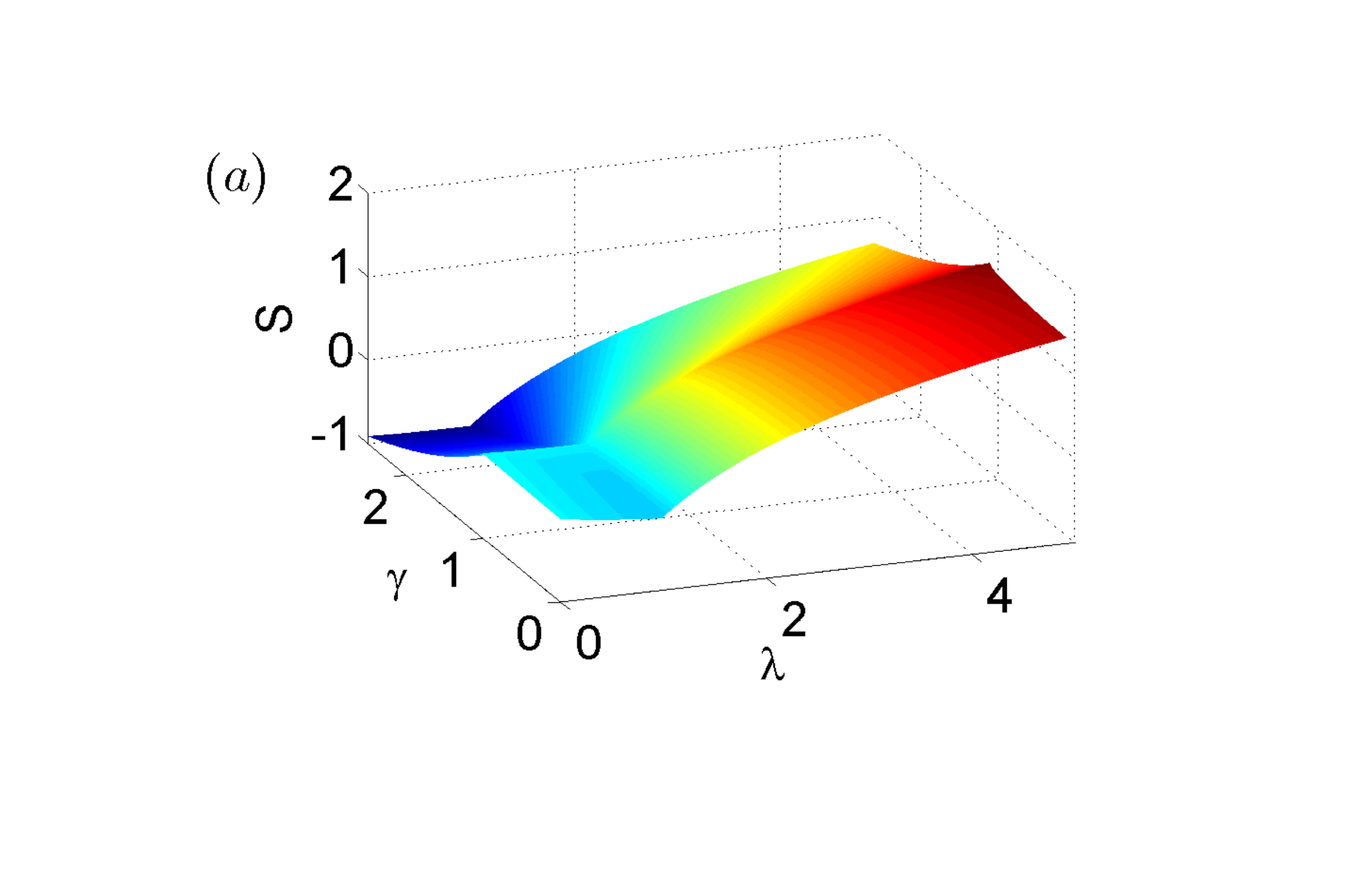}
     \includegraphics[angle=0,totalheight=4.5cm]{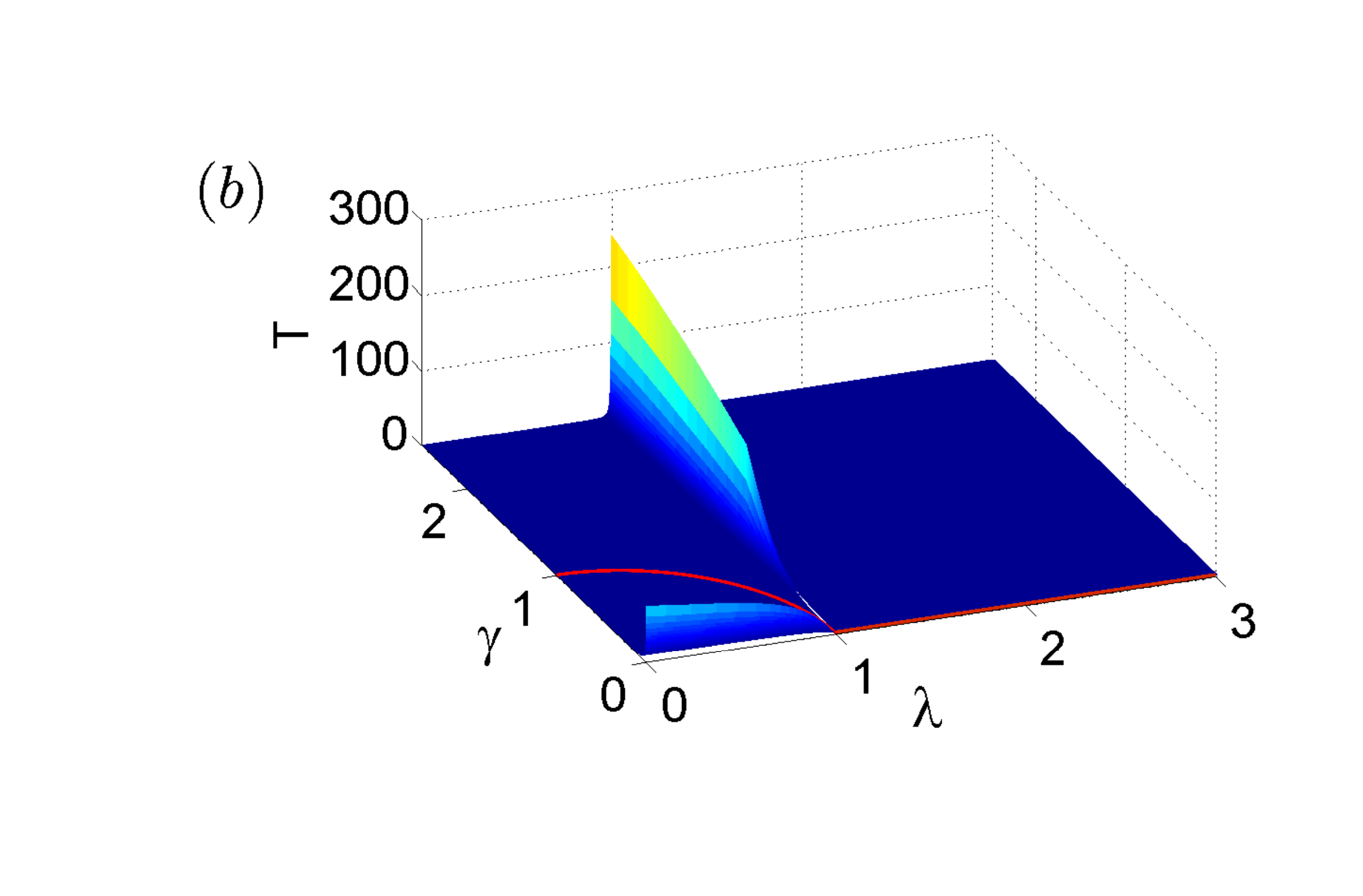}
     \includegraphics[angle=0,totalheight=4cm]{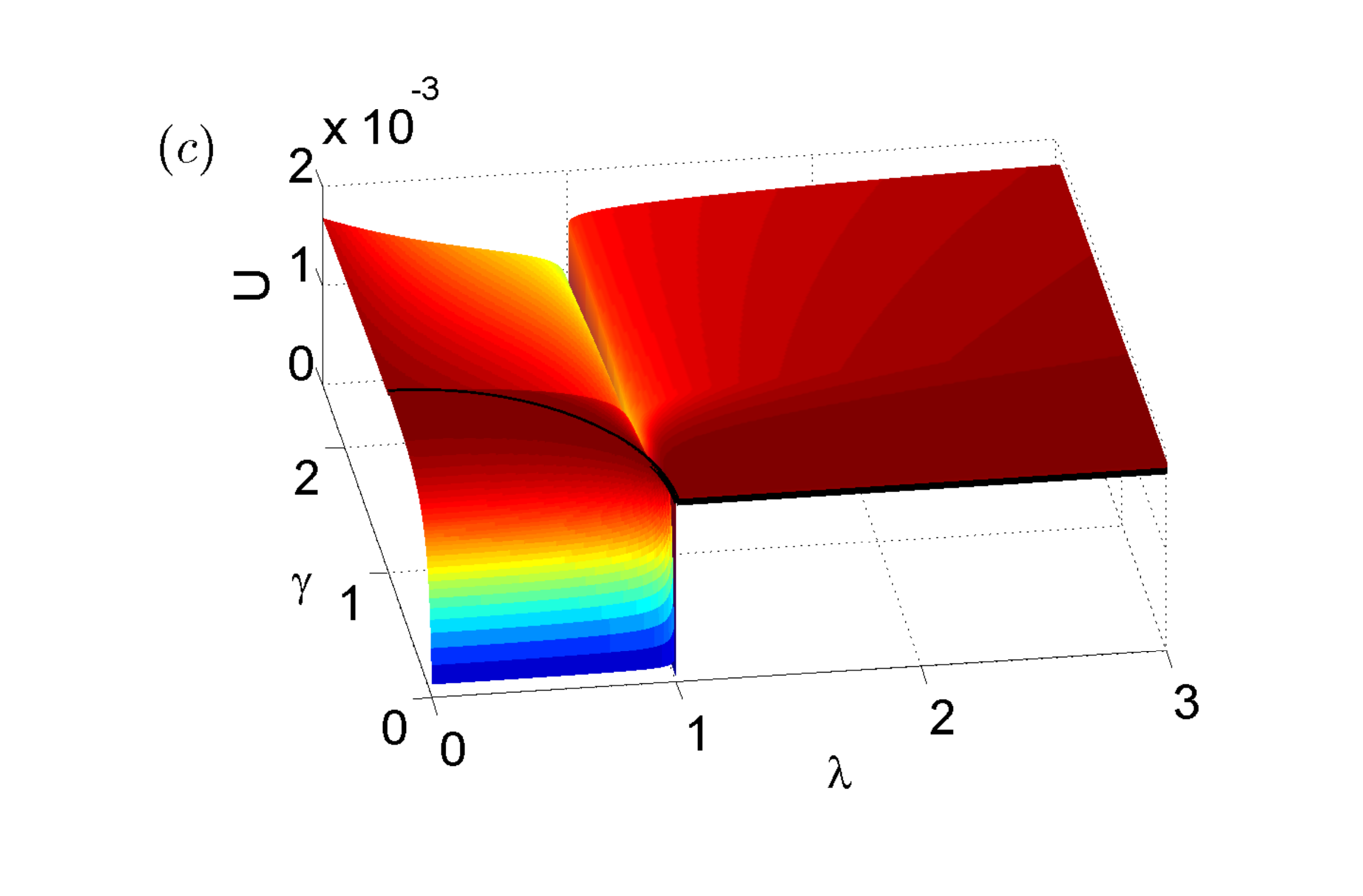}
      \caption{ Fidelity entropy $S_f(\lambda, \gamma)$,  fidelity temperature $T_f(\lambda, \gamma)$  and fidelity internal energy $U_f(\lambda, \gamma)$ for the quantum XY chain.  Here, we restrict ourselves to $\lambda \geq0$ and $\gamma \geq 0$, due to the symmetry of  the Hamiltonian (\ref{xyham}) with respect to $\gamma \leftrightarrow - \gamma$  and $\lambda \leftrightarrow - \lambda$.
      (a) Fidelity entropy $S_f(\lambda, \gamma)$ exhibit singularities at two dual lines: $\gamma= 1$ and $\lambda=0$ and the disorder circle: $\lambda^2+\gamma^2=1$, in addition to two critical lines at $\gamma =0, \lambda \in (-1,1)$ and $\gamma \neq 0, \lambda =1$.  One might view such singularities as ``phase transitions" in fidelity mechanics.  Note also that fidelity entropy $S_f(\lambda, \gamma)$ reaches its local maximum at $(0, 1)$.
       (b) Fidelity temperature $T_f(\lambda, \gamma)$ diverges at two lines of critical points $\gamma =0, \lambda \in (-1,1)$ and $\gamma \neq 0, \lambda =1$, but  is zero on the disorder circle: $\lambda^2+\gamma^2=1$, as well as at a characteristic point, representing a factorizing field, when $\lambda$ is infinite in value.
       (c) Fidelity internal energy $U_f(\lambda, \gamma)$ takes the same value at all stable fixed points and on the disorder circle: $\lambda^2+\gamma^2=1$,  including a characteristic point at infinity, representing a factorizing field, when $\lambda$ is infinite in value.
      }\label{xyentropy}
 \end{figure}

Once a dominant control parameter is chosen, fidelity entropy $S_f(\lambda, \gamma)$ may be determined straightforwardly. Accordingly, fidelity temperature $T_f (\lambda, \gamma)$ and fidelity internal energy $U_f(\lambda, \gamma)$ are determined by solving a singular first-order differential equation for $V(\lambda, \gamma)$.  In five different principal regimes as well as on the boundaries between different regimes, the explicit expressions for fidelity entropy $S_f(\lambda, \gamma)$, fidelity temperature $T_f(\lambda, \gamma)$ and fidelity internal energy $U_f(\lambda, \gamma) $ may be found in Appendix~\ref{xychain}.  As a result,
we plot fidelity entropy $S_f(\lambda, \gamma)$, fidelity temperature $T_f(\lambda, \gamma)$  and fidelity internal energy $U_f(\lambda, \gamma)$ as a function of $\lambda$ and $\gamma$ in Fig.~\ref{xyentropy} (a), (b) and (c), respectively.
A contribution to fidelity entropy from rescaling due to dualities has been taken into account.

In addition to unstable fixed points, which are identified as critical points,  there are three stable fixed points, identified as characteristic points located at $(0,1), (\infty, 1)$ and $(\infty,0)$.
Note that, at unstable fixed points,  fidelity temperature $T_f(\lambda, \gamma)$  diverges, indicating strong quantum fluctuations, whereas at stable fixed points, fidelity temperature $T_f(\lambda, \gamma)$ is zero, indicating the absence of quantum fluctuations. This also happens on the disorder circle:  $\lambda^2+\gamma^2=1$, with factorized states as ground states. Therefore,  zero fidelity temperature is a feature of factorizing fields in fidelity mechanics.  However, at  a PT transition point (1, 0), fidelity temperature $T_f(\lambda, \gamma)$ is not well-defined. In fact, it takes any value, ranging from 0 to $\infty$, depending on how it is approached, since all fidelity isothermal lines, defined as lines with the same constant values of fidelity temperature,  converge at a PT transition point. This bears a resemblance to a previous result~\cite{korepin} that entanglement entropy is not well-defined at the PT transition point (1, 0); its value depends on how the PT transition point (1, 0) is approached.

We remark that, in addition to QPTs detected through singularities in ground state fidelity per site, fidelity mechanical state functions exhibit singularities at the two dual lines: $\gamma= 1$ and $\lambda=0$, and on the disorder circle: $\lambda^2+\gamma^2=1$. One might view such singularities as ``phase transitions" in fidelity mechanics.  This interpretation resolves a confusing point raised in Ref.~\cite{Illuminati}; as claimed, long-range entanglement driven order exists inside the disorder circle, suggesting a
QPT occurs on the disorder circle.  However, the same long-range entanglement driven order also exists  on the line $\lambda =0$, $\gamma \geq 1$, due to the presence of duality between $\lambda =0$, $\gamma \geq 1$ and $\lambda =0$, $0< \gamma \leq 1$.  This indicates that no QPT occurs on the disorder circle.  Otherwise, QPTs should also occur at the line $\lambda =0$, $\gamma \geq 1$.

We have to bear in mind that there are different choices of a dominant control parameter in each regime, yielding different fidelity mechanical state functions.  However, a connection  exists between different choices, as discussed in Appendix~\ref{xychain}. A crucial point is that both stable and unstable fixed points remain the same, regardless of  choices of a dominant control parameter in a given regime.

\section{Transverse field quantum Ising chain in a longitudinal field - a typical example for discontinuous quantum phase transitions}~\label{tfismodel}

 \begin{figure}
    \centering
     \includegraphics[angle=0,totalheight=4cm]{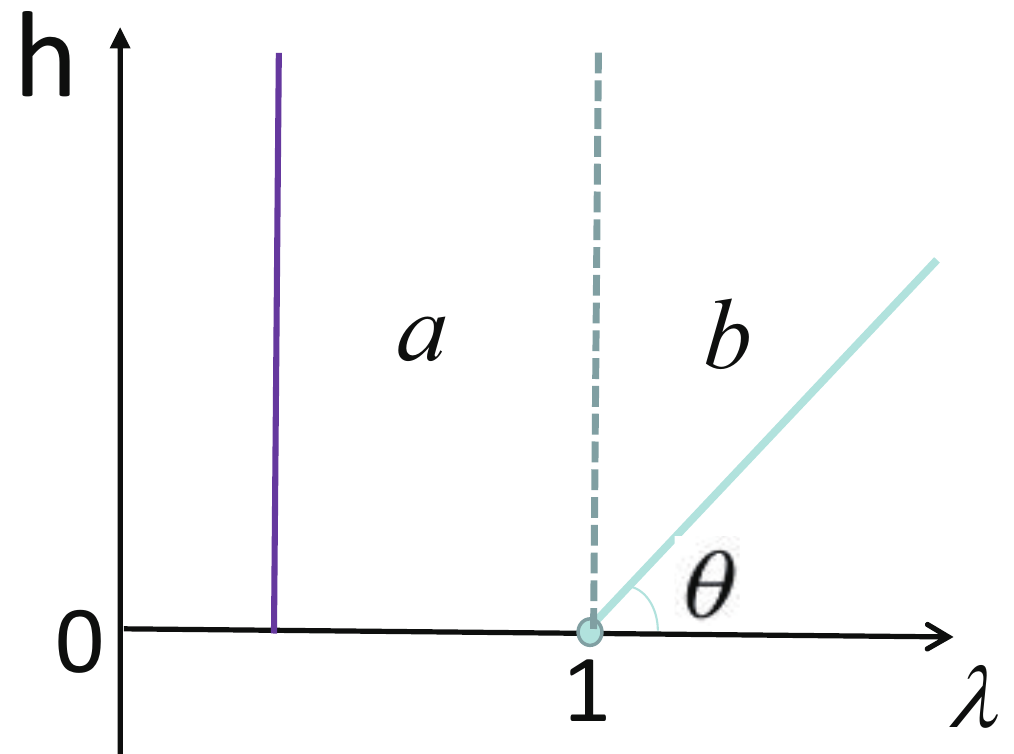}
     \caption{ Choices of a dominant control parameter  for the transverse field quantum Ising chain in a longitudinal field. There are two principal regimes: regime $a$: $0 \leq \lambda<1$ and $h \in (0, \infty)$, and  regime $b$: $\lambda \geq 1$ and $h \in (0, \infty)$.     
 In regime $a$, we rescale the ground state energy density: $e(\lambda, h) =k'(h') e(\lambda, h')$, with
 $ h'=h/(1+h)$, and $k'(h') =1/(1-h')$.  We choose $h'$  as a dominant control parameter, for a fixed $\lambda$. Note that $h$ ranges from $h=h_d=0$ to $h=\infty$, but $h'$ ranges from $h'=h'_d=0$ to $h'=1$.  In regime $b$, we  define a radius $r$ and an azimuthal  angle $\theta$: $r =\sqrt{(\lambda -1)^2 + h^2}$ and $\theta =\arctan h/(\lambda-1)$, and rescale the ground state energy density: $e(r,\theta) = k'(r') \; e'(r', \theta)$, with $r' = r/(1+r)$, and $k'(r')=1/(1-r')$.  We choose $r'$ as a dominant control parameter, for a fixed $\theta$.  Here,  $r$ ranges from $r=r_c=0$ to $r=\infty$, but $r'$ ranges from $r'=r'_c=0$ to $r'=1$.}\label{hlam}
  \end{figure}

As an illustrative example for first-order QPTs, we consider the transverse field quantum Ising chain in a longitudinal field. A canonical form of the
Hamiltonian takes the form
  \begin{equation}
     H(\lambda, h)=-\sum_{i}{(\sigma_i^x\sigma_{i+1}^x+\lambda \sigma_{i}^{z}+h\sigma_i^x)} ,
     \label{tfimham}
  \end{equation}
where $\sigma_i^{\beta}$ are the Pauli matrices at site $i$, with $\beta=x,y,z$,  $\lambda$ is the transverse field, and $h$ is the longitudinal field.  When $h=0$, the model becomes  the transverse field quantum Ising chain, and possesses the $Z_2$ symmetry. It exhibits a second-order QPT at $\lambda_c =1$, characterized by the $Z_2$ symmetry breaking order for $\lambda<1$.   When $h\neq 0$ and $\lambda \neq 0$, no symmetry exists in the Hamiltonian. However,  a $U(1)$ symmetry occurs when $\lambda=0$, as well as when $\lambda$ and/or $h$ are infinite in value, with factorized states as ground states.

The ground state  phase diagram is simple:  there exists a first-order QPT line: $0 \leq \lambda<1$ and $h=0$, which ends at a critical point $(1, 0)$.  The first-order QPTs occur from a phase with spin polarization in $-x$  to a phase with spin polarization in $x$, when $h$ changes its sign.  As already mentioned, duality occurs in  the transverse field quantum Ising chain. Taking the symmetry and duality into account, we may divide the parameter space into two principal regimes, as shown in Fig.~\ref{hlam}:  regime $a$, defined as $0 \leq \lambda<1$, $h \in (0, \infty)$ and regime $b$, defined as $\lambda \geq 1$, $h \in (0, \infty)$.  In regime $a$, we rescale the ground state energy density: $e(\lambda, h) =k'(h') e(\lambda, h')$, with $ h'=h/(1+h)$, and $k'(h') =1/(1-h')$.  We choose $h'$  as a dominant control parameter, for a fixed $\lambda$. Note that  $h$ ranges from $h=h_d=0$ to $h=\infty$, but $h'$ ranges from $h'=h'_d=0$ to $h'=1$.  In regime $b$, we  define a radius $r$ and an azimuthal  angle $\theta$: $r =\sqrt{(\lambda -1)^2 + h^2}$ and $\theta =\arctan h/(\lambda-1)$, and rescale the ground state energy density: $e(r,\theta) = k'(r') \; e'(r', \theta)$, with $r' = r/(1+r)$, and $k'(r')=1/(1-r')$.  We choose $r'$ as a dominant control parameter, for a fixed $\theta$.  Here,  $r$ ranges from $r=r_c=0$ to $r=\infty$, but $r'$ ranges from $r'=r'_c=0$ to $r'=1$.  This choice is consistent with the requirement from duality when $\theta =0$.

It is numerically confirmed that fidelity entropy $S_{\phi}(r,\theta)$ scales as $r^{3/2}$ for $\theta\neq0$ and as $r^2$ for $\theta = 0$ near the critical point,  indicating that  critical exponent $\nu =1/2$ for $\theta\neq0$ and $\nu =1$  for $\theta = 0$, respectively (cf. Appendix~\ref{scaling}).

The explicit mathematical expressions for fidelity entropy $S_f(\lambda, h)$, fidelity temperature $T_f(\lambda, h)$ and fidelity internal energy $U_f(\lambda, h)$ for the transverse field quantum Ising chain in a longitudinal field may be found in Appendix~\ref{ishchain}.

\begin{figure}
     \includegraphics[angle=0,totalheight=5cm]{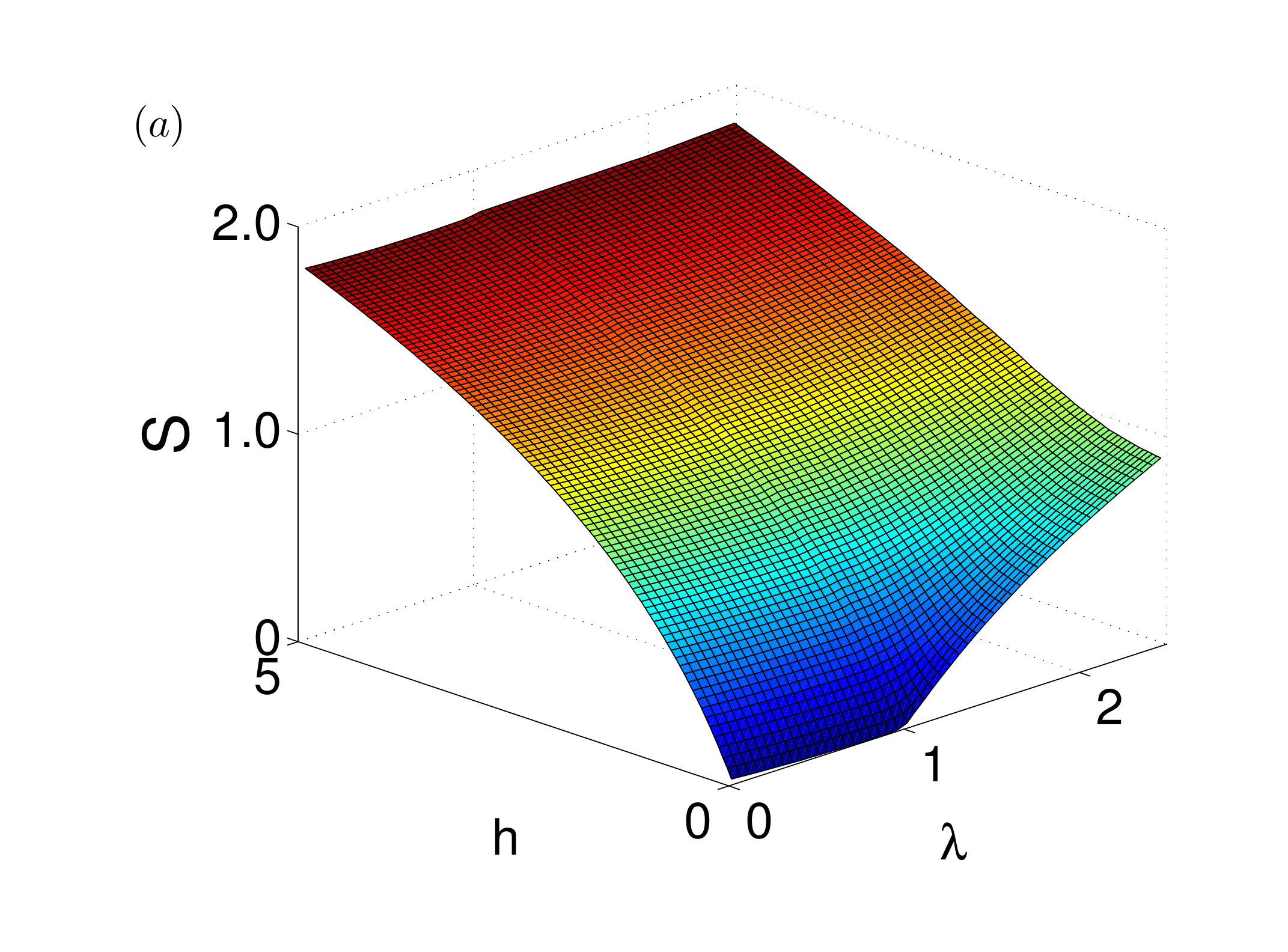}
     \includegraphics[angle=0,totalheight=5cm]{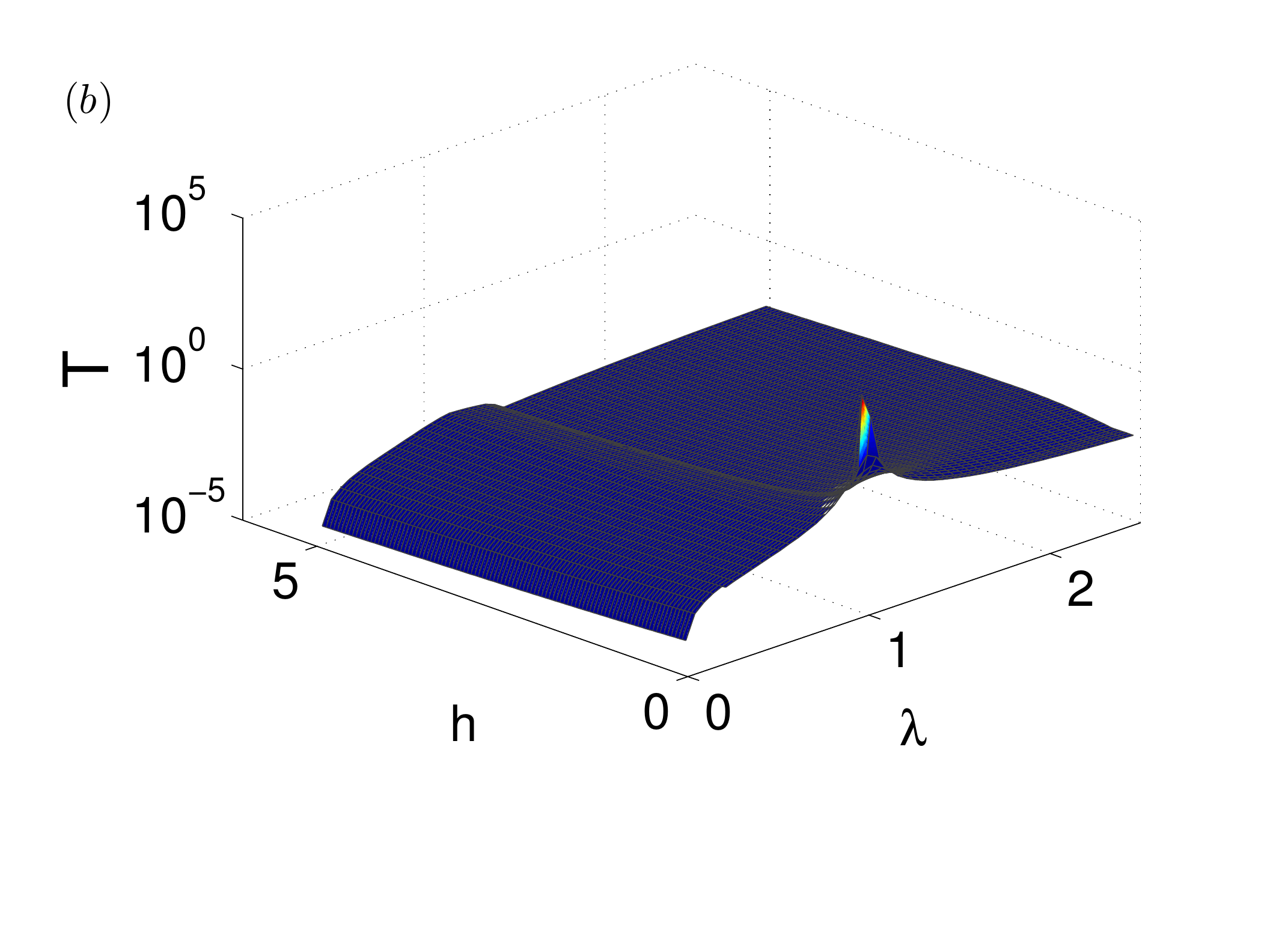}
     \includegraphics[angle=0,totalheight=5cm]{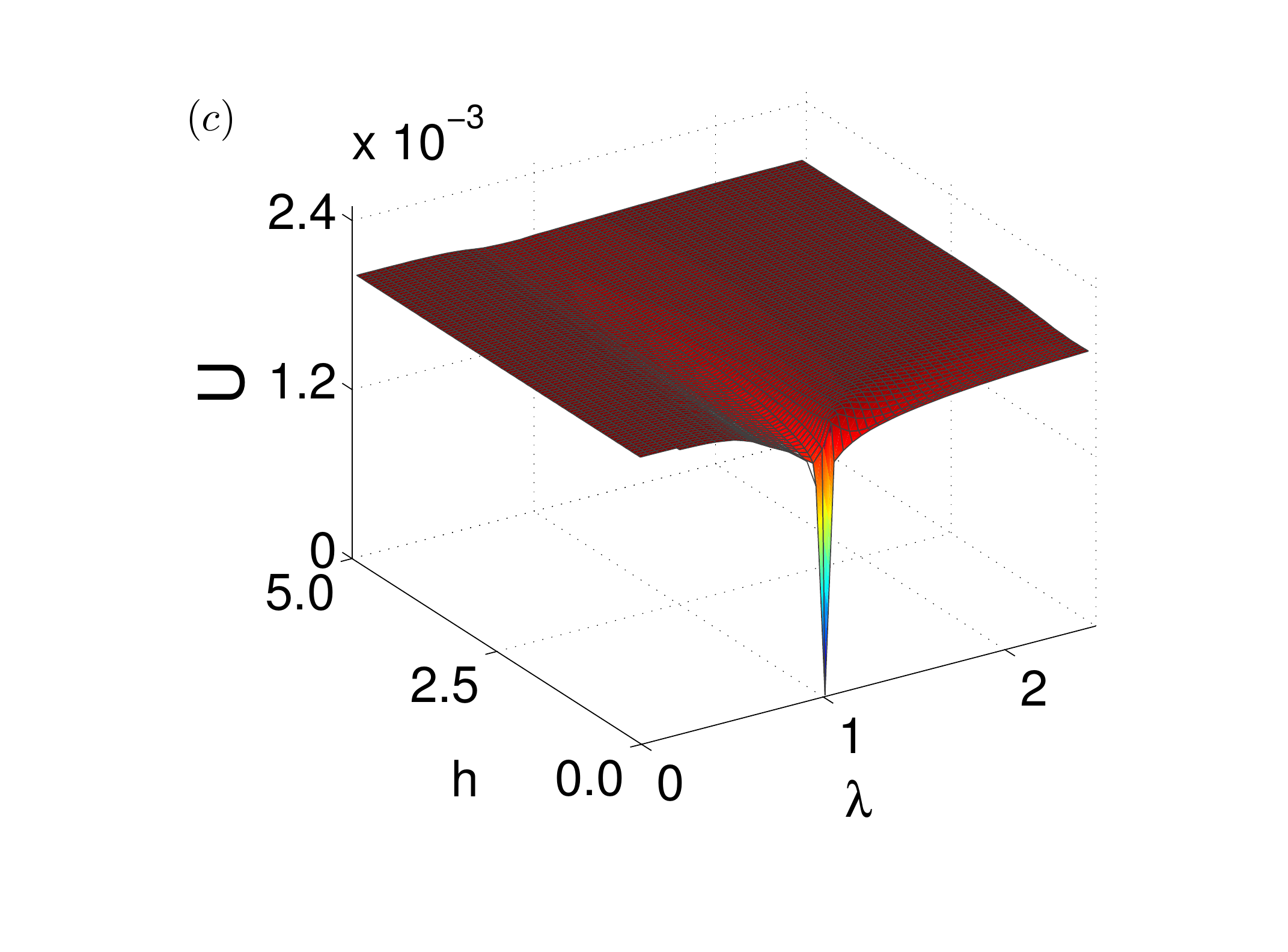}

       \caption{Fidelity entropy $S_f(\lambda,h)$, fidelity temperature $T_f(\lambda,h)$ and fidelity internal energy $U_f(\lambda,h)$  as a function of $\lambda$ and $h$ for the transverse field quantum Ising chain in a longitudinal field, with $h \geq 0$ and $\lambda \geq 0$.
       (a) There exist two singular lines, $h=0$ and $\lambda=1$, in fidelity entropy $S_f(\lambda,h)$.  Note that singularities on the line $h=0$ arise from duality and should be attributed to the $Z_2$ symmetry, whereas singularities on $\lambda=1$ reflects the fact that spins point towards $+ x$ direction for $0 \leq \lambda<1$ when $h$ is large and towards other directions for $\lambda>1$ when $\lambda$ and $h$ are large.
       (b) Fidelity temperature $T_f(\lambda,h)$ diverges at the critical point  (1,0), but vanishes when $\lambda=0$, and when $\lambda$ and/or $h$  are infinite in value.
       (c) Fidelity internal energy $U_f(\lambda,h)$ takes the same value at all stable fixed points, as well as at factorizing fields, when $\lambda=0$, and when $\lambda$ and/or $h$ are infinite in value.
        }\label{sutis}

\end{figure}

We plot fidelity entropy $S_f(\lambda,h)$, fidelity temperature $T_f(\lambda,h)$ and fidelity internal energy $U_f(\lambda,h)$  as a function of  $\lambda$ and $h$ in Fig.~\ref{sutis}.  Fidelity entropy $S_f(\lambda, h)$ reaches a local maximum when $\lambda =1$  and reaches the maximum when $\lambda =0$, if scaling entropy $\ln (1+|h|) $ is excluded.  This is consistent with the existence of stable fixed points at $(0,0)$, $(0,\infty)$, $(\infty, 0)$, and $(1, \infty)$, which are seen as characteristic points in the parameter space. The existence of stable fixed points $(0,0)$ and $(\infty, 0)$ is protected by the $Z_2$ symmetry when $h=0$, whereas the existence of stable fixed points $(0, \infty)$ and $(1, \infty)$ may be interpreted as a consequence of the variation of the symmetry group with $\lambda$: $U(1)$ for $\lambda =0$, and none for  $\lambda \neq 0$, when $h\neq 0$.  In particular, the existence of a stable fixed point at $(1, \infty)$ might also be related to a well-known fact that, at nonzero $h$,  a massive excitation spectrum involves eight massive particles, which shows a deep relation with $E_8$ algebraic structure,  as predicted in perturbed conformal field theory~\cite{zamolodchikov}.  Fidelity temperature $T_f(\lambda,h)$ diverges at the critical point $(1,0)$ and reaches zero at stable fixed points, as well as at factorizing fields when $\lambda=0$, and when $\lambda$ and/or $h$ are infinite in value.

We remark that, in addition to a critical point ($1,0$) and a first-order transition line $h=0$, with $\lambda \in [0,1)$,  detected through singularities in ground state fidelity per site, fidelity mechanical state functions exhibit singularities at one dual line:  $h=0$,  a line with a $U(1)$ symmetry: $\lambda=0$ and a characteristic line $\lambda=1$.
One might view such singularities as ``phase transitions" in fidelity mechanics.  Note that singularities on the line $h=0$ arise from duality and should be attributed to the $Z_2$ symmetry, whereas singularities on $\lambda=1$ reflects the fact that spins point towards $+ x$ direction for $\lambda<1$ when $h$ is large and towards other directions for $\lambda>1$, when $\lambda$  and $h$ are large.  Note that the line $\lambda=1$, as a characteristic line, does depend on choices of a dominant control parameter.  Thus, it is different from characteristic lines arising from  symmetries, dualities and factorizing fields (cf.  Appendix~\ref{fictitious}).  Further, fidelity internal energy $U_f(\lambda,h)$  takes the same value at stable fixed points,  as well as  at factorizing fields, when $\lambda=0$, and when $\lambda$ and/or $h$ are infinite in value.

\section{XYZ chain - a typical example for dualities}~\label{xyzmodel}

\begin{figure}
     \includegraphics[angle=0,totalheight=4.5cm]{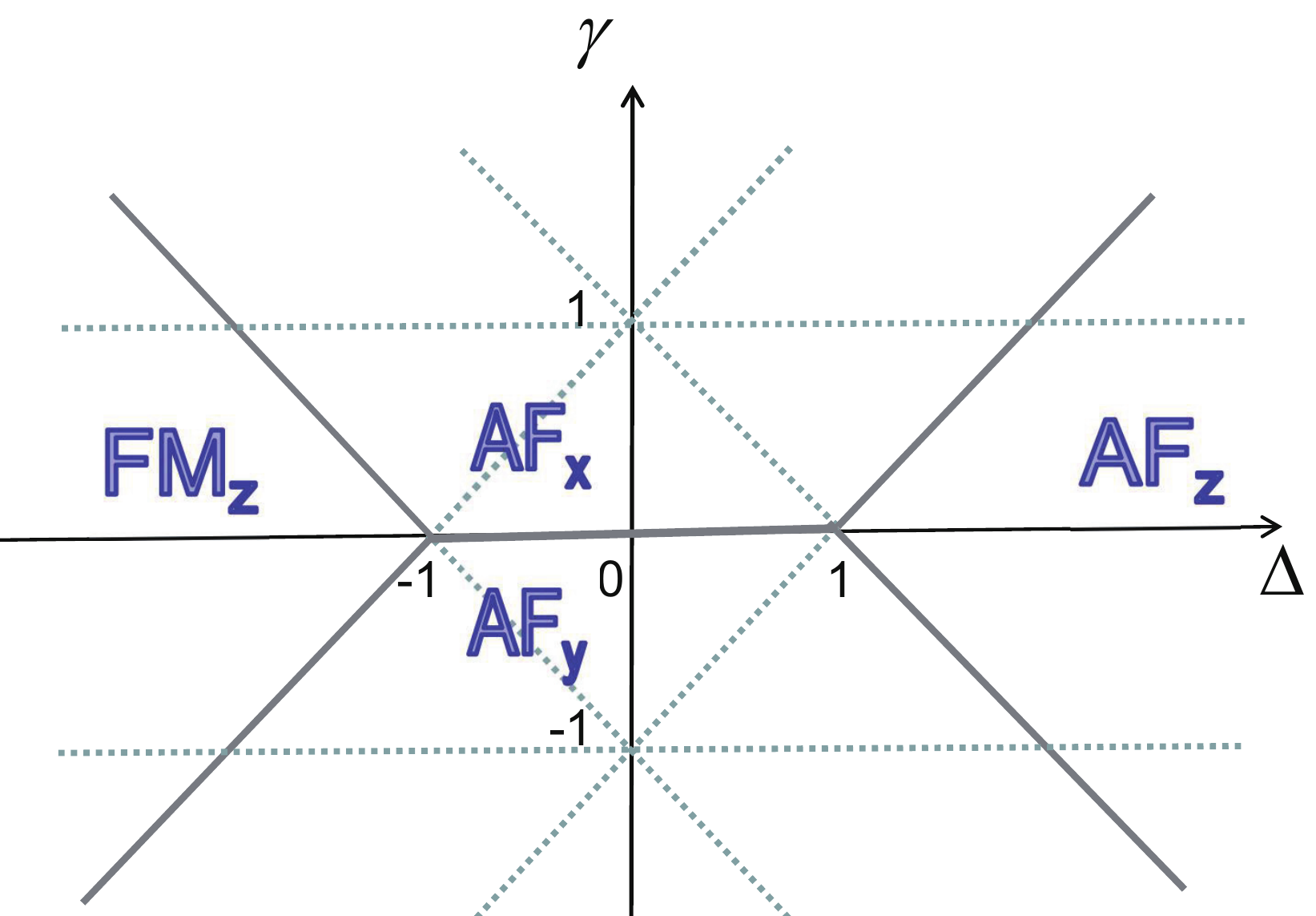}
       \caption{  Ground state phase diagram for the quantum spin-$1/2$ XYZ chain.  Solid lines, $\gamma=0~(-1<\Delta\leq 1)$, $\gamma=1+\Delta~(\Delta<-1)$, $\gamma =   1 -\Delta ~(\Delta \geq 1)$, $\gamma=-1-\Delta~(\Delta<-1)$ and $\gamma =  - 1 +\Delta ~(\Delta \geq 1)$, characterize its phase boundaries.  In addition, factorizing fields occur on two lines: $\gamma=1+\Delta$ and $\gamma=-1-\Delta$, with $\Delta>-1$, as well as when $|\Delta|$ is infinite in value. Moreover, a multi-critical point at infinity occurs when $|\gamma|$ is infinite in value.}\label{xyzphase}

\end{figure}

A canonical form of the Hamiltonian for the quantum spin-$1/2$ XYZ chain is
\begin{equation}
H(\gamma,\Delta)=\sum_{i}{(\frac{1+\gamma}{2}\sigma_i^x\sigma^x_{i+1}+\frac{1-\gamma}{2}\sigma_i^y\sigma^y_{i+1}+\frac{\Delta}{2}\sigma_i^z\sigma^z_{i+1})}, \label{xyz}
\end{equation}
where $\sigma^{x,y,z}_i$ are the Pauli matrices at site $i$, $\gamma$ and $\Delta$ are the anisotropic coupling constants.
This model is exactly solvable by Bethe ansatz via its equivalence to the eight-vertex model~\cite{Baxterxyz, Baxter8v, Baxterbook}.  Its ground state phase diagram is shown in
Fig.~\ref{xyzphase}.  There are four different phases, labeled as $\rm {AF}_x$, $\rm {AF}_y$, $\rm {AF}_z$, and $\rm {FM}_z$,
representing an antiferromagnetic phase in the $x$ direction, an antiferromagnetic phase in the $y$ direction, an antiferromagnetic phase in the $z$ direction, and a ferromagnetic phase in the $z$ direction, respectively.
In addition, there are five critical lines, $\gamma=0~(-1<\Delta\leq 1)$, $\gamma=1+\Delta~(\Delta<-1)$, $\gamma =   1 -\Delta ~(\Delta \geq 1)$, $\gamma=-1-\Delta~(\Delta<-1)$ and $\gamma =  - 1 +\Delta ~(\Delta \geq 1)$, depicted as five solid lines in Fig.\ref{xyzphase}.  For $\gamma=0$, a critical line exists between $-1<\Delta\leq 1$,  which is a Luttinger liquid with central charge $c=1$.  A KT phase transition occurs at $\Delta=1$,  protected by a $U(1)$ symmetry, from a critical phase to the $\rm {AF}_z$ phase for $\Delta>1$.  For $\gamma \neq 0$, four critical lines exist: $\gamma = -1+\Delta ~(\Delta \geq 1)$,  $\gamma = 1-\Delta ~(\Delta \geq 1)$,    $\gamma = -1-\Delta ~(\Delta < -1)$, and  $\gamma = 1+\Delta ~(\Delta < -1)$, which are Luttinger liquids with central charge $c=1$.  We note that, along the lines $\gamma = -1+\Delta$ and $\gamma = 1-\Delta$,  the KT transitions occur at $\Delta=1$,  protected by a $U(1)$ symmetry, from a critical phase to the $\rm {AF}_y$ phase and the $\rm {AF}_x$ phase for $\Delta<1$, respectively.  When $\Delta\rightarrow -\infty$, it yields a factorized ground state in the $\rm {FM}_z$ phase. When $\Delta\rightarrow\infty$, it yields a factorized ground state in the $\rm {AF}_z$ phase.
In addition, two lines: $\gamma=1+\Delta$ and $\gamma=-1-\Delta$ with $\Delta>-1$ represent factorizing fields~\cite{Illuminati}.  Moreover, a Gaussian critical point occurs when $\gamma$ is infinite in value for any fixed $\Delta$.

For $\gamma \neq 0$, the Hamiltonian (\ref{xyz}) possesses a  $Z_2$ symmetry group, defined by  $\sigma^x_i\leftrightarrow - \sigma^x_i$,
$\sigma^y_i\leftrightarrow - \sigma^y_i$ and $\sigma^z_i \leftrightarrow  \sigma^z_i$. For $\gamma = 0$, the symmetry group becomes $U(1)$.  In addition, a $U(1)$ symmetry  occurs at four characteristic lines: $\gamma = \pm 1\pm \Delta$, as well as when $\Delta$ or $\gamma$ is infinite in value.
Further, there are five different dualities, as discussed in Appendix \ref{dual}.  A remarkable fact is that  the ground state phase diagram for the quantum XYZ model, as presented above, may be exactly reproduced via dualities.

\begin{figure}
     \includegraphics[angle=0,totalheight=4.5cm]{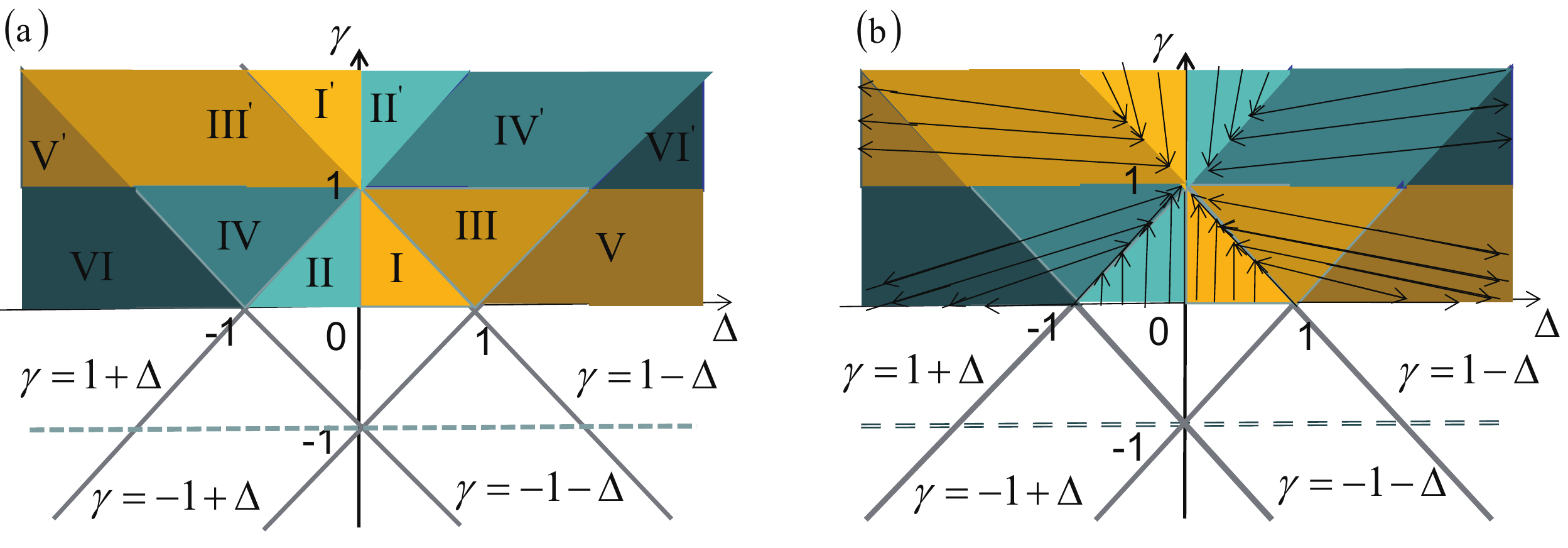}
       \caption{ (a) Twelve regimes for the quantum spin-$1/2$ XYZ chain with $\gamma>0$. Here, regimes $\rm{I}$, $\rm{III}$, $\rm{V}$, $\rm{I'}$, $\rm{III'}$ and $\rm{V'}$  are dual to each other,  and regimes $\rm{II}$, $\rm{IV}$, $\rm{VI}$, $\rm{II'}$, $\rm{IV'}$ and $\rm{VI'}$ are dual to each other. Therefore, there are only two principal regimes, with regime $\rm{I}$ and regime $\rm{II}$ as our choice.  (b) Choices of a dominant control parameter $x$ in  twelve different regimes for the quantum spin-$1/2$ XYZ chain. In regime $\rm{I}$ and  regime $\rm{II}$,  we choose $\gamma$ as a dominant control parameter.  Here, $\gamma$ ranges from $\gamma=\gamma_c=0$ to $\gamma=1-\Delta$ for a fixed $\Delta \in (0,1)$ in regime I and to $\gamma = 1+\Delta$ for a fixed $\Delta \in (-1,0)$ in regime II. Choices of a dominant control parameter in other regimes then follow from their respective dualities to regime $\rm{I}$ and  regime $\rm{II}$.}
\label{duality2}
\end{figure}

The Hamiltonian (\ref{xyz}) is symmetric under $\gamma \leftrightarrow - \gamma$.  Therefore, we may restrict ourselves to the region $\gamma \geq 0$.
Taking into account the symmetries, dualities and factorizing fields~\cite{factorizing}, we may divide the region $\gamma \geq 0$ into twelve different regimes, with five lines defined by $\gamma=1$ and $\gamma=\pm1\pm\Delta$ as boundaries. These twelve regimes are separated into two groups, with six regimes in each group dual to each other.  As shown in Fig.~\ref{duality2}\;(a), regimes $\rm{I}$, $\rm{III}$, $\rm{V}$, $\rm{I'}$, $\rm{III'}$ and $\rm{V'}$  are dual to each other, while regimes $\rm{II}$, $\rm{IV}$, $\rm{VI}$, $\rm{II'}$, $\rm{IV'}$ and $\rm{VI'}$ are dual to each other.   Therefore, there are only two principal regimes, which represent the physics underlying the quantum spin-$1/2$ XYZ model.  Here, we choose regime $\rm{I}$ and regime $\rm{II}$ as two principal regimes.

In regime $\rm{I}$ and regime $\rm{II}$,  it is natural to choose $\gamma$ as a dominant control parameter, given that $\gamma=0$, with $-1<\Delta \leq 1$ is a line of critical points. Here, $\gamma$ ranges from $\gamma=\gamma_c=0$ to $\gamma=1-\Delta$ for a fixed $\Delta \in (0,1)$ in regime I and to $\gamma = 1+\Delta$ for a fixed $\Delta \in (-1,0)$ in regime II.  However, other choices are possible,
as long as such a choice is consistent with the constraints imposed by the symmetries, dualities and factorizing fields.
If a dominant control parameter is chosen in regime $\rm{I}$ and regime $\rm{II}$,  then the choices of a dominant control parameter in other regimes simply follow from their respective dualities to regime  $\rm{I}$ and regime $\rm{II}$.

It is numerically confirmed that fidelity entropy $S(\Delta,\gamma)$ scales as $\gamma ^{ \nu(\Delta) +1}$ near a critical line $\gamma =0$, with $\Delta \in (-1, 1]$. Here,  $\nu(\Delta)$ is the critical exponent for the correlation length (cf. Appendix \ref{scaling}).

Now we move to discuss fidelity mechanical state functions for the quantum XYZ model. In a given regime, it is straightforward to compute fidelity entropy $S_f(\Delta,\gamma)$, fidelity temperature $T_f(\Delta,\gamma)$ and fidelity internal energy $U_f(\Delta,\gamma)$.  Mathematical details about their explicit expressions may be found in Appendix \ref{xyzchain}.

\begin{figure}
\centering
\includegraphics[angle=0,totalheight=5cm]{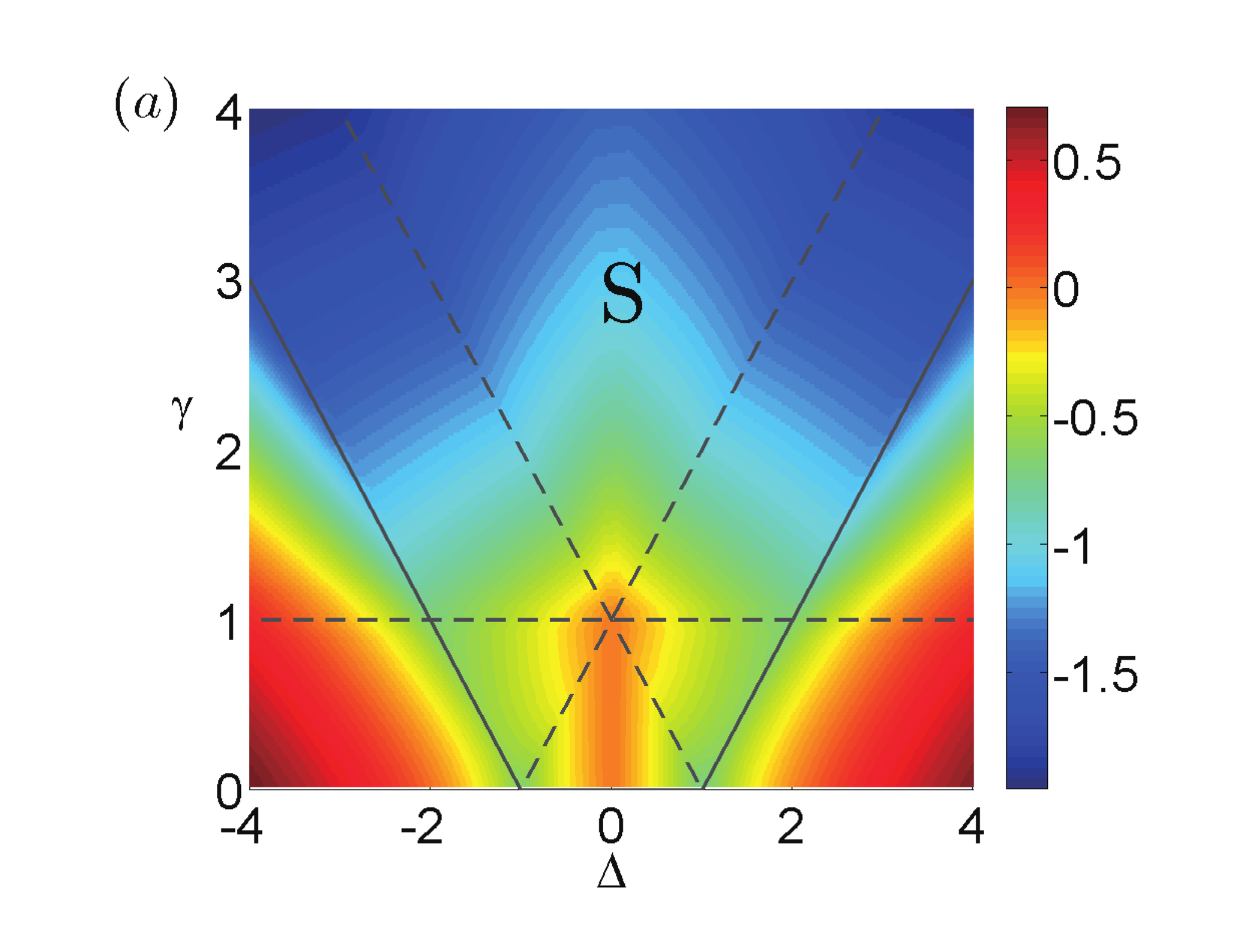}
     \includegraphics[angle=0,totalheight=5cm]{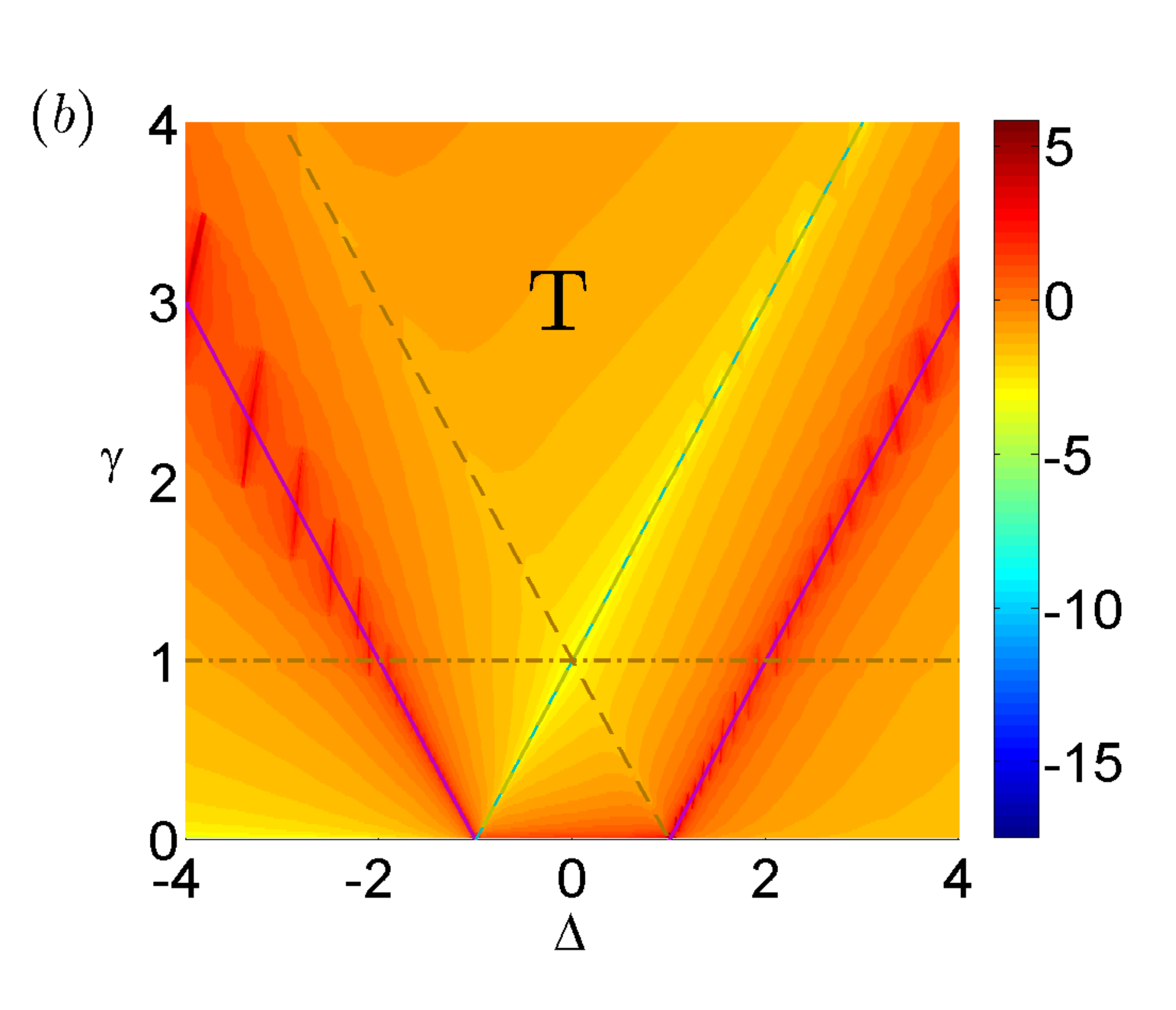}
     \includegraphics[angle=0,totalheight=5cm]{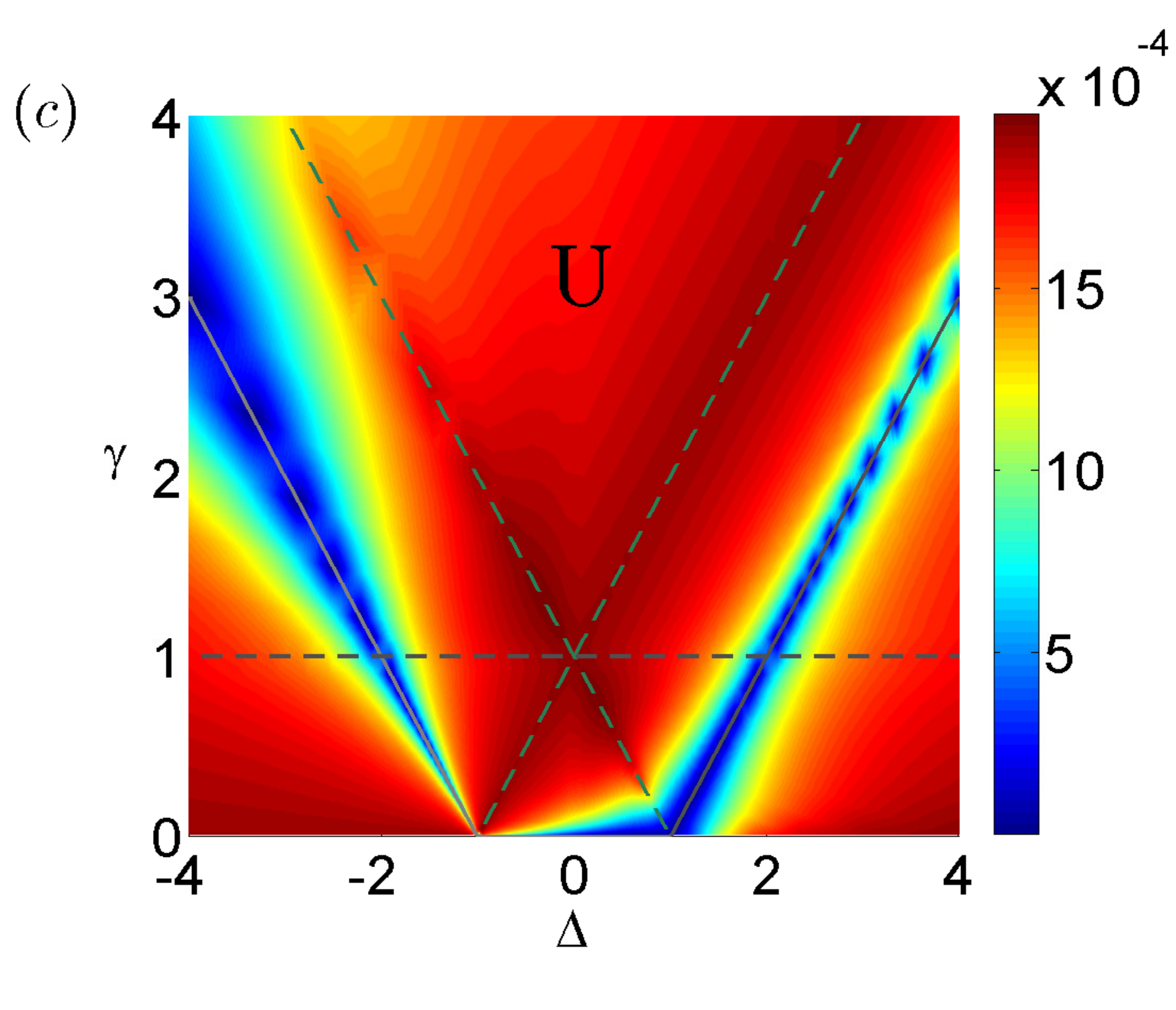}

     \caption{Fidelity entropy $S_f(\Delta,\gamma)$,  fidelity temperature $T_f(\Delta,\gamma)$ and fidelity internal energy $U_f(\Delta,\gamma)$ for the quantum spin-$1/2$ XYZ chain.
                    (a)  Fidelity entropy $S_f(\Delta,\gamma)$ exhibits a local maximum at a stable fixed point $(1, 0)$. Singularities occur at three lines of critical points: $\gamma=0~(-1<\Delta\leq 1)$,  $\gamma =  -1 + \Delta ~(\Delta \geq 1)$,  and  $\gamma = -1 -\Delta~(\Delta < -1)$, and at a line: $\gamma=1+\Delta~(\Delta>-1)$, representing the factorizing fields, and at two lines: $\gamma=1-\Delta ~(\Delta\leq 1)$ and $\gamma=1$, representing the self-dualities.
                      (b)  Fidelity temperature $T_f(\Delta,\gamma)$ diverges at three lines: $\gamma=0~(-1<\Delta\leq 1)$,  $\gamma =  -1 + \Delta ~(\Delta \geq 1)$,  and  $\gamma = -1 -\Delta~(\Delta < -1)$, representing lines of critical points, and is zero at a line: $\gamma=1+\Delta~(\Delta > -1)$, as well as at factorizing fields, when $|\Delta|$ is infinite in value.
                    (c)  Fidelity internal energy $U_f(\Delta,\gamma)$ takes the same value at all stable fixed points: $(0,1)$, $(\pm \infty, 0)$ and $(\pm \infty, 1)$, as well as at a line: $\gamma=1+\Delta~(\Delta>-1)$, and at factorizing fields, when $|\Delta|$ is infinite in value.}
    \label{fmsxyz}
 \end{figure}

A contour plot is depicted in  Fig.~\ref{fmsxyz}\;(a) for fidelity entropy $S_f(\Delta,\gamma)$ in the  parameter space for $\gamma \geq 0$. Fidelity entropy $S_f(\Delta,\gamma)$ takes a local maximum at $(0, 1)$.
Fidelity temperature $T_f(\Delta,\gamma)$ and fidelity internal energy $U_f(\Delta,\gamma)$ are shown as contour plots in Fig.~\ref{fmsxyz}\;(b) and (c), respectively.  As we see, fidelity temperature $T_f(\Delta,\gamma)$ diverges at three lines of critical points: $\gamma=0~(-1<\Delta\leq 1)$,  $\gamma =  -1 + \Delta ~(\Delta \geq 1)$,  and  $\gamma = -1 -\Delta~(\Delta < -1)$, and vanishes at a line: $\gamma=1+\Delta$, with $\Delta>-1$, where factorized states occur, in addition to two characteristic lines at infinity: $\Delta = \pm \infty$.  We remark that there are five stable fixed points identified as characteristic points in the region $\gamma \geq 0$: $(0,1)$, $(\pm \infty, 0)$ and $(\pm \infty, 1)$.

Also note that fidelity entropy $S_f(\Delta,\gamma)$, fidelity temperature $T_f(\Delta,\gamma)$ and fidelity internal energy $U_f(\Delta,\gamma)$ exhibit singular behaviors at three lines of critical points: $\gamma=0~(-1<\Delta\leq 1)$,  $\gamma =  -1 + \Delta ~(\Delta \geq 1)$,  and  $\gamma = -1 -\Delta~(\Delta < -1)$, and at a line: $\gamma=1+\Delta~(\Delta>-1)$, representing the factorizing fields, and at two lines: $\gamma=1-\Delta ~(\Delta\leq 1)$ and $\gamma=1$, representing the self-dualities.  This singular behavior may be recognized as ``phase transitions" in fidelity mechanics.  In addition, fidelity internal energy $U_f(\Delta,\gamma)$ takes the same value at all stable fixed points $(0,1)$, $(\pm \infty, 0)$ and $(\pm \infty, 1)$, as well as at factorizing fields.

\section{Quantum XXZ model in a magnetic field - intermediate cases between Kosterlitz-Thouless and Pokrovsky-Talapov transitions}~\label{xxzhmodel}

\begin{figure}
     \includegraphics[angle=0,totalheight=3.2cm]{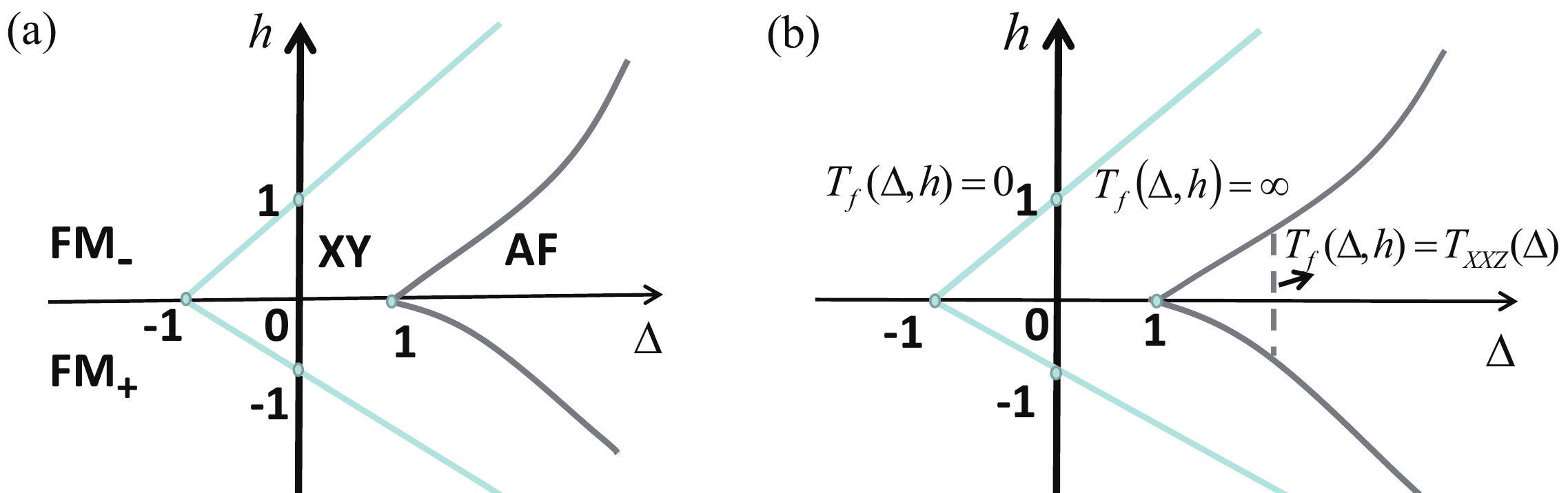}
       \caption{(a) Ground state phase diagram for the quantum XXZ model in a magnetic field.  There are four phases, labeled  AF, ${\rm FM}_-$, ${\rm FM}_+$, and XY, representing an antiferromagnetic phase, a ferromagnetic phase with all spin down,  a ferromagnetic phase with all spin up, and a critical phase with central charge $c=1$, respectively. Here, the phase boundary between the XY phase and the AF phase asymptotically approaches the PT transition line between the ${\rm FM}_-$/${\rm FM}_+$ phase and the XY phase for $h>0$/$h<0$, respectively.
       (b) Fidelity temperature $T_f(\Delta, h)$ for the quantum XXZ model in a magnetic field.  In the phases ${\rm FM}_-$ and ${\rm FM}_+$, it is zero; in the XY phase, it diverges, and in the AF phase, it takes the same value as fidelity temperature $T_{XXZ}(\Delta)$ for the quantum XXZ model. }\label{xxz}
\end{figure}

For the quantum XXZ chain in a magnetic field, a canonical form of the Hamiltonian  is
\begin{equation}
H(\Delta,h)=\sum_{i}(\sigma_i^x\sigma^x_{i+1}+\sigma_i^y\sigma^y_{i+1}+\Delta\sigma_i^z\sigma^z_{i+1}+h\sigma^z_i), \label{xxzh}
\end{equation}
where $\sigma^{x,y,z}_i$ are the Pauli matrices at site $i$, $\Delta$ is the anisotropic coupling constant and $h$ is the magnetic field strength. The model is exactly solvable by Bethe ansatz~\cite{yangcn,luther,essler}, with its ground state phase diagram shown in Fig.~\ref{xxz}. The Hamiltonian (\ref{xxzh}) is symmetric with respect to $h \leftrightarrow -h$.  There are four phases, labeled as AF, ${\rm FM}_-$, ${\rm FM}_+$, and XY, representing an antiferromagnetic phase, a ferromagnetic phase with all spin down,  a ferromagnetic phase with all spin up, and a critical phase with central charge $c=1$, respectively.

In the entire ${\rm FM}_-$ and ${\rm FM}_+$ phases, up to the phase boundary between the FM phase and the XY phase, the ground state remains to be the same: a spin polarized state with all spins down for $h>0$ and a spin polarized state with all spins up for $h<0$.
These two states coexist at $h=0$.  Therefore, there is a first-order QPT if the line $h=0$ is crossed, where a $Z_2$ symmetry, defined
as $\sigma^ x_i  \leftrightarrow \sigma^ y_i$ and $\sigma^ z_i \leftrightarrow - \sigma^ z_i$, is spontaneously broken.   In these two phases, we may rescale the ground state energy density: $e(\Delta, h) = k(\Delta, h)  e' (\Delta', h')$, with $\Delta'= \Delta/(|h|-\Delta)$, $h'=h/(|h|-\Delta)$, and $k(\Delta, h)= |h|-\Delta$. Then, we have $e'(\Delta', h')=-1$.  Therefore, fidelity entropy $S_f(\Delta, h)$ takes the form: $S_f(\Delta, h) = \ln (|h|-\Delta)$, fidelity temperature $T_f(\Delta, h)$
is zero, and fidelity internal energy  $U_f(\Delta, h)$ is a constant, which should be equal to fidelity internal energy for the XXZ model at $\Delta =\infty$, since the phase boundary between the XY phase and the AF phase asymptotically approaches the PT transition line  between the ${\rm FM}_-$/${\rm FM}_+$ phase and the XY phase for $h>0$/$h<0$, respectively.

In the critical XY phase, fidelity temperature $T_f(\Delta, h)$ diverges, indicating strong quantum fluctuations. Fidelity internal energy is, by convention, zero, and the only contribution to fidelity entropy is residual fidelity entropy $S_0$. Since residual fidelity entropy is extrinsic in nature, it can only be determined if the model is embedded into a more general model,  such as the quantum XYZ model in a magnetic field, which accommodates a cone on which factorized states occur~\cite{factorizing}. For our purpose, we simply set residual fidelity entropy $S_0$, thus fidelity entropy $S_f(\Delta, h)$ itself, to be zero in this phase. This makes sense, since no information close to a critical point is available, if we restrict ourselves to the quantum XXZ chain in a magnetic field.  At the boundary between the ${\rm FM_-}$ phase and the XY phase, continuous QPTs occur when $\Delta >-1$, belonging to the PT transitions. Hence, fidelity temperature is not well-defined at this boundary; in fact, it ranges from 0 to $\infty$.  At $\Delta =-1$, a first-order QPT occurs.  This transition is special, as an intersection point between two PT transition lines.  Indeed, fidelity temperature is also not well-defined, ranging from 0 to $\infty$.

In the AF phase, the ground state wave functions do not depend on $h$.  Hence, fidelity entropy $S_f(\Delta, h)$, fidelity temperature $T_f(\Delta, h)$ and fidelity internal energy $T_f(\Delta, h)$ do not depend on $h$. Hence,  we only need to determine fidelity entropy $S_f(\Delta, 0)$, fidelity temperature $T_f(\Delta, 0) $ and fidelity internal energy $U_f(\Delta, 0) $ at $h=0$. In this limit, a continuous QPT occurs from the critical XY phase to the AF phase at $\Delta=1$, belonging to the KT phase transitions.  Then, fidelity mechanical state functions follow from that for the quantum XXZ model, a special case of the quantum XYZ model, which has been discussed in Section~\ref{xyzmodel}.  Therefore,
we have $S_f(\Delta, 0)=S_{XXZ}(\Delta)$, $T_f(\Delta, 0)=T_{XXZ}(\Delta)$,  and $U_f(\Delta, 0)=U_{XXZ}(\Delta)$.

Now we are  ready to discuss the phase boundary between the XY phase and  the AF phase.  Along this boundary, fidelity temperature $T_f(\Delta_p, h_p)$ is monotonically decreasing from $\infty$ at $\Delta_p=1$ to zero at $\Delta_p=\infty$, where $\Delta_p$ and $h_p$  represent the corresponding values of $\Delta$ and $h$ at the phase boundary.  Therefore, at a specific point $(\Delta_p, h_p)$, fidelity temperature is not well-defined for $\Delta_p >1$, ranging from $T_{XXZ}(\Delta_p)$ to $\infty$.  That is,  a QPT at this phase boundary is an intermediate case interpolating between a KT transition  and a PT transition, which represents a new universality class, different from the KT transition  and the  PT transition.

\section{ Analogues of the four thermodynamic laws, fidelity flows and miscellanea}~\label{answer}

\subsection{Analogues of the four thermodynamic laws}~\label{4laws}

Let us now address whether or not there is any formal similarity between QPTs and  black holes, which has been raised as the first question in Section~\ref{intro}. The answer is affirmative.  Here, we note that, in this section, a subscript $f$ will be dropped from fidelity mechanical state functions to keep notation simple.  As shown in Table~\ref{tab1}, there is a dictionary that translates each notion in one theory to its counterparts in other theories, among fidelity mechanics, black hole thermodynamics~\cite{Bardeen}  and standard thermodynamics~\cite{Kittel}. 

Actually, we may formulate analogues of the four thermodynamic laws in fidelity mechanics:\\

(i) Zeroth law - For a given fidelity mechanical system, which is in equilibrium with its environment, fidelity temperature $T(x_1,x_2)$ quantifies quantum fluctuations. \\

(ii) First law - Fidelity internal energy may be transferred from a fidelity mechanical system, as  fidelity work or fidelity heat (defined via fidelity entropy), into  its environment, or vice versa. Mathematically, we have  $dU(x_1,x_2)=T(x_1,x_2)dS(x_1,x_2)+ \dbar W(x_1,x_2)$.\\

(iii) Second law - The total  fidelity entropy of  a fidelity mechanical system and its environment never decreases. Physically, this amounts to stating that {\it information gain we are able to recover from the environment never exceeds information loss incurred due to information erasure in a fidelity mechanical system}. Mathematically, we have $\Delta S(x_1,x_2) + \Delta S_e(x_1,x_2) \geq 0$. Generically, $\Delta S(x_1,x_2)\geq0$ and $\Delta S_e(x_1,x_2) \leq 0$. Therefore, $k(x_1,x_2)\leq 1$, with $k(x_1,x_2)$  being defined by  $\Delta S_e(x_1,x_2) = -k(x_1,x_2)\Delta S(x_1,x_2)$.  \\

 (vi) Third law - For a fidelity mechanical system,  fidelity entropy $S(x_1,x_2)$  approaches a (local) maximum and fidelity temperature $T(x_1,x_2)$ approaches zero, as a stable fixed point is approached. However, the probability for getting access to a stable fixed point is zero.\\

\begin{table}
\begin{tabular}{cc@{\hspace{3pt}}c}
\hline
Thermodynamics & Black Hole Thermodynamics & Fidelity Mechanics\\
\hline
Temperature $T$ & Surface gravity $\kappa$  &  Fidelity temperature $T(x)$\\
$dU=TdS+ \dbar W$ & $dE=\frac{\kappa}{8\pi}dA+\Omega dJ+\Phi dQ$ & $dU(x)=T(x)dS(x)+\dbar W(x)$  \\
\tabincell{c}{$S=k\ln Z$ \\increasing  monotonically} &  \tabincell{c}{ $S_{BH}=\frac{\kappa A}{4\ell_p^2}$ \\ increasing  monotonically }  &\tabincell{c}{$S(x)=-2\int_{x_c}^{x}{\ln{d(x,y)}dy}$ + $S_0$\\ increasing  monotonically  }\\
 Probability for $T=0$ is zero & Probability for $\kappa=0$ is zero & Probability for getting access to a stable fixed point is zero\\
Equilibrium states & Static black holes & Ground states \\
Non-equilibrium states & Dynamic black holes & Low-lying states\\
Quasi-static & Slowly-evolving  & Adiabatic\\
\hline

\end{tabular}
\caption{ \label{tab1} A dictionary for thermodynamics, black hole mechanics and fidelity mechanics.  Here, $S_{BH}$ is the Bekenstein-Hawking entropy, $A$ is the horizon area, $\ell_p$ is the Plank length, $E$ is the energy, $\kappa$ is the surface gravity, $\Omega$ is the angular velocity, $J$ is the angular momentum, $\Phi$ is the electrostatic potential, and $Q$ is the electric charge. In fidelity mechanics, fidelity internal energy $U(x)$ is defined as $U(x)=\mp \ln(e(x)/e(x_c))V+U_0$,  where $x$ is a dominant control parameter, $e(x)$ is the ground state energy density, $U_0$ is an additive constant, and $V(x)$ is an unknown function of $x$, determined from a singular first-order differential equation (\ref{alpha}), with fidelity temperature $T(x)=-V_x(x)$ quantifying quantum fluctuations. Here, a dominant control parameter is introduced via a one-to-one correspondence between $(x_1,x_2)$ and $(x, \tau)$, with $\tau$ being an auxiliary control parameter, and $-/+$ in fidelity internal energy $U(x)$ corresponds to monotonically increasing/decreasing $e(x)$ with increasing $x$, respectively. For brevity, no contribution from scaling entropy is taken into account.}
\end{table}

\subsection{Fidelity flows as an alternative form of RG flows}~\label{fidelityflows}

In real space RG theories,  a number of high energy degrees of freedom are discarded  during the construction of an effective Hamiltonian. This results in a reduction of the number of degrees of freedom, leading to an apparent irreversibility and so causing complications around this issue. However, fidelity mechanics offers us new insights into our understanding of the irreversibility of RG flows.  This is achieved by introducing an alternative form of RG flows - fidelity flows.

A fidelity mechanical system, which is in equilibrium with its environment,  is {\it unstable} under a random perturbation. That is, it is spontaneous for such a fidelity mechanical system to flow away.  Therefore, a trajectory is traversed in the parameter space, along which we formally treat $x_1$ and $x_2$ as a function of time $t$: $x_1=x_1(t), x_2=x_2(t)$.
Then there is a quantum state $\psi(t)$ attached to a point $(x_1(t), x_2(t))$ on the trajectory, according to the (time-dependent) Schr\"{o}dinger equation, with a time-dependent Hamiltonian  $H(x_1(t),x_2(t))$.
Apparently, there are two possibilities, given fidelity heat capacity $C(x_1,x_2)=T(x_1,x_2) {\Delta S(x_1,x_2)}/{\Delta T(x_1,x_2)}<0$ is generically negative: (i) if fidelity temperature $T(x_1,x_2)$ decreases, then fidelity  entropy $S(x_1,x_2)$ increases, due to information erasure; (ii) if fidelity temperature $T(x_1,x_2)$ increases, then fidelity entropy $S(x_1,x_2)$ decreases, due to information gain.  However,  the second possibility is forbidden:  if it happened, then the future would be remembered.  This contradicts the psychological/computational arrow  (cf. Appendix~\ref{timearrows}). Therefore, fidelity entropy $S(x_1,x_2)$ is monotonically increasing and fidelity temperature $T(x_1,x_2)$ is monotonically decreasing along a trajectory.   Although such an evolution is time-reversal invariant in quantum mechanics, a corresponding evolution in fidelity mechanics is irreversible.  Here, it is proper to remark that, in contrast to quantum mechanics, there are no equations of motion in fidelity mechanics, a situation exactly the same as in thermodynamics.  As a consequence, irreversibility  is stronger than time-reversal non-invariance in fidelity mechanics.  In other words, $t$, as a microscopic time, appears in the
Schr\"{o}dinger equation. However,  a macroscopic time emerges in a fidelity mechanical system. That is, an arrow of time emerges, resulted from information storage via recording information encoded in the past states in media - a key ingredient in a fidelity mechanical system (cf. Appendix~\ref{timearrows} for a definition of both microscopic and macroscopic time). Here, we note that, for a generic trajectory traversed by a fidelity mechanical system in the parameter space, the past states, recorded in media, differ from the past states really occurred. Actually, the past states, recorded in media, are subject to changes as time goes by.  This is a consequence of the fact that an increment in fidelity internal energy is separated into an increment in fidelity heat and an increment in fidelity work.  However, only the increment in fidelity heat due to an increment in fidelity entropy is attributed to changes in information storage. Physically, this is plausible, given the difference between the past states recorded in media and the past states really occurred may be attributed to a difference in the same type of irrelevant information encoded in ground state wave functions in the same regime. That is, such a trajectory never crosses any boundary between different regimes even in the same phase.

We define  such a trajectory transversed by a fidelity mechanical system in the parameter space as a fidelity flow.  As argued, fidelity flows are irreversible.  Following from the second law, fidelity entropy $S(x_1,x_2)$ is monotonically increasing  and fidelity temperature $T(x_1,x_2)$ is monotonically decreasing along any fidelity flow: it starts from a point close to an unstable fixed point  and ends at a point close to a stable fixed point  in the parameter space, with fidelity temperature $T(x_1,x_2)$ being divergent at an unstable fixed point and fidelity entropy being a (local) maximum and fidelity temperature being zero at a stable fixed point.  Here, we emphasize that, only in this sense, does it make sense to speak of fidelity flows from an unstable fixed point to a stable fixed point.  This offers us a characterization of  both unstable and stable fixed points.

Fidelity flows, as defined above, may be regarded as an idealized form of RG flows in real space RG theories. Indeed, an effective Hamiltonian may be kept in the same form as the original Hamiltonian, if any irrelevant coupling constants are ignored. In addition, relevant information encoded in ground states is retained and irrelevant information encoded in ground states is discarded during the construction of an effective Hamiltonian, according to a prescribed criterion (cf. Appendix~\ref{pinch} for the notions of irrelevant and relevant information). Note that different real space RG schemes adopt different criteria, according to which high energy degrees of freedom are distinguished from low energy degrees of freedom~\cite{notunique}.  In this sense, there is a correspondence between fidelity flows and  RG flows in real space RG theories.  Hence, the irreversibility of RG flows in real space RG theories is the manifestation of the second law in fidelity mechanics.  This answers the second question concerning  the irreversibility of RG flows from an unstable fixed point to a stable fixed point, as raised in Section~\ref{intro}.  However, we emphasize that there is a subtle difference between RG flows in real space RG theories and RG flows in Zamolodchikov's $c$-theorem: the former only concern discarding a certain type of irrelevant information in a given regime, thus they never cross any boundary between different regimes, but the latter involve different critical points, due to the existence of a monotonically decreasing function interpolating between central charges of the ultraviolet and infrared conformal field theories. Therefore, it is necessary to extend the current definition of fidelity flows to accommodate this type of RG flows in fidelity mechanics.

\begin{figure}
     \includegraphics[angle=0,totalheight=3.6cm]{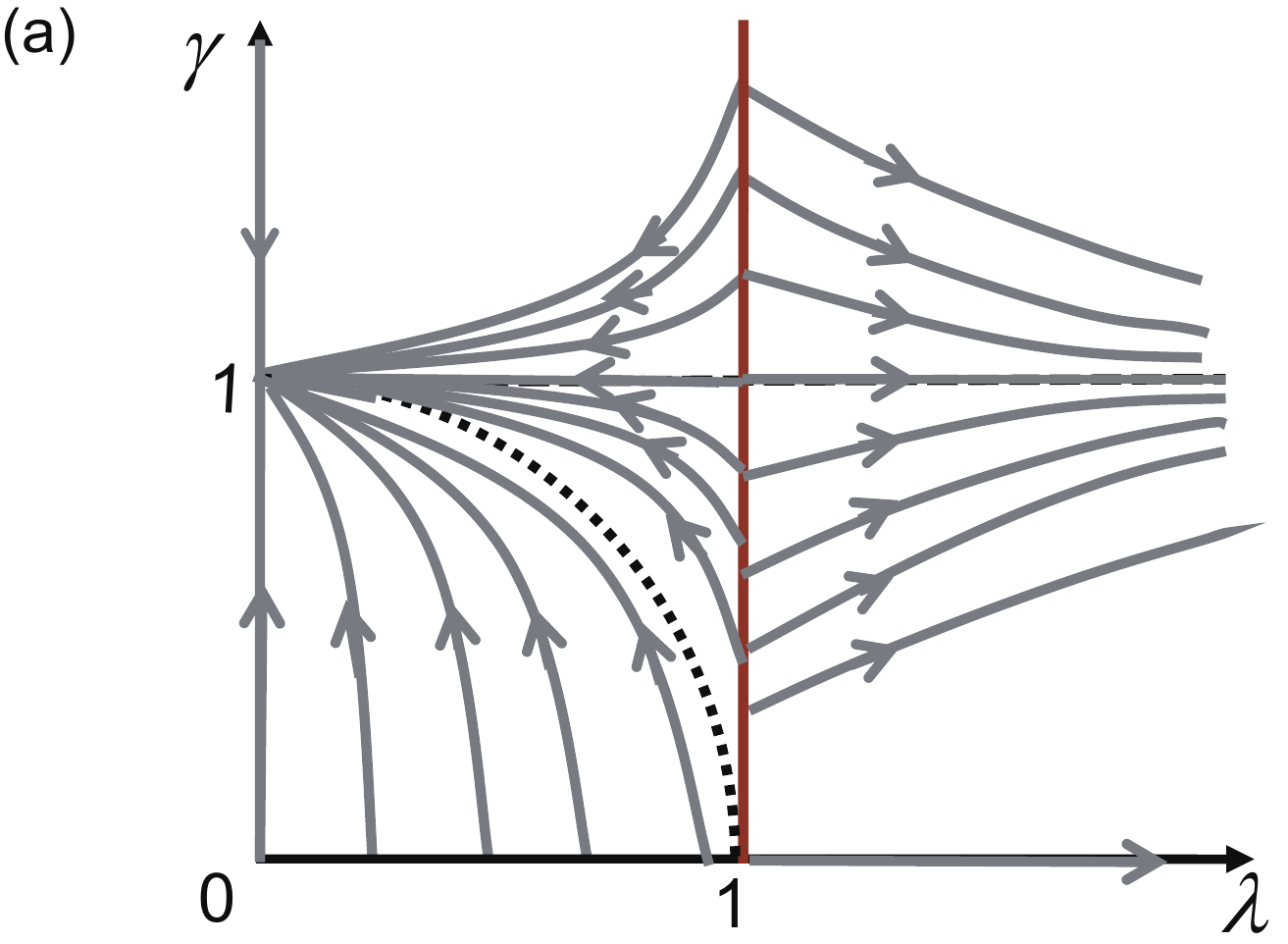}
        \includegraphics[angle=0,totalheight=3.6cm]{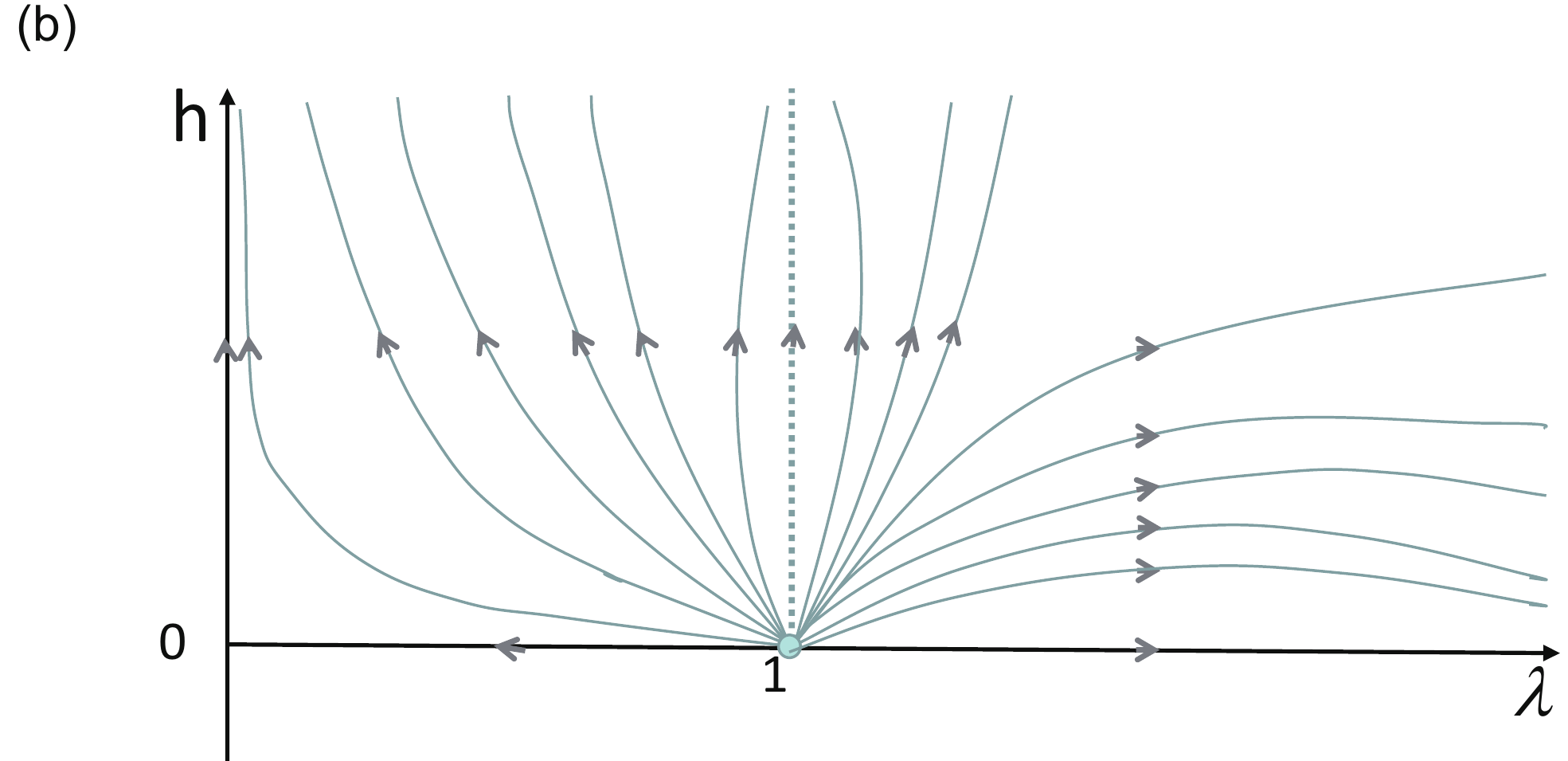}
    \includegraphics[angle=0,totalheight=4.2cm]{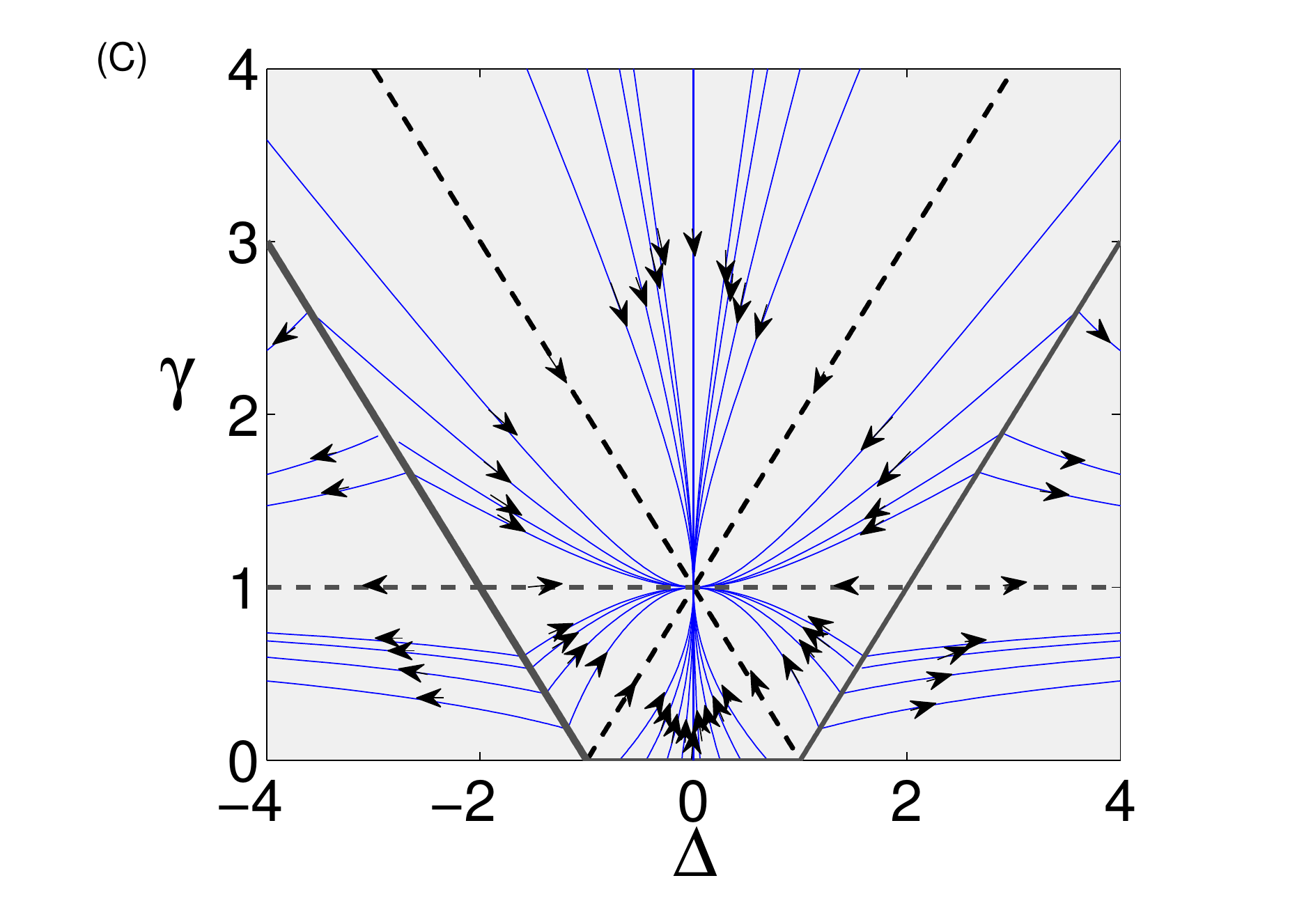}

       \caption{Typical fidelity flows for (a) the quantum XY model, (b) the transverse field quantum Ising chain in a longitudinal field, and (c) the spin-$1/2$ XYZ model.
       (a) For the quantum XY model, two stable fixed points are identified for the Ising universality class at (0,1) and ($\infty$, 1), which is protected by the $Z_2$ symmetry, and one stable fixed point for the PT universality class at ($\infty$, 0), which is protected by the $U(1)$ symmetry.  (b) For the transverse field quantum Ising chain in a longitudinal field, two stable fixed points are identified at (0, 0) and ($\infty$,0) for the Ising university class, which is protected by the $Z_2$ symmetry, and two stable fixed fixed points are identified at (1, $\infty$) and ($\infty$,0) for the universality class without any symmetry, when $\lambda \neq 0$ and $h\neq 0$. In addition, there is one stable fixed point at (0,$\infty$), which is protected by the $U(1)$ symmetry when $\lambda=0$, characterized by zero fidelity temperature and divergent fidelity entropy.
       (c) For the quantum spin-$1/2$ XYZ model,  three stable fixed points are identified at (0,1), ($\pm \infty$, 1) for the Gaussian universality class, and two stable fixed points are identified at (0,1) and ($\infty$, 0) for the KT university class. In addition, a stable fixed point at (-$\infty$,0) originates from a first-order QPT at $\Delta =-1$, protected by the $U(1)$ symmetry. }\label{flowxy}

\end{figure}

In Fig.~\ref{flowxy},  we sketch typical fidelity flows for the quantum XY chain, the transverse field quantum Ising chain in a longitudinal field, and the quantum spin-$1/2$ XYZ chain.  Here, we remark that fidelity flows for the quantum XXZ chain in a magnetic field are merely adapted  from fidelity flows for the quantum XXZ chain, a special case of the quantum XYZ model with $\gamma = 0$.   This follows from the fact that  the ground state wave functions do not depend on $h$ in the AF phase, whereas  the ground state wave function remains to be the same in the ${\rm FM}_-$ and ${\rm FM}_+$ phases, respectively.

(1) For the quantum XY model ($\lambda \geq 0,\gamma \geq 0$), two stable fixed points are identified for the Ising universality class at (0,1) and ($\infty$, 1), which is protected by the $Z_2$ symmetry, and one stable fixed point for the PT universality class at ($\infty$, 0), which is protected by the $U(1)$ symmetry.  For  the Ising universality class, a $U(1)$ symmetry emerges at (0,1) and ($\infty$, 1), in addition to the $Z_2$ symmetry, whereas for the PT universality class, a $Z_2$ symmetry, defined by  $\sigma^x_{2i} \leftrightarrow - \sigma^x_{2i}$,
$\sigma^y_{2i} \leftrightarrow - \sigma^y_{2i}$, $\sigma^z_{2i} \leftrightarrow  \sigma^z_{2i}$ and  $\sigma^x_{2i+1} \leftrightarrow \sigma^x_{2i+1}$,
$\sigma^y_{2i+1} \leftrightarrow \sigma^y_{2i+1}$, $\sigma^z_{2i+1} \leftrightarrow  \sigma^z_{2i+1}$,  emerges at ($\infty$, 0), in addition to the $U(1)$ symmetry. Generically, it is the emergence of such an extra symmetry at a stable fixed point that justifies why it is not accessible. On the other hand,
given two lines of critical points belonging to two different universality classes,  we interpret the disorder circle as a separation line between two different types of fidelity flows, with one type of fidelity flows starting from unstable fixed points with central charge $c=1$, and the other type of fidelity flows starting from unstable fixed points with central charge $c=1/2$. Note that both types of fidelity flows end at the same stable fixed point $(0,1)$, at which fidelity entropy $S(\lambda, \gamma)$ reaches a local maximum.

(2) For the transverse field quantum Ising chain in a longitudinal field ($h \geq 0$), two stable fixed points at (0, 0) and ($\infty$,0) are identified for the Ising universality class, which is protected by the $Z_2$ symmetry, and two stable fixed points at (1, $\infty$) and ($\infty$,0) are identified for the universality class without any symmetry, corresponding to $\lambda \neq 0$ and $h\neq 0$. In addition, there is one stable fixed point at (0,$\infty$), protected by the $U(1)$ symmetry when $\lambda =0$. Note that an extra $U(1)$ symmetry emerges at stable fixed points (0, 0),  ($\infty$,0), and (1, $\infty$), and an extra $Z_2$ symmetry, defined by $\sigma^x_{2i} \leftrightarrow  \sigma^x_{2i}$,
$\sigma^y_{2i} \leftrightarrow - \sigma^y_{2i}$, $\sigma^z_{2i} \leftrightarrow  - \sigma^z_{2i}$ and  $\sigma^x_{2i+1} \leftrightarrow \sigma^x_{2i+1}$,
$\sigma^y_{2i+1} \leftrightarrow \sigma^y_{2i+1}$, $\sigma^z_{2i+1} \leftrightarrow  \sigma^z_{2i+1}$, emerges at a stable fixed point (0,$\infty$). This justifies why a stable fixed point is not accessible.

(3) For the quantum spin-$1/2$ XYZ model ($\gamma \geq 0$),  three stable fixed points are identified for the Gaussian universality class at (0,1) and ($\pm \infty$, 1), and two stable fixed points are identified for the KT universality class at (0,1) and ($\infty$, 0).   In addition, a stable fixed point at (-$\infty$,0) originates from a first-order QPT at $\Delta =-1$, protected by the $U(1)$ symmetry.  The fact that ($\infty$, 1) and ($\infty$, 0) represent two different stable fixed points may be understood from both symmetry-breaking order and RG flows.  In fact,  a $Z^\sigma_2\times Z^\tau_2$ symmetry exists on the line  $\gamma=0$ for $\Delta>1$, where  $Z^\sigma_2$ is generated by $\sigma$: $\sigma^x_i \leftrightarrow \sigma^y_i$ and  $\sigma^z_i \leftrightarrow -\sigma^z_i$, and  $Z^\tau_2$ is generated by the one-site translation $\tau$: $\sigma ^\alpha_i \rightarrow \sigma ^\alpha_{i+1}$, with $\alpha =x, y, z$. However, only two-fold degeneracies exist, with each degenerate ground state invariant under the combined action $\sigma \; \tau$, which generates another $Z_2^{\sigma \tau}$. Thus, the symmetry group, which is spontaneously broken, is $Z^\sigma_2 \times Z^\tau_2/Z^{\sigma \tau}_2$. This is different from the cases with non-zero $\gamma$, in which the spontaneously broken symmetry group is $Z^\tau_2$. This also matches an observation that, for $\gamma =0$, there is a $U(1)$ symmetry, which protects the KT transition. Once $\gamma$ becomes nonzero,  the $U(1)$ symmetry is lost , and a continuous QPT changes from the KT universality class to the Gaussian  universality class.  In addition, it is the emergence of an extra symmetry at a stable fixed point, such as a $U(1)$ symmetry at  (0,1) and ($\pm \infty$, 1), and a $Z_2$ symmetry, defined by  $\sigma^x_{2i} \leftrightarrow - \sigma^x_{2i}$,
$\sigma^y_{2i} \leftrightarrow - \sigma^y_{2i}$, $\sigma^z_{2i} \leftrightarrow  \sigma^z_{2i}$ and  $\sigma^x_{2i+1} \leftrightarrow \sigma^x_{2i+1}$,
$\sigma^y_{2i+1} \leftrightarrow \sigma^y_{2i+1}$, $\sigma^z_{2i+1} \leftrightarrow  \sigma^z_{2i+1}$, at ($\infty$, 0) that justifies why a stable fixed point is not accessible.

\subsection{Miscellanea}~\label{mixeditems}

In our formulation of fidelity mechanics, an analogue of Landauer's principle at zero temperature has been assumed to keep internal logical consistency (cf. Appendix~\ref{connectionUT}), which states that,
in a fidelity mechanical system, to erase one bit of information at zero temperature, we need to do the minimum fidelity work $w(x)$: $w(x)=\pm T(x) \; \ln2 $.  Here, $T(x)$ characterizes quantum fluctuations at zero temperature, and $+/-$ corresponds to increasing/decreasing $e(x)$ with increasing $x$, respectively. This answers the third question raised in Section~\ref{intro}.

The fourth question, as raised in Section~\ref{intro}, concerns an observation that, during the construction of an effective Hamiltonian along any RG flow, an unlimited number of irrelevant coupling constants proliferate. In practice, this prevents access to a stable fixed point. According to fidelity mechanics, this simply follows from the third law in fidelity mechanics.  In fact, the third law may be rephrased as follows. {\it It is impossible to completely erase irrelevant information encoded in ground states in any given regime.}
Indeed, at a stable fixed point, there exists a singularity in fidelity mechanical state functions for all the models under investigation. In addition, such inaccessibility is also reflected in
the conventional Landau's SSB theory, since an extra symmetry always emerges at a stable fixed point, as discussed in the previous subsection.

In our opinion, the traditional definition based on a singularity in the ground state energy density is under-descriptive,  since it fails to signal QPTS in many quantum many-body systems~\cite{Wolf}. Moreover, if one defines QPTs as a singularity in any physical quantities, then such a definition is over-descriptive. In fact, according to this definition, factorizing fields should be mistakenly treated as QPTs.  In contrast, ground state fidelity per site offers us a proper criterion to detect QPTs, regardless of internal order arising from symmetry-breaking order and/or topological order (cf. Appendix~\ref{pinch}). Hence, a singularity in ground state fidelity per site is a proper criterion to define QPTs,  thus offering us an answer to the fifth question raised in Section~\ref{intro}.

Therefore, fidelity mechanics offers  a systematic framework to investigate QPTs in quantum many-body systems. It not only provides a characterization of unstable fixed points and stable fixed points, but also clarifies in what sense  a quantum many-body system flows from an unstable fixed point to a stable fixed point in the coupling parameter space by  erasing irrelevant information encoded in ground state wave functions along a fidelity flow. In Table~\ref{tab2}, we list basic notions in fidelity mechanics, with their counterparts in the conventional theories of local order parameters and RG flows.

\begin{table}

\begin{tabular}{lrlll}
\hline\hline
~ & Orders and Fluctuations	& Renormalization Group& 	Fidelity  Mechanics& \\
\hline
&Orders	&Low energy degrees of freedom  & Relevant information   &  \\
&Fluctuations	& 	High energy degrees of freedom   & Irrelevant information  &\\
&Local order parameters	& 	Effective Hamiltonians   & Fidelity mechanical quantities  & \\
&Transition points		& Unstable fixed points   & Divergent fidelity temperature & \\
&Ordered (disordered) states	&Stable fixed points   & Zero fidelity temperature &\\
&                                                &                               &and maximal fidelity entropy\\
&Not available                                          &RG flows                  &Fidelity flows\\
\hline\hline
\end{tabular}
\caption{\label{tab2} Fidelity mechanics offers a systematic framework to investigate  quantum critical phenomena. Here, we list basic notions in fidelity mechanics, with their counterparts in the conventional theories of local order parameters and RG flows.}
\end{table}

Fidelity mechanics might also offer a novel perspective to understand a long-standing mystery in physics: why should the thermodynamic,  psychological/computational and cosmological arrows of time align with each other? As discussed in  Appendix~\ref{timearrows}, only for a macroscopic time does it make sense to speak of an arrow of time. In fact, for any macroscopic time, there must exist a physical process which can, in principle, serve as a clock to track and record it. Therefore, one may single out the psychological/computational arrow of time as a master arrow of time. Then, it is necessary to develop a systematic theory to describe the psychological/computational arrow of time.  In fact, the psychological/computational arrow of time is to fidelity mechanics as the thermodynamical arrow of time is to thermodynamics. The fact that both entropy and fidelity entropy are monotonically increasing underlies why the thermodynamic arrow of time aligns with the psychological/computational arrow of time.  As for the cosmological arrow of time, we examine the universe from a fidelity mechanical perspective.  Since the universe itself is a perfect example of naturally occurring physical systems that act as memories or records, it is a fidelity mechanical system~\cite{thermalFM}. However, a peculiar feature arises, when one treats the universe as a fidelity mechanical system: there is no outside observer. That is, the universe itself is its own observer. Nevertheless, in contrast to classical and quantum mechanics, cosmology is a historical science~\cite{Spergel}.
As we have learned from cosmology, the universe may be traced back to a big bang, through different thresholds, such as the formation of solar systems, the formation of galaxies, the formation of stars, the formation of atoms, and the formation of subatomic particles. One may attribute these thresholds to dynamic phase transitions at different time scales during the evolution of the universe. Then, at each scale, a macroscopic time emerges, associated with a non-equilibrium physical process which can, in principle, serve as a clock. However, if one traces back further, the universe is so hot and so dense that it dissolves entirely into fluctuations at the Planck scale, with no regular oscillations left, thus no clock is available. As such, any macroscopic time ceases to exist, but a microscopic time remains, due to fluctuations.  Therefore, in the universe, fidelity entropy is monotonically increasing, and so entropy is monotonically increasing.  If one interprets dark energy as a result of  Landauer's principle~\cite{gough}, then the universe kepdf expanding since the big bang~\cite{darkmatter}.  In this sense, one may speculate that the cosmological arrow of time results from the psychological/computational arrow of time.

\section{Conclusion}~\label{conclusion}

In this work, fidelity mechanics has been formalized as a systematic framework to investigate QPTs in quantum many-body systems. Fidelity temperature has been introduced to properly quantify quantum fluctuations, which, together with fidelity entropy and fidelity internal energy, constitute three basic state functions in fidelity mechanics, thus enabling us to formulate analogues of the four thermodynamic laws and Landauer's principle at zero temperature. It is the notion of information storage that makes it possible to address a novel aspect of quantum information-information extractible by comparing the current state with the past states, both of which are stored in media. In fact,  for a given fidelity mechanical system,  we are able to quantify what amount of information may be recovered, due to information storage,  in terms of fidelity entropy. In addition, the importance of duality in fidelity mechanics has been clarified. Indeed, it plays a defining role in the determination of  a canonical form of the Hamiltonian for quantum many-body systems in fidelity mechanics. Fidelity flows have been defined, which are irreversible, as follows from the second law in fidelity mechanics.  On the other hand,  fidelity flows may be interpreted as an alternative form of RG flows, and allow us to characterize both stable and unstable fixed points: divergent fidelity temperature for unstable fixed points and zero fidelity temperature and maximal fidelity entropy for stable fixed points.

A  detailed analysis of fidelity mechanical state functions has been presented for a number of models, these are the quantum XY model, the transverse field quantum Ising chain in a longitudinal field, the spin-$1/2$ XYZ model and the XXZ model in a magnetic field, with extensive simulations of quantum many-body systems in terms of a tensor network algorithm in the context of matrix product states.
Rich physics has been unveiled even for  these well-studied models. First, for the quantum XY chain, we resolved a confusing point raised in Ref.~\cite{Illuminati}; as claimed, the so-called long-range entanglement driven order exists in the disordered regime, suggesting a QPT occurs on the disorder circle $\lambda^2+\gamma^2=1$. However,  the same long-range entanglement driven order also exists for $\gamma \geq 1$ at $\lambda =0$, due to the presence of duality between $\gamma \geq 1$ and $\gamma \leq 1$ at $\lambda =0$.
In our opinion, no QPT occurs on the disorder circle, but  a fidelity mechanical ``phase transition" does occur, since fidelity mechanical state functions exhibit singularities on the disorder circle, which has been interpreted as a separation line between two different types of fidelity flows, with one type of fidelity flows starting from unstable fixed points with central charge $c=1$, and the other type of fidelity flows starting from unstable fixed points with central charge $c=1/2$.  Both types of fidelity flows end at the same stable fixed point $(0,1)$, at which fidelity entropy $S(\lambda, \gamma)$ reaches its local maximum.  Another remarkable feature is that fidelity temperature is zero on the disorder circle, as it should be, since no quantum fluctuations exist in a factorized state. However, at  the PT transition point (1, 0), fidelity temperature $T$ is not well-defined. In fact, it takes any value, ranging from 0 to $\infty$, depending on how it is approached. This bears a resemblance to a previous result~\cite{korepin} that entanglement entropy is not well-defined at the PT transition point (1, 0); its value depends on how the PT transition point (1, 0) is approached.  Second, for the transverse field quantum Ising chain in a longitudinal field,  there are stable fixed points at $(0,0)$, $(0,\infty)$, $(\infty, 0)$, and $(1, \infty)$. The existence of stable fixed points $(0,0)$ and $(\infty, 0)$ is protected by the $Z_2$ symmetry when $h=0$, whereas the existence of stable fixed points $(0, \infty)$ and $(1, \infty)$ may be interpreted as a consequence of the variation of the symmetry group with $\lambda$: $U(1)$ for $\lambda =0$, and none for  $\lambda \neq 0$, when $h\neq 0$. In particular, the presence of a stable fixed point at $(1, \infty)$ might also be related to a well-known fact that, at $\lambda =1$ but nonzero $h$,  a massive excitation spectrum involves eight massive particles, which shows a deep relation with $E_8$ algebraic structure~\cite{zamolodchikov}.  Third, for the quantum spin-$1/2$ XYZ model, five different dualities have been identified, which enable us to reproduce the ground state phase diagram. Fourth, for the quantum XXZ model in a magnetic field, at the phase boundary between the XY phase and  the AF phase,  fidelity temperature is not well-defined, ranging from a finite value to $\infty$.  That is, a QPT at this phase boundary is an intermediate case interpolating between a KT transition  and a PT transition, which represents a new universality class.

Fidelity mechanics not only characterizes quantum critical phenomena arising from symmetry-breaking order, as illustrated in this work, but also characterizes quantum critical phenomena arising from topological order. The latter will be a topic of forthcoming publications, addressing the Haldane phase - a symmetry protected topological phase for the quantum XYZ chain with integer spin~\cite{qqshi}  and  a quantum spin liquid for the quantum spin-$1/2$ Kitaev model on a honeycomb lattice~\cite{qqshi1}, respectively.
We anticipate that fidelity flows always start from a point close to an unstable fixed point  and end at a point close to a stable fixed point, and never cross any characteristic lines as boundaries between different regimes in the parameter space, regardless of symmetry-breaking and/or topological order. In particular, the Hamiltonians are unitarily equivalent at all stable fixed points, even in a topologically ordered phase, with the only difference lying in the fact that the unitary operator involved is local for symmetry-breaking order and non-local for topological order. This is consistent with a heuristic argument that, for quantum many-body systems, ground state wave functions may be represented in terms of the multi-scale entanglement renormalization ansatz~\cite{mera}, with a top tensor being a (unentangled) product state for a SSB ordered state and with a top tensor being an entangled state (characterized by a non-local unitary operator) for a topologically ordered state.

Fidelity mechanics has been formalized as an analogue of black hole thermodynamics. In addition to the formal similarity discussed in the text,  they share one more common feature: both fidelity heat capacity in fidelity mechanics and heat capacity in  black hole thermodynamics are negative. Nevertheless, the formal  similarity between critical points and black holes, as unveiled, is not surprising, in the sense that both QPTs and black holes share singularities as their key ingredients. Moreover,  a brief speculative discussion has been presented,  justifying why the thermodynamic, psychological/computational and cosmological arrows of time should align with each other in the context of fidelity mechanics, with the psychological/computational arrow of time  being singled out as a master arrow of time.

{\it Acknowledgements.}
 We acknowledge enlightening discussions with Murray Batchelor,  Vladimir Bazhanov, Daniel Braak,  Anthony Bracken, Gavin  Brennen, Sam Young Cho, Fabian Essler, Andreas Kl\"umper, Vladimir Mangazeev, German Sierra, Javier Rodriguez-Laguna and Ming-Wei Wu. Talks based on this work have been delivered at The 6th workshop on Quantum Many-Body Computation at Beijing Computational Science Research Center, Beijing, China (April 2016), International Workshop on Ultra-Strong Light-Matter Interactions: Theory and Applications to Quantum Information, Bilbao, Spain (September 2016), Coogee'17 Sydney Quantum Information Theory Workshop, Sydney, Australia (February 2017), The 7th Workshop on Quantum Many-Body Computation, University of Chinese Academy of Sciences, Beijing, China (May, 2017) and at University of Science and Technology of China, Hefei, China (April 2016), Universidad Aut\'onoma de Madrid, Madrid, Spain (September 2016),  Scuola Internazionale Superiore di Studi Avanzati, Trieste, Italy (October 2016), University of Oxford, Oxford, United Kingdom (November 2016), and Australian National University, Canberra, Australia (February and May 2017).

 \appendix
  \renewcommand{\appendixname}{Appendix~}

\section{Relevant and irrelevant information via ground state fidelity per site}\label{pinch}

Fidelity, a basic notion in quantum information science, is a measure of the similarity between two quantum states $|\psi (x)\rangle$ and $|\psi(y)\rangle$.  Mathematically, it is defined as the absolute value of the overlap between two pure states $F(x,y)=|\langle\psi(y)|\psi(x)\rangle|$. It should be emphasized that, as a convention, we use  $x$ and $y$ to denote two different values of the {\it same} control parameter.

For quantum many-body systems, two ground states are always distinguishable (orthogonal) in the thermodynamic limit. As such,  fidelity between these two states vanishes.
For a large but finite lattice,  fidelity $F_N(x,y)$ scales as $d^N(x,y)$, with $N$ being the total number of lattice sites, and $d(x,y)$  being a scaling parameter. Physically, $d(x,y)$ may be interpreted as ground state fidelity per site:  $d(x,y)=\lim_{N\rightarrow\infty}{F_N^{1/N}(x,y)}$,
which is well-defined even in the thermodynamic limit.  Ground state fidelity per site $d(x,y)$ enjoys some properties inherited from fidelity $F(x,y)$: (i)
symmetry under interchange $d(x,y)=d(y,x)$;
(ii) normalization $d(x,x)=1$;
(iii) range $0\leq d(x,y)\leq 1$.

\begin{figure}
  \centering
\includegraphics[angle=0,totalheight=3.2cm]{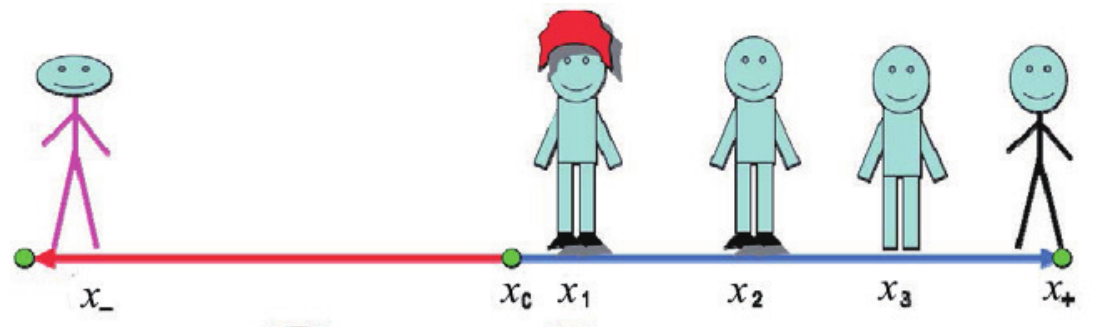}
\caption{A cartoon picture to illustrate the notions of relevant and irrelevant information in fidelity mechanics.  Here, information encoded in bare bodies is relevant, whereas information encoded in hats, clothes and shoes is irrelevant.  RG flows from an unstable fixed point $x_c$ to two stable fixed points $x_-$ and $x_+$ are also depicted, to justify that the notions of irrelevant and relevant information are introduced as counterparts of high energy degrees of freedom and low energy degrees of freedom in real space RG theories.  We remark that irrelevant information originated from the same unstable fixed point is identical, but relevant information at the two stable fixed points is different. }
\label{relevant}
\end{figure}

As demonstrated in Refs.~\cite{Zhou1,Zhou2}, QPTs may be detected through singularities exhibited in ground state fidelity per site $d$, regardless of internal order arising from symmetry breaking order and/or topological order.  The reason why ground state fidelity per site $d(x,y)$ may be used to signal QPTs is due to the fact that it distinguishes  {\it relevant information}  from {\it irrelevant information} encoded in ground state wave functions for a quantum many-body system. Here,  relevant information is defined to be a counterpart of orders in Landau's SSB theory. That is, any information encoded in a ground state wave function corresponding to an ordered (disordered) state is relevant. In contrast, irrelevant information is defined to be a counterpart of fluctuations in Landau's SSB theory. Therefore, any information encoded in a ground state wave function that makes it {\it deviate} from a ground state wave function at an ordered (disordered) state is irrelevant. A remarkable fact is that such a deviation may be quantified by ground state fidelity per site. In this scenario, a critical point is simply characterized as follows.  {\it At a critical point, relevant information is covered up by irrelevant information}.  In addition to Landau's SSB theory,  RG flows may also be used to justify the introduction of irrelevant and relevant information as counterparts of high energy degrees of freedom and low energy degrees of freedom in the context of fidelity approach to QPTs (see a cartoon picture in Fig.~\ref{relevant}).

Two typical examples are the transverse field quantum Ising chain and the two-dimensional spin-$1/2$  Kitaev model on a honeycomb lattice~\cite{Kitaev}. The former exhibits a continuous symmetry breaking QPT and the latter exhibits a topological QPT.

\begin{figure}
  \centering
\includegraphics[angle=0,totalheight=4.5cm]{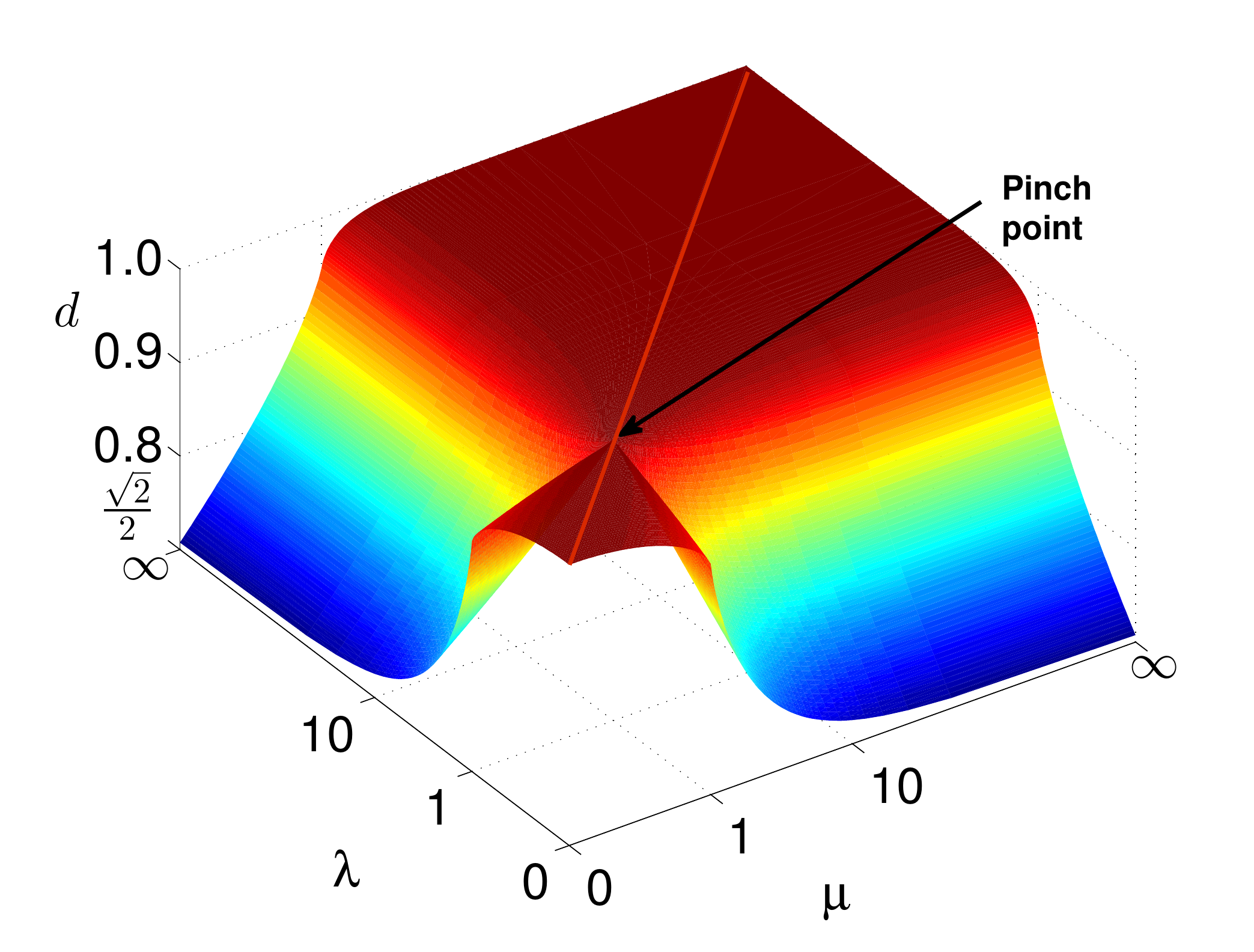}
\caption{Ground state fidelity per site $d(\lambda,\mu)$ is shown as a function of $\lambda$ and $\mu$ for the transverse field quantum Ising chain, which exhibits a pinch point at  $(1,1)$. A pinch point is defined to be an intersection point of two singular lines $\lambda=1$ and $\mu=1$.}\label{is2d}
\end{figure}

The transverse field quantum Ising chain is a special case of the quantum XY model, described by the Hamiltonian (\ref{xyham}), with $\gamma =1$. We choose $\lambda$ as a control parameter.  As shown in Fig.~\ref{is2d}, a critical point $\lambda_c = 1$ is characterized as a pinch point $(1, 1)$~\cite{Zhou1}.  Generically, a pinch point is defined as an intersection point between two singular lines $\lambda = \lambda_c$ and $\mu = \lambda_c$ .

For the two-dimensional spin-$1/2$ Kitaev model on a honeycomb lattice,  the Hamiltonian takes the form
\begin{equation}
H(J_x,J_y,J_z)=-J_x\sum_{x-{\rm bonds}}{\sigma_i^x\sigma^x_{j}}-J_y\sum_{y-{\rm bonds}}{\sigma_i^y\sigma^y_{j}}-J_z\sum_{z-{\rm bonds}}{\sigma_i^z\sigma^z_{j}},
\label{kitaevham}
\end{equation}
where $J_\alpha$ are interaction parameters and $\sigma_J^{\alpha}$
are the Pauli matrices at the site $j$, with $\alpha=x,y,z$.

\begin{figure}
  \centering
\includegraphics[angle=0,totalheight=4.5cm]{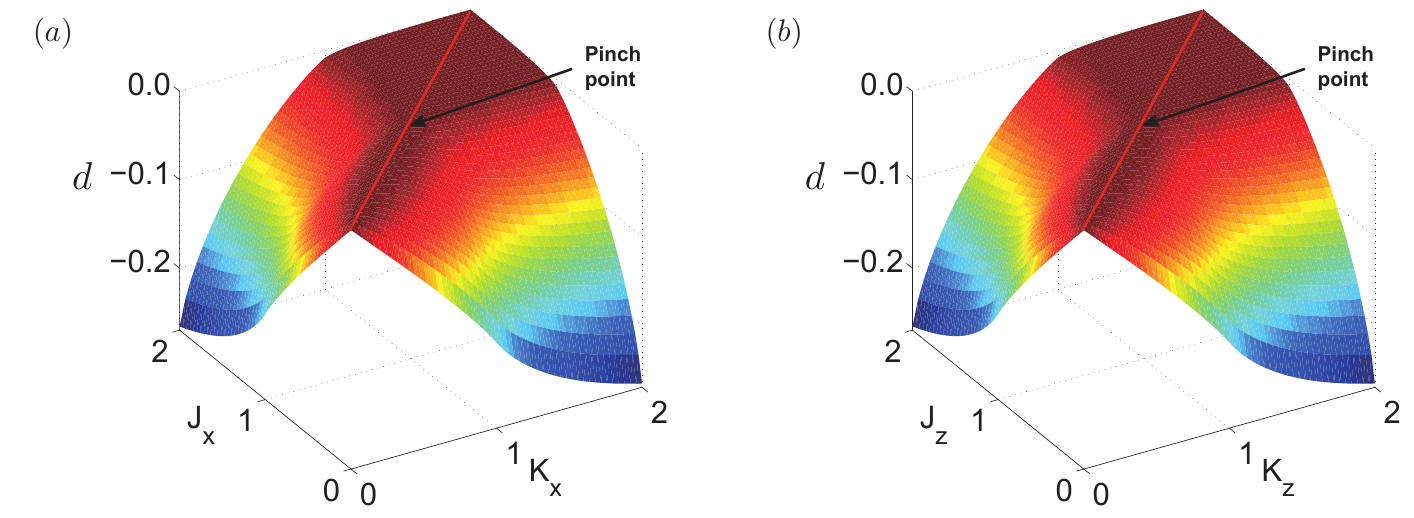}
\caption{
(a) Ground state fidelity per  site
$d(J_x, K_x)$ is shown as a function of $J_x$ and $K_x$, with $J_y =K_y =
J_z = K_z = 1/2$,   for the spin-$1/2$ Kitaev model on a honeycomb lattice.  It exhibits a pinch point at $(J_{xc}, J_{xc}) = (1, 1)$. (b)
Ground state fidelity per site $d(J_z, K_z)$ is shown as a
function of $J_z$ and $K_z$,
with $J_x = K_x = J_y = K_y = 1/2$, for the spin-$1/2$ Kitaev model on a honeycomb lattice. It exhibits
a pinch point at $(J_{zc}, J_{zc}) = (1, 1)$.  Here,  $d(J_x, K_x)$  and $d(J_z, K_z)$ have been plotted in the logarithmic scale.
}\label{kt2d}
\end{figure}

For this model, we choose $J_x$ or $J_z$ as a control parameter.  A critical point  at $J_{xc}=1$ is reflected as a pinch point in  ground state fidelity per site, $d(J_x, K_x)$,  at ($J_{xc}=1, J_{xc}=1$) for fixed $J_y =K_y = J_z = K_z = 1/2$ and  as a pinch point in  ground state fidelity per site, $d(J_z, K_z)$,  at ($J_{zc}=1, J_{zc}=1$)   for fixed $J_x = K_x = J_y = K_y = 1/2$, as shown in Fig.~\ref{kt2d}\;(a) and Fig.~\ref{kt2d}\;(b), respectively~\cite{Zhao}. We emphasize that the two plots are essentially the same, due to the symmetry under exchanges: $(x,y,z) \leftrightarrow (y,z,x) \leftrightarrow (z,x,y) $.

 \section{Ground state fidelity from tensor networks: matrix product states}\label{fidelityfromTN}

In this Appendix, we describe an efficient way to evaluate ground state fidelity per site for quantum many-body systems in the context of tensor network representations.  Here, we
restrict ourselves to matrix product states, which is suitable to quantum many-body systems in one spatial dimension.

\begin{figure}
\includegraphics[angle=0,totalheight=5.8cm]{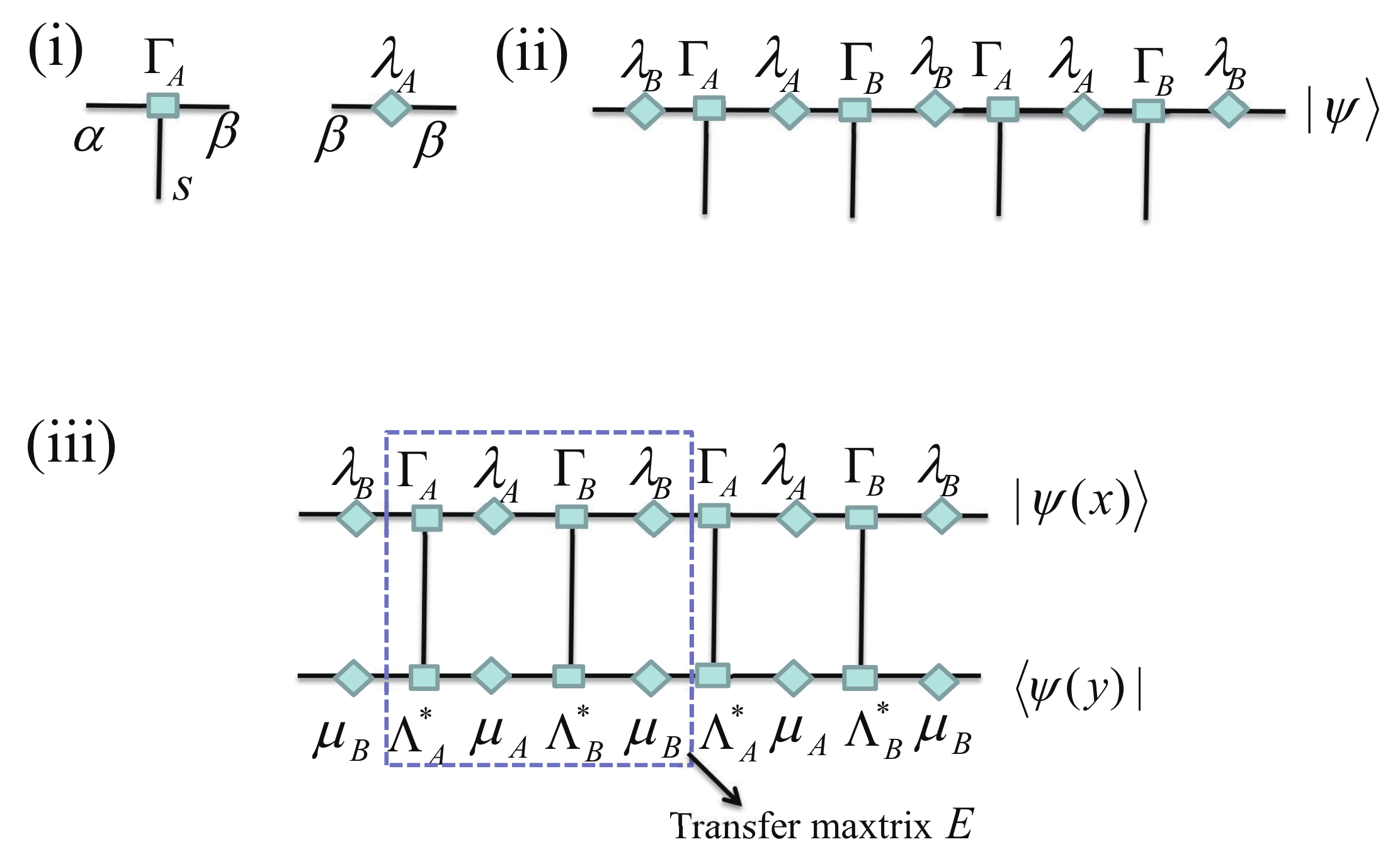}
\caption{(i)  A three-index tensor $\Gamma_A$ is labeled
by two bond indices,  denoted as $\alpha$ and $\beta$, and one physical index $s$, and $\lambda_A$ is a
singular value diagonal  matrix,  taking nonzero values only when two bond indices $\alpha$ and $\beta$ are the same. (ii) A two-site translation-invariant matrix product state representation for a ground state wave function $|\psi\rangle$,
consisting of alternating copies of the
tensors $\Gamma_A$, $\lambda_A$, $\Gamma_B$, and $\lambda_B$, with each tensor
connected through two bond indices.
(iii) A matrix product state representation for ground state fidelity between  $|\psi(x)\rangle$ and $|\psi(y)\rangle$. The former is represented by $\Gamma_A$, $\lambda_A$,
$\Gamma_B$, and $\lambda_B$, and the latter is represented by $\Lambda_A$, $\mu_A$,
$\Lambda_B$, and $\mu_B$,  respectively. Here, $E$ is the transfer matrix. }\label{imps}
\end{figure}

Consider a quantum many-body system characterized by a translation-invariant
Hamiltonian $H$, with the nearest-neighbor interactions:
$H=\sum_i {h_{i i+1}}$, where $h_{i i+1}$ is the Hamiltonian density.
In the context of the infinite time-evolving block
decimation algorithm~\cite{TN2}, a ground state wave function $|\psi\rangle$ is
translation-invariant under two-site shifts.  Then one only needs two
three-index tensors $\Gamma_A$ and $\Gamma_B$ and two singular value diagonal matrices $\lambda_A$ and
$\lambda_B$ to represent a ground state wave function $|\psi\rangle$, as shown in Fig.~\ref{imps}.
Here,  three-index tensors $\Gamma_A$ and $\Gamma_B$ are labeled by one physical
index $s$ and two bond indices $\alpha$ and $\beta$,  and $\lambda_A$ and $\lambda_B$ are real and diagonal matrices.
Note that the physical index $s$ runs over
$1,\ldots,\mathrm{d}$, and each bond index takes $1,\ldots,\chi$, with
$\mathrm{d}$ being the physical dimension, and $\chi$ being  the bond
dimension.

Hence, for two ground states, $|\psi(x)\rangle$ and $|\psi(y)\rangle$,  ground state fidelity $F(x,y)$ between $|\psi(x)\rangle$ and $|\psi(y)\rangle$
is represented as a tensor network,  with $E$ being a transfer matrix, which is shown in Fig.~\ref{imps}(iii).  Here,
$|\psi(x)\rangle$ is represented by $\Gamma_A$, $\lambda_A$,
$\Gamma_A$, and $\lambda_B$, and $|\psi(y)\rangle$ is represented by
tensors $\Lambda_A$, $\mu_A$, $\Lambda_B$, and $\mu_B$, respectively.
Therefore, ground state fidelity per site $d(x,y)$ is, by definition, related to  the dominant eigenvector $\lambda_{max}(x,y)$ of the transfer matrix $E$:
$d(x,y)=\sqrt {\lambda_{max}(x,y)}$.

The argument may be extended to projected-entangled pair states~\cite{TN1}, suitable to represent ground state wave functions for quantum many-body systems in two and higher spatial dimensions~\cite{Zhou2}.

\section {The infinite time-evolving block decimation algorithm}~\label{itebd}

In this Appendix, we briefly recall the infinite time-evolving block decimation algorithm~\cite{TN2}. The algorithm is based on a matrix product state representation of ground state wave functions, to simulate infinite-size translation-invariant quantum many-body systems in one spatial dimension.

Consider a quantum many-body system described by the Hamiltonian $H$
\begin{equation}
H=\sum_i {h_{i,i+1}},
\end{equation}
where $h_{i,i+1}$ is the Hamiltonian density describing the nearest-neighbor interactions.  A two-site translation-invariant ground state wave function takes the form
\begin{equation}
\ket{\psi}=\sum_{s}(...\Gamma_A^{s_{2i-1}}\lambda_A\Gamma_B^{s_{2i}}\lambda_B\Gamma_A^{s_{2i+1}}\lambda_A\Gamma_B^{s_{2i+2}}\lambda_B...)\ket{...s_{2i-1}s_{2i}s_{2i+1}s_{2i+2}...}.
\label{impswf}
\end{equation}
Here, $s$ is a physical index, $\Gamma_A$ and $\Gamma_B$
are three-index tensors on odd and even sites, $\lambda_A$ and $\lambda_B$  are $\chi \times \chi$  singular value
diagonal matrices on odd and even bonds, respectively.
If a random matrix product state $\ket{\psi_{0}}$ in the form (\ref{impswf}) is chosen as
an initial state, then  the imaginary
time evolution yeilds
$\ket{\psi_{\tau}}$ at imaginary time $\tau$, which takes the form
\begin{equation}
\label{initialstate}
\ket{\psi_{\tau}}=\frac{\exp(-H\tau)\ket{\psi_{0}}}{\|\exp(-H\tau)\ket{\psi_{0}}\|}.
\end{equation}
If $\tau\rightarrow\infty$, then a matrix product state representation of a ground-state wave function is
projected out, as long as the initial state is not orthogonal to the genuine ground state. The algorithm is efficient, with the computational costs being proportional to $\chi^3$.

\begin{figure}
     \includegraphics[angle=0,totalheight=4cm]{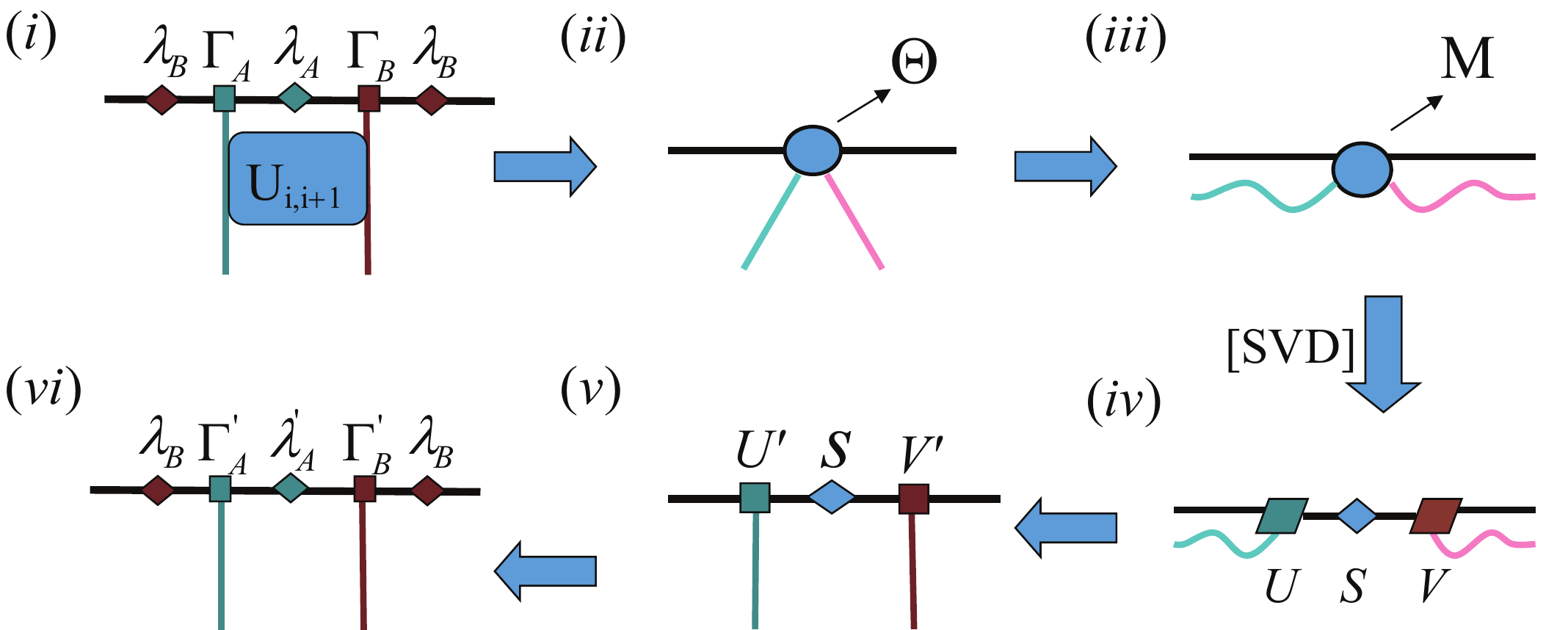}
\caption{ The procedure to update three-index tensors $\Gamma_A$ and $\Gamma_B$ and two singular value diagonal matrices $\lambda_A$ and $\lambda_B$
via absorbing a two-site gate $U_{i,i+1}$. (i) A two-site gate $U_{i,i+1}$ is applied
onto a matrix product state, represented in terms of $\Gamma_A$, $\lambda_A$, $\Gamma_B$ and $\lambda_B$. (ii) A four-index tensor $\Theta$ is formed by contracting the tensors $\Gamma_A$, $\lambda_A$, $\Gamma_B$ and $\lambda_B$ with
the two-site gate $U_{i,i+1}$. (iii) Reshape the four-index tensor into a matrix $M$. (iv) A singular value decomposition
is performed for the matrix $M$,  which yields $U$, $V$ and $\lambda'_A$. The latter is formed from the $\chi$ largest singular
values of $M$, due to truncation.  That is, $\lambda_A$ is updated to $\lambda'_A$. (v) Reshape two matrices $U$ and $V$ into two three-index
tensors $U'$ and $V'$. (vi) Recover the singular value diagonal matrix $\lambda_B$, thus $\Gamma_A$ and $\Gamma_B$ are updated to $\Gamma'_A$ and $\Gamma'_B$, respectively.}
\label{svdupdate}
\end{figure}

The imaginary time evolution operator $\exp(- H\tau)$ for $\tau\rightarrow \infty$ can be implemented by a series of
operators $ \exp(-H\delta \tau)$ over a time slice $\delta \tau$, where $\tau = M
\delta \tau$, with $\delta \tau \rightarrow 0$ and $M \rightarrow \infty$.  When
$\delta \tau$ is infinitesimal, the evolution operator $\exp(- H\tau)$ can be decomposed into a sequence
of the two-site gates $U_{i,i+1} = \exp(-h_{i,i+1}\delta \tau)$, as a result of the Trotter-Suzuki decomposition.
The translational invariance of the Hamiltonian allows us to consider two different types of the two-site gates $U_e$ and $U_o$, corresponding to even and odd sites, where $U_e=\exp(-h_{2i,2i+1}\delta \tau)$ and $U_o=\exp(-h_{2i+1,2i}\delta \tau)$, respectively.
A peculiar feature of such a decomposition is that all the two-site gates in $U_e$ and $U_o$ are commutative
with each other. Therefore, the problem to implement the imaginary time evolution reduces to absorb a
two-site gate acting on a matrix product state.  This is achieved in terms of the singular value decomposition, as described in Fig.~\ref{svdupdate}.
Following the procedure,  $\Gamma_A$, $\lambda_A$, $\Gamma_B$ and $\lambda_B$ are updated to $\Gamma'_A$, $\lambda'_A$, $\Gamma'_B$ and $\lambda'_B$ via absorbing a two-site gate $U_{i,i+1}$.  In the algorithm, the initial imaginary time slice $\delta \tau$ may be set as, e.g., $10^{-1}$, and then gradually decrease to a relatively small value.  During the simulation,
$\Gamma_A$, $\lambda_A$, $\Gamma_B$ and $\lambda_B$ are updated repeatedly until the singular value  diagonal matrices $\lambda_A$ and $\lambda_B$ converge.

\begin{figure}
     \includegraphics[angle=0,totalheight=3cm]{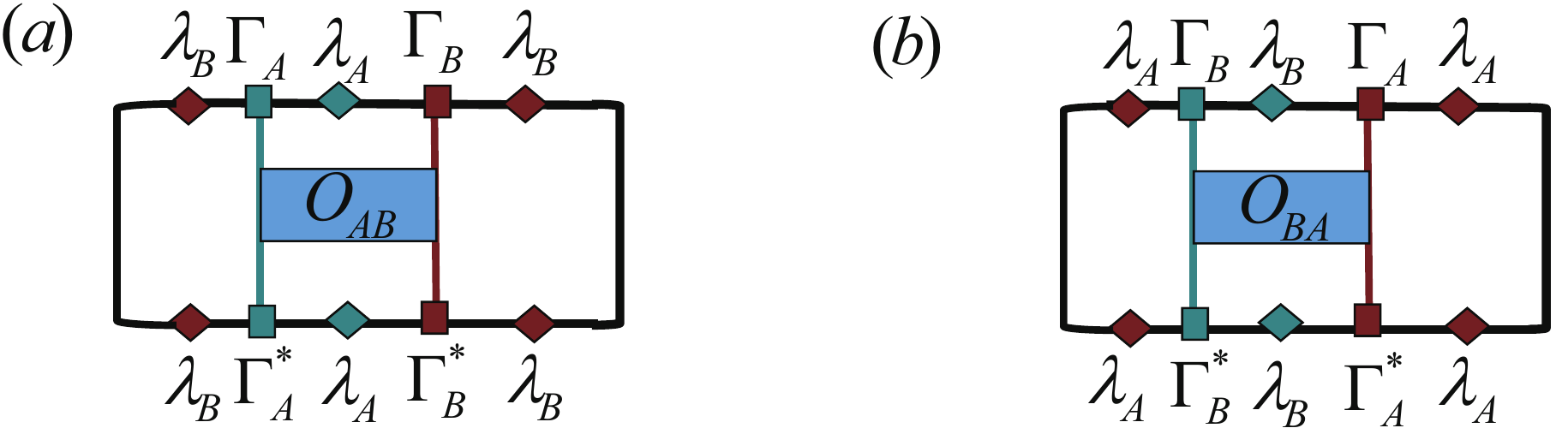}
\caption{The expectation value of a two-site operator $\langle O_{i,i+1}\rangle$ for a two-site translation-invariant matrix product state.
 (a) The expectation value of a two-site operator $\langle O_{AB}\rangle$ is computed by contracting tensors $\Gamma_A$, $\Gamma_B$,  $\Gamma^*_A$, $\Gamma^*_B$, $\lambda_A$, $\lambda_B$, and
a two-site operator $O_{AB}$. (b) The expectation value of two-site operator $\langle O_{BA}\rangle$ is computed by contracting tensors $\Gamma_A$, $\Gamma_B$, $\Gamma^*_A$, $\Gamma^*_B$, $\lambda_A$, $\lambda_B$, and
a two-site operator $O_{BA}$.}
\label{impsenergy}
\end{figure}

Once a matrix product state representation of a ground state wave function is generated, one may compute the expectation value of  any two-site operators $O_{AB}$ and $O_{BA}$ by contracting the tensors, as described in Fig.~\ref{impsenergy} (a) and Fig.~\ref{impsenergy} (b), respectively.
Specifically, for $O_{AB} = O_{BA}\equiv h_{i,i+1}$, this yields the ground-state energy density $e=(e_{AB} + e_{BA})/2$, where $e_{AB}$ and $ e_{BA}$ represents the expectation values of the Hamiltonian density  $h_{i,i+1}$ by contracting the tensors in Fig.~\ref{impsenergy} (a) and Fig.~\ref{impsenergy} (b), respectively.

\section {Geometric entanglement}~\label{gedensity}

Geometric entanglement has been introduced as a measure of multi-partite entanglement present in a quantum state~\cite{tcwei},  As it turns out,
it serves as an alternative indicator to locate critical points for quantum many-body  systems undergoing QPTs.  Moreover, geometric entanglement may be used as an indicator to identify factorized states, given that it must vanish for any unentangled states~\cite{geometricGE}.

For a pure quantum state $|\psi\rangle$ with $N$ parties, geometric entanglement $E(|\psi\rangle)$
takes the form
\begin{equation}
 E(|\psi\rangle)=- 2 \log_2 \Lambda_{{\rm max}}, \label{ge}
\end{equation}
where $\Lambda_{\rm max}$ is the maximum fidelity between $|\psi\rangle$ and all possible
separable (unentangled) and normalized states $|\phi\rangle$:
\begin{equation}
 \Lambda_{{\rm max} }={\rm \max_{|\phi\rangle }} \; |\langle\psi|\phi\rangle|.
\end{equation}

Physically, this amounts to identifying the closest separable (unentangled) state to  $|\psi\rangle$.
Then, geometric entanglement per party ${\mathcal E}_{N}(|\psi\rangle)$ is defined as
\begin{equation}
 {\mathcal E}_{N}(|\psi\rangle)=N^{-1}E(|\psi\rangle).
\end{equation}
Or, equivalently, we have
\begin{equation}
{\mathcal E}_{N}(|\psi\rangle)=- 2 \log_2 {\lambda^{{\rm
max}}_{N}},
\end{equation}
where $\lambda^{{\rm  max}}_{N}$ is the maximum fidelity per party, which is defined as
\begin{equation}
 \lambda^{{\rm max}}_{N}=\Lambda_{{\rm max}}^{\frac{1}{N}}.
\end{equation}

For a quantum many-body system,  one may introduce ground state geometric entanglement per unit cell ${\mathcal E}_{N}(|\psi\rangle)$, which is well-defined even in the thermodynamic limit. Our aim is to find an efficient way to compute the maximum fidelity between a ground state wave function $|\psi\rangle$ and all possible
separable (unentangled) and normalized states $|\phi\rangle$ in the context of tensor networks.  Here, we restrict ourselves to a matrix product state representation of ground state wave functions for quantum many-body systems in one spatial dimension, although the extension to projected-entangled pair states is straightforward for quantum many-body systems in two and higher spatial dimensions.
\begin{figure}
\includegraphics[angle=0,totalheight=5cm]{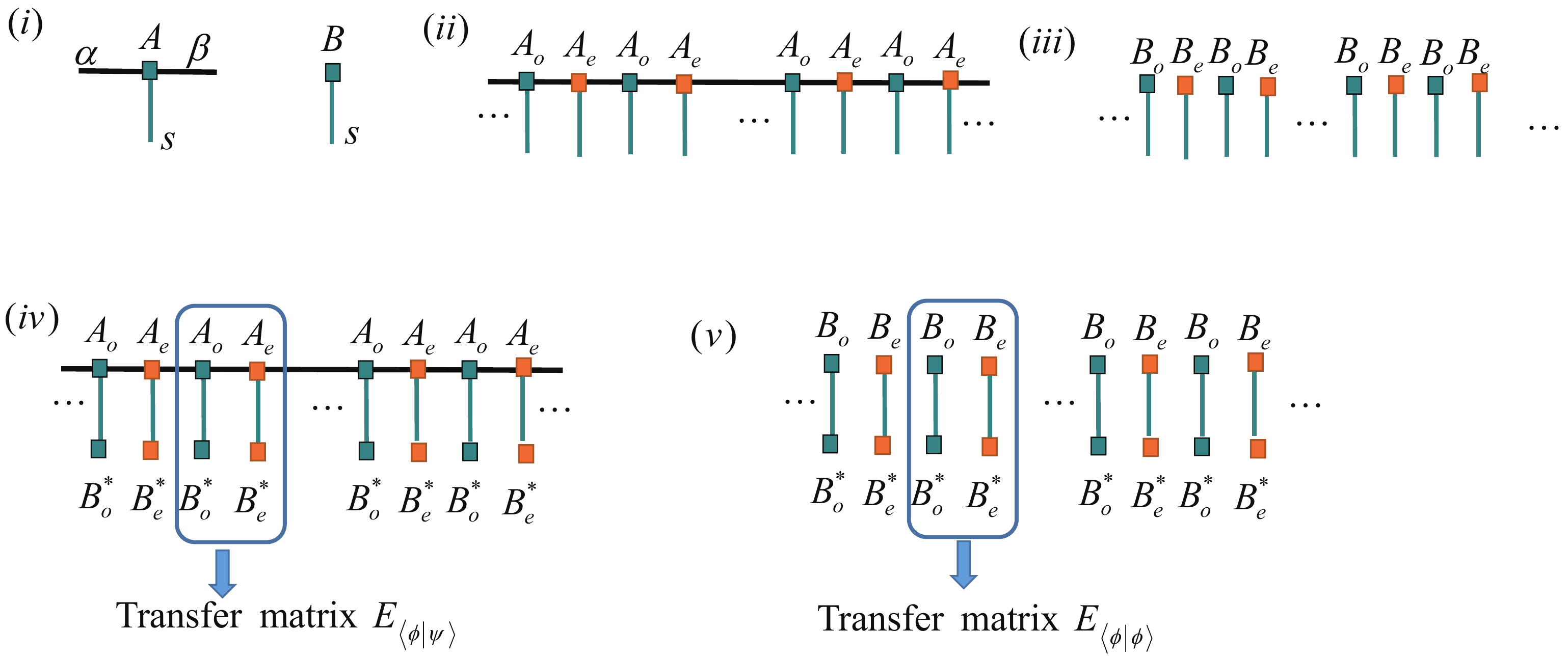}
\caption{(color online) (i) A three-index tensor $A$  and a one-index tensor $B$. Here, $s$ is the physical
index, $\alpha$ and $\beta$ are the inner bond indices.  (ii) A matrix product state
representation for a ground-state wave function $|\psi\rangle$. Here, two three-index tensors $A_{o}$
and $A_{e}$ are attached to odd and even sites, respectively.
(iii) A matrix product state representation for a separable state
$|\phi\rangle$. Here, two one-index tensors
$B_{o}$ and $B_{e}$ are attached to odd and even sites,
respectively.  (iv)  The fidelity between a ground-state wave
function $|\psi\rangle$ and a separable state $|\phi\rangle$.  The
transfer matrix $E_{\langle\phi|\psi\rangle}$ is
constructed from the tensors $A_{o}$, $A_{e}$, $B_{o}^{*}$ and
$B_{e}^{*}$. (v) The norm for a separable state $|\phi\rangle$,
where the transfer matrix $E_{\langle\phi|\phi\rangle}$ is
constructed from $B_{o}$, $B_{e}$ and their conjugates. }
  \label{gepersite}
 \end{figure}

A crucial step in evaluating geometric entanglement per site is how to maximize
$|\langle\phi|\psi\rangle|$ over all the possible separable states
$|\phi\rangle$.  In this regard,
a gradient-directed method turns to be an efficient way.  Specifically,
consider a two-site translation-invariant matrix product state, represented in terms of three-index tensors $A_o$ and $A_e$, as shown in Fig.~\ref{gepersite}\;(i) and (ii).
Then, the closest separable state may be represented in terms of one-index
tensors $ B_o$ and $B_e$, as shown in Fig.~\ref{gepersite}\;(i) and (iii).  Here,
subscripts $o$ and $e$ represent odd and even sites, respectively.
In Fig.~\ref{gepersite} (iv), we introduce the transfer
matrix $E_{\langle\phi|\psi\rangle}$ for fidelity between a ground state wave
function $|\psi\rangle$ and a separable state $|\phi\rangle$, which is constructed from two three-index
tensors, $A_{o}$ and $A_{e}$, and two one-index tensors, $B_{o}^{*}$
and $B_{e}^{*}$.  In Fig.~\ref{gepersite} (v), we introduce the transfer
matrix $E_{\langle\phi|\phi\rangle}$ for the norm of a separable state $|\phi\rangle$,  which is constructed from  two one-index tensors, $B_{o}$
and $B_{e}$, together with their conjugates.  Then,
the fidelity per unit cell $ \lambda$ between a ground state wave function $|\psi\rangle$ and a
separable state $|\phi\rangle$ takes the form
\begin{equation}
  \lambda=\frac{|\eta_{\langle\phi|\psi\rangle}|}
  {\sqrt{\eta_{\langle\psi|\psi\rangle}\eta_{\langle\phi|\phi\rangle}}},
\end{equation}
where $\eta_{\langle\phi|\psi\rangle}$, $\eta_{\langle\phi|\phi\rangle}$ and
$\eta_{\langle\phi|\phi\rangle}$ are the dominant
eigenvalues of the transfer matrix $E$ for the
matrix product state representations of $\langle\phi|\psi\rangle$, $\langle\psi|\psi\rangle$ and
$\langle\phi|\phi\rangle$, respectively.  For a normalized $|\psi\rangle$, we have $\eta_{\langle\psi|\psi\rangle}=1$.

We then proceed to compute geometric entanglement per
unit cell, which involves the optimization over all the separable states.
For our purpose, we define $F=\lambda^{2}$. The optimization amounts
to computing the logarithmic derivative of $F$ with respect to
$B^{*}$, which is expressed as
\begin{equation}
G \equiv \frac{\partial \ln F}{\partial
B^{*}}=\frac{1}{\eta_{\langle\phi|\psi\rangle}}\frac{\partial\eta_{\langle\phi|\psi\rangle}}{\partial
B^{*}}-\frac{1}{\eta_{\langle\phi|\phi\rangle}}\frac{\partial\eta_{\langle\phi|\phi\rangle}}{\partial
B^{*}}.
\end{equation}
Here, $B^{*}$ is either $B_{o}^{*}$ or $B_{e}^{*}$.
The problem therefore reduces to the computation of $G$ in the
context of the tensor network representations.  Once
$G$ is determined, we update the real and imaginary parts of $B^s$
separately:
\begin{eqnarray*}
\Re(B^s)&=&\Re(B^s)+\delta \Re(G^s),\\
\Im(B^s)&=&\Im(B^s)+\delta \Im(G^s).
\end{eqnarray*}
Here, $\delta\in[0,1)$ is the step size,
which is tuned to be decreasing during the optimization process.
In addition, we need to normalize the real and imaginary parts of the
gradient $G$ such that their respective largest entry remains to be unity.
If $\lambda$ converges, then the closest separable state
$|\phi\rangle$ is achieved.  Therefore, ground state geometric entanglement per
unit cell for a ground state wave function $|\psi\rangle$ follows.

 \section{Thermodynamic arrow of time, psychological/computational arrow of time and cosmological arrow of time}\label{timearrows}

 There are at least three arrows of time~\cite{tenarrows}: the thermodynamic arrow of time, the psychological arrow of time and the cosmological arrow of time~\cite{vaas}.

As our everyday experience shows, we remember the past, but not the future. This defines the psychological arrow of time.  The psychological arrow of time may be rephrased as the computational arrow of time, if cognitive processes are regarded as computational.
But how do we distinguish the future from the past, given the interchangeability of past and future with respect to the laws of microscopic physics?  One possible answer is that the observed asymmetry of past and future arises from the second law of thermodynamics, which states that the entropy of an isolated thermodynamic system increases monotonically. This defines the thermodynamic arrow of time. The cosmological arrow of time arises from the observation that the universe has been expanding since the big bang.
In some sense, the present is an idealized point between the past and the future.
Many efforts have been made in attempt to understand why the thermodynamic arrow of time, the psychological/computational arrow of time and the cosmological arrow of time should align with each other (see, e.g., Hawking~\cite{Hawking}, Wolpert~\cite{Wolpert}, Hartle~\cite{Hartle} and Mlodinow and Brun~\cite{Mlodinow}).

However, time-reversal invariance and reversibility are not the same, but independent from each other~\cite{aiello}. Indeed, time-reversal invariance is a property of a dynamical equation, such as the Schr\"{o}dinger equation, thus involving a set of its solutions, whereas reversibility is a property of one single solution of the dynamical equation. Therefore, a plausible resolution to the apparent contradiction between the interchangeability of past and future with respect to the laws of microscopic physics and the irreversibility of physical processes observed in macroscopic phenomena is based on a conceptual distinction between a microscopic time and a macroscopic time~\cite{vaas1}: under the time-reversal operation, the former is symmetric, but the latter is asymmetric.  In this sense, the mystery regarding arrows of time solely concerns about a macroscopic time.  In addition, a microscopic time always exists, due to quantum fluctuations arising from the uncertainty principle, whereas a macroscopic time may be absent in certain circumstances.

According to Mlodinov and Brun~\cite{Mlodinow}, the key to unlock this mystery lies in the presence of a physical system that can function as a memory or a record, in the sense of preserving a record of the state of some other system. In our opinion, it is information storage involved in a memory or record that is a key ingredient underlying arrows of time, including the thermodynamic arrow of time, the psychological/computational arrow of time and the cosmological arrow of time. This is due to the fact that, for any macroscopic time, there must exist a physical process which can, in principle, serve as a clock to track and record it. With this observation in mind, we may single out the psychological/computational arrow of time as a master arrow of time.

However, no systematic theoretical description is available for the psychological/computational arrow of time,  in contrast to the thermodynamic arrow of time and the cosmological arrow of time. Actually, not only is the notion of entropy available to measure degree of disorder in a thermodynamic system, but also the whole machinery based on thermodynamics offers a full description of physical properties of the system. There is also a plethora of theories on the big bang to describe different scenarios for the cosmological arrow of time. In this aspect, fidelity mechanics may be regarded as an attempt to understand the psychological/computational arrow of time in the context of quantum many-body systems,  in a sense that a specific physical meaning has been attached to the present, the past and the future via information storage.

 \section{Three theorems in quantum information science}\label{threetheorems}

 We recall three theorems in quantum information science: the no-cloning theorem, the no-deleting theorem and the no-hiding theorem.

(a) {\it No-cloning theorem}: it is impossible to create an identical copy of an arbitrary unknown quantum state.  The theorem was  first articulated in Refs.~\cite{Zurek,Dieks}.  It has profound implications in quantum information processing.  Mathematically, the no-cloning theorem states that, for an arbitrary normalized state $|\psi\rangle_A$ on a system $A$ and an arbitrary normalized state $|e\rangle_B$ on a system $B$, there is no unitary operator $U$, satisfying
$U|\psi\rangle_A|e\rangle_B=\exp{i\alpha}|\psi\rangle_A|\psi\rangle_B$, with $\alpha$ depending on $|\psi\rangle$ and $|e\rangle$.

(b) {\it No-deleting theorem}: it appears as time-reversed dual to the no-cloning theorem.  Given two copies of some arbitrary quantum state, it is impossible to delete one of the copies~\cite{Pati}. Mathematically, suppose  $|\psi \rangle$ is an unknown quantum state in a Hilbert space. Then, there is no linear isometric transformation $U$ such that $U |\psi \rangle_A |\psi \rangle_B|A\rangle_C=|\psi \rangle_A |0 \rangle_B|A'\rangle_C$, with the final state of the ancilla being independent of $|\psi\rangle$.

(c) {\it No-hiding theorem}: If information is missing from a given system due to interaction with  the environment, then it is simply residing somewhere else. In other words, {\it the missing information cannot be hidden in the correlations between a system and its environment}. It was formalized in Ref.~\cite{Braunstein} and experimentally confirmed in Ref.~\cite{samal}.

The theorems follow from the linearity of quantum mechanics. In fact, the principle of superposition states that, when two evolving states solve the Schr\"odinger equation, any linear combination of the two is also a solution.  As a corollary, perfect copying can be achieved only when states involved are mutually orthogonal to each other~\cite{nocloning}.  That is,
for a collection of mutually orthogonal states,  it is possible to set up a quantum copier exclusively tailored to this set of mutually orthogonal states.

\section{Relation between an unknown function $V(x)$ and fidelity temperature $T(x)$}\label{connectionUT}

Consider a quantum many-body system  described by the Hamiltonian $H(x)$, with $x$ being a dominant control parameter.  An analogue of Landauer's principle at zero temperature states that, in a fidelity mechanical system, to erase one bit of information at zero temperature, we need to do the minimum fidelity work $w(x)=\pm T(x) \ln2$.  Here, $T(x)$ characterizes quantum fluctuations at zero temperature, and $+/-$ corresponds to increasing/decreasing ground state energy density $e(x)$ with increasing $x$, respectively.   Our task is to establish a relation between an unknown function $V(x)$  and fidelity temperature $T(x)$.

Assume that the Hamiltonian $H(x)$ is in a canonical form such that the ground state energy density $e(x)$ is generically negative.  We prepare a composite system consisting of two identical copies. That is, these two copies share an identical Hamiltonian $H_d(x)$
and $H_u(x)$, thus one bit of information has been encoded for each value of $x$. Therefore, the composite Hamiltonian $H_c(x)$ is $H_c(x)=H_u(x)+H_d(x)$.
If we denote the ground state energy density by $e(x)$ for $H_u(x)$ and $H_d(x)$, then the ground state energy density $e_c(x)$ for the composite Hamiltonian $H_c(x)$ is $e_c(x)=2 \; e(x)$.

\begin{figure}
  \centering
\includegraphics[angle=0,totalheight=4.5cm]{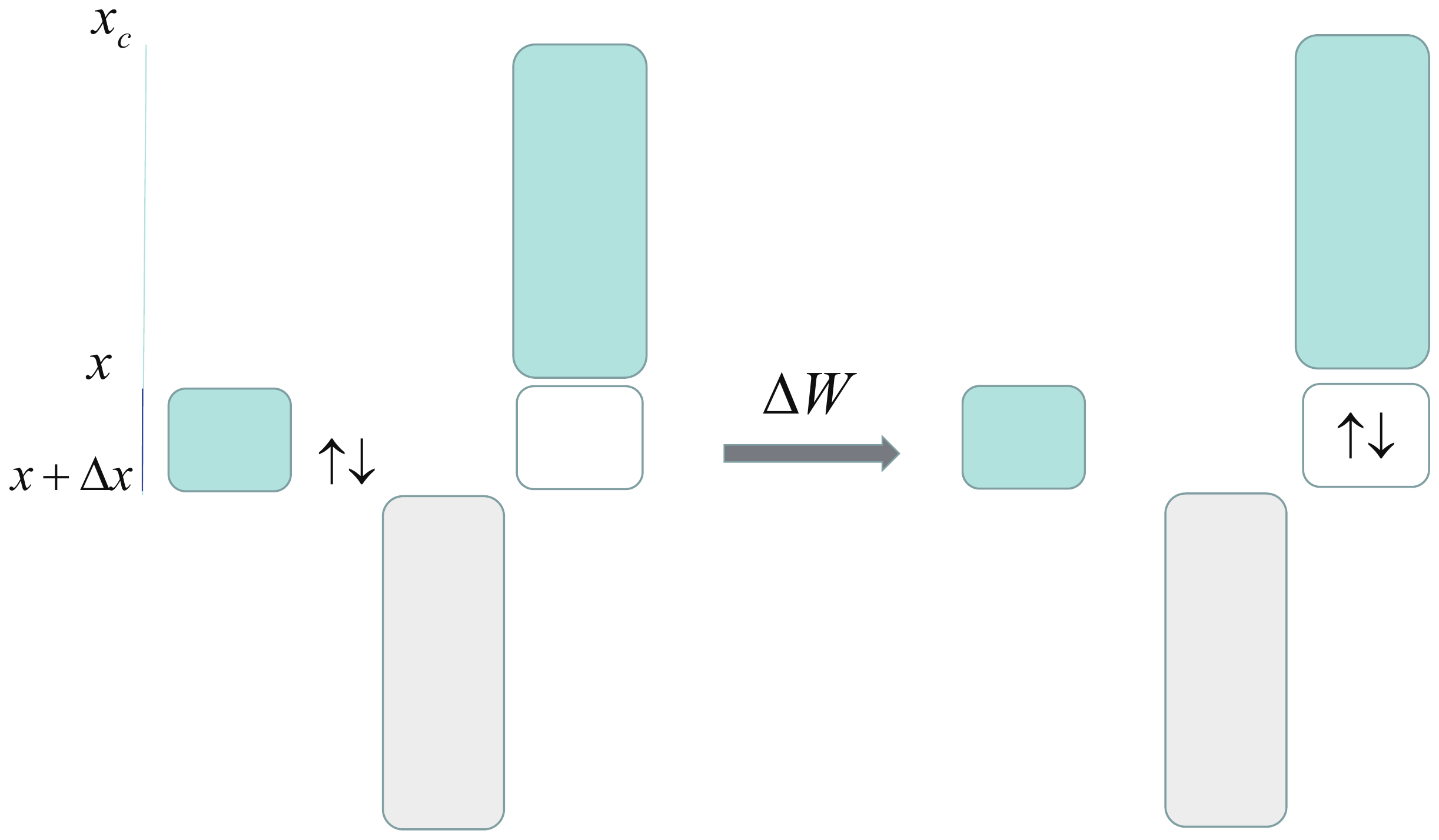}
\caption{One bit of information is erased from a fidelity mechanical  system for  each value of a dominant control parameter between $x$ and $x+\Delta x$. To do so,  a certain amount of fidelity work $\Delta W$ needs to be done, as required by an analogue of Landauer's principle at zero temperature, }\label{Vx}
\end{figure}

Suppose we erase one-bit of information from a fidelity mechanical system for each value of $x$ between $x$ and $x+\Delta x$,  as depicted in Fig.~\ref{Vx},  Then, a certain amount of fidelity work needs to be done, as required by an analogue of Landauer's principle at zero temperature.  To proceed further, we distinguish two cases:

(i) For a single copy system, if the ground state energy density $e(x)$ is monotonically decreasing  from a critical point $x_c$ to $x$, then fidelity internal energy $U(x)$ takes the form: $U(x)=\ln{(e(x)/e(x_c))}V(x)+U_0$, with $U_0$ an additive constant, and $V(x)$ being positive. However,  for a composite system, the increment of fidelity internal energy $\Delta U (x)$ due to the presence of  a bit of encoded information from $x$ to $x+\Delta x$
should be compensated by an extra amount of fidelity work  $\Delta W(x)=-T(x)\ln{2} \; \Delta \; x$:
\begin{equation}
\Delta (\ln{\frac{2e(x)}{e(x_c)} V(x)})=T (x)\Delta S(x)+ \Delta W (x).
\end{equation}
This  may be rewritten as
\begin{equation}
\Delta (\ln{\frac{e(x)}{e(x_c)} V(x)})=T(x) \Delta (S(x)+\ln 2 \;x)+ \Delta W(x),
\end{equation}
if we relate $T(x)$ with $V(x)$ as follows
\begin{equation}
T(x)=-\frac{\Delta V(x)}{\Delta x}.
\end{equation}
If $\Delta x \rightarrow 0$, then we have
\begin{equation}
T(x)=-V_x(x).
\end{equation}

(ii) For a single copy system, if the ground state energy density $e(x)$ is monotonically increasing  from a critical point $x_c$ to $x$, then fidelity internal energy takes the form: $U(x)=-\ln{(e(x)/e(x_c))}V(x)V(x) +U_0$ with $U_0$ an additive constant, and $V(x)$ being positive. However,  for a composite system, the increment of fidelity internal energy $\Delta U(x)$ due to the presence of  a bit of encoded information from $x$ to $x+\Delta x$
should be compensated by an extra amount of fidelity work  $\Delta W(x)=T(x)\ln{2}\; \Delta x$:
\begin{equation}
-\Delta (\ln{\frac{2e(x)}{e(x_c)} V(x)})=T(x) \Delta S(x)+ \Delta W(x).
\end{equation}
This may be rewritten as
\begin{equation}
-\Delta (\ln{\frac{e(x)}{e(x_c)} V(x)})=T(x) \Delta (S-\ln 2 \; x)+ \Delta W(x),
\end{equation}
if we relate $T(x)$ with $V(x)$ as follows
\begin{equation}
T(x)=-\frac{\Delta V(x)}{\Delta x}.
\end{equation}
If $\Delta x \rightarrow 0$, then we have
\begin{equation}
T(x)=-V_x(x).
\end{equation}

\section{Fidelity entropy, fidelity temperature and fidelity internal energy  for the quantum XY chain}\label{xychain}

In this Appendix,  we present mathematical details about fidelity entropy $S_f(\lambda, \gamma)$, fidelity temperature $T_f(\lambda, \gamma)$, and fidelity internal energy $U_f(\lambda, \gamma)$ for the quantum XY chain.

For this model, the Hamiltonian (\ref{xyham}) is symmetric under $\gamma \leftrightarrow - \gamma$  and $\lambda \leftrightarrow - \lambda$.  Therefore, we may restrict ourselves to the region defined by $\lambda \geq 0$ and $\gamma \geq 0 $.

In addition, dualities occur along the lines $\gamma=1$ and $\lambda=0$:

(i) If $\gamma=1$, then the Hamiltonian (\ref{xyham}) is reduced to the transverse field quantum Ising chain
$H(\lambda,1)=-\sum_{i}{(\sigma_i^x\sigma^x_{i+1}+\lambda \sigma^i_z)}$. Hence,
under a unitary transformation $U_1$: $\sigma_i^z \rightarrow \tau_i^x\tau_{i+1}^x$, and $\sigma^x_{i}\sigma^x_{i+1} \rightarrow \tau_i^z$,  we have
$H(\lambda,1)=k'(\lambda')  U_1 H(\lambda',1) U_1^{\dagger}$, with $\lambda'=1/\lambda$ and $k'(\lambda')=1/\lambda'$.

(ii) If $\lambda=0$, then the Hamiltonian (\ref{xyham}) is simplified to $H(0,\gamma)=-1/2\sum_{i}{[(1+\gamma)\sigma_i^x\sigma_{i+1}^x+(1-\gamma)\sigma_i^y\sigma_{i+1}^y]}$. Under a unitary transformation $U_2$:
$\sigma^x_{2i}\rightarrow \sigma^x_{2i}$, $\sigma^x_{2i+1}\rightarrow \sigma^x_{2i+1}$, $\sigma^y_{2i}\rightarrow \sigma^y_{2i}$, $\sigma^y_{2i+1}\rightarrow -\sigma^y_{2i+1}$, $\sigma^z_{2i}\rightarrow \sigma^z_{2i}$, and $\sigma^z_{2i+1}\rightarrow -\sigma^z_{2i+1}$,  we have  $H(0,\gamma)=k'(\gamma') U_2 H(0,\gamma') U_2^{\dagger}$,
with $\gamma'=1/\gamma$ and $k'(\gamma')=1/\gamma'$.

Therefore, we divide two dual lines into four parts:
(I) $0 \leq \lambda < \lambda_c = 1$ and $\gamma=1$; (II) $\lambda > \lambda_c = 1$ and $\gamma=1$; (III) $\lambda=0$ and $0 < \gamma \leq 1$; (IV) $\lambda=0$ and $\gamma\geq 1$.

Another interesting feature is the disorder circle $\gamma^2+\lambda^2=1$, characterized by the fact that ground states on the circle are factorized states~\cite{factorizing, factorizing1}.

The consideration of the dualities and factorizing fields allows us to separate the whole region, defined by $\gamma>0$ and $\lambda>0$, into five different principal regimes: (a) the regime inside the disorder circle, with $0<\lambda<1$ and $0<\gamma<\sqrt{1-\lambda^2}$; (b) the regime outside the disorder circle, with $0<\lambda<1$ and $\sqrt{1-\lambda^2}<\gamma<1$; (c) the regime with $\lambda<1$ and $\gamma>1$; (d) the regime with $\lambda>1$ and $0<\gamma<1$; (e) the regime with $\lambda>1$ and $\gamma>1$.
In each regime, we may choose a dominant control parameter, as long as such a choice is consistent with the constraints imposed by the symmetry groups, dualities and factorizing fields, meaning that any choice has to respect all the boundaries between different regimes. Here, our choice is: (1) for regime (a),
a dominant control parameter is chosen to be $\gamma$,  starting from $\gamma=\gamma_c = 0$ up to the disorder circle, with $\lambda$ fixed; (2)  for regime (b) or (c), a dominant control parameter is chosen to be $1-\lambda$, starting from $\lambda=\lambda_c = 1$ up to the disorder circle or $\lambda=0$, with $\gamma$ fixed; (3) for regime (d) or (e), we rescale the ground state energy density: $e(\lambda,\gamma)=k'(\lambda') e'(\lambda',\gamma)$ with $k'(\lambda')=1/\lambda'$ and $\lambda'=1/\lambda$ .
 A dominant control parameter is chosen to be  $1-\lambda'$, starting from $\lambda'=\lambda'_c = 1$ up to $\lambda'=0$, with $\gamma$ fixed. This choice is to keep consistency with duality for the transverse field quantum Ising chain, which corresponds to the quantum XY model with $\gamma =1$.

Let us start from fidelity entropy $S(\lambda, \gamma)$ for four parts on the two dual lines.

(1) In Part I ($0\leq\lambda<\lambda_c=1$, $\gamma=1$), a dominant control parameter is chosen to be $1-\lambda$.
From Eq.~(\ref{entropy}), fidelity entropy $S^{\rm I}(\lambda,1)$ takes the form:
\begin{equation}
S^{\rm I}(\lambda,1)=2\int_{1}^{\lambda}{\ln{d(\lambda,1;\mu,1)} \; d(\mu)}+S^{\rm I}_{0}.\label{sd1}
\end{equation}
Here, $S^{\rm I}_0$ is residual fidelity entropy at the critical point $\lambda_c=1$.
As  discussed in Section~\ref{fmsut},  we have $S^{\rm I}_f(\lambda,1) = S^{\rm I}(\lambda,1)$ .

(2) In Part II ($\lambda>\lambda_c= 1$, $\gamma=1$), the Hamiltonian $H(\lambda,1)$ is rescaled due to duality: $H(\lambda,1)=k'(\lambda') U H'(\lambda',1)U^{\dagger}$, with  $\lambda'=1/\lambda$ and $k'(\lambda')=1/\lambda'$.  This results in rescaling in the ground state energy density: $e(\lambda,1)=k'(\lambda') e'(\lambda',1)$. Here, a dominant control parameter is chosen to be $1-\lambda'$.  As discussed in Section~\ref{fmsut}, fidelity entropy $S^{\rm II}_f(\lambda,1)$ includes contributions from fidelity entropy $S'^{\rm II}_{\phi f}(\lambda',1)$ and scaling entropy $S'^{\rm II}_{\sigma f}(\lambda',1)$ arising from a rescaling factor $k'(\lambda')$, with $S'^{\rm II}_{\phi f}(\lambda',1) \equiv S'^{\rm II}_{\phi}(\lambda',1)$ and $S'^{\rm II}_{\sigma f}(\lambda',1) \equiv S'^{\rm II}_{\sigma}(\lambda',1)$.
Thus, we have
$S^{\rm II}_f(\lambda,1) = S^{\rm II}_{\phi f}(\lambda,1)+S^{\rm II}_{\sigma f}(\lambda,1)$,
with $S^{\rm II}_{\phi f}(\lambda,1)\equiv S'^{\rm II}_{\phi f}(\lambda',1)$ and $S^{\rm II}_{\sigma f}(\lambda,1)\equiv S'^{\rm II}_{\sigma f}(\lambda',1)$.
Here,
$S'^{\rm II}_\phi(\lambda',1)= S^{\rm I}(\lambda',1)$ and $S'^{\rm II}_{\sigma}(\lambda',1)=-\ln{\lambda'}$.

(3) In Part III ($\lambda=0$, $0<\gamma\leq 1$), a dominant control parameter is chosen to be $\gamma$.  From Eq.~(\ref{entropy}), fidelity entropy $S^{\rm III}(0,\gamma)$ takes the form:
\begin{equation}
S^{\rm III}(0,\gamma)=-2\int_{0}^{\gamma}{\ln{d(0,\gamma;0,\beta)} \; d{\beta}}+S^{\rm III}_{0}. \label{sgd}
\end{equation}
Here, $S^{\rm III}_0$ is residual fidelity entropy at the critical point $\gamma_c=0$ for $\lambda=0$.
As discussed in Section~\ref{fmsut},  we have $S^{\rm III}_f(0,\gamma) = S^{\rm III}(0,\gamma)$ .

(4) In Part IV ($\lambda=0$, $\gamma \geq 1$),  the Hamiltonian $H(\gamma)$ is rescaled due to duality: $H(0,\gamma)=k'(\gamma') U H'(0,\gamma')U^{\dagger}$, with  $\gamma'=1/\gamma$ and $k'(\gamma')=1/\gamma'$. This results in rescaling in the ground state energy density: $e(0,\gamma)=k'(\gamma') e'(0,\gamma')$. Here, a dominant control parameter is chosen to be $\gamma'$. As discussed in Section~\ref{fmsut}, fidelity entropy $S^{\rm IV}_f(0,\gamma)$ includes contributions from fidelity entropy $S'^{\rm IV}_{\phi f}(0,\gamma')$ and scaling entropy $S^{\rm IV}_{\sigma f}(0,\gamma')$ arising from a rescaling factor $k'(\gamma')$, with $S'^{\rm IV}_{\phi f}(0,\gamma') \equiv S'^{\rm IV}_{\phi}(0,\gamma')$ and $S'^{\rm IV}_{\sigma f}(0,\gamma') \equiv S'^{\rm IV}_{\sigma}(0,\gamma')$.
Thus, we have $S^{\rm IV}_f(0,\gamma) = S^{\rm IV}_{\phi f}(0,\gamma)+S^{\rm IV}_{\sigma f}(0,\gamma)$,
with $S^{\rm IV}_{\phi f}(0,\gamma)\equiv S'^{\rm IV}_{\phi f}(0,\gamma')$ and $S^{\rm IV}_{\sigma f}(0,\gamma)\equiv S'^{\rm IV}_{\sigma f}(0,\gamma')$.
Here, $S'^{\rm IV}_\phi(0,\gamma')= S^{\rm III}(0,\gamma')$ and $S'^{\rm IV}_{\sigma}(0,\gamma')=\ln{\gamma'}$.

On the disorder circle: $\lambda^2+\gamma^2=1$, we re-parameterize $\lambda$ and $\gamma$ by introducing a single parameter $\theta$ such that $\theta=\arctan{\gamma/\lambda}$.  Given an exotic QPT exists along the disorder circle, with $\theta_c=0$~\cite{Wolf}, we need to treat it separately.  Here, a dominant control parameter is chosen to be $\theta$. Then, fidelity entropy $S^{f}(\lambda,\gamma)$ takes the form
\begin{equation}
S^{f}(\lambda,\gamma)\equiv S(\cos{\theta},\sin{\theta})=-2\int_{0}^{\theta}{\ln{d(\cos{\theta},\sin{\theta};\cos{\eta},\sin{\eta})}} \; d{\eta}+S_0^f.
\label{s0}
\end{equation}
Here, $S_0^f$ is residual fidelity entropy at the transition point $\lambda_c=1$ and $\gamma_c=0$.
As discussed in Section~\ref{fmsut},  we have $S^f_f(\lambda,\gamma) = S^f(\lambda,\gamma)$ .

Now we move to five different regimes:

(a) In regime $a$ ($0<\lambda<1$ and $0<\gamma<\sqrt{1-\lambda^2}$), a dominant control parameter is chosen to be $\gamma$. Then, fidelity entropy $S^a(\lambda, \gamma)$ takes the form
\begin{equation}
S^a(\lambda,\gamma)=-2\int_{0}^{\gamma}{\ln{d(\lambda,\gamma;\lambda,\beta)} \; d{\beta}}+S^a_{0}(\lambda);\label{s1}
\end{equation}
Here, $S^a_{0}(\lambda)$ is residual fidelity entropy at a critical point $\gamma_c=0$ for a fixed $\lambda$.
As discussed in Section~\ref{fmsut}, we have $S^a_f(\lambda,\gamma) = S^a(\lambda,\gamma)$ .

(b) In regime $b$ ($0<\lambda<1$ and $\sqrt{1-\lambda^2}<\gamma<1$), a dominant control parameter is chosen to be $1-\lambda$. Then, fidelity entropy $S^b(\lambda, \gamma)$ takes the form
\begin{equation}
S^b(\lambda,\gamma)=2\int_{1}^{\lambda}{\ln{d(\lambda,\gamma;\mu,\gamma)} \; d{\mu}}+S^b_{0}(\gamma).\label{s2}
\end{equation}
Here, $S^b_{0}(\gamma)$ is residual fidelity entropy at a critical point $\lambda_c=1$ for a fixed $\gamma$.
As discussed in Section~\ref{fmsut}, we have $S^b_f(\lambda,\gamma) = S^b(\lambda,\gamma)$ .

(c) In regime $c$ ( $0<\lambda<1$ and $\gamma>1$), a dominant control parameter is chosen to be $1-\lambda$. Then, fidelity entropy $S^c(\lambda, \gamma)$ takes the form
\begin{equation}
S^c(\lambda,\gamma)=2\int_{1}^{\lambda}{\ln{d(\lambda,\gamma;\mu,\gamma)} \; d{\mu}}+S^c_{0}(\gamma).\label{s3}
\end{equation}

Here, $S^c_{0}(\gamma)$ is residual fidelity entropy at a critical point $\lambda_c=1$ for a fixed $\gamma$.
In this regime, the continuity requirement for $S_f(\lambda,\gamma)$ at the dual line $\lambda=0$, with $\gamma>1$, implies that $S^{c}_f(\lambda,\gamma)$ includes a contribution from residual scaling entropy $S^{\rm c}_{\sigma 0}(\gamma)$, due to a rescaling factor $k(\gamma)=\gamma$.  Hence, we have $S^c_f(\lambda,\gamma)=S^c(\lambda,\gamma)+S^c_{\sigma 0}(\gamma)$, with $S^c_{\sigma 0}(\gamma)=S^{\rm IV}_{\sigma f}(0,\gamma)=-\ln{\gamma}$.

(d) In regime $d$ ( $\lambda>1$ and $0<\gamma<1$), no duality exists. However,  the range of $\lambda$ is not finite. Thus, rescaling needs to be performed in the ground state energy density: $e(\lambda, \gamma)= k'(\lambda') e'(\lambda', \gamma)$, with $k'(\lambda')=1/\lambda'$.
As discussed in Section~\ref{fmsut}, $S^{d}_f(\lambda, \gamma)$ includes contributions from fidelity entropy $S'^d_{\phi f}(\lambda',\gamma)$ and scaling entropy $S'^d_{\sigma f}(\lambda',\gamma)$ with $S'^d_{\phi f}(\lambda',\gamma)\equiv S'^d_{\phi}(\lambda',\gamma)$ and $S'^d_{\sigma f}(\lambda',\gamma)\equiv S'^d_{\sigma}(\lambda',\gamma)$.  Here, $S'^d_{\sigma}(\lambda',\gamma)=-\ln{\lambda'}$.
 Thus,  we have
$S^d_f(\lambda,\gamma) =  S^d_{\phi f}(\lambda,\gamma)+S^d_{\sigma f}(\lambda,\gamma)$, with $S^d_{\phi f}(\lambda,\gamma)\equiv S'^d_{\phi f}(\lambda',\gamma)$ and $S^d_{\sigma f}(\lambda,\gamma) \equiv S'^d_{\sigma f}(\lambda',\gamma)$.
Hence, $S'^d_\phi(\lambda',\gamma)$ takes the form
\begin{equation}
S'^d_\phi(\lambda',\gamma)=2\int_{1}^{\lambda'}{\ln{d'(\lambda',\gamma;\mu',\gamma)}} \; d{\mu'}+S'^d_{\phi 0}(\gamma),
\label{sd2}
\end{equation}
where $d'(\lambda',\gamma;\mu',\gamma)\equiv d(\lambda,\gamma;\mu,\gamma)$, with $\mu$ denoting another value of the same control parameter as $\lambda$,  $\mu'=1/\mu$, and $S'^d_{\phi 0}(\gamma)$ is residual fidelity entropy at a critical point $\lambda_c=1$ for a fixed $\gamma$.

(e) In regime $e$ ( $\lambda>1$ and $\gamma>1$),
no duality exists.  However,  the range of $\lambda$ is not finite. Thus, rescaling needs to be performed in the ground state energy density: $e(\lambda, \gamma)= k'(\lambda') e'(\lambda', \gamma)$, with $k'(\lambda')=1/\lambda'$.
As discussed in Section~\ref{fmsut}, $S^{e}_f(\lambda, \gamma)$ includes contributions from fidelity entropy $S'^e_{\phi f}(\lambda',\gamma)$ and scaling entropy $S'^e_{\sigma f}(\lambda',\gamma)$, with $S'^e_{\phi f}(\lambda',\gamma)\equiv S'^e_{\phi}(\lambda',\gamma)$ and $S'^e_{\sigma f}(\lambda',\gamma)\equiv S'^e_{\sigma}(\lambda',\gamma)$.  Here, $S'^e_{\sigma}(\lambda',\gamma)=-\ln{\lambda'}+S'^e_{\sigma 0}(\gamma)$ with $S'^e_{\sigma 0}(\gamma)=-\ln{\gamma}$.
Thus, we have
$S^e_f(\lambda,\gamma) = S^e_{\phi f}(\lambda,\gamma)+S^e_{\sigma f}(\lambda,\gamma)$, with $S^e_{\phi f}(\lambda,\gamma)\equiv S'^e_{\phi f}(\lambda',\gamma)$ and $S^e_{\sigma f}(\lambda,\gamma) \equiv S'^e_{\sigma f}(\lambda',\gamma)$.
Hence, $S'^e_\phi(\lambda',\gamma)$ takes the form
\begin{equation}
S'^e_\phi(\lambda',\gamma)=2\int_{1}^{\lambda'}{\ln{d'(\lambda',\gamma;\mu',\gamma)}} \; d{\mu'}+S'^e_{\phi 0}(\gamma),
\label{sd2}
\end{equation}
where $d'(\lambda',\gamma;\mu',\gamma)\equiv d(\lambda,\gamma;\mu,\gamma)$, with $\mu$ denoting another value of the same control parameter as $\lambda$,  $\mu'=1/\mu$, and $S'^e_{\phi 0}(\gamma)$ is residual fidelity entropy at a critical point $\lambda_c=1$ for a fixed $\gamma$.

We turn to residual fidelity entropy on the two critical lines.  We remark that
fidelity entropy $S_f(\lambda,\gamma)$, generically, is {\it relative}, in the sense that it is  {\it only} determined up to a constant.
Here, as a convention, we choose fidelity entropy $S_f(\lambda,\gamma)$ to be zero at the critical point (0,0). This implies $S^{\rm III}_0=0$.
Then,
$S^{\rm III}(0,\gamma)$  follows from~(\ref{sgd}).   This in turn allows us to determine $S^{\rm IV}(0,\gamma)$ as a result of duality.
Further, residual fidelity entropy $S^f_0$ at the transition point (1,0) is determined from the continuity requirement for fidelity entropy  $S_f(\lambda,\gamma)$ at  (0,1): $S^f(0,1)=S^{\rm III}(0,1)$.  Hence, fidelity entropy $S^f(\lambda,\gamma)$ on the disorder circle follows from~(\ref{s0}). With this in mind, we are able to determine residual fidelity entropy $S^{a}_{0}(\lambda)$ and $S^{b}_{0}(\gamma)$, respectively, from the continuity requirements:  $S^{a}(\lambda,\sqrt{1-\lambda^2}) = S^f(\lambda,\sqrt{1-\lambda^2})$ and
$S^{b}(\sqrt{1-\gamma^2},\gamma)=S^{f}(\sqrt{1-\gamma^2},\gamma)$.  Similarly,  residual fidelity entropy $S^{c}_0(\gamma)$ are determined from the continuity requirement at the dual line $\lambda=0$, with $\gamma>1$: $S^{c}(0,\gamma)=S^{\rm IV}_{\phi f}(0,\gamma)$.  In addition, the continuity requirement for fidelity entropy $S_f(\lambda,\gamma)$ at a line of critical points $\lambda=1$, with $\gamma >0$,  implies that $S'^{d}_{\phi 0}(\gamma)=S^{b}_{0}(\gamma)$ and $S'^{e}_{\phi 0}(\gamma)=S^{c}_{0}(\gamma)$.

Once fidelity entropy $S_f(\lambda,\gamma)$ is determined, fidelity internal energy $U_f(\lambda,\gamma)$ and fidelity temperature $T_f(\lambda,\gamma)$ may be determined from solving the singular first-order differential equation (\ref{alpha}), as discussed for continuous QPTs in Section~\ref{fmsut}.

We start from four different parts on the two dual lines:

(1) In Part I ($0 \leq \lambda<\lambda_c = 1$ and $\gamma=1$), the ground state energy density $e(\lambda,1)$ is increasing with increasing  $1-\lambda$. Then, following Eq.~(\ref{internalenergy}), fidelity internal energy $U^{\rm I}(\lambda,1)$ takes the form
\begin{equation}
U^{\rm I}(\lambda,1)=-\ln{\frac{e(\lambda,1)}{e(1,1)}}V^{\rm I}(\lambda,1)+U^{\rm I}_0 \label{u1}
\end{equation}
Here, $U^{\rm I}_0$ is an additive constant, and $V^{\rm I}(\lambda,1)>0$ satisfies the differential equation

\begin{equation}
\partial{V^{\rm I}(\lambda,1)}/\partial{\lambda}=-\alpha^{\rm I} (\lambda,1) \; V^{\rm I}(\lambda,1),  \label{v1}
\end{equation}
with
\begin{equation}
\alpha^{\rm I}(\lambda,1)=\frac{\partial{\ln{(e(\lambda,1)/e(1,1))}}/\partial{\lambda}}{\partial{S^{\rm I}(\lambda,1)}/\partial{\lambda}+\ln{(e(\lambda,1)/e(1,1))}}.
\label{alphad1}
\end{equation}
Accordingly, fidelity temperature $T^{\rm I}(\lambda,1)$ follows from
\begin{equation}
 T^{\rm I}(\lambda,1)=\frac {\partial{V^{\rm I}(\lambda,1)}} {\partial{\lambda}}.
 \label{t1}
\end{equation}

(2) In Part II ($\lambda>\lambda_c = 1$ and $\gamma=1$),
the Hamiltonian $H(\lambda,1)$ is rescaled due to duality: $H(\lambda,1)=k'(\lambda') U H'(\lambda',1)U^{\dagger}$, with  $\lambda'=1/\lambda$ and $k'(\lambda')=1/\lambda'$.  This results in rescaling in the ground state energy density: $e(\lambda,1)=k'(\lambda') e'(\lambda',1)$. Here, a dominant control parameter is chosen to be $1-\lambda'$.  As discussed in Section~\ref{fmsut}, fidelity internal energy $U'^{\rm II}_\phi(\lambda',1)$ and fidelity temperature $T'^{\rm II}_\phi(\lambda',1)$ takes the form: $U'^{\rm II}_\phi(\lambda',1)= U^{\rm I}(\lambda',1)$ and $T'^{\rm II}_\phi(\lambda',1)=T^{\rm I}(\lambda',1)$, as follows from duality.

(3) In Part III ($0<\gamma \leq 1$ and $\lambda=0$), the ground state energy density $e(0,\gamma)$ is decreasing with increasing $\gamma$.  Then, fidelity internal energy $U^{\rm III}(0,\gamma)$ takes the form
\begin{equation}
U^{\rm III}(0,\gamma)=\ln{\frac{e(0,\gamma)}{e(0,0)}}V^{\rm III}(0,\gamma)+U^{\rm III}_0\label{u3}
\end{equation}
Here, $U^{\rm III}_0$ is an additive constant,
and $V^{\rm III}(0,\gamma)>0$ satisfies the differential equation

\begin{equation}
\partial{V^{\rm III}(0,\gamma)}/\partial{\gamma}=\alpha ^{\rm III}(0,\gamma) \; V^{\rm III} (0,\gamma), \label{v3}
\end{equation}
with
\begin{equation}
\alpha^{\rm III}(\gamma)=-\frac{\partial{\ln{(e(0,\gamma)/e(0,0))}}/\partial{\gamma}}{\partial{S^{\rm III}(\gamma)}/\partial{\gamma}+\ln{(e(0,\gamma)/e(0,0))}}.
\label{alphad3}
\end{equation}
Accordingly, fidelity temperature $T^{\rm III}(0,\gamma)$ follows from

\begin{equation}
 T^{\rm III}(0,\gamma)=-\frac{\partial{V^{\rm III}(0,\gamma)}} {\partial{\gamma}}.
 \label{t3}
\end{equation}

(4) In Part IV ($\lambda=0$ and $\gamma \geq 1$), the Hamiltonian $H(0,\gamma)$ is rescaled due to duality: $H(0,\gamma)=k'(\gamma') U H'(0,\gamma')U^{\dagger}$, with  $\gamma'=1/\gamma$ and $k'(\gamma')=1/\gamma'$. This results in rescaling in the ground state energy density: $e(0,\gamma)=k'(\gamma') e'(0,\gamma')$. Here, a dominant control parameter is chosen to be $\gamma'$.  As discussed in Section~\ref{fmsut}, fidelity internal energy $U'^{\rm IV}_\phi(0,\gamma')$ and fidelity temperature $T'^{\rm IV}_\phi(0,\gamma')$ takes the form: $U'^{\rm IV}_\phi(0,\gamma')= U^{\rm III}(0,\gamma')$ and $T'^{\rm IV}_\phi(0,\gamma')=T^{\rm III}(0,\gamma')$, as follows from duality.

On the disorder circle, the ground state energy density $e(\lambda,\gamma)$ is a constant.  Then, fidelity internal energy $U^{f}(\lambda,\gamma)$ is a constant, and fidelity temperature $T^{f}(\lambda,\gamma)$ is zero:
\begin{equation}
U^f(\lambda,\gamma) = U^{\rm III}(0,1),\qquad T^f(\lambda,\gamma) = 0.\label{utf}
\end{equation}
As discussed in Section~\ref{fmsut}, we have $U^f_f(\lambda,\gamma) = U^f(\lambda,\gamma)$ and $T^f_f(\lambda,\gamma) = T^f(\lambda,\gamma)$.

Now we move to five regimes $a$, $b$, $c$, $d$ and $e$:

(a) In regime $a$ ($0<\lambda<1$ and $0<\gamma<\sqrt{1-\lambda^2}$), for a fixed $\lambda$, the ground state energy density $e(\lambda,\gamma)$ is monotonically decreasing with  increasing $\gamma$.  Then, from Eq.~(\ref{internalenergy}), fidelity internal energy $U^a(\lambda,\gamma)$ takes the form
\begin{equation}
U^{a}(\lambda,\gamma)=\ln{\frac{e(\lambda,\gamma)}{e(\lambda,0)}}V^{a}(\lambda,\gamma)+U^{a}_0(\lambda). \label{uaxy}
\end{equation}
Here, $U^{a}_0(\lambda)$ is a function of $\lambda$,  and
$V^{a}(\lambda,\gamma)>0$ satisfies the differential equation
\begin{equation}
\frac{\partial{V^{a}(\lambda,\gamma)}}{\partial{\gamma}}=\alpha^{a}(\lambda,\gamma) \; V^{a} (\lambda,\gamma), \label{vaxy}
\end{equation}
with
\begin{equation}
\alpha^{a}(\lambda,\gamma)=-\frac{\partial{\ln{(e(\lambda,\gamma)/e(\lambda,0))}}/\partial{\gamma}}{\partial{S^{a}(\lambda,\gamma)}/\partial{\gamma}+\ln{(e(\lambda,\gamma)/e(\lambda,0))}}.
\label{alphaaxy}
\end{equation}
Accordingly, fidelity temperature $T^a(\lambda,\gamma)$ in this regime is given by
\begin{equation}
T^a(\lambda,\gamma)=-\frac{\partial{V^{a}(\lambda,\gamma)}}{\partial{\gamma}} \label{taxy}
\end{equation}

(b) In regime $b$ ($0<\lambda<1$ and $\sqrt{1-\lambda^2}<\gamma<1$ ), for a fixed $\gamma$, the ground state energy density $e(\lambda,\gamma)$ is monotonically increasing with  increasing $1-\lambda$.  Then, fidelity internal energy $U^b(\lambda,\gamma)$ takes the form
\begin{equation}
U^b(\lambda,\gamma)=-\ln{\frac{e(\lambda,\gamma)}{e(1,\gamma)}}V^b(\lambda,\gamma)+U^b_0(\gamma). \label{ubxy}
\end{equation}
Here, $U^{b}_0(\gamma)$ is a function of $\gamma$,  and
$V^b(\lambda,\gamma)>0$ satisfies the differential equation
\begin{equation}
\frac{\partial{V^{b}(\lambda,\gamma)}}{\partial{\lambda}}=\alpha^{b}(\lambda,\gamma) \; V^{b} (\lambda,\gamma), \label{vbxy}
\end{equation}
with
\begin{equation}
\alpha^b (\lambda,\gamma) =\frac{\partial{\ln{(e(\lambda,\gamma)/e(1,\gamma))}}/\partial{\lambda}}{\partial{S^{b}(\lambda,\gamma)}/\partial{\lambda}+\ln{e(\lambda,\gamma)/e(1,\gamma)}}.\label{alphabxy}
\end{equation}
Accordingly, fidelity temperature $T^b(\lambda,\gamma)$ in this regime is given by
\begin{equation}
 T^b(\lambda,\gamma)=\frac{\partial{V^{b}(\lambda,\gamma)}}{\partial{\lambda}} \label{tbxy}
\end{equation}

(c) In regime $c$ ($0<\lambda<1$ and $\gamma>1$ ), for a fixed $\gamma$,  the ground state energy density $e(\lambda,\gamma)$ is monotonically increasing with  increasing $1-\lambda$.  Then, fidelity internal energy $U(\lambda,\gamma)$ takes the form
\begin{equation}
U^c(\lambda,\gamma)=-\ln{\frac{e(\lambda,\gamma)}{e(1,\gamma)}}V^c(\lambda,\gamma)+U^c_0(\gamma). \label{uc}
\end{equation}
Here, $U^{c}_0(\gamma)$ is  a function of $\gamma$,  and
$V^{c}(\lambda,\gamma)>0$ satisfies the differential equation
\begin{equation}
\frac{\partial{V^{c}(\lambda,\gamma)}}{\partial{\lambda}}=\alpha^{c}(\lambda,\gamma) \; V^{c} (\lambda,\gamma),
\label{vc}
\end{equation}
with
\begin{equation}
\alpha^c(\lambda,\gamma)=\frac{\partial{{\ln{(e(\lambda,\gamma)/e(1,\gamma))}}}/\partial{\lambda}}{\partial{S^{c}(\lambda,\gamma)}/\partial{\lambda}+\ln{(e(\lambda,\gamma)/e(1,\gamma))}}.\label{alphac}
\end{equation}
Accordingly, fidelity temperature $T^c(\lambda,\gamma)$ in this regime is given by
\begin{equation}
 T^c(\lambda,\gamma)=\frac{\partial{V^{c}(\lambda,\gamma)}}{\partial{\lambda}}
 \label{tc}
\end{equation}

(d) In regime $d$ ($\lambda>1$ and $0<\gamma<1$),  the range of $\lambda$ is not finite.  Then, rescaling needs to be performed in the ground state energy density: $e(\lambda, \gamma)= k'(\lambda') e'(\lambda', \gamma)$, with $k'(\lambda')=1/\lambda'$.  Since  $e'(\lambda',\gamma)$ is monotonically increasing with increasing $1-\lambda'$, fidelity internal energy $U'^d_\phi(\lambda',\gamma)$ takes the form
\begin{equation}
U'^{d}_\phi(\lambda',\gamma)=-\ln{\frac{e'(\lambda',\gamma)}{e'(1,\gamma)}}V'^{d}(\lambda',\gamma)+U'^{d}_{\phi 0}(\gamma), \label{udxy}
\end{equation}
Here, $U'^{d}_{\phi 0}(\gamma)$ is a function of $\gamma$,  and
$V'^{d}(\lambda',\gamma)>0$ satisfies the differential equation

\begin{equation}
\frac{\partial{V'^{d}(\lambda',\gamma)}}{\partial{\lambda'}}=\alpha'^{d}(\lambda',\gamma)V'^{d} (\lambda',\gamma),
\label{vdxy}
\end{equation}
with
\begin{equation}
\alpha'^{d}(\lambda',\gamma)=
\frac{\partial{\ln{(e'(\lambda',\gamma)/e(1,\gamma))}}/\partial{\lambda'}}{\partial{S'^{d}_\phi(\lambda',\gamma)}/\partial{\lambda'}+
\ln{(e'(\lambda',\gamma)/ e'(1,\gamma))}}.\label{alphadxy}
\end{equation}
Accordingly, fidelity temperature $T'^d_\phi(\lambda',\gamma')$ in this regime is given by
\begin{equation}
 T'^d_\phi(\lambda',\gamma)=-\frac{\partial{V'^{d}(\lambda',\gamma)}}{\partial{\lambda'}}.
 \label{tdxy}
\end{equation}

(e) In regime $e$ ($\lambda>1$ and $\gamma>1$),  the range of $\lambda$ is not finite.  Then, rescaling needs to be performed in the ground state energy density: $e(\lambda, \gamma)= k'(\lambda') e'(\lambda', \gamma)$, with $k'(\lambda')=1/\lambda'$.   Since  $e'(\lambda',\gamma)$ is monotonically increasing with increasing $1-\lambda'$, fidelity internal energy $U'^e_\phi(\lambda',\gamma)$ takes the form
\begin{equation}
U'^{e}_\phi(\lambda',\gamma)=-\ln{\frac{e'(\lambda',\gamma)}{e'(1,\gamma)}}V'^{e}(\lambda',\gamma)+U'^{e}_{\phi 0}(\gamma), \label{ue}
\end{equation}
Here, $U'^{e}_{\phi 0}(\gamma)$ is a function of $\gamma$,  and
$V'^{e}(\lambda',\gamma)>0$ satisfies the differential equation

\begin{equation}
\frac{\partial{V'^{e}(\lambda',\gamma)}}{\partial{\lambda'}}=\alpha'^{e}(\lambda',\gamma) V'^{e} (\lambda',\gamma),
\label{ve}
\end{equation}
with
\begin{equation}
\alpha'^{e}(\lambda',\gamma)=
\frac{\partial{\ln{(e'(\lambda',\gamma)/e(1,\gamma))}}/\partial{\lambda'}}{\partial{S'^{e}_\phi(\lambda',\gamma)}/\partial{\lambda'}+
\ln{(e'(\lambda',\gamma)/ e'(1,\gamma))}}.\label{alphae}
\end{equation}
Accordingly, fidelity temperature $T'^e_\phi(\lambda',\gamma)$ in this regime is given by
\begin{equation}
 T'^e_\phi(\lambda',\gamma)=-\frac{\partial{V'^{e}(\lambda',\gamma)}}{\partial{\lambda'}}.
 \label{te}
\end{equation}

In order to solve a singular first-order differential equation in each regime, we perform a scaling analysis of $\alpha(\lambda,\gamma)$ near a critical point, which falls into two universality classes:
(i) the Gaussian universality class: $\gamma_c=0$ and (ii) the Ising universality class: $\lambda_c=1$.

(i) For a fixed $-1<\lambda<1$, when a Gaussian critical point $\gamma_c=0$ is approached,
 fidelity entropy $S(\lambda,\gamma)$ scales as $S(\lambda,\gamma) \sim (|\gamma-\gamma_c|)^{2}$, corresponding to the critical exponent $\nu=1$ (cf. Appendix~\ref{scaling} for details).  Combining with a scaling analysis of the ground state energy density $e(\lambda,\gamma)$ near a Gaussian critical point $\gamma_c=0$: $\ln(e(\gamma)/e(0))\sim \ln(\gamma)\; \gamma^2$ (cf. Appendix~\ref{gaussion} for details), we have
\begin{equation}
\alpha(\lambda,\gamma) \sim   \ln \; \gamma.
\label{alfags}
\end{equation}
Our numerical simulations confirm this analysis.

(ii) For a fixed $\gamma$, when an Ising critical point $\lambda_c=1$ is approached, we need to distinguish two cases:  $0\leq \lambda <1$ and $\lambda >1$.  When $0 \leq \lambda <1$, no rescaling in the Hamiltonian is needed.  Fidelity entropy $S(\lambda,\gamma)$ scales as $S(\lambda,\gamma) \sim (1-\lambda)^2$. This indicates that the  critical exponent $\nu=1$ (cf. Appendix~\ref{scaling} for details).
Taking into account the fact that the first-order derivative of $\ln{(e(\lambda,\gamma)/e(1,\gamma))}$ with respect to $\lambda$ at a critical point $\lambda_c = 1$ is nonzero,
we have
\begin{equation}
\alpha(\lambda,\gamma) \sim \frac{1}{1-\lambda}.
\end{equation}
To proceed further, we need to separate the divergent part from $\alpha(\lambda,\gamma)$.  To do so,
we set
\begin{equation}
\alpha(\lambda,\gamma)=\frac{p(\gamma)}{1-\lambda}+f(\lambda,\gamma).
\label{alfais}
\end{equation}
Here, $f(\lambda,\gamma)$ is a function of $\lambda$ and $\gamma$, which takes a finite value, and $p(\gamma)/(1-\lambda)$ is the divergent part of $\alpha(\lambda,\gamma)$, with $p(\gamma)$ determined by $p(\gamma)=\mu(\gamma)/(2w(\gamma)-\mu(\gamma))$, where $\mu(\gamma)$ is defined as $\mu(\gamma) = -\partial \ln{(e(\lambda,\gamma)/e(1,\gamma))} / \partial \lambda$ at $\lambda = 1$, and  $w(\gamma)$ is defined as $S(\lambda,\gamma) \approx w(\gamma)\; (1-\lambda)^2$ near a critical point $\lambda_c= 1$.
When  $\lambda >1$, rescaling in the Hamiltonian is necessary. Then, we only need to make the replacements: $\lambda \rightarrow \lambda'$,  $S(\lambda, \gamma) \rightarrow S'_\phi (\lambda', \gamma)$,  $\alpha (\lambda, \gamma) \rightarrow \alpha' (\lambda', \gamma)$, $f (\lambda, \gamma) \rightarrow f' (\lambda', \gamma)$, and $e(\lambda, \gamma) \rightarrow e'(\lambda', \gamma)$.

Our analysis enables us to solve a singular first-order differential equation, as shown below.

(1) In Part I ($0 \leq \lambda<\lambda_c=1$, $\gamma=1$),  a divergent part should be separated from $\alpha^{\rm I}(1,\lambda)$, as done in (\ref{alfais}). Then, the singular first-order differential equation (\ref{v1}) may be solved as follows
\begin{equation}
V^{\rm I}(\lambda,1)=V^{\rm I}_0V_1^{\rm I}(\lambda, 1),
\end{equation}
where $V^{\rm I}_0$ is a constant to be determined, and $V_1^{\rm I}(\lambda,1)$ takes the form
\begin{equation}
V_1^{\rm I}(\lambda,1)=(1-\lambda)^p \exp{(-\int_{1}^{\lambda}{f(\mu,1)d\mu})}.
\label{v1j}
\end{equation}

(2) In Part II ($\lambda>\lambda_c= 1$, $\gamma=1$), as mentioned above, $V'^{\rm II}(\lambda',1)$ follows from duality on the line $\gamma=1$.  That is, $V'^{\rm II}(\lambda',1)=V^{\rm I}(\lambda',1)$, with $\lambda'=1/\lambda$.

(3) In Part III ($\lambda=0$, $\gamma<1$), since the integration of $\alpha^{\rm III}(0,\gamma)$ with respect to $\gamma$ is finite,
the singular first-order differential equation (\ref{v3}) may be solved as follows
\begin{equation}
V^{\rm III}(0,\gamma)=V_0^{\rm III}V_1^{\rm III}(0,\gamma),
\label{vgd}
\end{equation}
where $V_0^{\rm III}$ is a constant to be determined, and $V_1^{\rm III}(0,\gamma)$ is defined as
\begin{equation}
V_1^{\rm III}(0,\gamma)=\exp{(\int_{0}^{\gamma}{\alpha^{\rm III} (0,\beta) d\beta})}.
\label{v1gd}
\end{equation}

(4) In Part IV ($\lambda=0$,$0< \gamma \leq 1$),  as mentioned above,  $V'^{\rm IV}(0,\gamma')$ follows from duality on the line $\lambda=0$.  That is, $V'^{\rm IV}(0,\gamma')=V^{\rm III}(0,\gamma')$, with $\gamma'=1/\gamma$ .

On the disorder circle: $\lambda^2+\gamma^2=1$,  $V(\lambda,\gamma)$ simply vanishes.

Similarly, we may determine $V(\lambda,\gamma)$ in five different regimes:

(a) In regime $a$ ($0<\lambda<1$ and $0<\gamma<\sqrt{1-\lambda^2}$),
since the integration of $\alpha(\lambda,\gamma)$ with respect to $\gamma$ for a fixed $\lambda$ is finite,
the singular first-order differential equation (\ref{vaxy}) may be solved as follows
\begin{equation}
V^{a}(\lambda,\gamma)=V_0^{a}(\lambda)V_1^a(\lambda,\gamma),
\end{equation}
where $V_0^{a}(\lambda)$ is a function of $\lambda$, and $V_1^{a}(\lambda,\gamma)$ is defined as
\begin{equation}
V_1^{a}(\lambda,\gamma)=\exp{(\int_{0}^{\gamma}{\alpha^a (\lambda,\beta) d\beta})}.
\label{v1axy}
\end{equation}

(b) In regime $b$ ($0<\lambda<1$ and $\sqrt{1-\lambda^2}<\gamma<1$),
 for a fixed $\gamma$, the singular first-order differential equation (\ref{vbxy}) may be solved as follows
\begin{equation}
V^{b}(\lambda,\gamma)=V^{b}_0(\gamma)V_1^{b}(\lambda,\gamma),
\end{equation}
where $V^{b}_0(\gamma)$ is a function of $\gamma$ , and $V_1^{b}(\lambda,\gamma)$ takes the form
\begin{equation}
V_1^{b}(\lambda,\gamma)=(1-\lambda)^p \exp{(-\int_{1}^{\lambda}{f(\mu,\gamma)d\mu})}.
\label{v1bxy}
\end{equation}

(c) In regime $c$ ( $0<\lambda<1$ and $\gamma>1$),
for a  fixed $\gamma$, the singular first-order differential equation (\ref{vc}) may be solved as follows
\begin{equation}
V^{c}(\lambda,\gamma)=V^{c}_0(\gamma)V_1^{c}(\lambda,\gamma),
\end{equation}
where $V^{c}_0(\gamma)$ is a function of $\gamma$ , and $V_1^{c}(\lambda,\gamma)$ takes the form
\begin{equation}
V_1^{c}(\lambda,\gamma)=(1-\lambda)^p \exp{(-\int_{1}^{\lambda}{f(\mu,\gamma)d\mu})}.
\label{v1c}
\end{equation}

(d) In regime $d$ ( $\lambda>1$ and $0<\gamma<1$),
for a fixed $\gamma$, the singular first-order differential equation (\ref{vdxy}) may be solved as follows
\begin{equation}
V'^{d}(\lambda',\gamma)=V'^{d}_0(\gamma)V_1'^{d}(\lambda',\gamma),
\end{equation}
where $V'^{d}(\gamma)$ is a function of $\gamma$, and $V_1'^{d}(\lambda',\gamma)$ takes the form
\begin{equation}
V_1'^{d}(\lambda',\gamma)=(1-\lambda')^p \exp{(-\int_{1}^{\lambda'}{f'(\mu',\gamma)d \mu'})};
\label{v1d}
\end{equation}

(e) In regime $e$ ( $\lambda>1$ and $\gamma>1$),
for a fixed $\gamma$, the singular first-order differential equation (\ref{ve}) may be solved as follows
\begin{equation}
V'^{e}(\lambda',\gamma)=V'^{e}_0(\gamma)V_1'^{e}(\lambda',\gamma),
\end{equation}
where $V'^{e}_0(\gamma)$ is a function of $\gamma$, and $V_1'^{e}(\lambda',\gamma)$ takes the form
\begin{equation}
V_1'^{e}(\lambda',\gamma)=(1-\lambda')^p \exp{(-\int_{1}^{\lambda'}{f'(\mu',\gamma)d \mu'})};
\label{v1e}
\end{equation}

As we have seen, the two dual lines and the disorder circle, as characteristic lines in the parameter space, divide a given phase into different regimes. In addition, we should also consider $\lambda'=0$ (or, equivalently, $\lambda=1/\lambda'=\infty$), with a finite $\gamma$, as a characteristic line, since a factorized ground state occurs there, thus leading to zero fidelity temperature.
Then, we need to ensure that fidelity mechanical state functions must be {\it continuous}  at boundaries between different regimes for each phase. Indeed, we have already taken into account  the continuity requirement for fidelity entropy, thus determining residual fidelity entropy at the two lines of critical points. The remaining task is to ensure the continuity requirements for fidelity temperature and fidelity internal energy.  As argued in Section~\ref{fmsut}, we need to determine  $T_0 \equiv T_m-T_t$, where $T_m$ represents fidelity temperature at a chosen point on a boundary, calculated from a dominant control parameter in one regime, whereas $T_t$ represents fidelity temperature at the same point, which is determined from a dominant control parameter along the boundary. Specifically, in Part I, we have $T^{\rm I}_m=T^{\rm I}_\phi(0,1)$, but $T^{\rm I}_t=0$.  In Part II, following the duality, we have $T^{\rm II}_m=T'^{\rm II}_\phi(0,1)$, but $T^{\rm II}_t=0$.  In Part III, we have $T^{\rm III}_m=T^{\rm III}(0,1)$, but $T^{\rm III}_t=0$.  In Part IV, following the duality, we have $T^{\rm IV}_m=T'^{\rm IV}_\phi(0,1)$, but $T^{\rm IV}_t=0$.
In regime $a$,  for a fixed $\lambda$,  we have $T^a_m(\lambda)=T^a(\lambda,\sqrt{1-\lambda^2})$, but $T^a_t(\lambda)=0$ on the disorder circle.   In regime $b$,  for a fixed $\gamma$, we have $T^b_m(\gamma)=T^b(\sqrt{1-\gamma^2},\gamma)$, but $T^b_t(\gamma)=0$ on the disorder circle.  In regime $c$, for a fixed $\gamma$, we have $T^c_m(\gamma)=T^c(0,\gamma)$, but $T^c_t(\gamma)=T^{\rm IV}_f(0,\gamma)$ at the dual line $\lambda=0$.  In regime $d$, for a fixed $\gamma$, we have $T^d_m(\gamma)=T'^d_\phi(0,\gamma)$, but $T^d_t(\gamma)=0$ at $\lambda'=0$.  In regime $e$,  for a fixed $\gamma$, we have $T^e_m(\gamma)=T'^e_\phi(0,\gamma)$, but $T^e_t(\gamma)=0$ at $\lambda'=0$.

We are free to choose $V_0$ on one of the characteristic lines, since fidelity internal energy $U_f(\lambda,\gamma)$ is only determined up to a constant factor. Here, we set $V^{\rm III}_0=1$ for the dual line $\lambda=0$.  Then, fidelity internal energy $U^{\rm III}(0,\gamma)$ on the dual line $\lambda=0$ is determined from (\ref{u3}).
Accompanied by a shift from $T^{\rm III}(0,\gamma)$ to $T^{\rm III}(0,\gamma)-T^{\rm III}_0$, fidelity internal energy in Part III is shifted to $U^{\rm III}(0,\gamma)-T^{\rm III}_0S^{\rm III}(0,\gamma)$. Here, $T^{\rm III}_0\equiv T^{\rm III}(0,1)$.
We refer to $T^{\rm III}(0,\gamma)-T^{\rm III}_0$ and
$U^{\rm III}(0,\gamma)-T^{\rm III}_0S^{\rm III}(0,\gamma)$ as $T^{\rm III}_f(0,\gamma)$ and $U^{\rm III}_f(0,\gamma)$, respectively.
That is, $T^{\rm III}_f(0,\gamma) \equiv T^{\rm III}(0,\gamma)-T^{\rm III}_0 $ and $U^{\rm III}_f(0,\gamma)\equiv U^{\rm III}(0,\gamma)-T^{\rm III}_0S^{\rm III}(0,\gamma)$.
Then, fidelity internal energy $U^{\rm IV}_f(0,\gamma)$ and $T^{\rm IV}_f(0,\gamma)$ follows from duality.
As shown in (\ref{utf}), fidelity internal energy $U^f(\lambda,\gamma)$ on the disorder circle is a constant, which is determined to be $U^{\rm III}_f(0,1)$ from the continuity requirement for fidelity internal energy $U_f(\lambda,\gamma)$.
On the dual line $\gamma=1$, for $0<\lambda<1$,  $T^{\rm I}(\lambda,1)$ is shifted to $T^{\rm I}(\lambda,1)-T_0^{\rm I}$, with $T_0^{\rm I}=T^{\rm I}(0,1)$.  Then,
$U^{\rm I}(\lambda,1)$ is shifted to $U^{\rm I}(\lambda,1)-T^{\rm I}_0 S^{\rm I}(\lambda,1)$.
As discussed in Section~\ref{fmsut}, we demand that (i) $U^{\rm I}(\lambda,1)-T^{\rm I}_0S^{\rm I}(\lambda,1)$ must be zero at a critical point. That is, $U^{\rm I}_0=T^{\rm I}_0 S^{\rm I}_0$.
(ii) Fidelity internal energy $U^{\rm I}(\lambda,1)-T^{\rm I}_0S^{\rm I}(\lambda,1)$ satisfies the continuity requirement at $\lambda=0$ and $\gamma=1$: $U^{\rm I}(0,1)-T^{\rm I}_0S^{\rm I}(0,1)=U^{\rm III}_f(0,1)$.  Therefore, $V^{\rm I}_0$ is determined as follows
 \begin{equation}
V^{\rm I}_0=\frac{U^{\rm III}_f(0,1)}{-\ln{(e(0,1)/e(1,1))}V_1^{\rm I}(0,1)+\alpha^{\rm I}(0,1) V_1^{\rm I}(0,1)(S^{\rm I}(0,1)-S^{\rm I}_0)},
\end{equation}
where $\alpha^{\rm I}(0,1)$ and $V_1^{\rm I}(0,1)$ are given by Eq.~(\ref{alphad1}) and  Eq.~(\ref{v1j}), respectively.
We refer to $T^{\rm I}(\lambda,1)-T^{\rm I}_0$ and
$U^{\rm I}(\lambda,1)-T^{\rm I}_0S^{\rm I}(\lambda,1)$ as $T^{\rm I}_f(\lambda,1)$ and $U^{\rm I}_f(\lambda,1)$, respectively.
That is, $T^{\rm I}_f(\lambda,1) \equiv T^{\rm I}(\lambda,1)-T^{\rm I}_0$ and $U^{\rm I}_f(\lambda,1)\equiv U^{\rm I}(\lambda,1)-T^{\rm I}_0S^{\rm I}(\lambda,1)$.
In Part II, fidelity internal energy $U^{\rm II}_f(\lambda,1)$ and $T^{\rm II}_f(\lambda,1)$ follows from duality.

(a)  In regime $a$ ($\lambda<1$ and $0<\gamma<\sqrt{1-\lambda^2}$ ), for a fixed $\lambda$, in order to ensure the continuity requirement for fidelity temperature   $T_f(\lambda,\gamma)$ on the disorder circle, $T^a(\lambda,\gamma)$ is shifted to $T^a(\lambda,\gamma)-T^a_0(\lambda)$, with $T^a_0(\lambda)=T^a(\lambda,\sqrt{1-\lambda^2})$.  Then,
$U^a(\lambda,\gamma)$ is shifted to $U^a(\lambda,\gamma)-T^a_0(\lambda)S^a(\lambda,\gamma)$, with $S^a(\lambda,\gamma)$ left intact.
Following our discussions in Section~\ref{fmsut}, fidelity entropy $S^a_f(\lambda,\gamma)$,  fidelity temperature $T^a_f(\lambda,\gamma)$, and fidelity internal energy $U^a_f(\lambda,\gamma)$ take the form: $S^a_f(\lambda,\gamma) = S^a(\lambda,\gamma)$,  $T^a_f(\lambda,\gamma) = T^a(\lambda,\gamma)-T^a_0(\lambda)$ and $U^a_f(\lambda,\gamma) = U^a(\lambda,\gamma)- T^a_0(\lambda) S^a(\lambda,\gamma)$, respectively.
In addition, we demand that (i) $U^a(\lambda,\gamma)$ must be zero at a critical point. That is, $U^a_0(\lambda)=T^a_0(\lambda) S^a_0(\lambda)$;
(ii) fidelity internal energy $U^a(\lambda,\gamma)$ satisfies the continuity requirement on the disorder circle: $U^a(\lambda,\sqrt{1-\lambda^2})-T^a_0(\lambda)S^a(\lambda,\sqrt{1-\lambda^2})=U^{\rm III}_f(0,1)$.
Therefore, $V^{a}_0(\lambda)$ is determined as follows
\begin{equation}
V^a_0(\lambda)=\frac{U^{\rm III}_f(0,1)}{\ln{(e(\lambda,\sqrt{1-\lambda^2})/e(\lambda,0))}V_1^a(\lambda,\sqrt{1-\lambda^2})+\alpha^a(\lambda,\sqrt{1-\lambda^2})V_1^a(\lambda,\sqrt{1-\lambda^2})(S^a(\lambda,\sqrt{1-\lambda^2})-S^a_0(\lambda))},
\end{equation}
where $\alpha^a(\lambda,\sqrt{1-\lambda^2})$ and $V_1^a(\lambda,\sqrt{1-\lambda^2})$ are determined from  (\ref{alphaaxy}) and (\ref{v1axy}).
Once $V^a_0(\lambda)$ and $U^a_{0}(\lambda)$ are determined, fidelity internal energy $U^a(\lambda,\gamma)$ and fidelity temperature $T^a(\lambda,\gamma)$ follow from (\ref{uaxy}) and (\ref{taxy}), respectively.
Hence, fidelity temperature $T^a_f(\lambda,\gamma)$ and fidelity internal energy $U^a_f(\lambda,\gamma)$ follow.

(b) In regime $b$ ( $\sqrt{1-\lambda^2}<\gamma<1$ and $0<\lambda<1$), for a fixed $\gamma$,
in order to ensure the continuity requirement for fidelity temperature $T_f(\lambda,\gamma)$ on the disorder circle,
$T^b(\lambda,\gamma)$ is shifted to $T^b(\lambda,\gamma)-T_0^b(\gamma)$, with $T_0^b(\gamma)=T^b(\sqrt{1-\gamma^2},\gamma)$.  Then,
$U^b(\lambda,\gamma)$ is shifted to $U^b(\lambda,\gamma)-T_0^b(\gamma) S^b(\lambda,\gamma)$, with $S^b(\lambda,\gamma)$ left intact.
Following our discussions in Section~\ref{fmsut}, fidelity entropy $S^b_f(\lambda,\gamma)$,  fidelity temperature $T^b_f(\lambda,\gamma)$, and fidelity internal energy $U^b_f(\lambda,\gamma)$ take the form: $S^b_f(\lambda,\gamma) = S^b(\lambda,\gamma)$,  $T^b_f(\lambda,\gamma) = T^b(\lambda,\gamma)-T^b_0(\gamma)$ and  $U^b_f(\lambda,\gamma) = U^b(\lambda,\gamma)- T^b_0(\gamma) S^b(\lambda,\gamma)$, respectively.
 In addition, we demand that (i) $U^{b}(\lambda,\gamma)$ must be zero at a critical point. That is, $U^b_0(\gamma)=T_0^b(\gamma) S^b_0(\gamma)$;
 (ii) fidelity internal energy $U^b(\lambda,\gamma)$ satisfies the continuity requirement: $U^b(\sqrt{1-\gamma^2},\gamma)-T_0^b(\gamma)S^b(\sqrt{1-\gamma^2},\gamma)=U^{\rm III}_f(0,1)$, on the disorder circle.
 Therefore, $V^b_0(\gamma)$ is determined as follows
 \begin{equation}
V^b_0(\gamma)=\frac{U^{\rm III}_f(0,1)}{-\ln{(e(\sqrt{1-\gamma^2},\gamma)/e(1,\gamma))}V_1^b(\sqrt{1-\gamma^2},\gamma)+\alpha^b(\sqrt{1-\gamma^2},\gamma)
V_1^b(\sqrt{1-\gamma^2},\gamma)(S^b(\sqrt{1-\gamma^2},\gamma)-S^b_0(\gamma))},
\end{equation}
where $\alpha^b(\sqrt{1-\gamma^2},\gamma)$ and $V_1^b(\sqrt{1-\gamma^2},\gamma)$ are determined from (\ref{alphabxy}) and  (\ref{v1bxy}).
Once $V^b_0(\gamma)$ and $U^b_{0}(\gamma)$ are determined, fidelity internal energy $U^b(\lambda,\gamma)$ and fidelity temperature $T^b(\lambda,\gamma)$ follow from (\ref{ubxy}) and (\ref{tbxy}), respectively.
Hence, fidelity temperature $T^b_f(\lambda,\gamma)$ and fidelity internal energy $U^b_f(\lambda,\gamma)$ follow.

(c)  In regime $c$ ( $0<\lambda<1$ and $\gamma>1$), for a fixed $\gamma$,
in order to ensure the continuity requirement for fidelity temperature $T_f(\lambda,\gamma)$ on a dual line $\lambda=0$,
$T^c(\lambda,\gamma)$ is shifted to $T^c(\lambda,\gamma)-T_0^c(\gamma)$, with $T^c_0(\gamma)=T^c(0,\gamma)-T^{\rm IV}_f(0,\gamma)$.  Then,
$U^c(\lambda,\gamma)$ is shifted to $U^{c}(\lambda,\gamma)-T^c_0(\gamma)S^c(\lambda,\gamma)$, with $S^c(\lambda,\gamma)$ left intact.
Following our discussions in Section~\ref{fmsut}, fidelity entropy $S^c_f(\lambda,\gamma)$,  fidelity temperature $T^c_f(\lambda,\gamma)$, and fidelity internal energy $U^c_f(\lambda,\gamma)$ take the form: $S^c_f(\lambda,\gamma) = S^c(\lambda,\gamma)$,  $T^c_f(\lambda,\gamma) =T^c(\lambda,\gamma)-T^c_0(\gamma)$ and  $U^c_f(\lambda,\gamma) = U^c(\lambda,\gamma)- T^c_0(\gamma) S^c(\lambda,\gamma)$, respectively.
 In addition,  we demand that (i) fidelity internal energy $U^{c}(\lambda,\gamma)$ at a critical point is zero.  That is, $U^{c}_0(\gamma)=T_0 S^c_0(\gamma)$;
(ii) fidelity internal energy $U^{c}(\lambda,\gamma)-T_0S^c(\lambda,\gamma)$ satisfies the continuity requirement: $U^{c}(0,\gamma)-T^c_0(\gamma) S^c(0,\gamma)=U^{\rm IV}_f(0,\gamma)$ at $\lambda=0$.
Therefore, $V^c_0(\gamma)$ is determined as follows
\begin{equation}
V^c_0(\gamma)=\frac{U^{\rm IV}_f(0, \gamma)-T^{\rm IV}_f(0, \gamma)(S^c(0,\gamma)-S^c_0(\gamma))}{-\ln{(e(0,\gamma)/e(1,\gamma))}
V_1^c(0,\gamma)+\alpha^c(0,\gamma)V_1^c(0,\gamma)(S^c(0,\gamma)-S^c_0(\gamma))},
\end{equation}
where $\alpha^c(\lambda,\gamma)$ and $V_1^c(\lambda,\gamma)$ are determined from (\ref{alphac}) and  (\ref{v1c}).
Once $V^c_0(\gamma)$ and $U^c_{0}(\gamma)$ are determined, fidelity internal energy $U^c(\lambda,\gamma)$ and fidelity temperature $T^c(\lambda,\gamma)$ follow from (\ref{uc}) and (\ref{tc}), respectively.
Hence, fidelity temperature $T^c_f(\lambda,\gamma)$ and fidelity internal energy $U^c_f(\lambda,\gamma)$ follow.

(d) In regime $d$ ( $\lambda>1$ and $0<\gamma<1$), for a fixed $\gamma$, in order to ensure the continuity requirement for fidelity temperature $T_f(\lambda,\gamma)$ on a dual line $\lambda'=0$, $T'^d_\phi(\lambda',\gamma)$ is shifted to $T'^d_\phi(\lambda',\gamma)-T^d_0(\gamma)$, with $T^d_0(\gamma)=T'^d_\phi(0,\gamma)$ and $\lambda'=1/\lambda$,
accompanied by a shift in  $U'^d_\phi(\lambda',\gamma)$: $U'^d_\phi(\lambda',\gamma)-T^d_0(\gamma) S'^d_\phi(\lambda',\gamma)$, with
$S'^{d}_\phi(\lambda',\gamma)$ left intact.
Following discussions in Section~\ref{fmsut}, we demand that (i) fidelity internal energy at a critical point is zero.  That is, $U'^d_{\phi 0}(\gamma)=T^d_0(\gamma) S'^d_{\phi 0}(\gamma)$;
(ii) fidelity internal energy $U'^d_{\phi}(\lambda',\gamma)-T^d_0(\gamma)S'^d_{\phi}(\lambda',\gamma)$ satisfies the continuity requirement at $\lambda'=0$: $U'^{d}_{\phi}(0,\gamma)-T^d_0(\gamma)S'^d_{\phi}(0,\gamma)=U^{\rm III}_f(0,1)$.
Therefore, $V'^d_{0}(\gamma)$ is determined as follows
\begin{equation}
V'^{d}_0(\gamma)=\frac{U^{\rm III}_f(0,1)}{-\ln{(e'(0,\gamma'))/(e'(1,\gamma'))}V_1'^{d}(0,\gamma)+\alpha'^{d}(0,\gamma')V_1'^{d}( 0,\gamma')(S'^{d}_\phi(0,\gamma')-S'^{d}_{\phi 0}(\gamma'))},
\end{equation}
where $\alpha'^d(\lambda',\gamma)$ and $V_1'^d(\lambda',\gamma)$ are determined from (\ref{alphadxy}) and (\ref{v1d}).
Once $V'^{d}_0(\gamma)$ and $U'^d_{\phi 0}(\gamma)$ are determined, fidelity internal energy $U'^d_\phi(\lambda',\gamma)$ and fidelity temperature $T'^d_\phi(\lambda',\gamma)$ follow from (\ref{udxy}) and (\ref{tdxy}), respectively.
We refer to  $T'^d_\phi(\lambda',\gamma)-T^d_0(\gamma) $ and
$U'^d_\phi(\lambda',\gamma)-T^d_0(\gamma)S'^d_\phi (\lambda',\gamma)$ as $T'^d_{\phi f}(\lambda',\gamma)$ and $U'^d_{\phi f}(\lambda',\gamma)$, respectively.
That is, $T'^d_{\phi f}(\lambda',\gamma) \equiv T'^d_\phi(\lambda',\gamma)-T^d_0(\gamma)$ and  $U'^d_{\phi f} (\lambda',\gamma) \equiv U'^d_\phi(\lambda',\gamma)-T^d_0(\gamma)S'^d_\phi (\lambda',\gamma)$.
Therefore, fidelity temperature $T^d_f(\lambda,\gamma)$ and fidelity internal energy $U^d_f(\lambda,\gamma)$ take the form:  $T^d_f(\lambda,\gamma) = T^d_{\phi f}(\lambda,\gamma)$ and $U^d_f(\lambda,\gamma) =
U^d_{\phi f} (\lambda,\gamma)$, with  $T^d_{\phi f}(\lambda,\gamma) \equiv T'^d_{\phi f}(\lambda',\gamma)$ and $U^d_{\phi f}(\lambda,\gamma) \equiv U'^d_{\phi f}(\lambda',\gamma)$.

(e) In regime $e$ ( $\lambda>1$ and $\gamma>1$), for a fixed $\gamma$,
in order to ensure the continuity requirement for fidelity temperature $T_f(\lambda,\gamma)$ on a dual line $\lambda'=0$,
$T'^e_\phi(\lambda',\gamma)$ is shifted to $T'^e_\phi(\lambda',\gamma)-T^e_0(\gamma)$, with $ T^e_0(\gamma) = T'^e_\phi(0,\gamma)$ and $\lambda'=1/\lambda$,
accompanied by a shift in
$U'^e_\phi(\lambda',\gamma)$ is shifted to $U'^e_\phi(\lambda',\gamma)-T^e_0(\gamma) S'^e_\phi(\lambda',\gamma)$, with
$S'^e_\phi(\lambda',\gamma)$ left intact.
Following discussions in Section~\ref{fmsut}, we demand that  (i) fidelity internal energy at a critical point is zero. That is, $U'^{e}_{\phi 0}(\gamma)=T^e_0(\gamma) S'^e_{\phi 0}(\gamma)$;
 (ii) fidelity internal energy $U'^{e}_\phi(\lambda',\gamma)-T^e_0(\gamma) S'^e_\phi(\lambda',\gamma)$ satisfies the continuity requirement at $\lambda'=0$: $U'^{e}_\phi(0,\gamma)-T^e_0(\gamma) S'^e_\phi(0,\gamma)=U^{\rm III}_f(0,1)$ .
 Therefore, $V'^e_0(\gamma)$ is determined as follows
\begin{equation}
V'^{e}_0(\gamma)=\frac{U^{\rm III}_f(0,1)}{-\ln{(e'(0,\gamma')/e'(1,\gamma'))}V_1'^{e}(0,\gamma)+\alpha'^{e}(0,\gamma')V_1'^{e}( 0,\gamma')(S'^{e}_\phi(0,\gamma')-S'^{e}_{\phi 0}(\gamma'))},
\end{equation}
where $\alpha'^e(\lambda,\gamma)$ and $V_1'^e(\lambda,\gamma)$ are determined from (\ref{alphae}) and (\ref{v1e}).
Once $V'^{e}_0(\gamma)$ and $U'^e_{\phi 0}(\gamma)$ are determined, fidelity internal energy $U'^e_\phi(\lambda',\gamma)$ and fidelity temperature $T'^e_\phi(\lambda',\gamma)$ follow from (\ref{ue}) and (\ref{te}), respectively.
We refer to  $T'^e_\phi(\lambda',\gamma)-T^e_0(\gamma)$ and
$U'^e_\phi(\lambda',\gamma)-T^e_0(\gamma)S'^e_\phi (\lambda',\gamma)$ as $T'^e_{\phi f}(\lambda',\gamma)$ and $U'^e_{\phi f}(\lambda',\gamma)$, respectively.
That is, $T'^e_{\phi f}(\lambda',\gamma) = T'^e_\phi(\lambda',\gamma)-T^e_0(\gamma) $ and  $U'^e_{\phi f} (\lambda',\gamma) = U'^e_\phi(\lambda',\gamma)-T^e_0(\gamma) S'^e_\phi (\lambda',\gamma)$.
Therefore, fidelity temperature $T^e_f(\lambda,\gamma)$ and fidelity internal energy $U^e_f(\lambda,\gamma)$ take the form: $T^e_f(\lambda,\gamma)\equiv T^e_{\phi f}(\lambda,\gamma)$ and $U^e_f(\lambda,\gamma)\equiv
U^e_{\phi f} (\lambda,\gamma)$, with  $T^e_{\phi f}(\lambda,\gamma) \equiv T'^e_{\phi f}(\lambda',\gamma)$ and $U^e_{\phi f}(\lambda,\gamma) \equiv U'^e_{\phi f}(\lambda',\gamma)$.

Numerical simulation results for fidelity entropy $S_f(\lambda, \gamma)$, fidelity temperature $T_f(\lambda, \gamma)$, and fidelity internal energy $U_f(\lambda, \gamma)$ for the quantum XY chain are shown in Fig.~\ref{xyentropy}~(a),~(b) and (c), respectively.

\section{Fidelity entropy, fidelity temperature and fidelity internal energy for the transverse field quantum Ising chain in a longitudinal field}\label{ishchain}

In this Appendix, we present mathematical details about fidelity entropy $S_f(\lambda,h)$, fidelity temperature $T_f(\lambda,h)$ and fidelity internal energy $U_f(\lambda,h)$ for the transverse field quantum Ising chain in a longitudinal field.

For this model,  the ground state phase diagram is simple.  A first-order QPT occurs at $h=0$, when $0 \leq \lambda<1$, which ends at a critical point $(1, 0)$.  At the first-order QPT points, the model is driven from a phase with spin polarization in $-x$  to a phase with spin polarization in $x$, when $h$ changes its sign.  When $h=0$, the model becomes the transverse field quantum Ising chain, which exhibits a second-order QPT at $\lambda_c =1$, characterized by the $Z_2$ symmetry breaking order for $\lambda<1$.
As mentioned in Appendix~\ref{xychain}, duality occurs for the transverse field quantum Ising chain. We remark that, when $h=0$, fidelity mechanical state functions have been determined as a special case of the quantum XY model, corresponding to $\gamma =1$. Therefore, we restrict ourselves to the region $h>0$.

With the symmetry and duality in mind,  we may divide the region $h>0$ into two principal regimes, as shown in Fig.~\ref{hlam}.  They are labeled as regime $a$, defined as $0 \leq \lambda<1$ and $h \in (0, \infty)$, and regime $b$, defined as $\lambda \geq 1$ and $h \in (0, \infty)$.

In each regime, we choose a dominant control parameter:

(a) In regime $a$ ($0 \leq \lambda<1$ and $h \in (0, \infty)$),  the range of $h$ is not finite.  Then, we choose $h'=h/(1+h)$ as a dominant control parameter for a fixed $\lambda$, which prevents from any divergence in the ground state energy density when $h \rightarrow \infty$.  In this regime, we rescale the ground state energy density: $e(\lambda,h) = k'(h') \; e'(\lambda, h')$, with $h' = h/(1+h)$ and $k'(h')=1/(1-h')$.

(b) In regime $b$ ($\lambda \geq 1$ and $h \in (0, \infty)$),  the ranges of $\lambda$ and $h$ are not finite.  Then, we re-parameterize $\lambda$ and $h$ by defining a radius $r$ and an azimuthal angle $\theta$: $r =\sqrt{(\lambda -1)^2 + h^2}$ and $\theta =\arctan h/(\lambda-1)$.  In this regime,  a dominant control parameter is chosen to be $r'=r/(1+r)$, with a fixed $\theta$.  Here,  $r$ ranges from $0$ to $\infty$, but $r'$ ranges from 0 to 1.  This choice is consistent with the requirement from duality, which occurs when $\theta =0$, corresponding to the transverse field quantum Ising chain.
We rescale the ground state energy density: $e(r,\theta) = k'(r') \; e'(r', \theta)$, with $r' = r/(1+r)$ and $k'(r')=1/(1-r')$.

Fidelity entropy $S_f(\lambda,h)$, fidelity temperature $T_f(\lambda,h)$ and fidelity internal energy $U_f(\lambda,h)$  may be determined following our formalism on
first-order QPTs  (regime $a$) and continuous QPTs (regime $b$), respectively, as discussed in Section ~\ref{fmsut}.

(a) In regime $a$ ($0 \leq \lambda<1$ and $h \in (0, \infty)$),  fidelity mechanical state functions for $h'=0$ has been known, as a special case of the quantum XY model, with $\gamma =1$.
We denote them as fidelity entropy $S^{\rm {Is}}(\lambda)$, fidelity temperature $T^{\rm {Is}}(\lambda)$ and fidelity $U^{\rm {Is}}(\lambda)$, respectively. Since rescaling in the ground state energy: $e(\lambda,h) = k'(h') \; e'(\lambda, h')$ is performed.  As discussed in Section ~\ref{fmsut}, $S^a_f(\lambda,h)$ includes contributions from fidelity entropy $S'^a_{\phi f} (\lambda,h')$ and from scaling entropy $S'^a_{\sigma f}(\lambda,h')$, with $S'^a_{\phi f} (\lambda,h')\equiv S'^a_{\phi} (\lambda,h')$ and $S'^a_{\sigma f}(\lambda,h')\equiv S'^a_{\sigma}(\lambda,h')$. Here, $S'^a_{\sigma}(\lambda,h')=\ln{k'(h')}$.
Thus, we have $S^a_f(\lambda,h) = S^a_{\phi f} (\lambda,h) +S^a_{\sigma f}(\lambda,h)$ with $S^a_{\phi f} (\lambda,h) \equiv S'^a_{\phi f} (\lambda,h')$ and  $S^a_{\sigma f}(\lambda,h) \equiv S'^a_{\sigma f}(\lambda,h')$.
For a fixed $\lambda$, following Eq.~(\ref{entropyd}), fidelity entropy $S'^a_\phi (\lambda, h')$ takes the form
\begin{equation}
S'^a_\phi (\lambda,h')=-2\int_{0}^{h'}{\ln{d'(\lambda,h';\lambda,l')}d{l'}} +S'^a_{\phi \; 0}(\lambda),
\label{sh}
\end{equation}
where $d'(\lambda,h';\lambda,l')\equiv d(\lambda,h;\lambda,l)$, with $l$ denoting another value of the same control parameter as $h$, and $l'=l/(1+l)$, and $S'^a_{\phi 0}(\lambda)$ is residual fidelity entropy at a transition point $h'_d=0$ for a fixed $\lambda$.
The continuity requirement for fidelity entropy $S_f(\lambda,h)$ at $h=0$ implies that $S'^a_{\phi 0}(\lambda)=S_{\rm {Is}}(\lambda)$.

Since $e'(\lambda, h')$ is monotonically increasing with increasing $h'$, fidelity internal energy $U'^a_\phi (\lambda,h')$ takes the form
\begin{equation}
U'^a_\phi (\lambda,h')= -[\ln \kappa + \ln(\frac{ e'(\lambda,h')}{e'(\lambda,0)})]V'^a(\lambda,h') + U'^a_{\phi 0}(\lambda).
\label{uaish}
\end{equation}
Here, $U'^a_{\phi 0}(\lambda)$ is a function of $\lambda$, and $V'^a(\lambda,h')$ takes the form
\begin{equation}
V'^a(\lambda,h')=V'^a_0(\lambda)\exp{\int_{0}^{h'}\alpha'^a(\lambda,l') dl'},
\label{vaish}
\end{equation}
with
\begin{equation}
\alpha'^a(\lambda,h')=\frac{\partial{\ln{(e'(\lambda,h')/e'(\lambda,0))}/\partial{h'}}}
{\partial{S'^a_\phi (\lambda,h')}/\partial{h'}-\ln\kappa-\ln{(e'(\lambda,h')/e'(\lambda,0))}}.
\label{alphaaish}
\end{equation}
Accordingly, fidelity temperature $T'^a_\phi (\lambda,h')$ takes the form
\begin{equation}
T'^a_\phi (\lambda,h')=-\partial{V'^a(\lambda,h')}/\partial{h'}.
\label{taish}
\end{equation}

In addition to the dual line: $h'=0$, there is another characteristic line: $h'=1$, with $0<\lambda<1$,
since a factorized ground state occurs at $h'=1$, thus leading to zero fidelity temperature, with fidelity internal energy being
$U^{\rm Is}(0)$.  In order to ensure the continuity requirements for fidelity temperature and fidelity internal energy,
we need to perform a shift: $T'^a_\phi (\lambda,h') \rightarrow T'^a_\phi (\lambda,h')-T^a_0(\lambda)$, accompanied by a shift: $U'^a_\phi (\lambda,h') \rightarrow U'^a_\phi (\lambda,h') - T^a_0(\lambda) S'^a_\phi (\lambda,h')$.  Here, $T^a_0(\lambda) \equiv T^a_m(\lambda)-T^a_t(\lambda)$,  where $T^a_m(\lambda)$ represents fidelity temperature at a characteristic line, evaluated from a dominant control parameter in one regime, whereas $T^a_t(\lambda)$ represents fidelity temperature at the characteristic line.
Specifically, in regime $a$,
for a fixed $\lambda$, we have $T^a_m(\lambda)=T'^a_\phi(\lambda,1)$, but $T^a_t(\lambda)=0$ at $h'=1$.
Therefore, $T'^a_\phi(\lambda,h')$ is shifted to $T'^a_\phi(\lambda,h')-T'^a_\phi(\lambda,1)$ and $U'^a_\phi(\lambda,h')$ is shifted to $U'^a_\phi(\lambda,h')-T'^a_\phi(\lambda,1)S'^a_\phi (\lambda,h')$.
At $h'=0$, the continuity requirements for fidelity temperature and fidelity internal energy imply that $T'^a_\phi(\lambda,0)-T'^a_\phi(\lambda,1)=T^{\rm{Is}}(\lambda)$ and $U'^a_\phi(\lambda,0)-T'^a_\phi(\lambda,1)S^{\rm Is}(\lambda)=U^{\rm{Is}}(\lambda)$, respectively.
On the other hand, at $h'=1$, the continuity requirement for fidelity internal energy implies that
$U'^a_\phi(\lambda,1)-T'^a(\lambda,1)S'^a_\phi(\lambda,1)=U^{\rm{Is}}(0)$.

Therefore, $\kappa$, $U'^a_{\phi 0}(\lambda)$ and $V'^a_0(\lambda)$ are determined from these continuity requirements for a fixed $\lambda$.
Thus, $U'^a_\phi(\lambda,h')$ and $T'^a_\phi(\lambda,h')$ follow from (\ref{uaish}) and (\ref{taish}), respectively.
We refer to $T'^a_\phi(\lambda,h)-T^a_0(\lambda)$ and
$U'^a_\phi(\lambda,h')-T^a_0(\lambda)S'^a_\phi (\lambda,h')$ as $T'^a_{\phi f}(\lambda,h')$ and $U'^a_{\phi f}(\lambda,h')$, respectively.
Here, $T^a_0(\lambda)=T'^a_\phi(\lambda,1)$.
That is, $T'^a_{\phi f}(\lambda,h) \equiv T'^{a}_\phi (\lambda,h')-T^a_0(\lambda)$ and  $U'^a_{\phi f} (\lambda,h') \equiv U'^a_{\phi }(\lambda,h')-T^a_0(\lambda)S'^a_{\phi} (\lambda,h')$.
Then, following our discussions in Section~\ref{fmsut}, fidelity temperature $T^a_f(\lambda,h)$ and fidelity internal energy $U^d_f(\lambda, h)$ take the form:  $T^a_f(\lambda, h) = T^a_{\phi f}(\lambda, h)$ and $U^a_f(\lambda, h) =
U^a_{\phi f} (\lambda,h)$, with $T^a_{\phi f}(\lambda,h) \equiv T'^a_{\phi f}(\lambda,h')$ and $U^a_{\phi f}(\lambda,h) \equiv U'^a_{\phi f}(\lambda,h')$.

 \begin{figure}
    \centering
     \includegraphics[angle=0,totalheight=4cm]{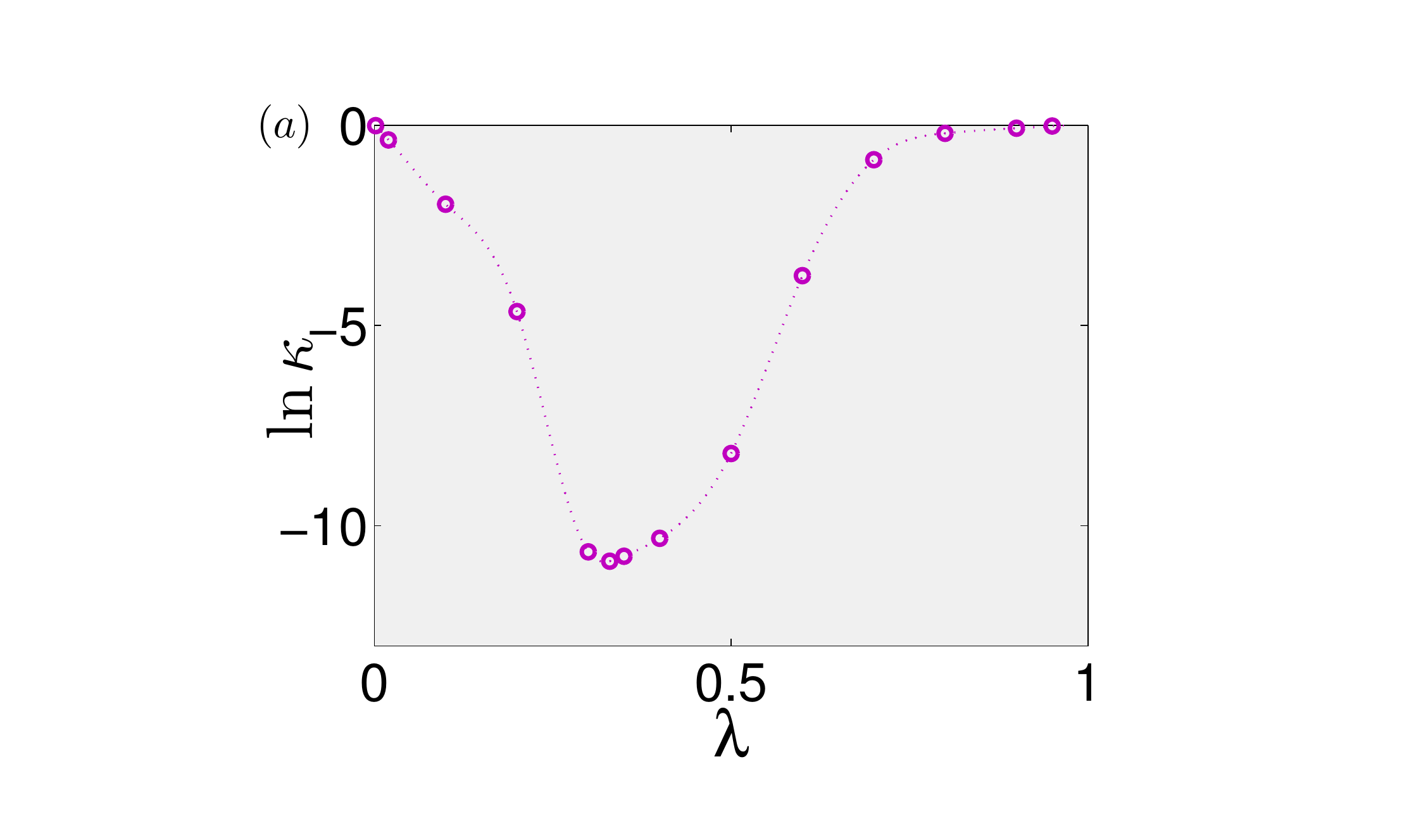}
        \includegraphics[angle=0,totalheight=4cm]{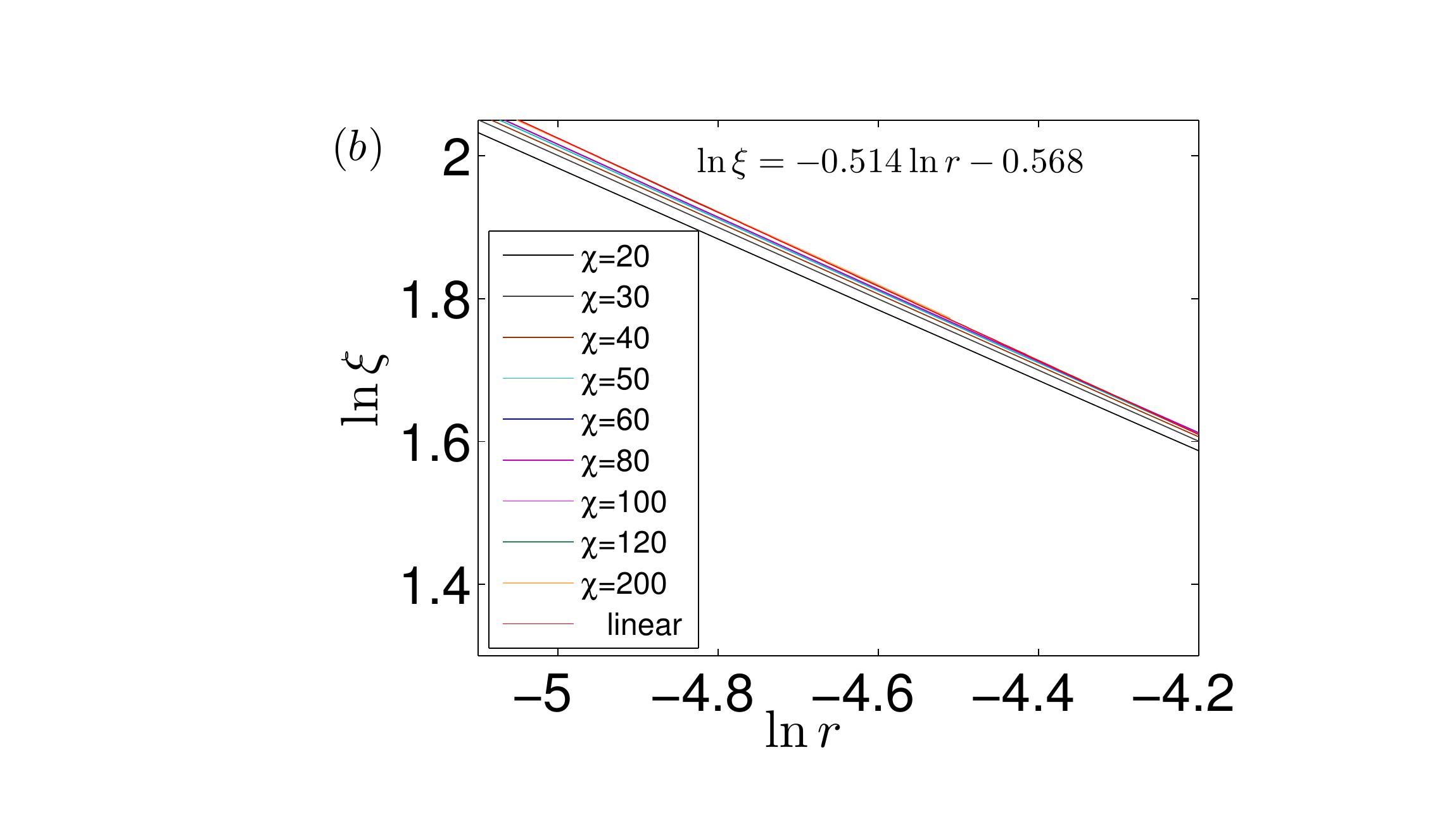}
      \caption{
   (a) The parameter $\ln \kappa$ for the transverse field quantum Ising chain in a longitudinal field at the first-order QPT points.  It approaches zero when $\lambda \rightarrow 0$ and $\lambda \rightarrow 1$, and exhibits a minimum around $\lambda =0.33$.
   (b) The critical exponent $\nu$ for the correlation length $\xi$ for the transverse field quantum Ising chain in a longitudinal field. The critical point $(1,0)$ is approached along a straight line with $\theta=\pi/4$. The  correlation length $\xi$ scales as $\xi\sim r^{-\nu}$. Our numerical simulation shows that $\nu \sim 0.51$.  For other choices of $\theta$,
it yields similar results, with the critical exponent $\nu$ ranging from $0.51$ to $0.52$.  Here, the model has been simulated by exploiting a tensor network  algorithm~\cite{TN1,TN2}, in the context of matrix product states, with the bond dimension $\chi$ ranging from 20 to 200.}\label{cn}
    \end{figure}

The model is not exactly solvable when $h \neq 0$. Instead, we simulate it numerically to compute  ground state wave functions by exploiting a tensor network  algorithm~\cite{TN1,TN2}, in the context of matrix product states, with the bond dimension $\chi$ ranging from 20 to 200.  This allows us to determine, among others,  the parameter $\ln \kappa$, as plotted in Fig.~\ref{cn}\;(a). It approaches zero when  $\lambda \rightarrow 0$ and $\lambda \rightarrow 1$, and exhibits a minimum around $\lambda =0.33$.

(b) In regime $b$ ($\lambda \geq 1$ and $h \in (0, \infty)$),
the critical point at  (1,0) controls the underlying physics.  In this regime, fidelity mechanical state functions are computed following our discussions in Section~\ref{fmsut} for continuous QPTs.  The ranges of $r'$ and $\theta$ in this regime are $0<r'<1$ and $0<\theta<\pi/2$, respectively.
Since rescaling in the ground state energy: $e(r,\theta) = k'(r') \; e'(r', \theta)$ is performed.  As discussed in Section ~\ref{fmsut},
 $S^b_f(r,\theta)$ includes contributions from fidelity entropy $S'^b_{\phi f} (r',\theta)$ and from scaling entropy $S'^b_{\sigma f}(r',\theta)$, with $S'^b_{\phi f} (r',\theta) \equiv S'^b_{\phi} (r',\theta)$ and $S'^b_{\sigma f}(r',\theta) \equiv S'^b_{\sigma}(r',\theta)$.  Here, $S'^b_\sigma(r',\theta)=\ln{k'(r')}$.
Thus, we have $S^b_f(r,\theta) = S^b_{\phi f} (r,\theta) +S^b_{\sigma f}(r,\theta)$, with $S^b_{\phi f} (r,\theta) \equiv S'^b_{\phi f} (r',\theta) $ and $S^b_{\sigma f}(r,\theta) \equiv S'^b_{\sigma f}(r',\theta)$.
For a fixed $\theta$, following Eq.~(\ref{entropy}), fidelity entropy $S'^b_\phi (r',\theta)$ takes the form

\begin{equation}
S'^{b}_\phi (r', \theta)=-2\int_{0}^{r}{\ln{(d'(r', \theta; w', \theta))} d w'}+S'^{b}_{\phi 0}(\theta),
\label{sr}
\end{equation}
where $d'(r', \theta; w', \theta)\equiv d(r, \theta; w, \theta)$,  with $w$ denoting another value of the same control parameter as $r$, $w'=w/(1+w)$, and $S'^{b}_{\phi 0} (\theta)$ is residual fidelity entropy at the critical point (1,0), which is equal to fidelity entropy for the transverse field quantum Ising chain, i.e., $S'^{b}_{\phi 0}(\theta)=S^{\rm Is}(1)$ for any $\theta$.

Since $e'(r',\theta)$ is monotonically increasing with increasing $r'$,  $U'^b_\phi (r',\theta)$ takes the form
\begin{equation}
U'^{b}_\phi (r',\theta)=-\ln{\frac{e'(r',\theta)}{e'(0,\theta)}}V'^{b}(r',\theta)+U'^{b}_{\phi 0}(\theta),
\label{ubth}
\end{equation}
Here, $U'^{b}_{\phi 0}(\theta)$ is a function of $\theta$, and
$V'^{b}(r',\theta)>0$ satisfies the singular first-order differential equation
\begin{equation}
\frac{\partial{V'^{b}(r',\theta)}}{\partial{r'}}=\alpha'^{b}(r',\theta) V'^{b} (r',\theta),
\label{vbth}
\end{equation}
with
\begin{equation}
\alpha'^{b}(r',\theta)=
\frac{\partial{\ln{(e'(r',\theta)/e(0,\theta))}}/\partial{r'}}{\partial{S'^{d}_\phi (r',\theta)}/\partial{r'}-
\ln{(e'(r',\theta)/ e'(0,\theta))}}.\label{alphath}
\end{equation}
Accordingly, fidelity temperature $T'^b_\phi (r',\theta)$ in this regime is given by
\begin{equation}
 T'^b_\phi (r',\theta)=-\frac{\partial{V'^{b}(r',\theta)}}{\partial{r'}}.
 \label{tbth}
\end{equation}

In order to solve the singular first-order differential equation, we analyze the scaling behavior of $\alpha'^{b}(r',\theta)$ near the critical point.
For $\theta\neq0$, we find that the critical exponent $\nu$ takes $\nu \simeq 0.5$, with $\theta=\pi/4$ as an example, shown in Fig.~\ref{cn}(b),  Thus, fidelity entropy $S'^{b}_\phi(r',\theta)$ scales as $r'^{3/2}$.  This is consistent with a general scaling analysis in
Appendix~\ref{scaling}, which predicts that  $S'^{b}_\phi(r',\theta)\sim r'^{\nu+1}$.
Therefore, for a fixed $\theta$, $\alpha'^{b}(r',\theta)$ diverges, when $r'$ goes to 0:

\begin{equation}
\alpha'^{b}(r',\theta)\propto \frac{1}{r'^{1/2}}.
\end{equation}

For a fixed $\theta$, the differential equation (\ref{vbth}) may be solved as follows
\begin{equation}
 V'^{b}(r',\theta)=V'^{b}_0(\theta)V'^{b}_1(r',\theta),
\end{equation}
where $V'^{b}_0(\theta)>0$ is a function of $\theta$,  and $V'^{b}_1(r',\theta)$ is defined as
\begin{equation}
V'^{b}_1(r',\theta)=\exp{(\int_{0}^{r'}{\alpha'^b(w',\theta) d w'})},
\label{v1th}
\end{equation}
Therefore, fidelity internal energy $U'^{b}_\phi (r', \theta)$ and fidelity temperature $T'^b_\phi (r', \theta)$
follow from (\ref{ubth}) and (\ref{tbth}), respectively.

In this regime, in order to ensure that fidelity temperature vanishes when $r'=1$, or, equivalently,  $r=r'/(r'-1)\rightarrow \infty$,
$T'^b_\phi(r',\theta)$ is shifted to $T'^b_\phi(r',\theta)-T^b_0(\theta)$, with $T^b_0(\theta)=T'^b_\phi(1,\theta)$,
accompanied by a shift in
$U'^{b}_\phi(r',\theta)$: $U'^{b}_\phi(r',\theta)-T^b_0(\theta)S'^b_\phi(r',\theta)$.
Following discussions in Section~\ref{fmsut}, we demand that (i) fidelity internal energy at a critical point is zero, i.e., $U'^{b}_{\phi 0}(\theta)=T^b_0(\theta) S'^b_{\phi 0}(\theta)$;  (ii)
fidelity internal energy at $r'=1$ satisfies the continuity requirement: $U'^{b}_\phi(1,\theta)-T^b_0(\theta) S'^b_\phi (1,\theta)=U^{\rm Is}(0)$.
Hence, $V'^b_0(\theta)$ is determined as follows

\begin{equation}
V'^{b}_0(\theta)=\frac{U_{\rm Is}(0)}
{\ln{(e'(r',\theta)/e'(0,\theta))}V'^{b}_1(1,\theta)+\alpha'^b(r',\theta) V'^{b}_1(r',\theta)
(S'^{b}_{\phi}(1,\theta)-S^{\rm Is}(1))},
\end{equation}
where $\alpha'^b(r',\theta)$ and $V'^b_1(r',\theta)$ are determined from (\ref{alphath}) and (\ref{v1th}), respectively.
Thus,  $U'^b_\phi(r',\theta)$ and $T'^b_\phi(r',\theta)$ follow from (\ref{ubth}) and (\ref{tbth}), respectively.   We refer to
$T'^b_\phi(r',\theta)-T^b_0(\theta)$ and
$U'^b_\phi(r',\theta)-T^b_0(\theta)S'^b_\phi (r',\theta)$ as $T'^b_{\phi f}(r',\theta)$ and $U'^b_\phi (r',\theta)$, respectively.  That is, $T'^b_{\phi f}(r',\theta) \equiv T'^b_\phi (r',\theta)-T^b_0(\theta)$ and  $U'^b_{\phi f} (r',\theta) \equiv U'^b_\phi(r',\theta)-T^b_0(\theta)S'^b_\phi (r',\theta)$.  Therefore, fidelity temperature $T^b_f (r,\theta)$ and fidelity internal energy $U^b_f(r,\theta)$ take the form  $T^b_f(r,\theta) = T^b_{\phi f}(r,\theta)$  and  $U^b_f(r,\theta) =
U^b_{\phi f} (r,\theta)$, with $T^b_{\phi f}(r,\theta) \equiv T'^b_{\phi f}(r',\theta)$ and  $U^b_{\phi f}(r,\theta) \equiv U'^b_{\phi f}(r',\theta)$.

Numerical simulation results for fidelity entropy $S_f(\lambda,h)$, fidelity temperature $T_f(\lambda,h)$ and fidelity internal energy $U_f(\lambda,h)$ for the transverse field quantum Ising chain in a longitudinal field are  shown in Fig.~\ref{sutis}~(a),~(b) and (c), respectively.

\section{Fidelity entropy, fidelity temperature and fidelity internal energy for the quantum spin-$1/2$ XYZ model}~\label{xyzchain}

In this Appendix, we present mathematical details about fidelity entropy $S_f(\Delta,\gamma)$, fidelity temperature  $T_f(\Delta,\gamma)$ and fidelity internal energy $U_f(\Delta,\gamma)$ for the quantum spin-$1/2$ XYZ model.

For this model, the Hamiltonian (\ref{xyz}) is symmetric under $\gamma \leftrightarrow - \gamma$.  Therefore, we may restrict ourselves to the region $\gamma \geq 0$.
As is shown in Fig.~\ref{xyzphase}, there are three different phases, labeled as $\rm {AF}_x$, $\rm {AF}_z$, and $\rm {FM}_z$,
representing an antiferromagnetic phase in the $x$ direction, an antiferromagnetic phase in the $z$ direction, and a ferromagnetic phase in the $z$ direction, respectively.
There are three critical lines: $\gamma=0~(-1<\Delta\leq 1)$, $\gamma = -1+\Delta ~(\Delta \geq 1)$ and $\gamma = -1-\Delta ~(\Delta < -1)$.  A remarkable feature of the quantum spin-$1/2$ XYZ model is that there are five different dualities.  The details may be found in Appendix~\ref{dual}.   In addition, there is a characteristic line $\gamma=1+\Delta$ with $\Delta>-1$, representing factorizing fields~\cite{factorizing}.

Taking into account the symmetries, dualities, and factorizing fields, we may divide the region $\gamma\geq 0$ into twelve different regimes, with the lines defined by $\gamma=0$, $\gamma=1$ and $\gamma=\pm1\pm\Delta$ as boundaries.
These twelve regimes are separated into two groups, with six regimes in each group dual to each other.  As shown in Fig.~\ref{duality2} (a), regimes $\rm{I}$, $\rm{III}$, $\rm{V}$, $\rm{I'}$, $\rm{III'}$ and $\rm{V'}$  are dual to each other, whereas regimes $\rm{II}$, $\rm{IV}$, $\rm{VI}$, $\rm{II'}$, $\rm{IV'}$ and $\rm{VI'}$ are dual to each other.
Therefore, there are only two principal regimes, representing the physics underlying the quantum spin-$1/2$ XYZ model.  Here, we choose regime $\rm{I}$ ($0<\Delta<1$ and $0<\gamma<1-\Delta$) and regime $\rm{II}$ ($-1<\Delta<0$ and $0<\gamma<1+\Delta$) as  two principal regimes.

In regime $\rm{I}$ and regime $\rm{II}$,  a dominant control parameter is chosen to be $\gamma$, given that $\gamma=0$ is a critical line.
Then, choices of a dominant control parameter in other regimes simply follow from their respective dualities to regime  $\rm{I}$ and regime $\rm{II}$.

Let us determine fidelity entropy $S^f(\Delta,\gamma)$ on a line of factorizing fields $\gamma=1+\Delta$ and fidelity entropy $S^d(\Delta,\gamma)$ on a dual line $\gamma=1-\Delta$, with $0<\gamma<1$:

(1) On the line $\gamma=1+\Delta$, with $0<\gamma<1$,  the same factorized state occurs as ground state wave functions, with the ground state energy density $e(\Delta, 1+\Delta)=-(\Delta+2)/2$. We rescale the ground state energy density: $e(\Delta, 1+\Delta)=k'(\Delta')e'(\Delta', 1+\Delta')$, with $k'(\Delta')=(\Delta'+2)/2$ and $\Delta'=\Delta$.
Note that the quantum spin-$1/2$ XYZ model becomes the quantum XY model, when $\Delta =0$. Therefore, fidelity mechanical state functions on the line $\Delta=0$ has been determined, as discussed in Appendix~\ref{xychain}.  In particular, fidelity entropy $S^f(0,1)$ at $\Delta=0$ and $\gamma=1$ is known.  With this in mind, fidelity entropy $S^f(\Delta, \gamma)$ on the line $\gamma=1+\Delta$, with $0<\gamma<1$, is identical to $S^f(0,1)$, up to scaling entropy $S'^f_\sigma(\Delta',1+\Delta')$. Thus, we have
$S^f(\Delta,1+\Delta) = S^f(0,1)+S^f_{\sigma f}(\Delta,1+\Delta)$ with $S^f_{\sigma f}(\Delta,1+\Delta)\equiv S^f_{\sigma}(\Delta,1+\Delta)$, Here,
$S^f_{\sigma}(\Delta,1+\Delta)\equiv S'^f_\sigma(\Delta',1+\Delta')=\ln((\Delta'+2)/2)$.
As discussed in Section~\ref{fmsut}, we have $S_f^f(\Delta,1+\Delta) = S^f(\Delta,1+\Delta)$.

(2) On the line $\gamma=1-\Delta$, with $0<\gamma<1$, the Hamiltonian $H(\Delta,1-\Delta)$ may be rescaled as $H(\Delta,1-\Delta)=k'(\Delta') H'(\Delta', 0)$, with $k'(\Delta')=\Delta'/(1+\Delta')$, $\Delta'=(2-\Delta)/\Delta$.  Here, $H'(\Delta',0)$ is unitarily equivalent to the quantum spin-$1/2$ XXZ model, which corresponds to $\gamma =0$ in the Hamiltonian (\ref{xyz}). Note that $\Delta' >1$. For $H'(\Delta',0)$, we choose $1-1/\Delta'$ as a dominant control parameter.
It should be emphasized that the ground state energy density $e(\Delta,1-\Delta)$ is not monotonic as a function of $\Delta$. However,  rescaling in the ground state energy: $e(\Delta,1-\Delta)=k'(\Delta') e'(\Delta', 0)$ ensures that both $k'(\Delta')$ and $e'(\Delta', 0)$ are monotonically increasing with increasing $1-1/\Delta'$. In particular, $\Delta'$ has been chosen to be consistent with duality between regime III and regime IV, as discussed in Appendix~\ref{dual}.
Then, $S^d_f(\Delta,1-\Delta)$ includes contributions from fidelity entropy $S'^d_{\phi f}(\Delta',0)$ and from scaling entropy $S'^d_{\sigma f}(\Delta',0)$, with $S'^d_{\phi f}(\Delta',0)\equiv S'^d_{\phi }(\Delta',0)$ and  $S'^d_{\sigma f}(\Delta',0)\equiv S'^d_{\sigma}(\Delta',0)$.
Here, $S'^d_{\sigma f}(\Delta',0)=\ln (\Delta')-\ln (1+\Delta')$.
That is, we have $S^d_f(\Delta,1-\Delta)=S^d_{\phi f}(\Delta,1-\Delta)+S^d_{\sigma f}(\Delta,1-\Delta)$, with  $S^d_{\phi f}(\Delta,1-\Delta)\equiv S'^d_{\phi f}(\Delta',0)$ and $S^d_{\sigma f}(\Delta,1-\Delta)\equiv S'^d_{\sigma f}(\Delta',0)$.
Here, $S'^d_\phi(\Delta',0)$ takes the form
\begin{equation}
S'^d_\phi(\Delta',0)=2\int_{1}^{\frac{1}{\Delta'}}{\ln{d'(\Delta',0;\Theta',0)}} \; d(\frac{1}{\Theta'})+S'^d_{\phi 0}.
\label{sd2}
\end{equation}
Here, $d'(\Delta',0;\Theta',0)\equiv d(\Delta,1-\Delta;\Theta,1-\Theta)$, with $\Theta$ denoting another value of the same control parameter as $\Delta$, $\Theta'=(2-\Theta)/\Theta$, and $S'^d_{\phi 0}$ is residual fidelity entropy at a critical point $\Delta'=1$  ( or, equivalently, $\Delta=1$) and $\gamma=0$, in addition to a contribution to residual fidelity entropy from scaling entropy $S'^d_{\sigma}(\Delta',0)$, which is $-\ln 2$.

From the continuity requirement for fidelity entropy $S^d_f(\Delta,1-\Delta)$ at $\Delta=0$ and $\gamma=1$, $S'^d_{\phi 0}$ is determined from $S^d_f(0,1)=S^f(0,1)$.

We move to regime ${\rm I}$ and regime ${\rm II}$:

(a) In regime $\rm{I}$ ($0<\Delta<1$ and $\gamma<1-\Delta$) , a dominant control parameter is chosen to be $\gamma$. Then, for a fixed $\Delta$, fidelity entropy $S^{\rm I}(\Delta,\gamma)$ takes the form
\begin{equation}
  S^{\rm I}(\Delta,\gamma)=-2\int_{0}^{\gamma}{\ln{d(\Delta,\gamma;\Delta,\beta)}d{\beta}}+S^I_{0}(\Delta).
  \label{sxyz1}
\end{equation}
Here, $S^{\rm I}_{0}(\Delta)$ is a function of $\Delta$, representing residual fidelity entropy at a critical point $\gamma=0$ for a fixed $\Delta \in (0,1)$.

(b) In regime $\rm{II}$ ($-1<\Delta<0$ and $\gamma<1+\Delta$), a dominant control parameter is chosen to be $\gamma$. Then, for a fixed $\Delta$, fidelity entropy $S^{\rm II}(\Delta,\gamma)$  takes the form

\begin{equation}
  S^{\rm II}(\Delta,\gamma)=-2\int_{0}^{\gamma}{\ln{d(\Delta,\gamma;\Delta,\beta)}d{\beta}}+S^{\rm II}_{0}(\Delta).
  \label{sxyz2}
\end{equation}
Here, $S^{\rm II}_{0}(\Delta)$ is a function of $\Delta$, representing residual fidelity entropy at a critical point $\gamma=0$ for a fixed $\Delta \in (-1,0)$.

We turn to residual fidelity entropy $S^{\rm I}_{0}(\Delta)$ and $S^{\rm II}_{0}(\Delta)$ on a line of critical points: $\gamma=0$, with $-1<\Delta\leq 1$.
Residual fidelity entropy $S^{\rm I}_{0}(\Delta)$ is determined from the continuity requirement for fidelity entropy at the dual line $\gamma = 1 - \Delta$: $S^{\rm I}(\Delta,1-\Delta)=S^d_{\phi f}(\Delta,1-\Delta)$, whereas residual fidelity entropy $S^{\rm II}_{0}(\Delta)$ is determined from the continuity requirement for fidelity entropy at a line of factorizing fields $\gamma = 1 + \Delta$: $S^{\rm II}(\Delta,1+\Delta)=S^f(0,1)$.
Moreover, the continuity requirements for $S_f(\Delta,\gamma)$ on these two characteristic lines $\gamma=1-\Delta$ and $\gamma=1+\Delta$, $S_f(\Delta,\gamma)$ in regime $\rm I$ and regime $\rm{II}$ should include a contribution from residual scaling entropy $S^{\rm I}_{\sigma 0}(\Delta)$ and $S^{\rm II}_{\sigma 0}(\Delta)$, due to rescaling in the ground state energy density, respectively.
Hence, we have $S^{\rm I}_f(\Delta,\gamma)=S^{\rm I}(\Delta,\gamma)+S^{\rm I}_{\sigma 0}(\Delta)$ and  $S^{\rm{II}}_f(\Delta,\gamma)=S^{\rm{II}}(\Delta,\gamma)+S^{\rm I}_{\sigma 0}(\Delta)$.
Here, $S^{\rm I}_{\sigma 0}(\Delta)=S^d_{\sigma f}(\Delta,1-\Delta)$ and $S^{\rm II}_{\sigma 0}(\Delta)=S^f_{\sigma f}(\Delta,1+\Delta)$.

As discussed in Section \ref{fmsut}, fidelity entropy for other regimes follows from their respective dualities to regime $\rm{I} $ and regime $\rm{II}$, with extra scaling entropy $S_{\sigma}(\Delta,\gamma)=\pm \ln{k(\Delta,\gamma)}$ up to a constant, in each regime.  Here, $\pm$ is determined to ensure that $\partial{ S_{\sigma}(\Delta,\gamma)}/\partial{\gamma}>0$, and
$k(\Delta,\gamma)$ follows from dualities in Appendix~\ref{dual}.  Specifically, regimes $\rm{III}$, $\rm{V}$, $\rm{I'}$, $\rm{III'}$ and $\rm{V'}$  are dual to regime $\rm I$, then
fidelity entropy $S_f(\Delta,\gamma)$ in  each of these regimes includes a contribution from fidelity entropy $S'^\beta_{\phi f}(\Delta',\gamma')$ and from scaling entropy $S'^\beta_{\sigma f}(\Delta',\gamma')$, with $S'^\beta_{\phi f}(\Delta',\gamma')\equiv S'^\beta_{\phi }(\Delta',\gamma')$ identical to fidelity entropy $S^{\rm I}(\Delta',\gamma')$ in regime $\rm{I}$ and $S'^\beta_{\sigma f}(\Delta',\gamma')\equiv S'^\beta_{\sigma}(\Delta',\gamma')$, with $\beta= \rm{III}$, $\rm{V}$, $\rm{I'}$, $\rm{III'}$ and $\rm{V'}$.  Here, scaling entropy $S'^\beta_{\sigma}(\Delta',\gamma')$ follows from a rescaling factor $k(\Delta, \gamma)$,  as discussed in Appendix~\ref{dual}:
(i) in regime $\rm{III}$, $S'^{\rm III}_{\sigma}(\Delta',\gamma')=-2\ln 2+\ln(2-\Delta')+\ln(1+\Delta'+\gamma')$;
(ii) in regime $\rm{V}$,  $S'^{\rm V}_{\sigma}(\Delta',\gamma')=-2\ln 2+\ln(2-\Delta') + 2\ln(1+\Delta' + \gamma' )-\ln(1+\Delta'-\gamma')$;
(iii) in regime $\rm{I'}$, $S'^{\rm I'}_{\sigma}(\Delta',\gamma')=-\ln 2+\ln(2-\Delta')+\ln{\gamma'}$;
(iv) in regime $\rm{III'}$,  $S'^{\rm III'}_{\sigma}(\Delta',\gamma')=-2\ln 2+\ln(2-\Delta') + \ln(1-\Delta'+\gamma')$;
(v) in regime $\rm{V'}$, $S'^{\rm V'}_{\sigma}(\Delta',\gamma')=-2\ln 2+\ln(2-\Delta') + 2 \ln(1-\Delta'+\gamma')-\ln(1-\Delta'-\gamma')$.
Following from our prescription,  we have $S^{\beta}_f(\Delta,\gamma)=S^\beta_{\phi f}(\Delta,\gamma)+S^\beta_{\sigma f}(\Delta,\gamma)$, with
$S^\beta_{\phi f}(\Delta,\gamma)\equiv S'^\beta_{\phi f}(\Delta',\gamma')$ and $S^\beta_{\sigma f}(\Delta,\gamma)\equiv S'^\beta_{\sigma f}(\Delta',\gamma')$.
Similarly, regimes $\rm{IV}$, $\rm{VI}$, $\rm{II'}$, $\rm{IV'}$ and $\rm{VI'}$ are dual to regime $\rm{II}$, then $S_f(\Delta,\gamma)$ in each of these regimes includes a contribution from fidelity entropy $S'^\beta_{\phi f}(\Delta',\gamma')$ and from scaling entropy $S'^\beta_{\sigma f}(\Delta',\gamma')$, with $S'^\beta_{\phi f}(\Delta',\gamma')\equiv S'^\beta_{\phi }(\Delta',\gamma')$ identical to fidelity entropy $S^{\rm II}(\Delta',\gamma')$ in regime $\rm{II}$ and $S'^\beta_{\sigma f}(\Delta',\gamma')\equiv S'^\beta_{\sigma}(\Delta',\gamma')$ with $\beta= \rm{IV}$, $\rm{VI}$, $\rm{II'}$, $\rm{IV'}$ and $\rm{VI'}$.  Here, scaling entropy $S'^\beta_{\sigma}(\Delta',\gamma')$  follows from a rescaling factor $k(\Delta, \gamma)$,  as discussed in Appendix~\ref{dual}:
(i) in regime $\rm{IV}$,  $S'^{\rm IV}_{\sigma}(\Delta',\gamma')=-2\ln 2+\ln(\Delta'+2)+\ln(1-\Delta'+\gamma')$;
(ii) in regime $\rm{VI}$, $S'^{\rm VI}_{\sigma}(\Delta',\gamma')=-2\ln 2 +\ln(\Delta'+2)+ 2\ln(1-\Delta' + \gamma')-\ln(1-\Delta'-\gamma')$;
(iii) in regime $\rm{II'}$,  $S'^{\rm II'}_{\sigma}(\Delta',\gamma')=-\ln 2+\ln(\Delta'+2)+\ln{\gamma'}$;
(iv) in regime $\rm{IV'}$,  $S'^{\rm IV'}_{\sigma}(\Delta',\gamma')=-2\ln 2+\ln(\Delta'+2) + \ln(1+\Delta'+\gamma')$;
(v) in regime $\rm{VI'}$, $S'^{\rm VI'}_{\sigma}(\Delta',\gamma')=-2\ln 2+\ln(\Delta'+2)+ 2 \ln(1+\Delta'+\gamma')-\ln(1+\Delta'-\gamma')$.
Following from our prescription,  we have  $S^\beta_f(\Delta,\gamma)=S^\beta_{\phi f}(\Delta,\gamma)+S^\beta_{\sigma f}(\Delta,\gamma)$, with
$S^\beta_{\phi f}(\Delta,\gamma)\equiv S'^\beta_{\phi f}(\Delta',\gamma')$ and $S^\beta_{\sigma f}(\Delta,\gamma)\equiv S'^\beta_{\sigma f}(\Delta',\gamma')$.

Once fidelity entropy $S_f(\Delta,\gamma)$ is determined, fidelity temperature $T_f(\Delta,\gamma)$ and fidelity internal energy $U_f(\Delta,\gamma)$ may be determined from solving the singular first-order differential equation (\ref{alpha}), as discussed for continuous QPTs in Section II.

We determine fidelity temperature $T_f(\Delta,\gamma)$ and fidelity internal energy $U_f(\Delta,\gamma)$  for two characteristic  lines: (1) the line of factorizing fields $\gamma=1+\Delta$, with $0<\gamma<1$ and (2) the dual line $\gamma=1-\Delta$, with $0<\gamma<1$.

(1) On the line of factorizing fields $\gamma=1+\Delta$, with $0<\gamma<1$, fidelity temperature $T^f(\Delta,1+\Delta)$ vanishes: $T^f(\Delta,1+\Delta)=0$.  Meanwhile,
fidelity internal energy $U^f(\Delta,1+\Delta)$ is a constant: $U^f(\Delta,1+\Delta)= U^f(0,1)$,
where $U^f(0,1)$ has been determined in Appendix~\ref{xychain}, since the quantum spin-$1/2$ XYZ model becomes the quantum XY model, when $\Delta =0$.
As discussed in Section~\ref{fmsut}, we have $T_f^f(\Delta,1+\Delta) = T^f(\Delta,1+\Delta)$ and $U_f^f(\Delta,1+\Delta) = U^f(\Delta,1+\Delta)$ .

(2) On the dual line $\gamma=1-\Delta$, with $0<\gamma<1$,  rescaling in the ground state energy density: $e(\Delta,1-\Delta)=k'(\Delta') e'(\Delta', 0)$ is performed.  Since $e'(\Delta',0)$ is monotonically increasing with increasing $1-1/\Delta'$,  $U'^d_\phi(\Delta',0)$ takes the form
\begin{equation}
U'^{d}_\phi(\Delta',0)=-\ln{\frac{e'(\Delta',0)}{e'(1,0)}}V'^{d}(\Delta',0)+U'^{d}_{\phi 0}.
\label{udxyz}
\end{equation}
Here, $U'^{d}_{\phi 0}$ is an additive constant, and
$V'^{d}(\Delta',0)>0$ satisfies the differential equation

\begin{equation}
\frac{\Delta'^2\partial{V'^{d}(\Delta',0)}}{\partial{\Delta'}}=\alpha'^{d}(\Delta',0) V'^{d} (\Delta',0),
\label{vdxyz}
\end{equation}
with
\begin{equation}
\alpha'^{d}(\Delta',0)=
\frac{\Delta'^2\partial{\ln{(e'(\Delta',0)/e(1,0))}}/\partial{\Delta'}}{\Delta'^2\partial{S'^{d}_\phi(\Delta',0)}/\partial{\Delta'}-
\ln{(e'(\Delta',0)/ e'(1,0))}}.\label{alphadxyz}
\end{equation}
Accordingly, fidelity temperature $T'^d(\Delta',0)$ on the dual line  is given by
\begin{equation}
 T'^d_\phi(\Delta',0)=-\frac{\Delta'^2\partial{V'^{d}(\Delta',0)}}{\partial{\Delta'}}.
 \label{tdxyz}
\end{equation}

Next, we move to fidelity temperature and fidelity internal energy in two regimes:

(a) In regime $\rm{I}$ ($0<\Delta<1$ and $0<\gamma<1-\Delta$), for a fixed $\Delta$, the ground state energy density $e(\Delta,\gamma)$ is monotonically decreasing with an increasing dominant control parameter $\gamma$.  Then, from Eq.~(\ref{internalenergy}), fidelity internal energy $U^{\rm I}(\Delta,\gamma)$ takes the form
\begin{equation}
U^{\rm I}(\Delta,\gamma)=\ln{\frac{e(\Delta,\gamma)}{e(\Delta,0)}}V^{\rm I}(\Delta,\gamma)+U^{\rm I}_0(\Delta). \label{uxyzi}
\end{equation}
Here, $U^{\rm I}_0(\Delta)$ is a function of $\Delta$,  and
$V^{\rm I}(\Delta,\gamma)>0$ satisfies the differential equation
\begin{equation}
\frac{\partial{V^{\rm I}(\Delta,\gamma)}}{\partial{\gamma}}=\alpha^{\rm I}(\Delta,\gamma) \; V^{\rm I} (\Delta,\gamma), \label{vxyzi}
\end{equation}
with
\begin{equation}
\alpha^{\rm I}(\Delta,\gamma)=-\frac{\partial{\ln{(e(\Delta,\gamma)/e(\Delta,0))}}/\partial{\gamma}}{\partial{S^{\rm I}(\Delta,\gamma)}/\partial{\gamma}+\ln{(e(\Delta,\gamma)/e(\Delta,0))}}.
\label{alphaxyzi}
\end{equation}
Accordingly, fidelity temperature $T^{\rm I}(\Delta,\gamma)$ in this regime is given by
\begin{equation}
T^{\rm I}(\Delta,\gamma) =-\frac{\partial{V_{\rm I}(\Delta,\gamma)}}{\partial{\gamma}} \label{txyzi}
\end{equation}

(b) In regime $\rm{II}$ ($-1<\Delta<0$ and $0<\gamma<1+\Delta$), for a fixed $\Delta$, the ground state energy density $e(\Delta,\gamma)$ is monotonically decreasing with an increasing dominant control parameter $\gamma$.  Then, from Eq.~(\ref{internalenergy}), fidelity internal energy $U^{\rm II}(\Delta,\gamma)$ takes the form
\begin{equation}
U^{\rm II}(\Delta,\gamma)=\ln{\frac{e(\Delta,\gamma)}{e(\Delta,0)}}V^{\rm II}(\Delta,\gamma)+U^{\rm II}_0(\Delta). \label{uxyzii}
\end{equation}
Here, $U^{\rm II}_0(\Delta)$ is a function of $\Delta$,  and
$V^{\rm II}(\Delta,\gamma)>0$ satisfies the differential equation
\begin{equation}
\frac{\partial{V^{\rm II}(\Delta,\gamma)}}{\partial{\gamma}}=\alpha^{\rm II}(\Delta,\gamma) \; V^{\rm II} (\Delta,\gamma), \label{vxyzii}
\end{equation}
with
\begin{equation}
\alpha^{\rm II}(\Delta,\gamma)=-\frac{\partial{\ln{(e(\Delta,\gamma)/e(\Delta,0))}}/\partial{\gamma}}{\partial{S^{\rm II}(\Delta,\gamma)}/\partial{\gamma}+\ln{(e(\Delta,\gamma)/e(\Delta,0))}}.
\label{alphaxyzii}
\end{equation}
Accordingly, fidelity temperature $T^{\rm II}(\Delta,\gamma)$ in this regime is given by
\begin{equation}
T^{\rm II}(\Delta,\gamma) =-\frac{\partial{V^{\rm II}(\Delta,\gamma)}}{\partial{\gamma}} \label{txyzii}
\end{equation}

\begin{figure}
     \includegraphics[angle=0,totalheight=5cm]{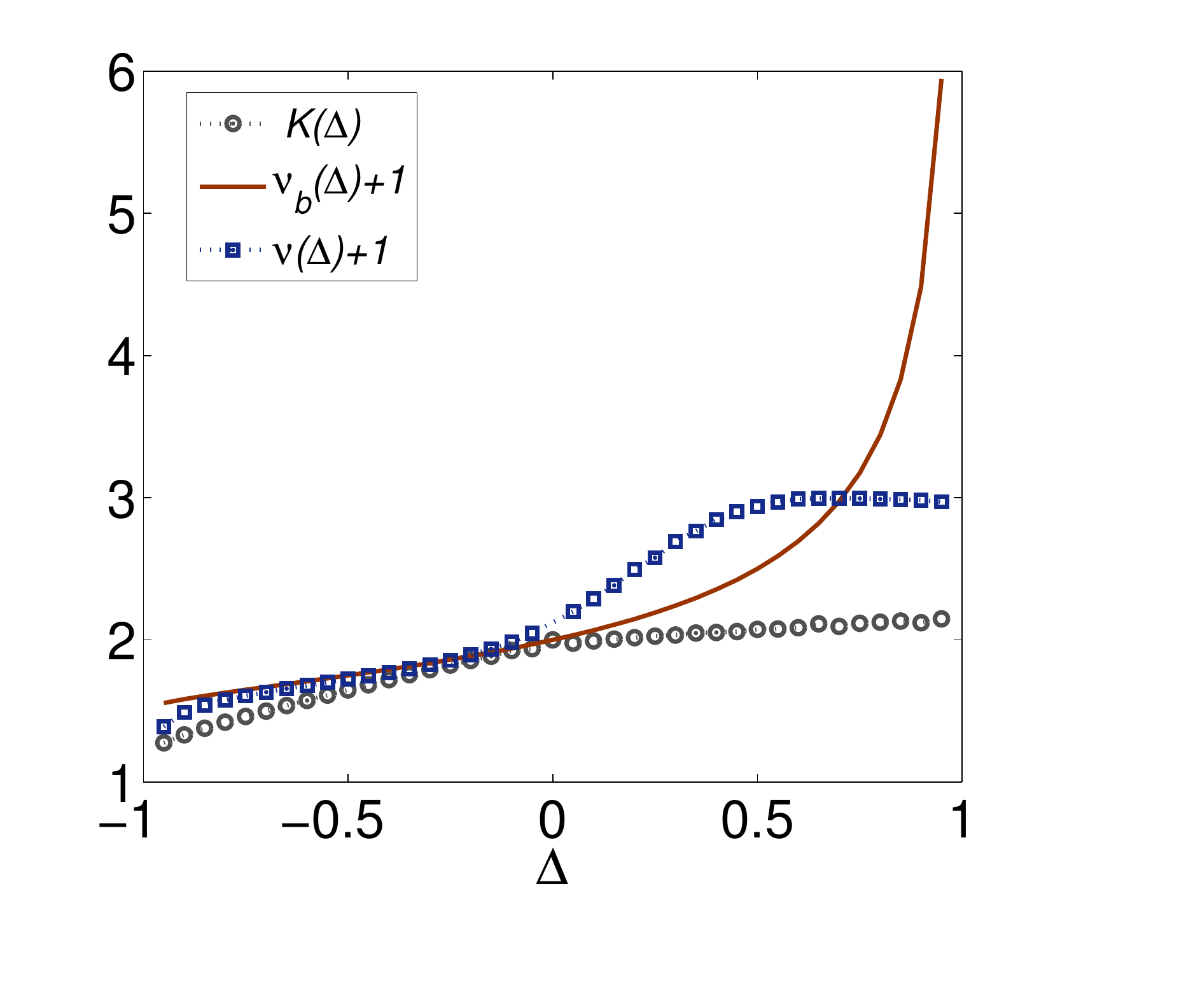}
       \caption{A parameter $K(\Delta)$, $1+\nu_{b}(\Delta)$ and $1+\nu(\Delta)$  as a function of $\Delta \in (-1,1)$.  Here, $K(\Delta)$ is defined via a scaling relation for the ground state energy density $e(\Delta,\gamma) \sim \gamma^{K(\Delta)} \; \ln \gamma$  for a fixed $\Delta$, and $\nu_b(\Delta)$ and $\nu(\Delta)$ represent, respectively, two different sets of the critical exponent $\nu$ for the correlation length:   one is $\nu (\Delta)$, which is extracted from the scaling  behavior of fidelity entropy $S^{\rm I/II}(\Delta,\gamma)$,  the other is $\nu_b(\Delta)$, extracted from the leading singular term via the exact solution. }\label{betanu}
   \end{figure}

To solve the singular first-order differential equation (\ref{vxyzi}) and (\ref{vxyzii}), we analyze the scaling behavior of $\alpha^{\rm I/II}(\Delta,\gamma)$ near a critical point $\gamma_c=0$ for $-1 <\Delta<1$.
As discussed in Appendix~\ref{scaling}, fidelity entropy $S^{\rm I/II}(\Delta,\gamma)$ scales as $S^{\rm I/II}(\Delta, \gamma) \sim (|\gamma-\gamma_c|)^{\nu (\Delta)+1}$, with $\nu (\Delta)$, as a function of $\Delta \in (-1,1)$, being the critical exponent for the correlation length.
In addition, for a fixed $\Delta \in (-1,1)$, our numerical simulation shows that the ground state energy density $e(\Delta,\gamma)$ near a critical point $\gamma_c=0$ scales as
 \begin{equation}
 \ln{\frac{e(\Delta,\gamma)}{e(\Delta,0)}}\sim \gamma^{K(\Delta)} \; \ln \gamma.
 \end{equation}
In regimes $\rm I$ and $\rm II$, as long as $\nu(\Delta)<K(\Delta) \leq \nu(\Delta)+1$,  if $\gamma \rightarrow 0$, then $\alpha (\Delta,\gamma)$ diverges as
\begin{equation}
 \alpha(\Delta,\gamma) \propto \gamma^{K(\Delta)-\nu(\Delta)-1}\ln{\gamma}.
 \label{caseii}
 \end{equation}
This is confirmed numerically, as shown in Fig.~\ref{betanu}.  Actually, two different sets of the critical exponent $\nu$ are plotted as a function of $\Delta \in (-1,1)$: one is $\nu (\Delta)$, which is extracted from the scaling  behavior of fidelity entropy $S^{\rm I/II}(\Delta,\gamma)$,  the other is $\nu_b(\Delta)$, extracted from the leading singular term via the exact solution in Ref.~\onlinecite{Baxterbook} (see, also Ref.~\cite{luther}).
For $-1<\Delta<0$, $\nu(\Delta)$ and $\nu_b(\Delta)$ matches, with accuracy up to $5\%$.
However, for $\Delta>0$,  a significant discrepancy arises between $\nu(\Delta)$ and $\nu_b(\Delta)$. One might attribute this discrepancy to the fact that only the leading singular term is taken into account to extract $\nu_b(\Delta)$, which also neglects the presence of a logarithmic factor $\ln \gamma$.  Indeed, the necessity to include this logarithmic factor
may be justified from a heuristic argument that  it exists, since  $\alpha(\Delta,\gamma)$ should be smooth along a line of critical points with $\Delta \in (-1,1)$, combining with the fact that it exists at an infinite number of discrete points between $\Delta =0$ and $\Delta=1$, if $\pi/ \mu$ is an even integer, with $\cos \mu = \Delta$~\cite{Baxterbook}.
Note that our numerical result for $\nu (\Delta)$, as $\Delta$ approaches 1, coincides with a previous observation that $\nu (1) \approx 2$~\cite{haiqing}.

Since the integration of $\alpha^{\rm I/\rm II}(\Delta,\gamma)$ with respect to $\gamma$ is finite,  the singular first-order differential equations  (\ref{vxyzi}) and (\ref{vxyzii}) for regime ${\rm I}$ and ${\rm II}$ can be solved in a straightforward way.
Let us first determine $V(\Delta,\gamma)$ on two characteristic lines.

(1) On the line of factorizing fields $\gamma=1+\Delta$, with $0<\gamma<1$, $V^f(\Delta,1+\Delta)$ vanishes: $V^f(\Delta,1+\Delta)=0$.

(2) On the dual line $\gamma=1-\Delta$, with $0<\gamma<1$, when a KT critical point $\Delta'_c = 1$ is approached, fidelity entropy $S'^d_{\phi}(\Delta',0)$ scales as $S'^d_{\phi}(\Delta',0) \sim (|1-1/\Delta'|)^3$.  This implies that the critical exponent for the correlation length $\nu \approx 2$ (cf. Appendix~\ref{scaling} for details).  Taking into account the fact that  the first-order derivative of $\ln{(e'(\Delta',0)/e'(1,0))}$ with respect to $\Delta'$ at a critical point $\Delta'_c = 1$ is nonzero, $\alpha'^d(\Delta',0)$ diverges as
\begin{equation}
\alpha'^d(\Delta',0) \sim -\frac{\Delta'}{\Delta'-1}.
\end{equation}
This is a situation similar to the dual line $\gamma =1$ for the quantum XY model. That is, in order to solve the singular first-order differential equation (\ref{vdxyz}), we need to separate the divergent part from $\alpha'^d(\Delta',0)$.
We set
\begin{equation}
\alpha'^d(\Delta',0)=-\frac{\Delta'}{\Delta'-1}+f(\Delta',0).
\label{alfaxyzdi}
\end{equation}
Here, $f(\Delta',0)$ is a function of $\Delta'$, which takes a finite value, and $-\Delta'/(\Delta'-1)$ is the divergent part of $\alpha'^d(\Delta',0)$.
Then, the singular first-order differential equation (\ref{vdxyz}) may be solved as follows
\begin{equation}
V'^{d}(\Delta',0)=V'^{d}_0V_1'^{d}(\Delta',0),
\end{equation}
where $V'^{d}_0$ is a constant to be determined, and $V_1'^{d}(\Delta',0)$ takes the form
\begin{equation}
V_1'^{d}(\Delta',0)=\frac{\Delta'}{\Delta'-1}\exp{(\int_{1}^{\Delta'}{\frac{f(\Theta',0)}{\Theta'^2}d \Theta'})}.
\label{v1dxyz}
\end{equation}

On the dual line $\gamma=1-\Delta$, in order to ensure that fidelity temperature vanishes at $\Delta'\rightarrow \infty$,  or, equivalently, $\Delta=2/(1+\Delta')\rightarrow 0$, $T'^d_\phi(\Delta',0)$ is shifted to $T'^d_\phi(\Delta',0)-T^d_0$ with $T^d_0\equiv T'^d_\phi(\infty,0)$,  accompanied by a shift in  $U'^d_\phi(\Delta',0)$: $U'^d_\phi(\Delta',0)-T^d_0S'^d_\phi(\Delta',0)$.
Following discussions in Section~\ref{fmsut}, we demand that (i) fidelity internal energy $U'^d_\phi(\Delta',0)$ at a critical point $\Delta'=1$ is zero, i.e., $U'^d_{\phi 0}=T^d_0 S'^d_{\phi 0}$;  (ii)
fidelity internal energy $U'^d_\phi(\Delta',0)$ at $\Delta'\rightarrow \infty$ satisfies the continuity requirement:  $U'^d_\phi(\infty,0)-T^d_0 S'^d_\phi(\infty,0)=U^f_f(0,1)$.
Therefore, $V'^d_0$ is determined as follows
\begin{equation}
V'^{d}_0=\frac{U^f_f(0,1)}{-\ln{(e'(\infty,0)/e'(1,0))}V'^d_1(\infty,0)+\alpha'^{d}(\infty,0)V'^{d}_1( \infty, 0)(S'^d_\phi(\infty,0)-S'^d_{\phi 0})},
\end{equation}
Once $V'^{d}_0$ and $U'^d_{\phi 0}$ are determined, $U'^d_\phi(\Delta',0)$ and $T'^d_\phi(\Delta',0)$ follow from (\ref{udxyz}) and (\ref{tdxyz}), respectively.
We refer to
$T'^d_\phi(\Delta',0)-T'^d_0$ and
$U'^d_\phi(\Delta',0)-T'^d_0S'^d_\phi(\Delta',0)$ as $T'^d_{\phi f}(\Delta',0)$ and $U'^d_{\phi f} (\Delta',0)$, respectively.  That is, $T'^d_{\phi f}(\Delta',0) \equiv T'^d_\phi(\Delta',0)-T'^d_0$ and  $U'^d_{\phi f}(\Delta',0)\equiv U'^d_\phi(\Delta',0)-T'^d_0S'^d_\phi(\Delta',0)$. Then, following our discussions in Section~\ref{fmsut}, fidelity temperature $T^d_f(\Delta,1-\Delta)$ and fidelity internal energy $U^d_f(\Delta,1-\Delta)$ take the form:  $T^d_f(\Delta,1-\Delta) = T^d_{\phi f}(\Delta,1-\Delta)$ and $U^d_f(\Delta,1-\Delta) =
U^d_{\phi f}(\Delta,1-\Delta)$, with  $T^d_{\phi f}(\Delta,1-\Delta) \equiv T'^d_{\phi f}(\Delta',0)$ and $U^d_{\phi f}(\Delta,1-\Delta) \equiv U'^d_{\phi f}(\Delta',0)$.

The singular first-order differential equations (\ref{vxyzi}) and (\ref{vxyzii}) for regime ${\rm I}$ and regime ${\rm II}$ can be solved in a similar way

(a) In regime ${\rm I}$ ($0<\Delta<1$ and $0<\gamma<1-\Delta$),
the singular first-order differential equation  (\ref{vxyzi}) may be solved as follows
\begin{equation}
V^{\rm I}(\Delta,\gamma)=V_0^{\rm I}(\Delta)V_1^{\rm I}(\Delta,\gamma),
\end{equation}
where $V_0^{\rm I}(\Delta)$ is a function of $\Delta$, and $V_1^{\rm I}(\Delta,\gamma)$ is defined as
\begin{equation}
V_1^{\rm I}(\Delta,\gamma)=\exp{(\int_{0}^{\gamma}{\alpha^{\rm I} (\Delta,\beta) d\beta})}.
\label{v1axyz}
\end{equation}

In regime ${\rm I}$, in order to ensure the continuity requirement for fidelity temperature $T_f(\Delta,\gamma)$ on a dual line $\gamma=1-\Delta$,
$T^{\rm I}(\Delta,\gamma)$ is shifted to $T^{\rm I}(\Delta,\gamma)-T_0^{\rm I}(\Delta)$,
accompanied by a shift in $U^{\rm I}(\Delta,\gamma)$: $U^{\rm I}(\Delta,\gamma)- T_0^{\rm I}(\Delta) S^{\rm I}(\Delta,\gamma)$, with
$S^{\rm I}(\Delta,\gamma)$ left intact.  Here, $T^{\rm I}_0(\Delta) \equiv T^{\rm I}(\Delta,1-\Delta)-T^{d}(\Delta,1-\Delta)$.  Following our discussions in Section~\ref{fmsut}, fidelity entropy $S^{\rm I}_f(\Delta,\gamma)$,  fidelity temperature $T^{\rm I}_f(\Delta,\gamma)$, and fidelity internal energy $U^{\rm I}_f(\Delta,\gamma)$ take the form: $S^{\rm I}_f(\Delta,\gamma) = S^{\rm I}(\Delta,\gamma)$,  $T^{\rm I}_f(\Delta,\gamma) = T^{\rm I}(\Delta,\gamma)-T^{\rm I}_0(\Delta)$ and  $U^{\rm I}_f(\Delta,\gamma) = U^{\rm I}(\Delta,\gamma)- T^{\rm I}_0(\Delta) S^{\rm I}(\Delta,\gamma)$, respectively.
In addition, we demand that (i) fidelity internal energy $U^{\rm I}(\Delta,\gamma)$ at a critical point is zero: $U^{\rm I}_0(\Delta)=T^{\rm I}_0(\Delta)S^{\rm I}_0(\Delta)$;  (ii) fidelity internal energy $U^{\rm I}(\Delta,\gamma)$ satisfies the continuity requirement at $\gamma=1-\Delta$: $U^{\rm I}(\Delta,1-\Delta)- T^{\rm I}_0(\Delta)S^{\rm I}(\Delta,1-\Delta)=U^d_f(\Delta,1-\Delta)$.
Therefore, $V^{\rm I}_0(\Delta)$  is determined as follows
\begin{equation}
V^{\rm I}_0(\Delta)=\frac{U^d_f(\Delta,1-\Delta)-T^{d}_f(\Delta,1-\Delta)(S^{\rm I}(\Delta,1-\Delta)-S^{\rm I}_0(\Delta))}{\ln{(e(\Delta,1-\Delta)/e(\Delta,0))}V_1^{\rm I}(\Delta,1-\Delta)+\alpha^{\rm I}(\Delta,1-\Delta)V_1^{\rm I}(\Delta,1-\Delta)(S^{\rm I}(\Delta,1-\Delta)-S^{\rm I}_0(\Delta))}.
\end{equation}
Once $V^{\rm I}_0(\Delta)$ and $U^{\rm I}_{0}(\Delta)$ are determined, fidelity internal energy $U^{\rm I}(\Delta,\gamma)$ and fidelity temperature $T^{\rm I}(\Delta,\gamma)$ follow from (\ref{uxyzi}) and (\ref{txyzi}), respectively.
 As such,  fidelity temperature $T^{\rm I}_f(\Delta,\gamma)$ and fidelity internal energy $U^{\rm I}_f(\Delta,\gamma)$ follow.

(b) In regime ${\rm II}$ ($-1<\Delta<0$ and $0<\gamma<1+\Delta$),
the singular first-order differential equation (\ref{vxyzii}) may be solved as follows
\begin{equation}
V^{\rm II}(\Delta,\gamma)=V_0^{\rm II}(\Delta)V_1^{\rm II}(\Delta,\gamma),
\end{equation}
where $V_0^{\rm II}(\Delta)$ is a function of $\Delta$, and $V_1^{\rm II}(\Delta,\gamma)$ is defined as
\begin{equation}
V_1^{\rm II}(\Delta,\gamma)=\exp{(\int_{0}^{\gamma}{\alpha^{\rm II} (\Delta,\beta) d\beta})}.
\label{v1bxyz}
\end{equation}

In regime ${\rm II}$,  in order to ensure the continuity requirements for fidelity temperature $T_f(\Delta,\gamma)$ on a dual line $\gamma=1-\Delta$,
$T^{\rm II}(\Delta,\gamma)$ is shifted to $T^{\rm II}(\Delta,\gamma)-T^{\rm II}_0(\Delta)$,
 accompanied by a shift in $U^{\rm II}(\Delta,\gamma)$: $U^{\rm II}(\Delta,\gamma)- T_0^{\rm II}(\Delta) S^{\rm II}(\Delta,\gamma)$, with $S^{\rm II}(\Delta,\gamma)$ left intact.  Here, $T^{\rm II}_0(\Delta)=T^{\rm II}(\Delta,1+\Delta)$.  Following our discussions in Section~\ref{fmsut}, fidelity entropy $S^{\rm II}_f(\Delta,\gamma)$,  fidelity temperature $T^{\rm II}_f(\Delta,\gamma)$, and fidelity internal energy $U^{\rm II}_f(\Delta,\gamma)$ take the form:  $ S^{\rm II}_f(\Delta,\gamma) = S^{\rm II}(\Delta,\gamma)$, $T^{\rm II}_f(\Delta,\gamma) = T^{\rm II}(\Delta,\gamma)-T^{\rm II}_0(\Delta)$ and  $U^{\rm II}_f(\Delta,\gamma) = U^{\rm II}(\Delta,\gamma)- T^{\rm II}_0(\Delta) S^{\rm II}(\Delta,\gamma)$, respectively.
In addition, we demand that (i) fidelity internal energy $U^{\rm II}(\Delta,\gamma)$ at a critical point is zero, i.e., $U^{\rm II}_0(\Delta)=T_0^{\rm II}(\Delta)S^{\rm II}_0(\Delta)$;  (ii) fidelity internal energy $U^{\rm II}(\Delta,\gamma)$ satisfies the continuity requirement at $\gamma=1+\Delta$: $U^{\rm II}(\Delta,1+\Delta)- T_0^{\rm II}(\Delta)S^{\rm II}(\Delta,1+\Delta)=U^f_f(0,1)$.
Therefore, $V^{\rm II}_0(\Delta)$  is determined as follows
\begin{equation}
V^{\rm II}_0(\Delta)=\frac{U^f_f(0,1)}{\ln{(e(\Delta,1+\Delta)/e(\Delta,0))}V_1^{\rm II}(\Delta,1+\Delta)+\alpha^{\rm II}(\Delta,1+\Delta)V_1^{\rm II}(\Delta,1+\Delta)(S^{\rm II}(\Delta,1+\Delta)-S^{\rm II}_0(\Delta))}.
\end{equation}
Once $V^{\rm II}_0(\Delta)$ and $U^{\rm II}_{0}(\Delta)$ are determined, fidelity internal energy $U^{\rm II}(\Delta,\gamma)$ and fidelity temperature $T^{\rm II}(\Delta,\gamma)$ follow from (\ref{uxyzii}) and (\ref{txyzii}), respectively.
Therefore,  fidelity temperature $T^{\rm II}_f(\Delta,\gamma)$ and fidelity internal energy $U^{\rm II}_f(\Delta,\gamma)$ follow.

If fidelity entropy $S_f(\Delta,\gamma)$, fidelity temperature $T_f(\Delta,\gamma)$ and fidelity internal energy $U_f(\Delta,\gamma)$ are determined in regime ${\rm I}$ and regime ${\rm II}$, then fidelity mechanical state functions in other regimes are determined from dualities in Appendix~\ref{dual}, taking into account a contribution from scaling entropy arising from dualities.

Numerical simulations for fidelity entropy $S_f(\Delta,\gamma)$, fidelity temperature $T_f(\Delta,\gamma)$ and fidelity internal energy $U_f(\Delta,\gamma)$ are shown in Fig.~\ref{fmsxyz}(a), Fig.~\ref{fmsxyz}(b) and Fig.~\ref{fmsxyz}(c), respectively.

\section{A fictitious parameter $\sigma$ relating to different choices of a dominant control parameter in fidelity mechanics}\label{fictitious}

For a quantum many-body system, a given phase in its ground state phase diagram is separated into different regimes, featuring characteristic lines as boundaries between them.  There are mainly four different types of characteristic lines:

(i) If the Hamiltonian possesses a distinct symmetry group when coupling constants take special values on a characteristic line in a given phase, then it separates this phase into different regimes in the parameter space.
An example to illustrate this observation is the quantum spin-$1/2$ XYZ model. In this model, the KT line: $\gamma=1-\Delta$, with $0<\Delta<1$,  is special,  since the Hamiltonian possesses a $U(1)$ symmetry on this line, which is lost, when the coupling parameters move away from this characteristic line.

(ii) Duality offers another type of characteristic lines, separating a given phase into different regimes in the parameter space. As we have seen in the quantum XY chain, dualities exist along two lines $\lambda =0$ and $\gamma = 1$.

(iii) Factorizing fields turn out to be a characteristic line between different regimes in a given phase. In fact, fidelity temperature is zero at factorizing fields.  This happens to
the quantum XY chain and the quantum spin-$1/2$ XYZ chain. An interesting feature for factorizing fields is that they alway occur in a symmetry broken phase, and in turn are associated with the PT transitions, given that fidelity temperature is undefined, ranging from 0 to $\infty$,  at a PT transition point.  We remark that factorizing fields also occur when one coupling parameter takes infinity in value, or when more than one coupling parameters are infinite in value.

(iv) There is a peculiar type of characteristic lines, originating from a multi-critical point and ending at a characteristic point - an intersection point between two characteristic lines arising from symmetries, dualities and factorizing fields, including those at infinity. This happens,  if no characteristic line, arising from symmetries, dualities and factorizing fields, originates from this multi-critical point.  In contrast to the rigid constraints imposed by symmetries, dualities and factorizing fields,  this type of characteristic lines does not impose any rigid constraints, in the sense that they do depend on choices of a dominant control parameter.  An illustrative example is a characteristic line: $\lambda = 1$, for the transverse field Ising chain in a longitudinal field.

For a given principal regime, there are different choices of a dominant control parameter,  as long as such a choice is consistent with the constraints imposed by symmetries, dualities and factorizing fields. Different choices result in different fidelity mechanical state functions.   Therefore, two points need to be addressed: first, it is necessary to relate different choices into each other; second, different choices have to produce the same stable and unstable fixed points.

\begin{figure}
     \includegraphics[angle=0,totalheight=4.5cm]{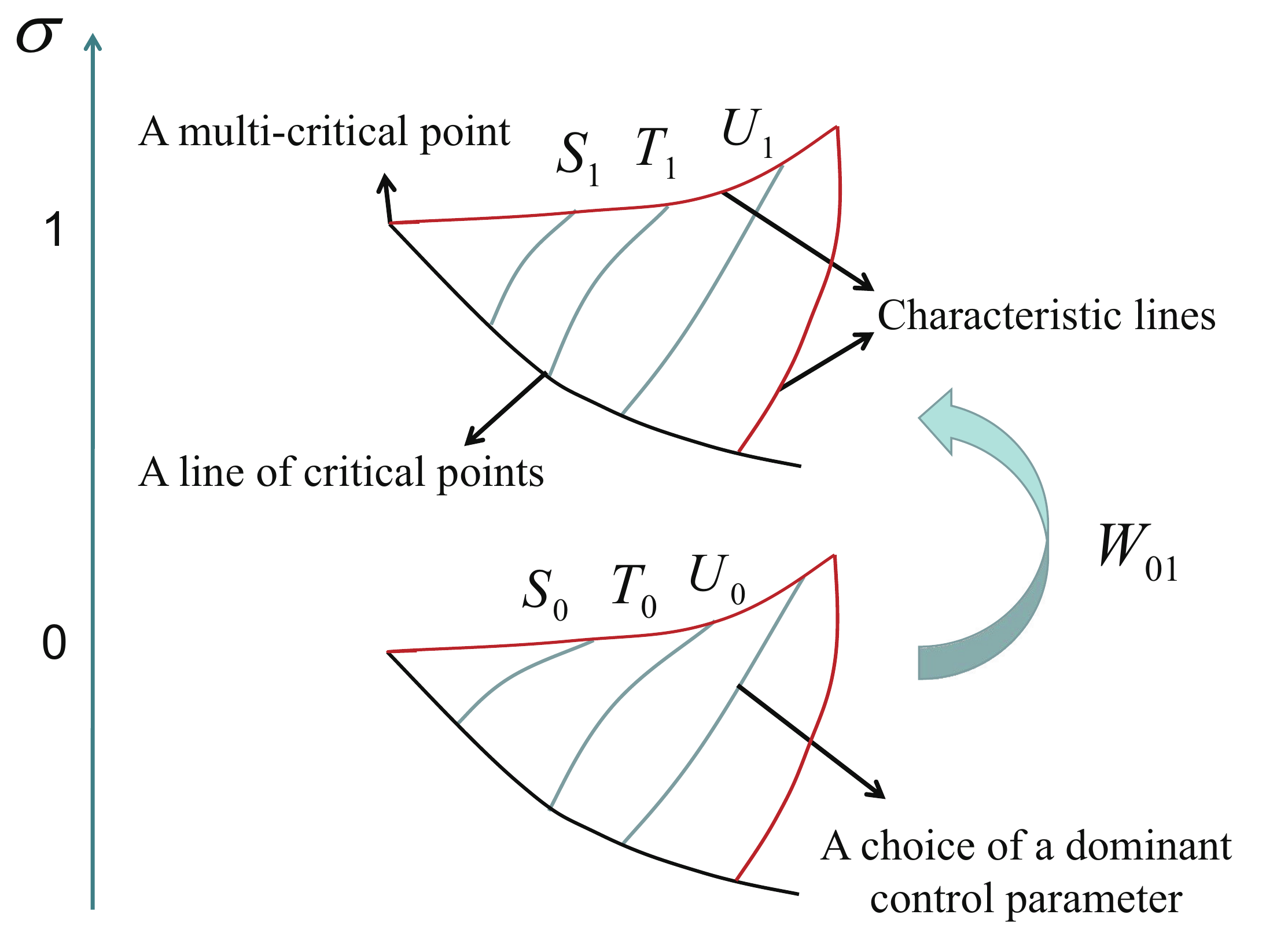}
       \caption{A fictitious parameter $\sigma$ connecting different choices of a dominant control parameter in a typical principal regime.}\label{twosets}
  \end{figure}

Let us start from the first point. Suppose we have made two different choices of a dominant control parameter in a given regime: one yields fidelity mechanical state functions  $U_{0}$, $S_{0}$, and $T_{0}$, and the other yields $U_{1}$, $S_{1}$, and $T_{1}$, as shown in  Fig.~\ref{twosets}.  Then, we may introduce a fictitious parameter $\sigma$, ranging from 0 to 1. Now it is legitimate to resort to a new set of fidelity mechanical state functions $U_{\sigma}$, $S_{\sigma}$, and $T_{\sigma}$,  which are some smooth functions of $\sigma$, such that $U_{\sigma}$, $S_{\sigma}$, and $T_{\sigma}$ interpolate between  $U_{0}$, $S_{0}$, $T_{0}$ and $U_{1}$, $S_{1}$, $T_{1}$, when $\sigma$ varies from 0 to 1.  This amounts to stating that we may smoothly deform one choice to the other, as depicted in  Fig.~\ref{twosets}.   Therefore, we are able to relate one choice to the other by doing a certain amount of fidelity work $W_{01}$:
\begin{equation}
W_{01}=\int{( dU-T_{\sigma} dS)}=\Delta U-T^*\Delta S,
\end{equation}
where $\Delta U=U_1-U_0$,  $\Delta S = S_1 - S_0$, and $T^*$ may be determined from the mean value theorem for a definite integral. Needless to say,
fidelity work to be  done $W_{01}$ depends on the way how we deform our choices into each other. Suppose $T_0 < T_1$, then we have  $T_0< T^*<T_1$. Simply putting $T^* = T_0$ or $T_1$, we may estimate an amount of fidelity work needed to be done.

Now we turn to the second point. Recall that different choices of a dominant control parameter have to be consistent with the constraints imposed by symmetries, dualities and factorizing fields. These constraints are {\it rigid}, in the sense that there is no flexibility in choosing a dominant control parameter on a characteristic line, arising from symmetries, dualities and factorizing fields, 
as depicted in Fig.~\ref{twosets}.  In other words,  
a dominant control parameter remains the same for any $\sigma \in [0,1]$ on these characteristic lines. However, a choice of different parametrization variables  is still possible on a characteristic line, subject to the condition that, for any two parametrization variables, one must be a monotonically increasing function of the other, and vice versa.  As for a characteristic line in type (iv),  the above argument implies that its starting and ending points, as a multi-critical point and a characteristic point, respectively, are rigid.
Although  fidelity mechanical state functions depend on both choices of a dominant control parameter and parametrization variables, this does not change where fidelity temperature diverges or becomes zero, and does not change where fidelity entropy takes a (local) maximum.
Therefore, both stable and unstable fixed points remain the same for any different choices of a dominant control parameter in a given regime.
In this sense, information encoded in a fictitious parameter $\sigma$ arising from different choices of a dominant control parameter is {\it irrelevant}.

\section{A canonical form of the Hamiltonian:  Definition of duality and a shift operation in the Hamiltonian}\label{shift}

We do not claim that a canonical form of the Hamiltonian, as defined in Section~\ref{duality}, occupies any {\it unique} position in fidelity mechanics.  To see this point, consider  a shifted Hamiltonian $H_b(x) = H(x) + b$. Here, we assume $H(x)$ is in a canonical form, as defined in Section~\ref{duality}. Actually, for a given Hamiltonian, its canonical form depends on the definition of duality. Our definition adopted in Section~\ref{duality}, is conventional (see, e.g.,  Ref.~\cite{ortiz}): $H(x) = k'(x') U H (x')U^{\dagger}$, where $U$ is a unitary operator performing duality.  Note that $k'(x')>0$,  with $x'$ being a function of  $x$.

If the definition of duality is changed, then a canonical form of the Hamiltonian follows. Suppose the definition of duality is changed to $H_b(x) = k'(x') U H_b (x')U^{\dagger} +\mu(x')$, with an extra parameter $\mu(x')$. The task is to find out a proper $\mu(x')$ to ensure that $H_b(x)$ is  in a canonical form according to this new definition of duality. As it turns out,  we have $\mu(x')=b(1-k'(x'))$.

Following our prescription in Section~\ref{fmsut}, fidelity entropy $S_b(x)$, fidelity temperature $T_b(x)$ and fidelity internal energy $U_b(x)$ are equally well-defined, as long as $e(x)+b$ is negative.  Given ground state wave functions remain the same,  we have $S_b(x)=S(x)$. If the same choice of a dominant control parameter is kept in a given regime, then the monotonicity of the shifted ground state energy density remains the same. Following our prescription,  both fidelity temperature $T_b(x)$ and fidelity internal energy $U_b(x)$ depend on $b$ explicitly.  However, we are able to relate fidelity mechanical  state functions for two different canonical forms of the Hamiltonian, via a shift operation: $H_b(x) = H(x) + b$, by doing a certain amount of fidelity work $W_{b}(x)$:
\begin{equation}
W_{b}(x)=\int{( dU_b-T_b dS_b)}=U_b(x)-U(x).
\end{equation}
Here, the fact that fidelity entropy $S_b(x)$ remains the same during the shift operation has been taken into account.

Combining this argument with the rigid constraints imposed by symmetries, dualities and factorizing fields, we conclude that, for a quantum many-body system,  both stable and unstable fixed points remain the same for any different definitions of a canonical form of the Hamiltonian, resulted from different definitions of duality, given all the characteristic lines remain the same. This means that, a canonical form of the Hamiltonian, as defined in Section~\ref{duality},  does not occupy any {\it unique}  position in fidelity mechanics.

\section{Dualities for the quantum XYZ models on different lattices}~\label{dual}

Let us start from the quantum spin-$1/2$ XYZ chain.  There are five different dualities for the Hamiltonian (\ref{xyz}):\\

(0) The Hamiltonian $H(\Delta, \gamma)$ for $\gamma \geq1$ is dual to the Hamiltonian $H(\Delta', \gamma')$ for $0<\gamma' \leq1$ under a local unitary transformation $U_0$:
$\sigma_{2i}^x\rightarrow\sigma_{2i}^x$, $\sigma_{2i}^y\rightarrow\sigma_{2i}^y$, $\sigma_{2i}^z\rightarrow\sigma_{2i}^z$, $\sigma_{2i+1}^x\rightarrow\sigma_{2i+1}^x$, $\sigma_{2i+1}^y\rightarrow-\sigma_{2i+1}^y$ and $\sigma_{2i+1}^z\rightarrow-\sigma_{2i+1}^z$: $H(\Delta, \gamma)=k(\Delta,\gamma) U_0 H(\Delta', \gamma') U^\dagger_0$,
with $\Delta'=-\Delta/\gamma$, $\gamma'=1/\gamma$, and $k(\Delta,\gamma) =\gamma$.
The Hamiltonian is invariant when $\Delta =0, ~\gamma= 1$.\\

(1) Under a local unitary transformation  $U_1$: $\sigma_{i}^x\rightarrow-\sigma_{i}^x$,
$\sigma_{i}^y\rightarrow\sigma_{i}^z$, $\sigma_{i}^z\rightarrow\sigma_{i}^y$, we have
  $H(\Delta, \gamma)=k(\Delta,\gamma) U_1 H(\Delta', \gamma') U^\dagger_1$, with
  $\Delta'=(2-2\gamma)/(1+\gamma+\Delta)$, $\gamma'=(1+\gamma-\Delta)/(1+\gamma+\Delta)$, and    $k(\Delta,\gamma)=(1+\gamma+\Delta)/2$.
The Hamiltonian on the line $\gamma=1-\Delta$ is invariant.\\

(2) Under a local unitary transformation $U_2$: $\sigma_{2i}^x\rightarrow-\sigma_{2i}^x$,  $\sigma_{2i}^y\rightarrow-\sigma_{2i}^z$, $\sigma_{2i}^z\rightarrow-\sigma_{2i}^y$, $\sigma_{2i+1}^x\rightarrow-\sigma_{2i+1}^x$, $\sigma_{2i+1}^y\rightarrow\sigma_{2i+1}^z$ and $\sigma_{2i+1}^z\rightarrow\sigma_{2i+1}^y$,
we have $H(\Delta, \gamma)=k(\Delta,\gamma) U_2 H(\Delta', \gamma') U^\dagger_2$, with $\Delta'=(2\gamma-2)/(1+\gamma-\Delta)$,
$\gamma'=(\gamma+\Delta+1)/(1+\gamma-\Delta)$, and  $k(\Delta,\gamma)=(1+\gamma-\Delta)/2$.
The Hamiltonian on the line $\gamma=1+\Delta$ is invariant.\\

(3) Under a local unitary transformation $U_3$:  $\sigma_{i}^x\rightarrow\sigma_{i}^z$, $\sigma_{i}^y\rightarrow-\sigma_{i}^y$, $\sigma_{i}^z\rightarrow\sigma_{i}^x$, we have
  $H(\Delta, \gamma)=k(\Delta,\gamma) U_3 H(\Delta', \gamma') U^\dagger_3$, with $\Delta'=(2+2\gamma)/(1-\gamma+\Delta)$, $\gamma'=(\gamma+\Delta-1)/(1-\gamma+\Delta)$, and  $k(\Delta,\gamma)=(1-\gamma+\Delta)/2$. The Hamiltonian on the line $\gamma=-1+\Delta$ is invariant.\\

(4) Under a local unitary transformation $U_4$: $\sigma_{2i}^x\rightarrow-\sigma_{2i}^z$,  $\sigma_{2i}^y\rightarrow-\sigma_{2i}^y$, $\sigma_{2i}^z\rightarrow-\sigma_{2i}^x$, $\sigma_{2i+1}^x\rightarrow\sigma_{2i+1}^z$, $\sigma_{2i+1}^y\rightarrow-\sigma_{2i+1}^y$ and $\sigma_{2i+1}^z\rightarrow\sigma_{2i+1}^x$,
we have $H(\Delta, \gamma)=k(\Delta,\gamma) U_4 H(\Delta', \gamma') U^\dagger_4$, with $\Delta'=(2\gamma+2)/(\gamma+\Delta-1)$,
 $\gamma'=(-1+\gamma-\Delta)/(1-\gamma-\Delta)$, and  $k(\Delta,\gamma)=(1-\gamma-\Delta)/2$.
The Hamiltonian on the line $\gamma=-1-\Delta$ is invariant.

\begin{figure}
     \includegraphics[angle=0,totalheight=3.5cm]{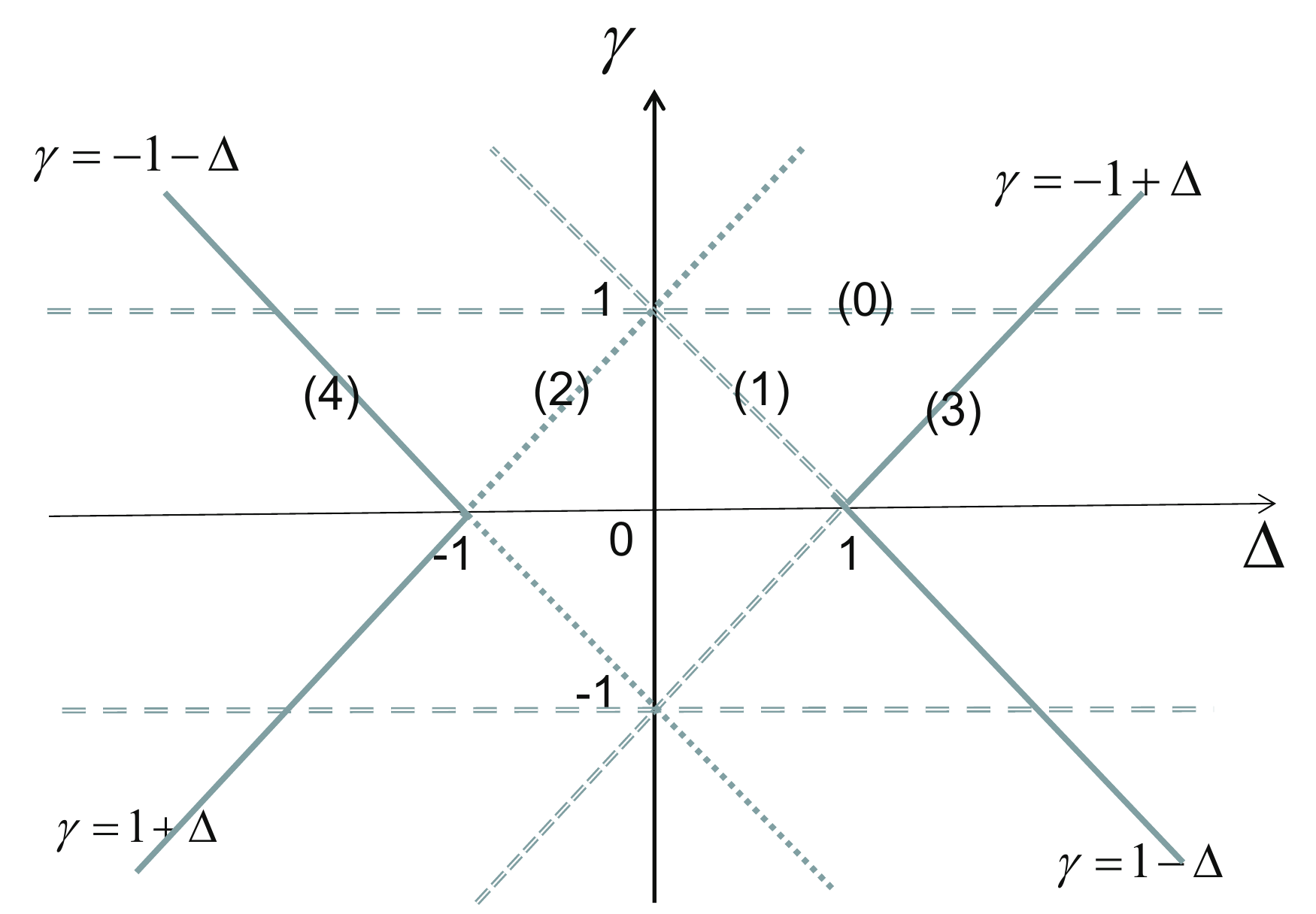}
       \caption{Dualities for the  quantum spin-$1/2$ XYZ chain. Note that the same labels are used to indicate dualities described in the text. }\label{xyzdual}

\end{figure}

We use the same labels to indicate dualities in Fig.~\ref{xyzdual}.  Here, we point out that the symmetries of  the Hamiltonian (\ref{xyz}) under the permutation of the three Pauli matrices $\sigma_{i}^x, \sigma_{i}^y$, and $\sigma_{i}^z$ have been discussed in Ref.~\cite{Baxterbook}, though not treated as dualities.  The presence of these dualities separates the parameter space $(\gamma>0,\Delta)$ into twelve different regimes, with five lines defined by $\gamma=1$ and $\gamma=\pm1\pm\Delta$ as boundaries.
And these twelve regimes are separated into two groups, with six regimes in each group dual to each other.  As shown in Fig.~\ref{duality2}, regimes $\rm{I}$, $\rm{III}$, $\rm{V}$, $\rm{I'}$, $\rm{III'}$ and $\rm{V'}$ are dual to each other, while regimes $\rm{II}$, $\rm{IV}$, $\rm{VI}$, $\rm{II'}$, $\rm{IV'}$ and $\rm{VI'}$ are dual to each other.  Therefore, only two regimes represent the physics underlying the quantum spin-$1/2$ XYZ model, with regime $\rm{I}$ and regime $\rm{II}$ as our choice.

As for the quantum XYZ models on different lattices, with arbitrary spin $s$, we restrict ourselves to make a few comments. First, dualities discussed for the quantum spin-$1/2$ XYZ chain are also valid for the quantum XYZ chain with arbitrary spin $s$, both integer and half integer~\cite{qqshi}. Second, dualities discussed for the quantum spin-$1/2$ XYZ chain may be extended to the quantum XYZ model with arbitrary spin $s$ on a bipartite lattice in any spatial dimensions, such as square and honeycomb lattices~\cite{lish}.  A few remarks are in order:  (1) self duality does not necessarily imply a critical point; (2) dualities yield exact results about critical points, but, generically, a quantum many-body system, which hosts dualities, is not integrable, in contrast to the quantum XY chain and the quantum spin-$1/2$ XYZ chain. In addition, the Haldane phase~\cite{haldane}, the so-called symmetry protected topological phase~\cite{wxg}, occurs in the quantum XYZ chain with integer spin $s$.  It is anticipated that fidelity mechanics furnishes a novel way to characterize a symmetry protected topological phase.

Duality defines a canonical form of the Hamiltonian, thus playing an inherently fundamental role in fidelity mechanics. Therefore, one may expect that duality has to occur for any specific quantum many-body system in some form. Here is an illustrative example for a non-bipartite lattice - the quantum XYZ model on a triangular lattice.  However, our discussion also applies to the quantum XYZ model on any non-bipartite lattice, such as the Kagom\'e  lattice. The Hamiltonian for the quantum XYZ model on a triangular lattice takes the form
\begin{equation}
  H =
  \sum_{( \vec r, \vec r')} \left(J_x S^{[\vec r]}_x S^{[\vec r']}_x
    +  J_y S^{[\vec r]}_y S^{[\vec r']}_y
    + J_z S^{[\vec r]}_z S^{[\vec r']}_z \right),
\end{equation}
where $S^{[\vec r]}_{\alpha} (\alpha=x,y,z)$ are spin-$s$ operators at site $\vec r$, and $( \vec r, \vec r')$
are nearest neighbor pairs on the triangular lattice.  The model accommodates the quantum frustrated spin-$1/2$ Heisenberg model as a special case: $J_x=J_y=J_z$. For simplicity, we restrict ourselves to $J_x \geq 0$, $J_y \geq 0$, and set $J_z=1$, and denote the Hamiltonian as $H(J_x,J_y)$.

The Hamiltonian $H(J_x,J_y)$ is symmetric under a local unitary transformation $U_0$: $S_{i}^x\rightarrow S_{i}^y$,
$S_{i}^y\rightarrow S_{i}^x$ and $S_{i}^z\rightarrow-S_{i}^z$. In addition,  there are two dualities:

(1) The Hamiltonian $H(J_x,J_y)$ is dual to the Hamiltonian $H(J'_x, J'_y) $
under a local unitary transformation $U_1$: $S_{i}^x\rightarrow -S_{i}^x$,
$S_{i}^y\rightarrow S_{i}^z$, $S_{i}^z\rightarrow S_{i}^y$:
 $H(J_x, J_y)=k'(J'_x, J'_y) U_1 H(J'_x, J'_y) U^\dagger_1$,
with $J_x=J'_x/J'_y$, $J_y=1/J'_y$, and $k'(J'_x,J'_y) =1/J'_y$.
The Hamiltonian is invariant when $J_y=1$.\\

(2) The Hamiltonian $H(J_x,J_y)$ is dual to the Hamiltonian $H(J'_x, J'_y) $
 under a local unitary transformation $U_2$: $S_{i}^x\rightarrow S_{i}^z$,
$S_{i}^y\rightarrow-S_{i}^y$, $S_{i}^z\rightarrow S_{i}^x$:
$H(J_x, J_y)=k'(J'_x, J'_y) U_2 H(J'_x, J'_y) U^\dagger_2$,
with $J_x=1/J'_x$, $J_y=J'_y/J'_x$, and $k'(J'_x,J'_y) =1/J'_x$.
The Hamiltonian is invariant when $J_x=1$.

In contrast to the quantum spin-$1/2$ XYZ chain, dualities are not enough to produce the ground state phase diagram for the quantum XYZ model on the triangular lattice.  However, the usefulness of dualities may be seen by combining with the tensor network algorithm in the context of the projected-entangled pair state representations~\cite{TN1},  which provides an efficient way to evaluate ground state fidelity per site~\cite{Zhou2}, thus mapping out the ground state phase diagram.

\section{Dualities for the quantum spin-$1/2$ Kitaev model on a honeycomb lattice}~\label{dualityinkitaev}

The two-dimensional spin-$1/2$ Kitaev model on a honeycomb lattice is described by the Hamiltonian (\ref{kitaevham}).
Here, we restrict ourselves to $J_x \geq 0$, $J_y \geq 0$, and  set $J_z=1$.  Then, we may simply use $H(J_x,J_y)$ to denote the Hamiltonian.

The Hamiltonian $H(J_x,J_y)$ is symmetric under a local unitary transformation $U_0$:  $\sigma_{i}^x\rightarrow\sigma_{i}^y$,
$\sigma_{i}^y\rightarrow\sigma_{i}^x$ and $\sigma_{i}^z\rightarrow-\sigma_{i}^z$,
accompanied by the lattice symmetry between $x$-bonds and $y$-bonds. In addition,  two dualities arise:

(1) The Hamiltonian $H(J_x,J_y)$ is dual to the Hamiltonian $H(J'_x, J'_y) $
 under a local unitary transformation $U_1$: $\sigma_{i}^x\rightarrow\sigma_{i}^z$,
$\sigma_{i}^y\rightarrow-\sigma_{i}^y$, $\sigma_{i}^z\rightarrow\sigma_{i}^x$,
 accompanied by the lattice symmetry between $x$-bonds and $z$-bonds. That is,
 $H(J_x, J_y)=k'(J'_x, J'_y) U_1 H(J'_x, J'_y) U^\dagger_1$,
with $J_x=J'_x/J'_y$, $J_y=1/J'_y$, and $k'(J'_x,J'_y) =1/J'_y$.
The Hamiltonian is invariant when $J_y=1$.\\

(2) The Hamiltonian $H(J_x,J_y)$ is dual to the Hamiltonian $H(J'_x, J'_y) $
 under a local unitary transformation $U_2$: $\sigma_{i}^x\rightarrow-\sigma_{i}^x$,
$\sigma_{i}^y\rightarrow\sigma_{i}^z$, $\sigma_{i}^z\rightarrow\sigma_{i}^y$,
accompanied by the lattice symmetry between $y$-bonds and $z$-bonds. That is,
$H(J_x, J_y)=k'(J'_x, J'_y) U_2 H(J'_x, J'_y) U^\dagger_2$,
with $J_x=1/J'_x$, $J_y=J'_y/J'_x$, and $k'(J'_x,J'_y) =1/J'_x$.
The Hamiltonian is invariant when $J_x=1$.\\

\begin{figure}
     \includegraphics[angle=0,totalheight=4.3cm]{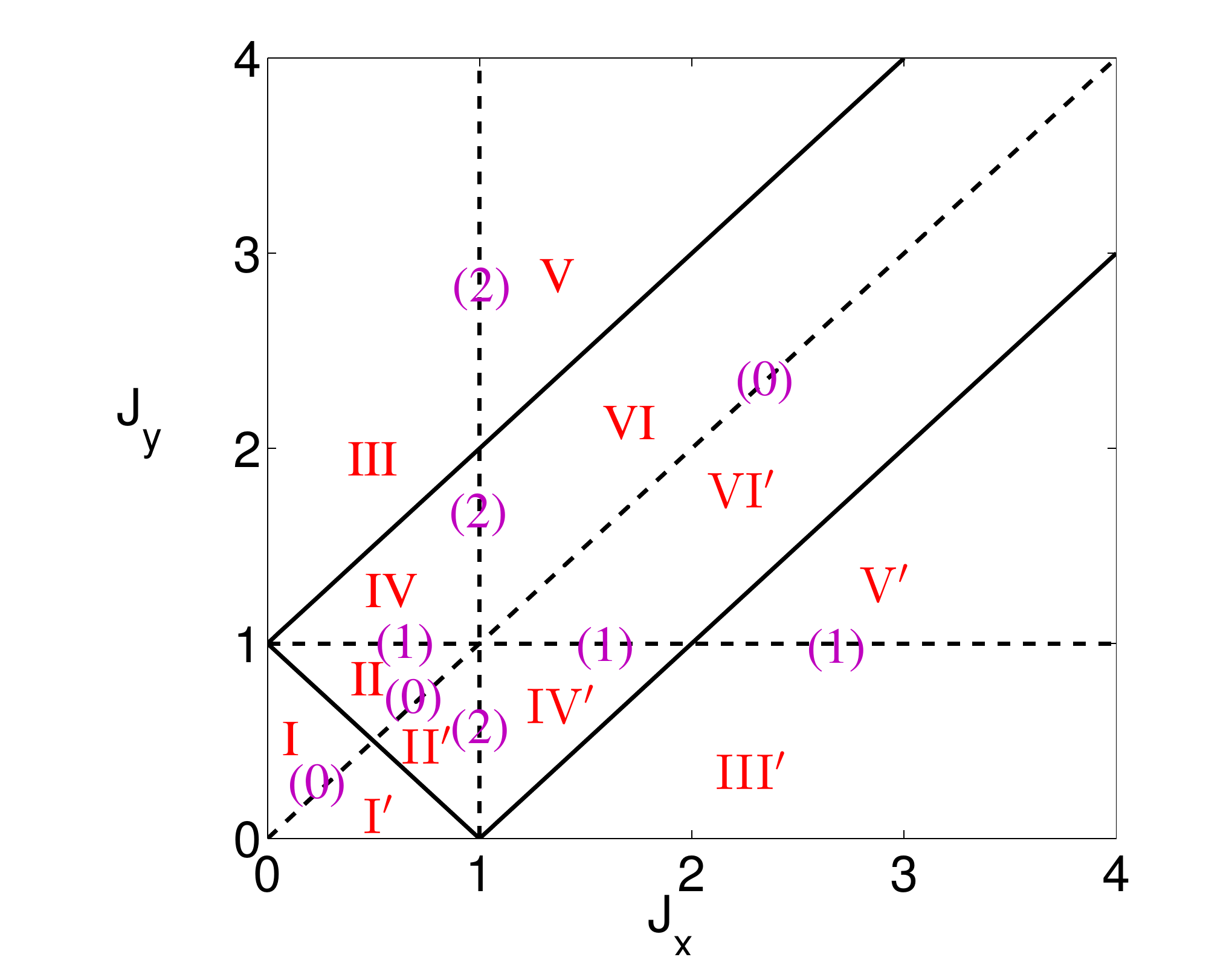}
       \caption{Dualities for the quantum spin-$1/2$ Kitaev model on a honeycomb lattice. Note that the same labels are used to indicate dualities described in the text. }\label{kitaevdual}
\end{figure}

As shown in Fig.~\ref{kitaevdual}, the presence of these dualities separates the region $J_x>0$ and $Jy>0$ into twelve different regimes, if critical lines are supposed to be determined. Six lines, defined by $J_x=1$, $J_y=1$, $J_y=J_x$, $J_y=1-J_x$ and $J_y=\pm1+J_x $, constitutes the boundaries between different regimes.  These twelve regimes are separated into two groups, with six regimes in each group dual to each other.  As shown in Fig. ~\ref{kitaevdual}, regimes $\rm{I}$, $\rm{III}$, $\rm{V}$, $\rm{I'}$, $\rm{III'}$ and $\rm{V'}$ are dual to each other, whereas regimes $\rm{II}$, $\rm{IV}$, $\rm{VI}$, $\rm{II'}$, $\rm{IV'}$ and $\rm{VI'}$ are dual to each other.  Therefore, only two regimes represent the physics underlying the quantum spin-$1/2$ Kitaev model on a honeycomb lattice.

We remark that dualities discussed for the quantum spin-$1/2$ Kitaev model on a honeycomb lattice are also valid for the quantum Kitaev model on a honeycomb lattice with arbitrary spin $s$~\cite{chenxh}.

\section{A scaling analysis of fidelity entropy $S(x)$ near a critical point $x_c$}\label{scaling}

In this Appendix, we perform a scaling analysis of fidelity entropy $S(x)$ near a critical point.

For simplicity, we denote $f(x,x')=\ln{d(x,x')}$, with $d(x,x')$ being ground state fidelity per site. Then,  fidelity entropy takes the form:  $S(x)=-2\int_{x_c}^{x}{f(x,x')d{x'}}+S_0$.
We expand $f(x, x')$ into a Taylor series at $x'=x$. Keeping up to the second-order term, we have
\begin{equation}
f(x,x')=f(x,x)+\frac{\partial f(x,x')}{\partial x'}\mid_{x'=x}(x'-x)+\frac{1}{2}\frac{\partial^2{f(x,x')}}{\partial x'^2}\mid_{x'=x}(x'-x)^2+O((x'-x)^3).
\end{equation}
Since $f(x,x) = 0$ and $\frac{\partial f(x,x')}{\partial x'}\mid_{x'=x} = 0$, we have
\begin{equation}
f(x,x')=\frac{1}{2}  \frac{\partial^2{f(x,x')}}{\partial x'^2} \mid_{x'=x}(x'-x)^2+O((x'-x)^3).
\end{equation}
Therefore,  the leading contribution to  fidelity entropy $S(x)$ is from the second-order derivative of $f(x,x')$ with respect to $x'$ near a critical point $x=x_c$:
\begin{equation}
S(x) =-\frac{1}{3} \frac{\partial^2{f(x,x')}}{\partial x'^2}\mid_{x'=x}(x-x_c)^3.
\end{equation}
Generically, the second-order derivative of $f(x,x')$  with respect to $x'$ at $x'=x$ is related with the critical exponent for the correlation length $\nu$~\cite{Marek}:

\begin{equation}
      \frac{\partial^2{f(x,x')}}{\partial x'^2} \mid_{x'=x}  \sim (x-x_c)^{d\nu-2},
\end{equation}
with $d$ being the dimension of a quantum many-body system under investigation. Therefore, when $x$ is close to a critical point $x_c$, fidelity entropy $S(x)$ scales as
\begin{equation}
S(x)  \sim (x-x_c)^{d\nu+1}.
\end{equation}
As an illustrative example, we consider the transverse field quantum Ising chain.  For this model,  we have $d=1$ and $\nu=1$.
Then,  $S (x) \sim (x-x_c)^2$, which is in agreement with our numerics.\\

\section{A scaling  analysis of the ground state energy density $e(\lambda,\gamma)$ close to a Gaussian critical point for the quantum XY chain}\label{gaussion}

In this Appendix, we perform a scaling  analysis of the  ground state energy $e(\lambda,\gamma)$ close to a Gaussian critical point for the quantum XY chain. For this model,  Gaussian phase transitions
occur  at $\gamma=0$ and $-1<\lambda<1$.  As an exactly solvable model,  the ground state energy density $e(\lambda,\gamma)$ is known to be~\cite{Lieb, pfeuty}

\begin{equation}
e(\lambda,\gamma)=-1/\pi\int_{0}^{\pi}{\sqrt{(\cos{\alpha}-\lambda)^2+(\gamma\sin{\alpha})^2}d\alpha}.
\end{equation}
For brevity, we denote $x\equiv \cos{\alpha}$. Then, the second-order derivative of $e(\lambda, \gamma)$ with respect to $\gamma$ takes the form:
\begin{equation}
\frac{\partial^2 e(\lambda,\gamma)}{\partial\gamma^2}=
1/\pi\int_{1}^{-1}{\frac{\sqrt{1-x^2}}{\sqrt{(x-\lambda)^2+(\gamma \sqrt{1-x^2})^2}} d x}. \label{int}
\end{equation}
As it turns out, the integral diverges when $x \rightarrow \lambda$ and $\gamma \rightarrow 0$.  To determine its leading divergent behavior,  we divide this integral into four parts, i.e., $\int_{1}^{-1}=\int_{1}^{\lambda+\delta}+\int_{\lambda+\delta}^{\lambda}+\int_{\lambda}^{\lambda-\delta}+\int_{\lambda-\delta}^{-1}$. Given the first and last parts are regular, we only need to consider the
second and third parts.  Since $\delta\rightarrow 0$, for these two parts, the integrand in the integral (\ref{int}) is simplified to
 \begin{equation}
 \frac{\sqrt{1-x^2}}{\sqrt{(x-\lambda)^2+(\gamma \sqrt{1-x^2})^2}}\rightarrow \frac{\sqrt{1-\lambda^2}}{\sqrt{(x-\lambda)^2+(\gamma \sqrt{1-\lambda^2})^2}}.
 \end{equation}
Hence, the leading contribution from the second and third parts  is
\begin{equation}
\frac{\partial^2 e(\lambda,\gamma)}{\partial\gamma^2} \sim  \frac{2\sqrt{1-\lambda^2}}{\pi}\ln{\gamma}+\cdots.
\end{equation}
We have numerically confirmed this scaling analysis.


\begin{thebibliography}{10}
\bibitem{note1} This is even valid for thermal phase transitions, if we extend
the notion of fidelity from pure states to mixed states to accommodate
thermal fluctuations (see the first reference in \cite{Zhou1}). In fact,
the logarithmic function of finite temperature fidelity per site for two
thermal mixed states corresponding to two different temperatures
reduces to nothing but the free energy if other non-thermal
control parameters are kept fixed. This implies that the singularities
in finite temperature fidelity per site coincide with those in the
free energy, thus showing the equivalence between the fidelity
approach and the conventional one to thermal phase transitions.
\bibitem{folklore} This usually refers to the fact that the effects from a critical point at zero temperature may be observed in a critical regime at low but finite temperature~\cite{pines}. However, we attempt to explore a formal similarity between them, given both of them arise from singularities.
\bibitem{twoton} For simplicity, we restrict ourselves to consider a Hamiltonian depending on two coupling parameters. However, it is straightforward to extend our discussion to a Hamiltonian with an arbitrary number of coupling parameters.
\bibitem{orthogonality} The orthogonality follows from the translational invariance, combining with the fact that any quantum state may be represented in terms of a matrix product state in one spatial dimension and a projected entangled pair state in two and higher spatial dimensions, as long as the bond dimension is large enough~\cite{TN2}. In fact, as shown in Ref.~\cite{Zhou2}, this is valid for any two translation-invariant states in the thermodynamic limit. Physically, this is related to the fact that more degrees of freedom are involved in a quantum system, more distinguishable its states are.  This might explain why ``more is different''~\cite{andersonmore}.
\bibitem{difference} In a sense, one may regard  time itself as  a fictitious information eraser, which constantly deletes  quantum states from a system during a quantum evolution. These states may be recovered from the environment, according to the no-hiding theorem (cf. Appendix~\ref{threetheorems}). However, there is a marked difference between quantum many-body systems and quantum few-body systems: for the former, these states are orthogonal in the thermodynamic limit; for the latter, they are not.
\bibitem{fidtem} Here, we simply denote fidelity temperature  by $T(x)$, without concern about any confusion with physical temperature $T$, which is zero for QPTs. However, we note that, even at a finite temperature $T$,  fidelity temperature $T(x)$ may also be defined to quantify fluctuations.
\bibitem{separation} As defined, fidelity temperature is associated with a variation of a chosen dominant control parameter, whereas fidelity work is associated with a variation of an auxiliary control parameter.
\bibitem{lowerbound} This statement is only valid for fidelity internal energy, determined from a specific Hamiltonian with the ground state energy density negative. As discussed in Section~\ref{duality}, that means the Hamiltonian is in a canonical form with respect to a specific definition of duality. However, as discussed there, for a Hamiltonian with the ground state energy density changing its sign, fidelity internal energy diverges when the ground state energy density is zero.
\bibitem{notunique} Note that both fidelity flows and RG flows are not unique.  In addition, fidelity flows depend on the choices of a dominant control parameter, whereas RG flows depend on the choices of a criterion to distinguish high energy degrees of freedom from low-energy degrees of freedom.
\bibitem{thermalFM} In this work, fidelity mechanics is formalized for QPTs at zero temperature. However, it may be extended to finite temperature. For this purpose, it is necessary to develop finite temperature tensor network algorithms~\cite{thermalTN,dai}. In fact, thermal state fidelity per site has been introduced in Ref.~\cite{dai} for quantum many-body systems at finite temperatures, which has been used to signal thermal phase transitions.
\bibitem{darkmatter} A possible interpretation for dark matter is that the analogue of Landauer's principle at zero temperature is responsible for dark matter, given the fact that galaxies are traces from quantum fluctuations in the early universe.
\bibitem{tenarrows} Actually, ten arrows of time have been listed in Ref.~\cite{vaas}.  It remains to be controversial whether or not there is a single arrow of time, which governs all physical processes.
\end{thebibliography}
\end{document}